\documentclass[intlimits,twoside,a4paper]{article}
\usepackage{amssymb,amsmath,amsfonts}
\usepackage{graphicx,epsfig}
\usepackage[T2A]{fontenc}
\usepackage[cp1251]{inputenc}
\usepackage[english]{babel}

\usepackage{cmpj}
\issue{2011}{14}{4}{42701}

\doinumber{10.5488/CMP.14.42701}

\articletype{Review article}

\title[Formation of charge and spin ordering in strongly correlated electron systems]
{Formation of charge and spin ordering  in strongly correlated electron systems}

\author[H. {\v C}en{\v c}arikov\'a, P. Farka{\v s}ovsk\'y]%{H. Cencarikova, P. Farkasovsky}%
{H. {\v C}en{\v c}arikov\'a, P. Farka{\v s}ovsk\'y}

\authorcopyright{H. {\v C}en{\v c}arikov\'a, P. Farka{\v s}ovsk\'y, 2011}

\address{Institute of Experimental Physics, Slovak Academy of Sciences,
Watsonov\'a~47, 040~01~Ko\v sice, Slovakia}
\date{Received July 7, 2011, in final form November 7, 2011}%

\begin{document}

\maketitle

\begin{abstract}
In this review we present results of our theoretical study of charge and spin ordering in strongly correlated electron systems obtained within various generalizations of the Falicov-Kimball model.  The primary goal of this study was to identify crucial interactions that lead to the stabilization of various types of charge ordering in these systems, like the axial striped ordering, diagonal striped ordering, phase-separated ordering, phase-segregated ordering, etc. Among the major interactions that come into account, we have examined the effect of local Coulomb interaction between localized and itinerant electrons,   long-range and correlated hopping of itinerant electrons,  long-range Coulomb interaction between localized and itinerant electrons, local Coulomb interaction between itinerant electrons,  local Coulomb interaction between localized electrons,  spin-dependent interaction between localized and itinerant electrons, both for zero and nonzero temperatures, as well as for doped and undoped systems. Finally, the relevance of resultant solutions for  a description of rare-earth and transition-metal compounds is discussed.
\keywords charge and spin ordering, metal-insulator transitions, valence transitions, Falicov-Kimball model,  strongly correlated systems
\pacs 75.10.Lp, 71.27.+a, 71.28.+d, 71.30.+h, 05.30.Fk
\end{abstract}

\tableofcontents

\section{Introduction}
\label{Introduction}

The problem of inhomogeneous charge and magnetic ordering in strongly interacting electron systems is certainly one of the most intensively studied problems of the  contemporary \linebreak solid state physics. The reason is that the inhomogeneous charge ordering (e.g., the striped phases)  were experimentally observed in many  rare-earth and transition-metal compounds~\cite{Cu1, Mook, Kajimoto, Howald}, like\linebreak \mbox{La$_{1.6}$Nd$_{0.4}$Sr$_x$CuO$_4$}, YBa$_2$Cu$_3$O$_{6+x}$, Bi$_2$Sr$_2$Cu$_2$O$_{8+x}$, La$_{1.5}$Sr$_{0.5}$NiO$_4$, Na$_x$CoO$_2$, some of which exhibit a high temperature superconductivity. This phenomenon was most frequently studied  in the literature within  the Hubbard  and   $t-J$ model~\cite{stripes_HM1, stripes_HM2, stripes_TJ1, stripes_TJ2, stripes_TJ3, stripes_TJ4, stripes_TJ5, stripes_TJ6}. Theoretical studies based on these models pointed to two possible mechanisms of forming the  inhomogeneous charge ordering in these materials:   (i) Kivelson, Emery, and co-workers~\cite{stripes_TJ1} proposed that strongly correlated systems have a natural tendency toward phase separation and the inhomogeneous spatial charge ordering arises from a competition between this tendency to phase separate and the long-range Coulomb interaction which does not allow the electron density to stray too far from its average; and (ii) White and Scalapino~\cite{stripes_TJ3, stripes_TJ4, stripes_TJ5, stripes_TJ6}  proposed that the stripe order arises from a competition between kinetic and exchange energies in a doped
antiferromagnet which does not require long-range Coulomb forces to stabilize the stripes.

Moreover, shortly after the introduction of the spinless Falicov-Kimball model in 1986 by Kennedy and Lieb~\cite{Kennedy_Lieb} and Brandt and Schmidt~\cite{Brandt_Schmidt} it was found that this probably the simplest model of correlated electrons on a lattice is capable of describing various types of charge ordering including the periodic as well as phase-separated and segregated configurations~\cite{rev1, rev2, Watson_Lemanski, Gajek_Jedrzejewski, Gajek_Jedrzejewski2, Kennedy94, Kennedy98, Haller_Kennedy01, Gruber_Iwanski, Gruber, Gruber_Lemberger, Gruber_Ueltschi, Gruber_Macris, Freericks_Falicov}. This opened a new route of studying this important phenomenon. The model describes a system of itinerant and localized particles on the lattice, which interact only through  the local Coulomb interaction. The Hamiltonian of the spinless Falicov-Kimball model
can be written in the form
\begin{eqnarray}
H=\sum_{ij}t_{ij}c^+_ic_j + U\sum_{i}w_ic^+_ic_i\, ,
\label{eq2.2}
\end{eqnarray}
where $c_i^+$ ($c_i$) are the creation (anihilation) operators of the itinerant spinless particles at site $i$, $t_{ij}$ describes the hopping probability from site $i$ to $j$  and $w_i$ is the occupation number of the localized particles taking the value 1 or 0 according to whether the site $i$ is occupied or unoccupied by the localized particle.

Accordingly, as the itinerant and localized particles are interpreted, we get different interpretations of the model. Historically, the first interpretation of the model: itinerant particles = electrons with spin up and localized particles = electrons with spin down, was already used in the original work by Hubbard~\cite{HM} as the approximative solution of the original Hubbard model, in which one type of particles, e.g., electrons with spin down, are immobilized. Therefore, this approximation is  sometimes referred to as the static Hubbard model. The second interpretation: itinerant particles = itinerant electrons and localized particles = immobile ions, comes from Kennedy and Lieb~\cite{Kennedy_Lieb} and represents a very simple model of crystallization in solids. With respect to the fact that  the Falicov-Kimball model was originally proposed to describe valence transitions and metal-insulator transitions in rare-earth compounds, one of the most frequent interpretations is to consider the conduction $d$ electrons instead of itinerant particles and the valence $f$ electrons instead of localized particles.

The greatest advantage of the spinless Falicov-Kimball model is its relative simplicity, which makes the model more accessible to analytical and numerical studies compared with e.g., the Hubbard or periodic Anderson model. Two diametrically different ways have been used to solve the spinless Falicov-Kimball model. The first direction represents analytical studies in the limit of infinite dimensions ($D=\infty$) and other analytical and numerical studies in reduced dimensions ($D=$ 1 and $D=$ 2).

Regarding the works devoted to the study of the Falicov-Kimball model in the infinite dimensional limit, it is worth noting that there are two good reasons to study the model in this limit. The first reason is that for $D\rightarrow \infty$ the self-energy becomes local~\cite{Metzner_Vollhardt}, which simplifies the analytical calculations to the extent that they permit to solve many problems exactly. The second reason for studying the model in the limit of infinite dimension is the fact that some physical quantities calculated for $D=\infty$ reproduce the three-dimensional results better than the one-dimensional solutions. A milestone in this direction is the work by Brandt and Mielsch~\cite{Brandt_Mielsch}, in which the exact solution of the spinless Falicov-Kimball model is presented for the symmetric case (the number of $d$ electrons = the number of $f$ electrons = $L/2$, where $L$ is the number of lattice sites). The main result of this study was a determination of  the critical transition temperature for a transition from the high temperature disordered phase to the low-temperature ordered (checkerboard) phase. In subsequent articles, Brandt et al.~\cite{Brandt_Mielsch, Brandt_Mielsch2, Brandt_Mielsch3, Brandt_Fledderjohann, Brandt_Fledderjohann2} built the foundations of physics of the Falicov-Kimball model in the limit of infinite dimensions, on which many other theoretical physicists built both the spinless and spin-one-half Falicov-Kimball model~\cite{vDongen_Vollhart, vDongen1, vDongen2, Freericks1, Freericks2, Chen, Freericks_Gruber, Letfulov, Freericks_Lemanski, Gruber_Macris2, Chung}. The results of these studies are summarized in an excellent review article by Freericks and Zlatic~\cite{Freericks_Zlatic} devoted to exact solutions of the Falicov-Kimball model in the dynamic mean field theory.

Regarding the second direction, the analytical solutions of the Falicov-Kimball model in the limit of reduced dimensions, it should be noted that despite the huge efforts of theorists and relative simplicity of the model, so far only very few exact results for the ground state and thermodynamics of the spinless model have been obtained. In addition to the aforementioned evidence of the long-range arrangement at low temperatures and dimensions $D\geqslant 2$, the following has been proven: (i) the absence of spontaneous hybridization at finite temperatures~\cite{Subrahmanyam_Barma}, (ii) the phase separation and periodic arrangement in the limit of strong Coulomb interactions and $D=$ 1~\cite{Lemberger92}, (iii) the phase separation in the two-dimensional model for selected values of electron concentrations and sufficiently large Coulomb interactions~\cite{Kennedy94, Kennedy98, Haller2000, Haller_Kennedy01}, (iv) the phase separation in the limit of strong Coulomb interactions for all dimensions~\cite{Freericks02a, Freericks02b}.

Under these circumstances,  numerical calculations seem to be very valuable.  Although they do not always provide definitive answers to the questions searched,  they are capable of pointing out the fundamental trends in the system, and thus help to understand the physics of the Falicov-Kimball model. Basically, there are two main directions of the numerical study of the Falicov-Kimball model. The first is an exact diagonalization of the model Hamiltonian on finite lattices over a complete set of accessible configurations of localized particles. As the number of configurations increases as $2^L$, this method is severely limited by the size of clusters, which are eligible for the numerical study ($L\sim$ 36--40). In the one dimensional case such cluster sizes are sufficient to extrapolate the results obtained to the thermodynamic limit ($L\rightarrow \infty$), yielding definite results on the ground state or thermodynamics of the model, at least for a certain area of model parameters (e.g., strong interactions). In the two-dimensional and three-dimensional cases, however, such cluster sizes are insufficient and, therefore, it is necessary to use approximate methods that somehow reduce the number of investigated configurations. This procedure was first used in the work by Freericks and Falicov~\cite{Freericks_Falicov} in the study of one-dimensional phase diagram of the spinless Falicov-Kimball model for selected concentrations of localized particles. Since the same formalism with minor modifications was used later by many other authors in the study of the ground states of the  one-dimensional and two-dimensional models, we briefly summarize the main points of this algorithm.
Consider some configuration $w=\{w_1, w_2, \dots, w_L\}$, which corresponds to a specific distribution of particles localized on the lattice consisting of $L$ sites, while the classical variable $w_i=$ 1, 0 indicates whether or not the site $i$  is  occupied by a localized particle. The energy of the system with $N_d$ itinerant electrons corresponding to the selected configuration is then given by
\begin{eqnarray}
E(U,N_d,w)=\int_{-\infty}^{E_{\rm F}}n(E,w)E\rd E ,
\label{eq2.3}
\end{eqnarray}
where $n(E,w)$ is the density of states  corresponding to the configuration $w$ and $E_{\rm F}$
is the Fermi energy,  which  can be determined from the condition
\begin{eqnarray}
N_d=\int_{-\infty}^{E_{\rm F}}n(E,w)\rd E .
\label{eq2.4}
\end{eqnarray}
The density of states $n(E,w)$ in the aforementioned expression is given by
\begin{eqnarray}
n(E,w)=-\frac{1}{\pi L}\Im\lim_{\varepsilon\to 0}\sum_jG_j(E+i\varepsilon)\, ,
\label{eq2.5}
\end{eqnarray}
where $G_j(E)$ are the local Green's functions for which Freericks and Falicov~\cite{Freericks_Falicov} found the following
expression
\begin{eqnarray}
G_j(E)=\frac{1}{E-Uw_j-\Delta^+_j(E)-\Delta^-_j(E)}\, ,
\label{eq2.6}
\end{eqnarray}
where
\begin{eqnarray}
\Delta^{\pm}_j=\displaystyle{\frac{1}
{E-Uw_{j\pm 1}-\displaystyle\frac{1}{E-Uw_{j\pm 2}-\displaystyle\frac{1}{E-Uw_{j\pm 3}-\dots}}}}\, .
\label{eq2.7}
\end{eqnarray}
The problem is thus reduced to the calculation of infinite fractions $\Delta^{\pm}_j$~\cite{Lyzwa}.
For the concentration of localized particles equal to $n_f=0.5$, Freericks and Falicov~\cite{Freericks_Falicov} investigated ten periodic configurations of ions (their list is given in table~\ref{tab1}),
\begin{table}[h!]
\begin{center}
\caption{ The complete list of all physically different periodic configurations of ions with the period $p<9$ and  $n_f=1/2$~\cite{Freericks_Falicov}.}
\vspace{2ex}
\begin{tabular}{|c|l|c|l|}
\hline
\mbox{k} & \mbox{configuration period} & \mbox{k} & \mbox{configuration period}\\
\hline\hline
1 & 10 & 6 & 11101000\\
2 & 1100 & 7 & 11100100\\
3 & 111000 & 8 & 11011000\\
4 & 110100 & 9 & 11010100\\
5 & 11110000 & 10 & 11010010\\
\hline
\end{tabular}
\end{center}
\label{tab1}
\end{table}
which represent all different physical configurations of localized particles with the period less than 9 and a segregated phase (an incoherent mixture of completely empty and fully occupied lattice) with a density of states
\begin{eqnarray}
n^{\rm seg}(E)=(1-n_f)n^{\rm empty}(E)+n_fn^{\rm fully}(E)\ .
\label{eq2.8}
\end{eqnarray}

\begin{figure}[h!]
\begin{center}
\mbox{\includegraphics[width=8.5cm,angle=0]{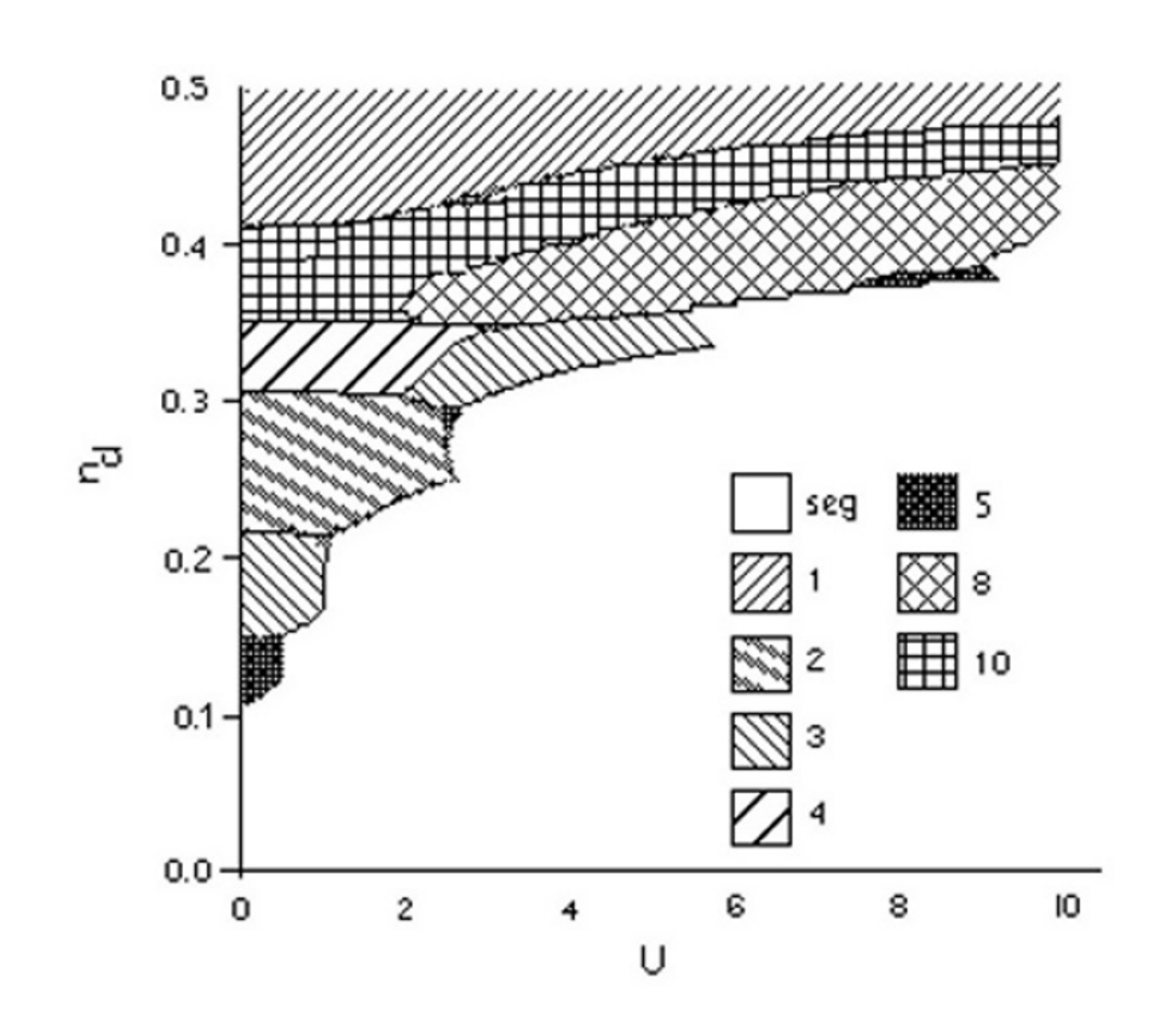}}
\end{center}
\vspace*{-0.5cm}
\caption{ The phase diagram of the one-dimensional spinless Falicov-Kimball  model in the  $n_d-U$ plane calculated for $n_f=1/2$~\cite{Freericks_Falicov}. The set of ground-state configurations consists of the segregated configuration and the periodic configurations with the smallest periods (see  table~\ref{tab1}).}
\label{fazy10}
\end{figure}

Using the numerical analysis they obtained (for the case $n_f=1/2$) the coherent ground-state phase diagram (see figure~\ref{fazy10})
of the model that exhibits some general features:
(a) The alternating phase $w_1=\{1010\ldots10\}$ is the ground state at $n_d=1/2$ for all values of  $U$ as stated by previous investigations~\cite{Kennedy_Lieb}.  (b) The phase diagram tends to be simplified as the interaction strength increases indicating that many-body effects stabilize the system (this is a consequence of the segregation principle). (c)~There is a trend for phases that disappear from the phase diagram  as $U$ increases to reappear as phase islands at even larger values of $U$ (e.g., the configuration No.~3). (d) Phase islands of configurations not present at $U=0$ may be formed at larger values of $U$ (e.g., the configuration No.~8). (e) Some configurations are not the ground state for any value of $U$ or  $n_d$ (e.g., the configurations  $\{11101000\ldots\}$ and $\{11010100\ldots\}$  do not appear).

A similar method was used later on by Gruber et al.~\cite{Gruber_Ueltschi} for the study of the one-dimensional phase diagram of the spinless Falicov-Kimball model for the so-called neutral case, where the concentration of itinerant electrons $n_d$ is equal to the concentration of localized electrons/ions $n_f$.
Unlike the previous work~\cite{Freericks_Falicov}, Gruber et al.  did not focus on studying the model only for selected values of $n_f$, but they studied the comprehensive  phase diagram of the model in the $n_f-U$ plane.  Similarly to the previous case, the set of input configurations is not complete, since only periodic configurations with small periods and mixtures of these periodic configurations with an empty lattice have been taken into account\footnote{The difference between the methods presented in~\cite{Freericks_Falicov} and~\cite{Gruber_Ueltschi} is that in reference~\cite{Freericks_Falicov} the canonical phase diagrams were constructed, where only the simplest periodic configurations  and the segregated phase are taken into account, but not a mixture thereof. However, in reference~\cite{Gruber_Ueltschi} the grand canonical phase diagrams were constructed first and only then they were transformed into the canonical phase diagrams. This procedure ensures both the simplest periodic configurations, the segregated phase and all possible mixtures thereof are included.}. The main result of their numerical and analytical studies is that the ground states of the spinless Falicov-Kimball model ($n_f\leqslant 0.5$) are either the most homogeneous distributions of localized particles (for $U$ and $n_f$ large enough) or  mixtures of periodic configurations and the empty lattice ($U$ and $n_f$ small).
Since the Fermi level for the mixtures of periodic configurations and the empty lattice lies in the conduction band, while for the most homogeneous configurations lie in the energy gap,  the boundary between these two domains, is a boundary of  metal-insulator transitions, which may be induced either by the Coulomb interaction or by changing the concentration of localized electrons.

\begin{figure}[h!]
 \begin{center}
\includegraphics[width=8cm]{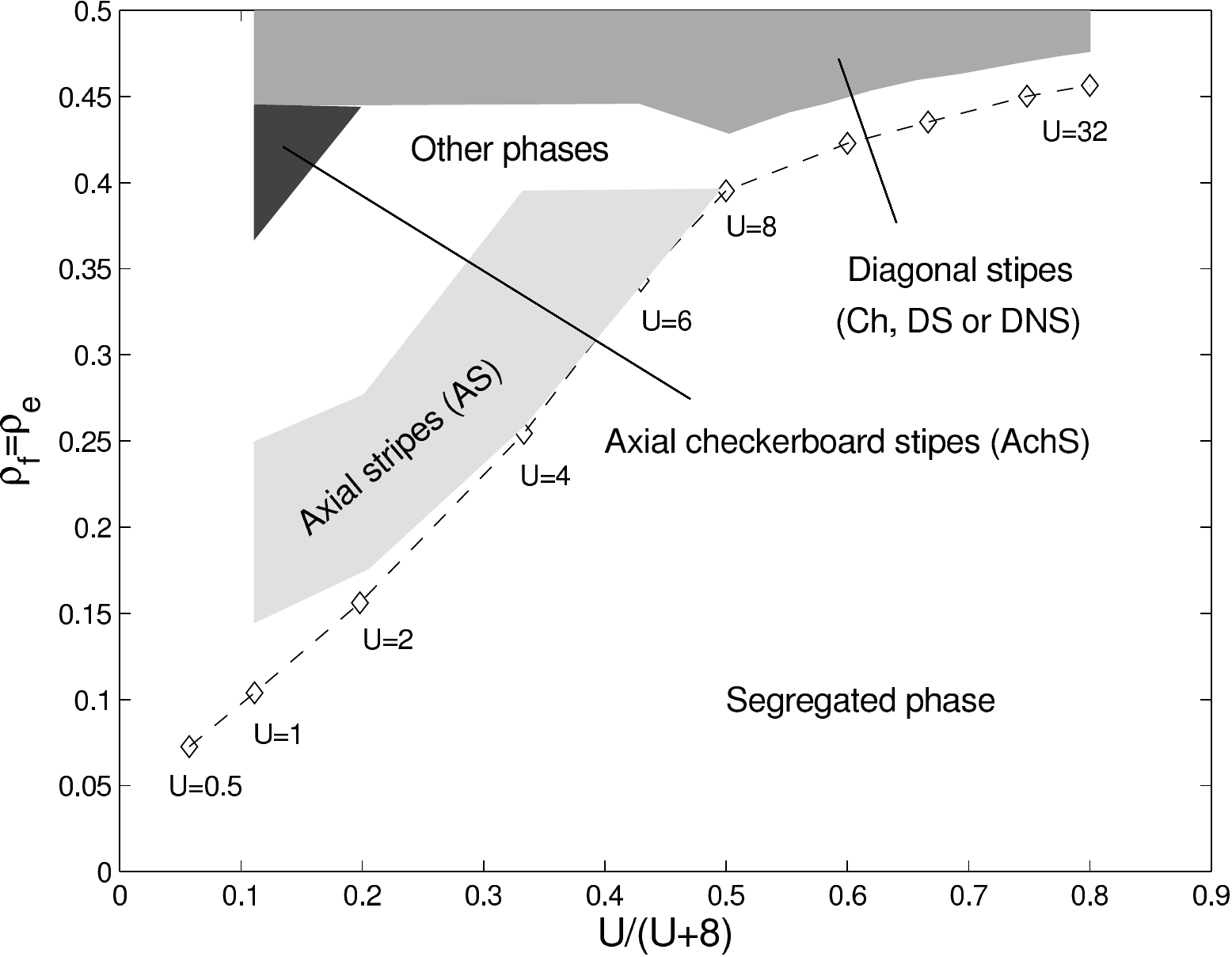}\hspace*{7mm}
\includegraphics[width=5.60cm]{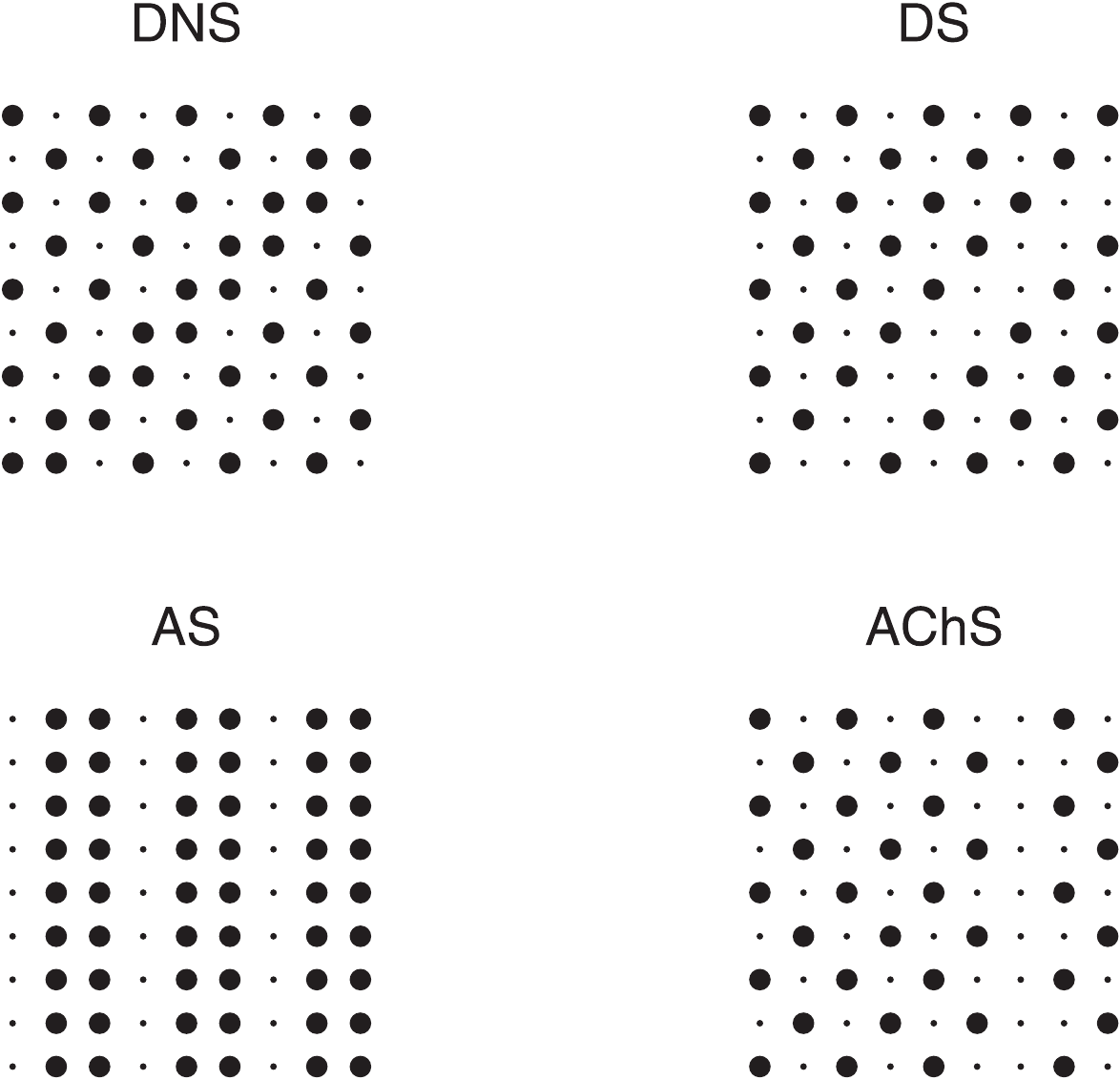}
\end{center}
%\vspace{-0.5cm}
\caption{ The ground-state phase diagram of the spinless Falicov-Kimball model in $D=2$ calculated by the method of restricted set diagrams~\cite{Lemanski_Freericks}. The typical ground-state configurations detected in the phase diagram. The large dots correspond to the occupied sites and the small dots correspond to the vacant sites.}
\label{o_Lemanski_Freericks}
\end{figure}

For the two-dimensional case, this method was generalized by Watson and Lemanski~\cite{Watson_Lemanski} and later it was used by Lemanski, Freericks and Banach to study the ground-state phase diagram of the model in  $D=2$~\cite{Lemanski_Freericks}. In this case,  the set of input configurations includes all periodic configurations with a unit cell having a smaller number of sites  than the selected critical value $N_{\mathrm c}$ and all possible mixtures thereof. The most interesting result of these studies was the observation of axial and diagonal striped phases of localized particles (see figure~\ref{o_Lemanski_Freericks}), suggesting that in the system of itinerant and localized particles, a sufficient mechanism leading to the formation of inhomogeneous charge arrangement is the local Coulomb interaction between these two electron subsystems.
This is a much simpler mechanism of forming inhomogeneous charge stripes than the one considered earlier within the Hubbard model~\cite{stripes_HM1, stripes_HM2}, respectively, within the  $t-J$ model~\cite{stripes_TJ1, stripes_TJ2, stripes_TJ3, stripes_TJ4, stripes_TJ5, stripes_TJ6}. Due to the incomplete input set of investigated configurations, an open question remains whether these results persist if the complete set of configurations will be considered, and especially, whether they persist for the more realistic three-dimensional case.

\begin{figure}[!b]
\begin{center}
\includegraphics[width=8.20cm]{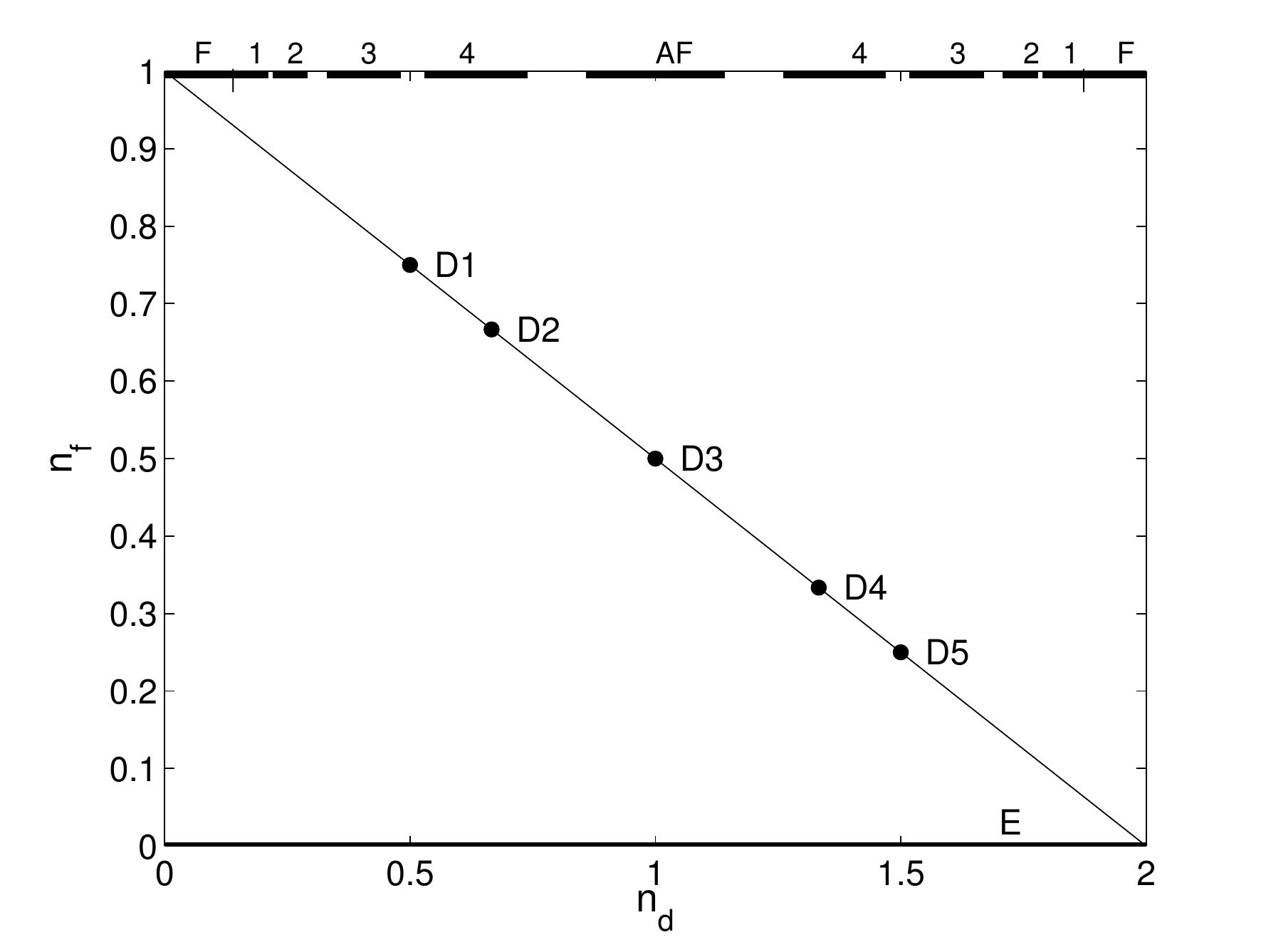}\hspace*{-0.8cm}
\includegraphics[trim = 2cm 1cm 1cm 0cm, clip,width=7.80cm]{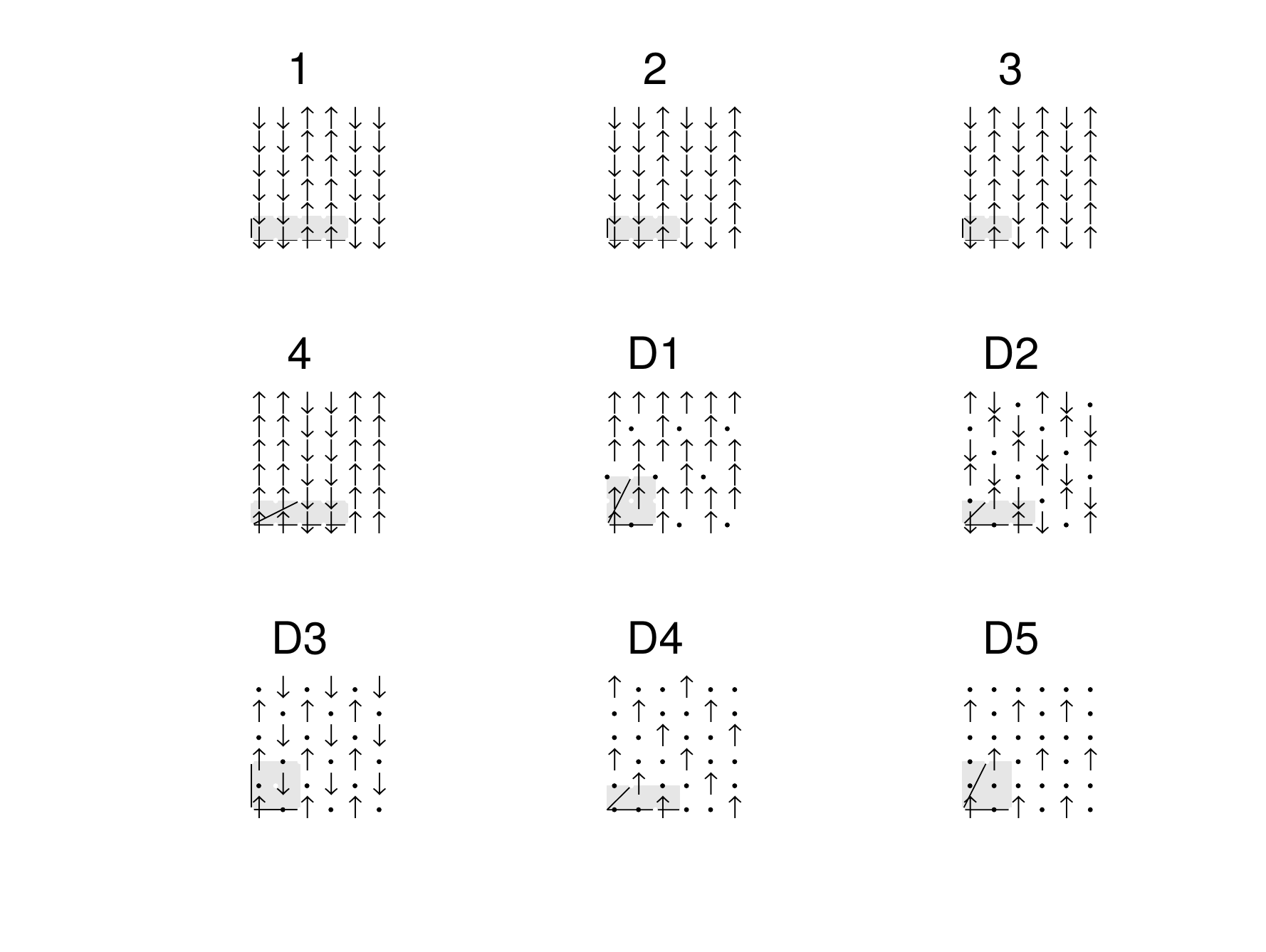}
\end{center}
\vspace{-0.3cm}
\caption{ The restricted ground-state phase diagram of the spin-1/2 Falicov-Kimball model with the Ising interaction for the Coulomb interaction $U = 8$ and the Ising interaction $J = 0.5$ (the canonical phase diagram in the plane
$n_f-n_d$). F, AF and E refer to the ferromagnetic, antiferromagnetic and empty  phases, respectively. Periodic phases with $n_f$=1 are denoted by the numbers 1--4 (right hand panel), and those with $n_f+n_d/2$= 1 by the symbols D1--D5.  The phases are ground states only on the bold straight-line segments or at single points. Outside these straight line segments or points there are mixtures of periodic phases that have lower energy than any periodic phase taken from the restricted set. The small vertical
straight line segments, crossing the $n_f$= 1 line, mark the limits of F. The diagonal line $n_f + n_d/2$ = 1 is only a visual guide~\cite{Lemanski05}.}
\label{o_Lemanski}
\end{figure}

One of the major shortcomings of the Falicov-Kimball model is that it does not include any spin interactions between electrons, and, therefore, it is not capable of describing the magnetic superstructures, which in many of rare-earth and transition-metal compounds coexist with  charge ordering. This phenomenon was observed,  for example, not only in nickelates~\cite{Ni1, Ni2, Ni3}, manganates~\cite{Mn}, cobaltates~\cite{Co1, Co2}, but also in materials exhibiting high-temperature superconductivity~\cite{Cu1, Cu2, Cu3}. At present, there are still intensive discussions on possible mechanisms of the formation of inhomogeneous charge and spin ordering and its relation to the physical properties of the systems, e.g., high-temperature superconductivity. The easiest way of introducing spin interactions in the system of itinerant and localized electrons is to bind them by the Ising interaction. This idea was first used by Lemanski~\cite{Lemanski05}, who found that turning on the Ising interaction between itinerant and localized electrons leads to the stabilization of different types of charge and spin arrangement, including the axial and diagonal striped phases (see figure~\ref{o_Lemanski}). Moreover, a number of simple rules of formation of various sorts of ground-state phases have been presented in reference~\cite{Lemanski_Wrzodak}.
Since these results were obtained on a restricted set of configurations, an open challenge for further theoretical studies was whether the nature of the ground state would remain unchanged after taking into account the complete set of configurations.
For this reason we have decided to perform a systematic study (within an exact diagonalization and well-controlled approximate method described in the next section) of the ground state of the spinless as well as  generalized spin-1/2 Falicov-Kimball model, in order to find the fundamental mechanisms of the formation of  inhomogeneous charge and spin ordering in strongly correlated systems. Apart from  the above mentioned local Coulomb interaction between $d$ and $f$ electrons, we have also investigated the effect of nonlocal Coulomb interaction~\cite{Farky41, Farky44}, the correlated hopping~\cite{Farky16, Farky21, Farky28, Farky31, Farky34}, the  lattice geometry~\cite{Farky36, Farky39},  the dimension of the system~\cite{Farky17, Farky32}, the anisotropic spin-dependent interaction~\cite{Farky29, Farky37, Farky40} and the Hubbard  interaction~\cite{Farky29}. In what follows, we state a brief overview of the main results we have reached in our numerical studies.

\section{Methods}
\label{Methods}
To study the ground-state properties of the model Hamiltonians based on the spinless/ spin-1/2 Falicov-Kimball model we have used the method of exact diagonalization on finite clusters, where  diagonalizations are performed over all possible distributions of localized particles, as well as the approximate method developed by us, in which the acceptance of  configuration is realized using the principle of reducing the total energy of the system. Periodic boundary conditions are used in all the examined cases, since the fastest convergence of numerical results to the thermodynamic limit is observed for this type of boundary conditions.

\subsection{Exact diagonalization technique}
\label{Exact diagonalization}

Although in the next major steps, the exact diagonalization method (EDM) will be illustrated  for the spinless  Falicov-Kimball model, the applicability of the method is much broader and with minor modifications it can be directly extended to the spin-1/2 Falicov-Kimball model, as well as the spin-1/2 Falicov-Kimball model with the Ising interaction between localized and itinerant electrons. The method is flexible with regard to the changes of the hopping  matrix $t_{ij}$ and so it may be used, without any additional numerical complications, to study the effects of long-range and correlated hopping of electrons on the ground-state properties of the model.

Hereinafter we will use solely  the interpretation of the spinless  Falicov-Kimball model in which  the itinerant particles are $d$ electrons and the localized particles are $f$ electrons from localized  $4f$ or $5f$ states of rare-earth ions. Then, spinless Falicov-Kimball model can be written in the form
\begin{eqnarray}
H=\sum_{ij}t_{ij}d_i^+d_j + U\sum_iw_id_i^+d_i\, ,
\label{eq3.1.1.1}
\end{eqnarray}
where $w_i$ ($w_i=1$, 0) now describes the occupancy  of the $f$ orbital at lattice site $i$.

It is important to note that for any distribution of $f$ electrons  $w=\{w_1, w_2 ,\ldots, w_L\}$,  the Hamiltonian (\ref{eq3.1.1.1}) is a single particle Hamiltonian in the representation of the second quantization
\begin{eqnarray}
H=\sum_{ij}h_{ij}(w)d_i^+d_j\, ,
\label{eq3.1.1.2}
\end{eqnarray}
where $h_{ij}(w)=t_{ij}+Uw_i\delta_{ij}$.
Thus, the solution of the model~(\ref{eq3.1.1.1}) reduces to the problem of determining the spectra of matrix $h(w)$ for different distributions of $f$ electrons on the lattice of the  size $L$. Since the problem is analytically solvable only for special types of configurations (e.g., the periodic configurations with the smallest periods), the only way to exactly solve this problem  is to use the  numerical diagonalization on finite clusters. Then, a fundamental task is to find a distribution of $f$ electrons, for which the system has the lowest energy. The numerical algorithm for finding the configuration $w^0$, which minimizes the energy of the system is as follows:
(i) Having $w$, $U$ and $t_{ij}$ fixed, we find all eigenvalues  $\lambda_k$ of $h(w)$. (ii) For a given  $N_f=\sum_iw_i$ we determine the ground-state energy $E(w)=\sum_{k=1}^{N_d}\lambda_k$ of  a particular $f$-electron configuration $w$ by filling in the lowest $N_d$ one-electron levels. (iii) We find $w^0$ (examining all possible distributions of localized  electrons), for which  $E(w,U)$ has a minimum.  Repeating this procedure for different values of model parameters one can immediately study the ground-state phase diagrams of the model.

Such exact calculations can be performed at present up to $L\sim$~36--40 sites, which in some cases (the one-dimensional case and strong Coulomb interactions) is sufficient for an extrapolation of the results  to the thermodynamic limit. In general, however, such  cluster sizes are insufficient to obtain reliable conclusions about the behaviour of macroscopic systems in higher dimensions. Under these circumstances, the only way  seems to be to use approximate methods. When selecting an appropriate approximate method one should have in mind  the fact that charge and spin ordering as well as valence  and metal-insulator transitions  are very sensitive to the type of approximation used~\cite{Farky1, Farky2}, and thus their description can only be successful within the approximations that introduce  only small simplifications of the model system. Instead of searching for an appropriate method among the existing approximations, comparing them and excluding the least accurate candidates, we decided to develop a new numerical method, which would be sewn from the beginning on the Falicov-Kimball model, while retaining some degree of variability due to possible generalizations of this model.

\subsection{Approximate method based on the reduction of the total energy}
\label{Approximate method based on the reduction of the total energy}

The natural starting point in building a new approximate method (AM) seemed to us to be the method of exact diagonalization. As stated above, within this method the single particle Hamiltonian $h(w)$ is exactly diagonalized over all possible ($2^L$) distributions of localized particles in order to find the only configuration  $w^0$, which minimizes the total energy of the system. This procedure is necessary, but not efficient. Much more efficient than passing through the complete set of configurations could be regulating the choice of configurations from the initial configuration $w$ to the final configuration  $w^0$, under some criterion that would significantly reduce the number of configurations that should be examined. The most  natural criterion seems to be a reduction of the total  energy in a sequence of configurations from $w$ to $w^0$. This is the basic idea of our AM, the algorithm of which can be described as follows~\cite{Farky17}:
(i) Choose a trial configuration $w=\{w_1,w_2,\dots,w_L\}$. (ii) Having $w$, $U$  fixed, find all eigenvalues $\lambda_k$ of $h(w)$. (iii) For a given $N_f=\sum_iw_i$ determine the ground-state energy $E(w)=\sum_{k=1}^{N_d}\lambda_k$ of a particular $f$-electron configuration $w$ by filling in the lowest $N_d$ one-electron levels. (iv) Generate a new configuration $w'$ by moving a randomly chosen electron to a new position which is also chosen as random. (v) Calculate the ground-state energy $E(w')$. If $E(w')<E(w)$, the new configuration is accepted, otherwise $w'$ is rejected. Then, the steps (ii)--(v) are repeated until the convergence (for a given $U$) is reached.

Of course, one can move instead of one electron (in step (iv)) simultaneously two or more electrons, thereby improving the convergence of method. Indeed, the tests that we have performed for a wide range of model parameters showed that the subsequent implementation of the method, in which  $1 < p < p_{\rm max}$ electrons ($p$ should be chosen at random) are moved to new positions, better overcomes the local minima of the ground-state energy. As usual, we have performed calculations with $p_{\rm max}=N_f$. The main advantage of this implementation is that in any iteration step the system has a chance to lower its energy (even if it is in a local minimum), thereby the problem of local minima is strongly reduced (in principle, the method becomes exact if the number of iteration steps goes to infinity). On the other hand, a disadvantage of this selection is that the method converges slower than for $p_{\rm max}=2$ and $p_{\rm max}=3$. To speed up the convergence of the method (for $p_{\rm max}=N_f$) and still to hold its advantage we generate instead the random number $p$ (in step (iv)) the pseudo-random number $p$ that probability of choosing decreases (according to the power law) with increasing~$p$. Such a modification considerably improves the convergence of the method.

Apart from the number of the moved electrons,  the method was also  tested on  the optimum length of the iteration cycle $M$. It is obvious that if $M\rightarrow \infty$, the method is exact. Unfortunately, with respect to the time factor, such a choice is not possible and we must consider only a finite number of iteration steps. The test process was realized  on different finite clusters  in the  one-, two- and three-dimensional cases. We have observed that it is very convenient to divide the whole iteration process into several smaller independent cycles (from  5 to 10), among which the ground-state configuration with the lowest energy is selected. This also minimizes the problem of local minima.
\begin{table}[h!]
\begin{center}
\caption{ The difference of the ground-state energies of the one-dimensional spinless Falicov-Kimball model calculated exactly and by our numerical method ($\Delta=|E_{\rm exact}-E_{\rm met.}|$) for three different Coulomb  interactions  ($U=2$, $U=4$ and $U=8$) for $M=200$, 400 and 600 iterations.}
\label{tab2}
\vspace*{2ex}
\scriptsize{
\begin{tabular}{|c||c|c|c||c|c|c||c|c|c|}
\hline
\mbox{$N_f$} &
\multicolumn{3}{c||}{M=200} &
\multicolumn{3}{c||}{M=400} &
\multicolumn{3}{c|}{M=600}\\
\cline{2-10}
 & \mbox{$\Delta(U$=2$)$} &\mbox{$\Delta(U$=4$)$} &\mbox{$\Delta(U$=8$)$} &
\mbox{$\Delta(U$=2$)$} &\mbox{$\Delta(U$=4$)$} &\mbox{$\Delta(U$=8$)$} &
\mbox{$\Delta(U$=2$)$} &\mbox{$\Delta(U$=4$)$} &\mbox{$\Delta(U$=8$)$} \\
\hline\hline%\cline{1-10}
1& 0 &0 &0 &0 &0 &0 &0 &0 &0\\
2& 0 &0 &0 &0 &0 &0 &0 &0 &0\\
3& 0 &0 &0 &0 &0 &0 &0 &0 &0\\
4& 0 &0 &0 &0 &0 &0 &0 &0 &0\\
5& 0 &0 &0 &0 &0 &0 &0 &0 &0\\
6& 0 &0 &0 &0 &0 &0 &0 &0 &0\\
7& 0 &0 &0 &0 &0 &0 &0 &0 &0\\
8& $1,5.10^{-8}$ &0 &0 &0 &0 &0 &0 &0 &0\\
9& 0 &0 &0 &0 &0 &0 &0 &0 &0\\
10&$8,7.10^{-4}$ &0 &$5,7.10^{-8}$ &0 &0 &$1,2.10^{-7}$& 0 &0 &0\\
11& 0 &0 &0 &0 &0 &0 &0 &0 &0\\
12& 0 &0 &$3,0.10^{-6}$ &0 &0 &0 &0 &0 &0\\
13& 0 &0 &0 &0 &0 &0 &0 &0 &0\\
14& 0 &0 &0 &0 &0 &0 &0 &0 &0\\
15& 0 &0 &0 &0 &0 &0 &0 &0 &0\\
16& $4,3.10^{-7}$ &$5,5.10^{-9}$ &0 &0 &0 &0 &0 &0 &0\\
17& 0 &0 &0 &0 &0 &0 &0 &0 &0\\
18& 0 &$5,2.10^{-8}$ &0 &0 &0 &0 &0 &0 &0\\
19& 0 &0 &0 &0 &0 &0 &0 &0 &0\\
20& 0 &0 &0 &0 &0 &0 &0 &0 &0\\
21& $1,6.10^{-7}$ &0 &0 &0 &0 &0 &0 &0 &0\\
22& 0 &0 &0 &0 &0 &0 &0 &0 &0\\
23& $6,4.10^{-6}$ &0 &0 &0 &0 &0 &0 &0 &0\\
24& 0 &$9,1.10^{-4}$ &$1,8.10^{-5}$ &0 &0 &$1,8.10^{-5}$ &0 &0 &0\\
25& $1,6.10^{-5}$ &0 &0 &0 &0 &0 &0 &0 &0\\
26& 0 &0 &0 &0 &0 &0 &0 &0 &0\\
27& $9,7.10^{-9}$  &0 &0 &0 &0 &0 &0 &0 &0\\
28& $6,6.10^{-7}$ &0 &0 &0 &0 &0 &0 &0 &0\\
29& 0 &0 &0 &0 &0 &0 &0 &0 &0\\
30& 0 &0 &0 &0 &0 &0 &0 &0 &0\\
\hline
\end{tabular}}
\end{center}
\end{table}
 In table~\ref{tab2} we  compare our numerical results on the  one-dimensional cluster of  $L=60$ sites, obtained  with 10 iteration cycles for $M=200$, $M=400$ and $M=600$ iterations per site with the exact results. It should be noted that the exact results were obtained by the EDM on a set of the most homogeneous  configurations, which are the ground states of the one-dimensional Falicov-Kimball model for all $n_f$ and $U> 1.2 $. Therefore, the comparative tests were made for $ U = $ 2, $ U = $ 4 and $ U = $ 8. As seen in table~\ref{tab2} already  relatively small number of iterations ($ M =600$) was sufficient to obtain the exact ground states  for every $N_f$ on a sufficiently large cluster of $L=60$. The same test was repeated on a twice larger lattice, where at 500 iterations, there were still observed small differences in the ground-state energy. These differences ranged in the order of about $10^{-4}-10^{-9}$, and their number decreased with the increasing  number of iteration steps $M$. For $M=2000$, we have found full consistency between our and exact results.

\newpage

\section{Charge ordering in the spinless Falicov-Kimball model}
\label{Charge ordering in the spinless Falicov-Kimball model}

\subsection{The effect of local Coulomb interaction}
\label{Influence of local Coulomb interaction}

{\bf One-dimensional case}

We have started the study  of ground-state properties of the spinless Falicov-Kimball model,   in the limit $U\geqslant 1$~\cite{Farky1}. This case was chosen for the reason that the analytical calculations showed~\cite{Lemberger92} that the ground states of the model (at sufficiently large $U$) can be only the most homogeneous  distributions, when $f$ electrons are as far apart as possible, taking into account the periodic boundary conditions. The knowledge of the ground states in the limit of strong interactions make possible  a direct comparison between our results obtained on finite clusters with the results obtained in the thermodynamic limit ($L\rightarrow \infty$) and at the same time it permits to specify more precisely the area of stability of these configurations. Numerical calculations were performed using the EDM, which permits to find the ground state of the model on the finite cluster of size $L$ for any model parameters: $n_d$, $n_f$ and $U$. Although $n_d$  and $n_f$ can be considered as independent parameters, we have bound them with the condition $n_f+n_d=1$, since we are also interested in valence transitions, i.e., transitions induced by migration of valence $f$ electrons to the conduction band. To minimize the finite-size effects, the  model was studied on finite clusters from  $L=4$  to $L=24$ sites.  We have chosen the values of the Coulomb interaction from 1  to 10 with a unit step. The main result of our theoretical study  was the finding  that the most homogeneous configurations are the ground states of the spinless  Falicov-Kimball model not only in the limit of strong Coulomb interactions, but for all the investigated values of $U>$ 1, and for all $f$-electron concentrations. Since the Fermi level for the most homogeneous configurations always lies within the energy gap~\cite{Gruber_Ueltschi}, all ground states  for $U>$ 1 are insulating, and thus the spinless  Falicov-Kimball model  is not capable of describing the metal-insulator transitions in this limit.

For this reason we have turned our attention to the case $U<1$. Using the same method, we have performed a systematic study of the model for the selected value of Coulomb interaction ($U=0.6$) and all even lattices from $L=16$ to $L=48$  lattice sites~\cite{Farky2}. The main result of our study was the finding that for every $L$ there exists a critical value of the $f$-electron occupation number $N_{\mathrm c}$ below which the ground states are no longer the most homogeneous configurations,  but the phase-separated configurations that may be formally presented as an incoherent mixture of a  configuration $w$ and the empty lattice ($w\&w_0$). A complete list of such configurations together with critical values $N_{\mathrm c}$ is summarized in table~\ref{tab3}.
\begin{table}[h!]
\begin{center}
\caption{ The critical values of the $f$-electron occupation number $N_{\rm c}$,  $f$-level position $E_{\rm c}$, and the ground-state configurations below $N_{\rm c}$ calculated for $U=0.6$ and different values of $L$. Here the lower index denotes the number of consecutive sites occupied (unoccupied) by $f$ electrons~\cite{Farky2}.}
\label{tab3}
\vspace*{2ex}
\begin{tabular}{|cccc|}
\hline
$L$ & $N_{\rm c}$ & $E_{\rm c}$ & GSC\\
\hline\hline
16 & 2 & 1.4149 &  $1_20_{14}$\\
18 & 2 & 1.2922 &  $1_20_{16}$, $1_20_41_10_{11}$\\
20 & 3 & 1.3440 &  $1_20_{18}$, $1_20_51_10_{14}$\\
22 & 3 & 1.3711 &  $1_20_{20}$, $1_20_41_10_{15}$\\
24 & 3 & 1.3957 &  $1_20_{22}$, $1_20_51_10_{16}$\\
26 & 4 & 1.3675 &  $1_20_{24}$, $1_20_61_10_{17}$, $1_20_31_20_{19}$\\
28 & 5 & 1.3091 & $1_20_{26}$, $1_20_61_10_{19}$, $1_20_41_20_{20}$, $1_20_31_20_31_10_{17}$\\
30 & 5 & 1.3399 & $1_20_{28}$, $1_20_71_10_{20}$, $1_20_41_20_{22}$, $1_20_31_20_41_10_{18}$\\
32 & 5 & 1.3532 & $1_20_{30}$, $1_20_71_10_{22}$, $1_20_41_20_{24}$, $1_20_31_20_41_10_{20}$\\
34 & 5 & 1.3678 & $1_20_{32}$, $1_20_71_10_{24}$, $1_20_41_20_{26}$, $1_20_41_20_41_10_{21}$\\
36 & 6 & 1.3485 & $1_20_{34}$, $1_20_81_10_{25}$, $1_20_51_20_{27}$, $1_20_41_20_51_10_{22}$, $1_20_31_20_31_20_{24}$\\
48 & 8 & 1.3434 & $1_20_{46}$, $1_30_{45}$, $1_20_61_20_{38}$, $1_20_51_20_61_10_{32}$, $1_20_41_20_41_20_{34}$\\
 & & & $1_20_41_20_41_20_{34}$, $1_20_41_20_41_20_41_10_{29}$, $1_20_31_20_31_20_31_20_{28}$\\
\hline
%\label{tab5}
\end{tabular}
\end{center}
\end{table}
Thus, in accordance with the results by Gruber~et~al.~\cite{Gruber_Ueltschi} obtained on an incomplete set of configurations (the periodic configurations and the  mixtures of periodic configurations with the empty lattice), we have found that the ground states of the spinless  Falicov-Kimball model for $U<1$ may be incoherent mixtures of the type $w\& w_0$, with a small difference, and namely, that the configuration $w$ is not necessarily periodic.

\begin{figure*}[t]
\begin{center}
\vspace*{-0.3cm}
\centerline{\mbox{\includegraphics[width=11cm,angle=0]{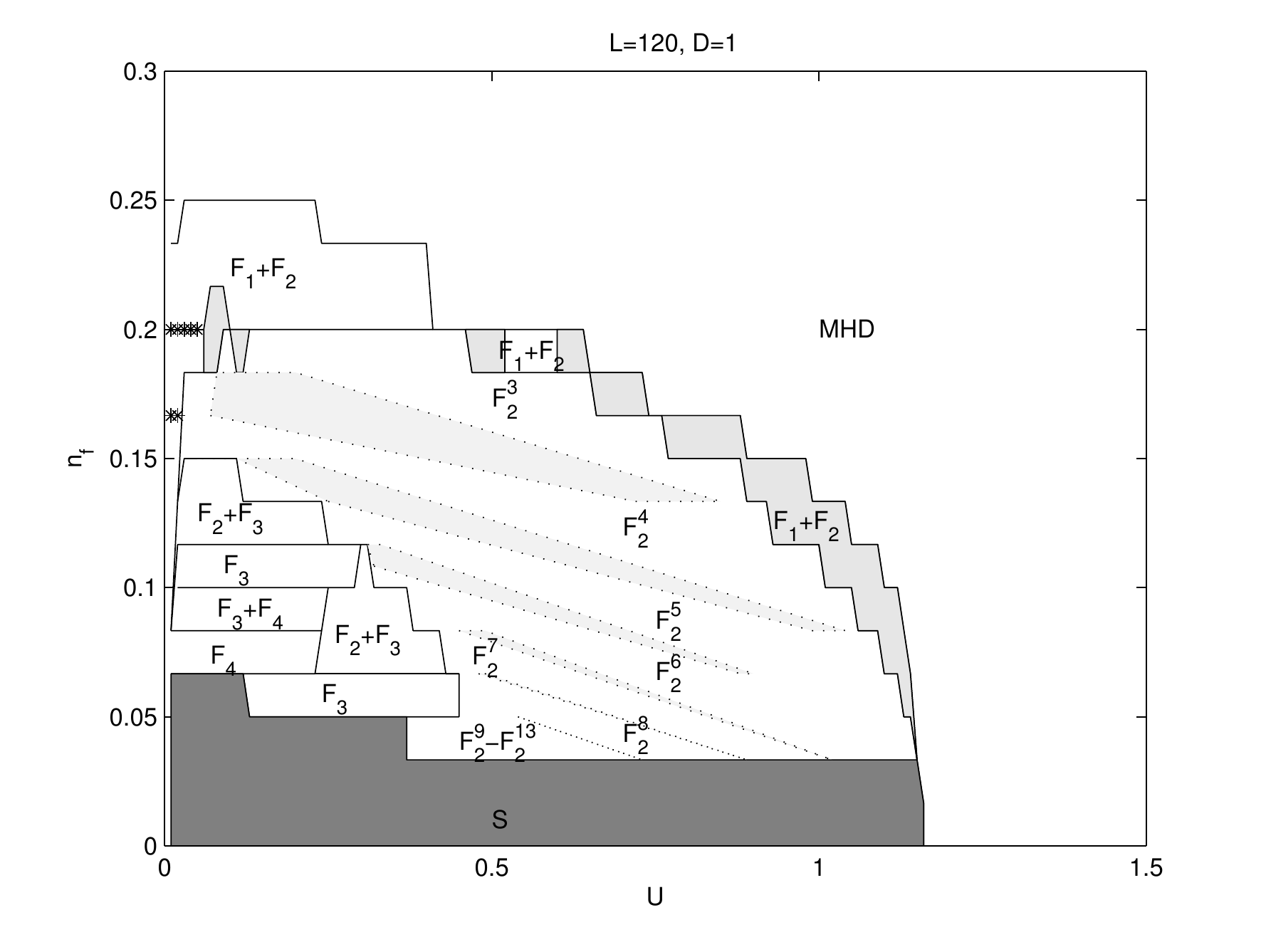}}}
\vspace*{-0.3cm}
\caption{ Phase diagram of the one-dimensional spinless Falicov-Kimball model. Different phases are discussed in detail in the text.}
\label{fd120}
\end{center}
\end{figure*}

To reveal the effect of Coulomb interaction $U$ on the formation of charge ordering in these phase-separated configurations, we have  performed an exhaustive numerical study of the model for a wide range of Coulomb interactions ($U=0.01, 0.02,\dots, 2$) using our new AM that permits to treat much larger clusters. Thereby the finite size effects are considerably reduced.
In figure~\ref{fd120}
\begin{figure*}[!b]
\hspace*{-0.6cm}
\centerline{\hspace{1cm}
\mbox{\includegraphics[width=0.48\textwidth,angle=0]{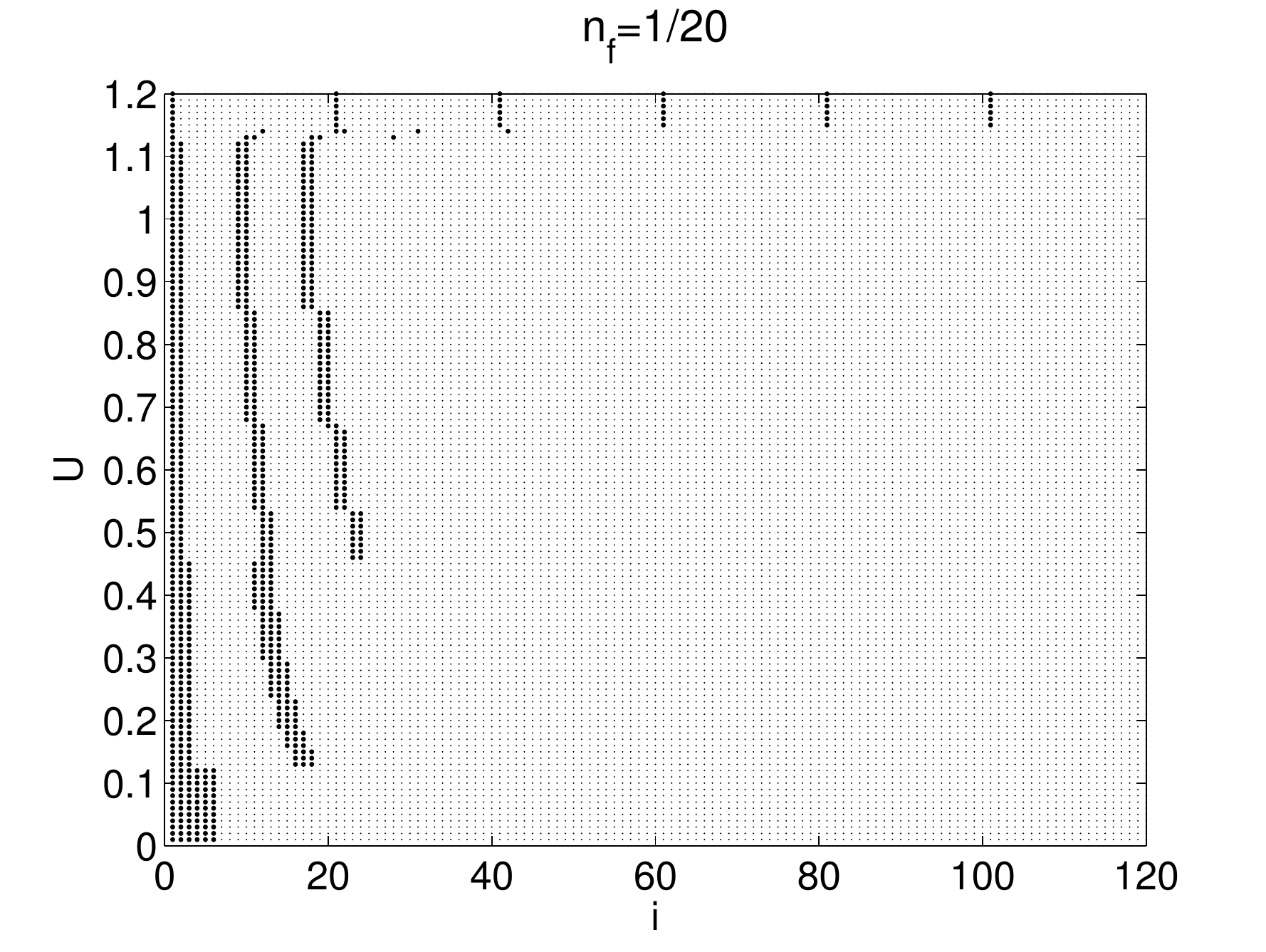}}\hfill%\hspace*{-0.8cm}
\mbox{\includegraphics[width=0.48\textwidth,angle=0]{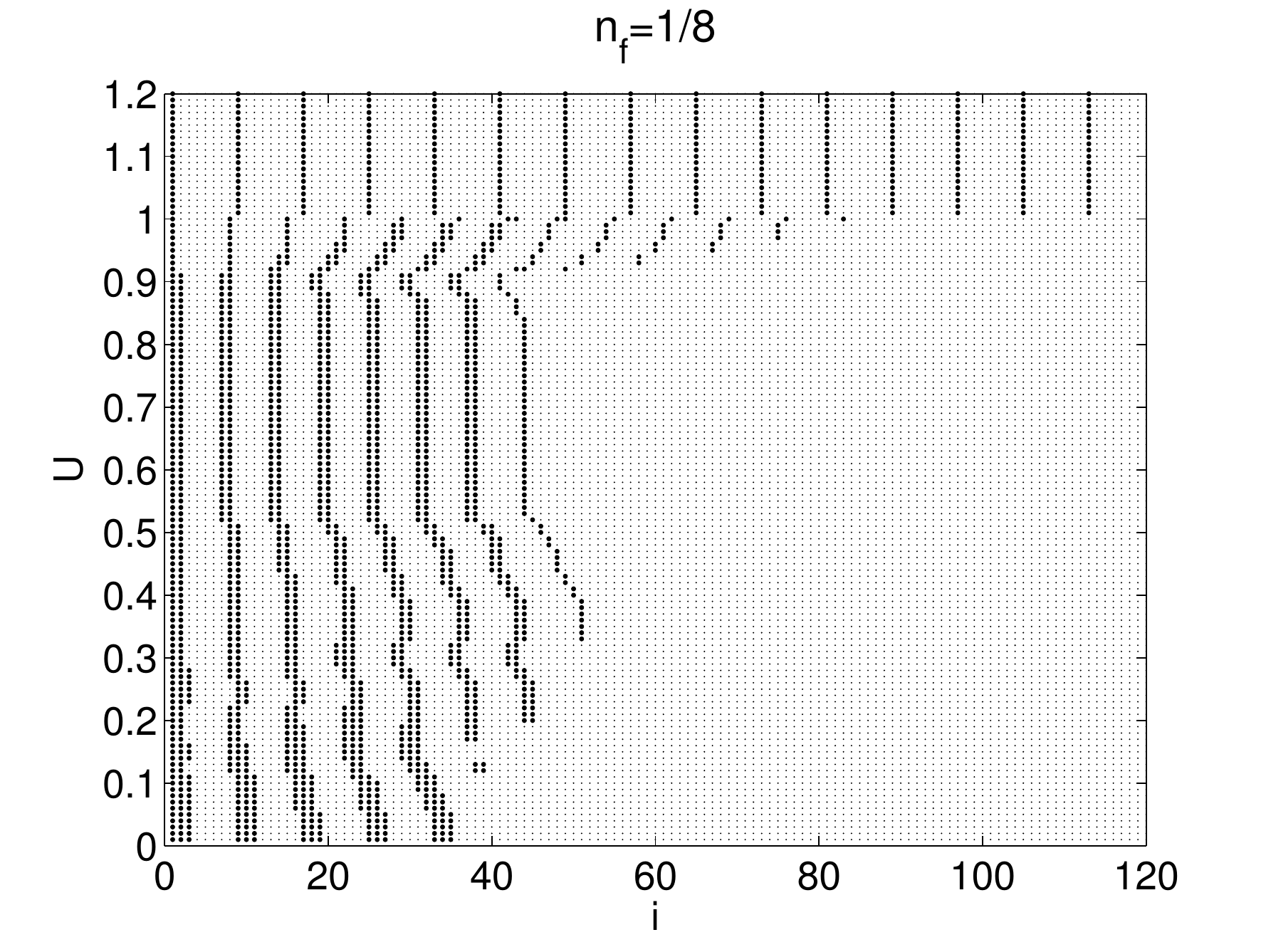}}}
%\vspace*{-0.5cm}
\caption{ Distributions of $f$ electrons on the finite cluster of $L=120$ sites calculated for $U=0.01$, 0.02, $\dots$, 1.2 and two different values of $n_f$.}
\label{konf}
\end{figure*}
we present numerical results obtained for the ground-state phase diagram in the $n_f-U$ plane on finite clusters of $L=120$ sites.
We have found that the phase separation takes place for all Coulomb interaction $U\lesssim 1.2$ for both small ($n_f<0.25$) and large ($n_f>0.75$) electron filling (the $n_f-U$ phase diagram is symmetric around $n_f=0.5$ line, and thus here we present  only the results for $n_f\leqslant 0.5$) and that this domain, except a few isolated points/lines (denoted by $\ast$) is continuous. This is an obvious difference between our phase diagram and one obtained by Gruber et al.~\cite{Gruber_Ueltschi} for periodic configurations and mixtures of these configurations with empty lattices, where large islands of the most homogeneous phase are observed in the phase-separated region. The second important difference is the existence of a narrow intermediate region (in our phase diagram) between the phase-separated configurations with regular (quasi-regular) distributions of $f$ electrons (within $w$) and the most homogeneous domain (MHD). In this region (the medium gray area) the ground states are mixtures of an empty lattice and the aperiodic configurations $w$ with two-molecule distributions in the middle of $w$ and atomic distributions (a single occupied site) at the beginning and at the end  of $w$ ($F_1+F_2$). This fact is clearly demonstrated in figure~\ref{konf}, where all ground-state configurations for $U$ from 0.01 to 1.2 are displayed for two selected values of $f$-electron concentration $n_f=1/20$ and $n_f=1/8$.
Analysing
our numerical results we have found the following trends in the system: (i) In the weak coupling and low concentration limit there is an obvious tendency to form  phase-segregated configurations ($S$) or mixtures of regularly (quasi-regularly) distributed $n$-molecules with the empty lattice (the phases $F_n$ with $n=3$ and~4). (ii) With increasing $U$ and $n_f$, large $n$-molecules split into smaller ones, but their regular distribution persists. (iii) The largest region of the phase-separated domain corresponds to two-molecule distributions $F_2$. This domain exhibits a relatively simple internal structure. There are pure regions $F_{2}^m$ ($m=3,4,\dots,13$), where the ground states are regular distributions of two-molecules with the distance between them being equal to $m$ and the mixtured regions (the light gray areas) where the ground states are mixtures of $F_2^m$ and $F_2^{m+1}$ ($m=3,4,\dots,8$) phases with regular (quasi-regular) distribution of distances $m$ and $m+1$, and an obvious tendency to reduce the distance $m$ with an increasing $n_f$.

\begin{figure*}[h!]
\begin{center}
%\hspace*{-0.6cm}
\includegraphics[width=8cm]{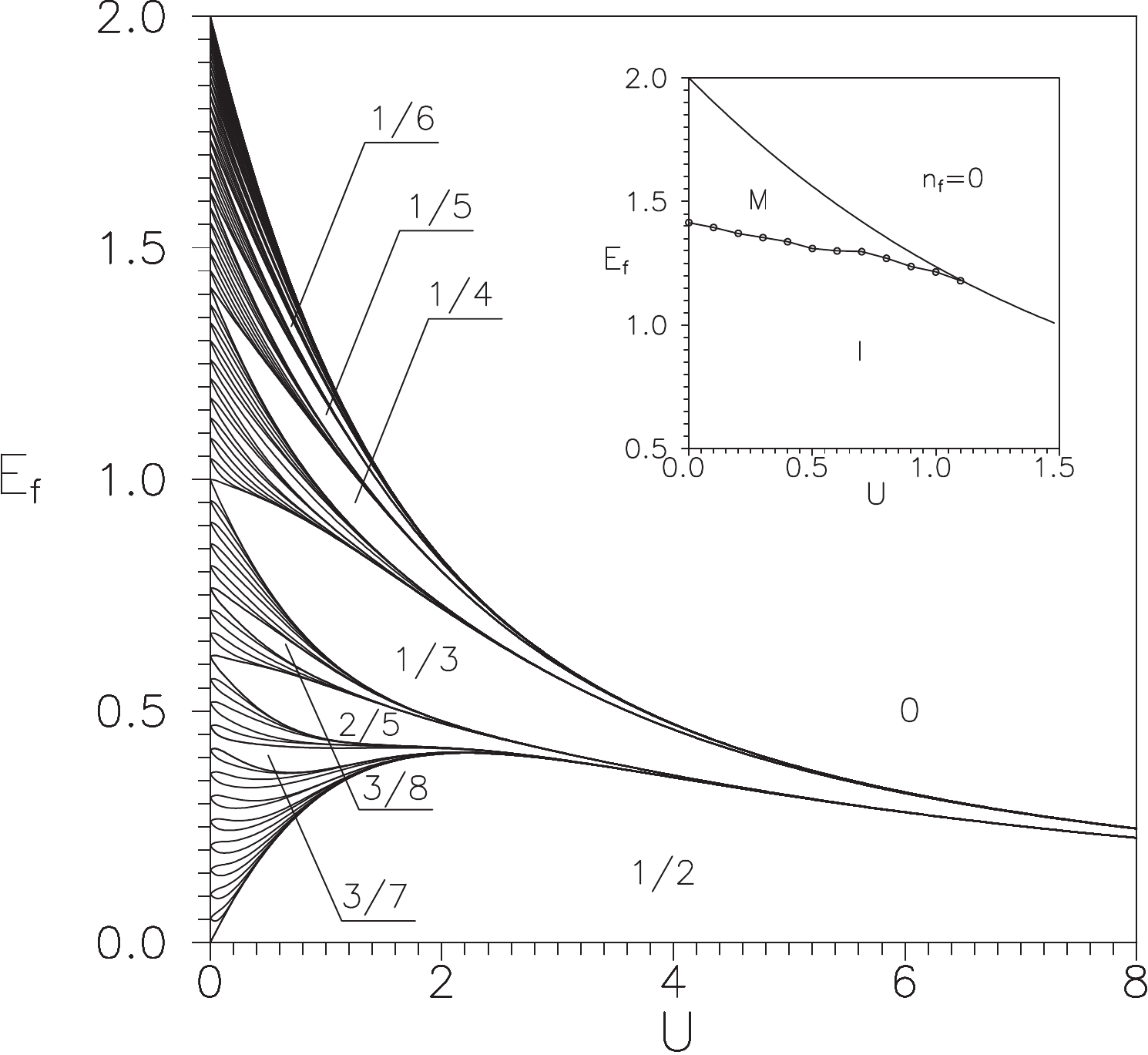}
%\vspace*{-0.5cm}
\caption{ The ground-state phase diagram of the one-dimensional spinless Falicov-Kimball model in the $E_f-U$ plane obtained for $L=240$. All 120 phases corresponding to $f$-electron densities between $n_f=0$ and $n_f=1/2$ are displayed. The largest regions of stability correspond to the periodic configurations with the smallest periods and the rational $f$-electron densities: $n_f=1/2,3/7,3/8,2/5,1/3,1/4,1/5,1/6$ and $n_f=0$. The 3/7 phase has been calculated for $L=420$. The inset shows the regions of stability of the metallic (M) and insulating (I) phase for $n_f>0$~\cite{Farky17}.}
\label{epj1}
\end{center}
\end{figure*}

\begin{figure*}[b!]
\begin{center}
%\hspace*{-0.6cm}
\centerline{
\mbox{\includegraphics[width=0.45\textwidth,angle=0]{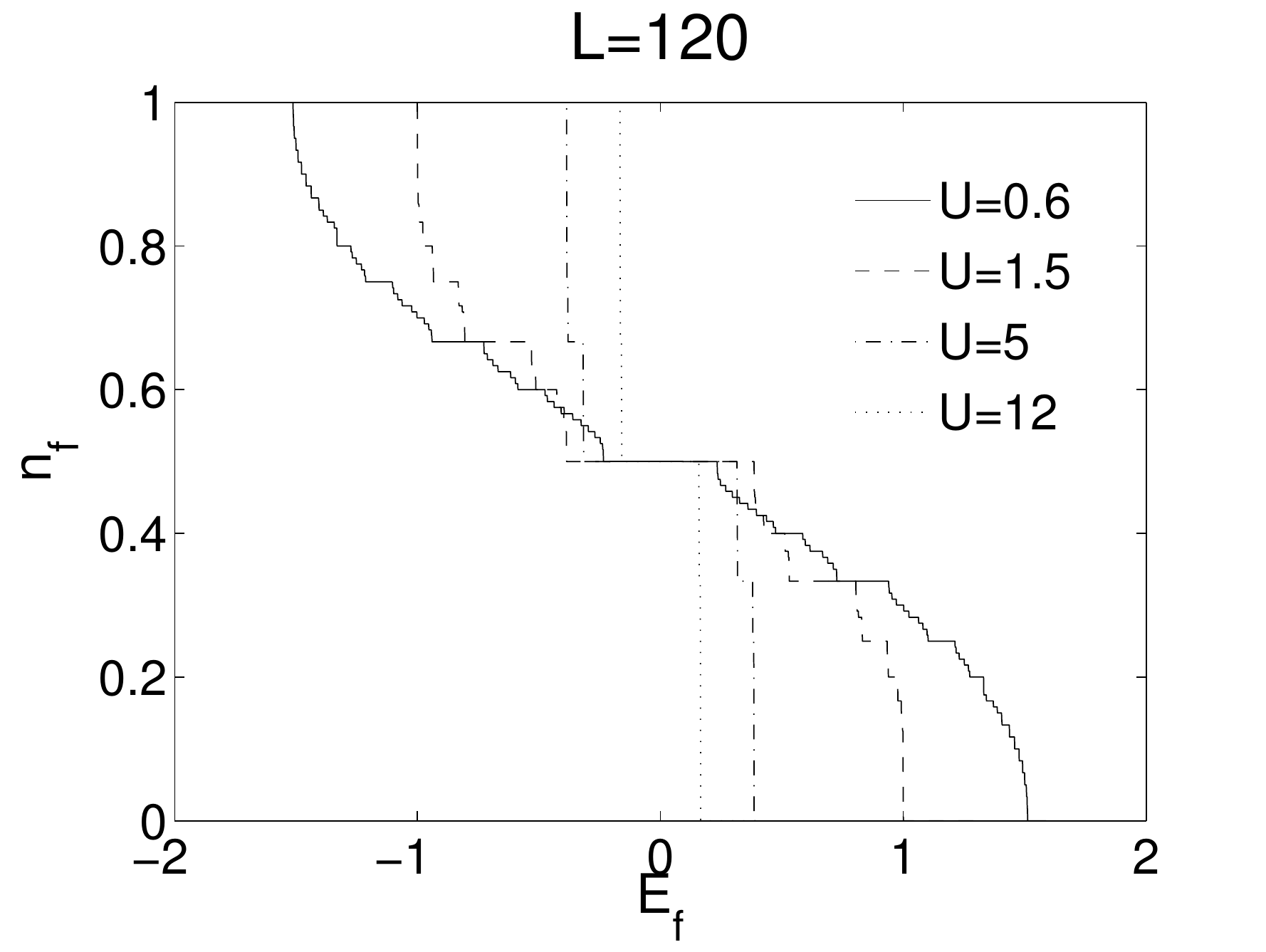}}
\hfill
\mbox{\includegraphics[width=0.45\textwidth,angle=0]{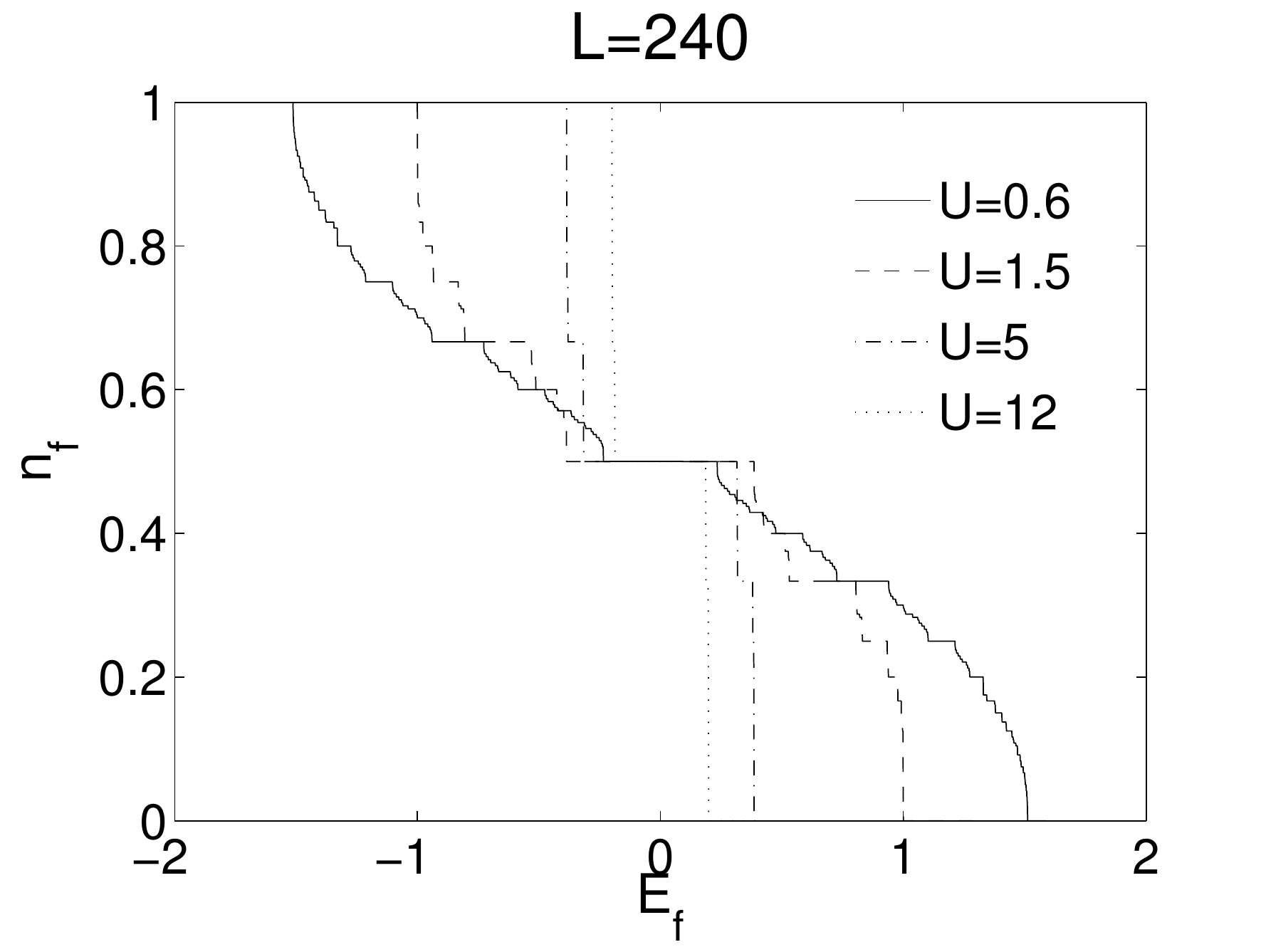}}}
%\vspace*{-0.5cm}
\caption{ Dependence of the $f$-electron occupation number $n_f$  on the  $f$-level position $E_f$ for $U=$ 0.6, 1.5, 5, 12 and $L=120$ (left panel) and $L=240$ (right panel).}
\label{valp}
\end{center}
\end{figure*}

From the viewpoint of rare-earth compounds where the role of itinerant particles is played by the $d$ electrons and the role of immobile particles is played by the $f$ electrons localized on the energy level $E_f$ (the term $E_fN_f$ should be added to the model Hamiltonian~(\ref{eq3.1.1.1})) it is interesting to transform the $n_f-U$ phase diagram into $E_f-U$ coordinates since the cuts of the $E_f-U$ phase diagram in the $n_f$ direction represent the valence transition at a given $U$. Since there is a direct parametrization between $E_f$ and the external pressure $p$~\cite{Goncalves}, the $n_f(E_f)$ behaviour is described (at least qualitatively) by the pressure induced valence changes in rare-earth compounds. The results of our numerical calculations for the $E_f-U$ phase diagram  are summarized in figure~\ref{epj1}.
One can see that the phases with the largest area of stability in the $E_f-U$ phase diagram correspond to the periodic configurations with the smallest periods ($p<9$) and the rational $f$-electron concentrations. The number of phases with the relevant width is strongly reduced with increasing $U$, and thus only a few relevant phases (with $p\leqslant 5$) form the basic structure of the phase diagram in the strong coupling limit. A detailed analysis of the model performed for $U=10$ ($L=240$ and $L=420$) showed that some periodic phases with larger periods also persist in the strong-coupling regime, but their width is considerably smaller. A complete set of phases (with width $w_D>10^{-10}$) that have been determined numerically as the ground states of the model for $U=10$ is shown in table~\ref{tab6}.
\begin{table}[h!]
\begin{center}
\caption{ A complete set of phases (with width $w_D > 10^{-10}$) that have been determined numerically as the ground states of the one-dimensional Falicov-Kimball model for $U=10$~\cite{Farky17}. }
\label{tab6}
\vspace*{0.5cm}
\begin{tabular}{|c|c|}
\hline
$n_f$  & $w_D$ \\
\hline\hline
$1/2$ & $1.8712578 \times 10^{-1}$\\
$1/3$ & $1.0677360 \times 10^{-2}$\\
$1/4$ & $2.1048106 \times 10^{-4}$\\
$2/5$ & $2.1852806 \times 10^{-5}$\\
$1/5$ & $3.4509125 \times 10^{-6}$\\
$2/7$ & $1.1864030 \times 10^{-8}$\\
$3/7$ & $3.7646341 \times 10^{-8}$\\
$1/6$ & $5.0851783 \times 10^{-8}$\\
$3/8$ & $1.0913297 \times 10^{-9}$\\
$1/7$ & $6.9578354 \times 10^{-10}$\\
\hline
\end{tabular}
\end{center}
\end{table}
The phases with the smallest periods also persist in the weak coupling limit, but with decreasing $U$ they are gradually suppressed by the periodic configuration with $p\geqslant 10$ for $E_f<E_f^{\rm c}(U)$ and by phase-separated configurations for  $E_f>E_f^{\rm c}(U)$.
The corresponding picture of valence transition based on this phase  diagram is displayed in figure~\ref{valp}.
It is seen that the valence transitions have a staircase structure, where different phases correspond to different areas of stability. The largest stability regions correspond to the most homogeneous configurations with the smallest periods ($n_f$=1/2, 1/3, 1/4, \ldots). These phases form a  primary structure of the valence transition. Its characteristic feature is that it does not change  with an increasing lattice size and therefore it can be used to represent the behaviour of macroscopic systems. The remaining phases form the secondary structure that unlike the previous one is very sensitive to the lattice size. This secondary structure is observed only for small values of the Coulomb interaction $U$ and  with an increasing $U$ it rapidly disappears. Consequently, the valence transitions for intermediate values of  Coulomb interactions consist of only a few valence steps, whose number is further reduced with an increasing~$U$. For example,  for $5<U<10$ the valence transition consists of only four relevant transitions, and namely,  from $n_f=1$ to $n_f=2/3$, from $n_f=2/3$ to $n_f=1/2$, from $n_f=1/2$ to $n_f=1/3$ and from $n_f=1/3$ to $n_f=0$  and for $U>10$ there are  even two relevant valence transitions: the first from $n_f=1$ to $n_f=1/2$ and the second  from $n_f=1/2$ to $n_f=0$. Thus, we can conclude that the spinless Falicov-Kimball model is capable of describing two basic types of valence transitions, and namely, the transition from the integer-valence ground state into the inhomogeneous intermediate-valence state and transition from one inhomogeneous intermediate-valence state into another inhomogeneous intermediate-valence state. Moreover, our numerical results confirmed that the crucial role in the mechanism of valence transitions is played by the Coulomb interaction between itinerant and localized electrons.
\\

{\bf Two-dimensional case}

The combination of EDM and AM has been  also used for the study of the charge ordering in a two-dimensional spinless Falicov-Kimball model~\cite{Farky17}. In this case we were forced to limit ourselves only to the area of intermediate and strong Coulomb interactions ($U>1$), whereas in the opposite limit $U<1$  lattice effects were still strong, even on lattices with $L=400$ sites.
\begin{figure*}[h!]
\begin{center}
\mbox{\includegraphics[width=5.8cm,angle=0]{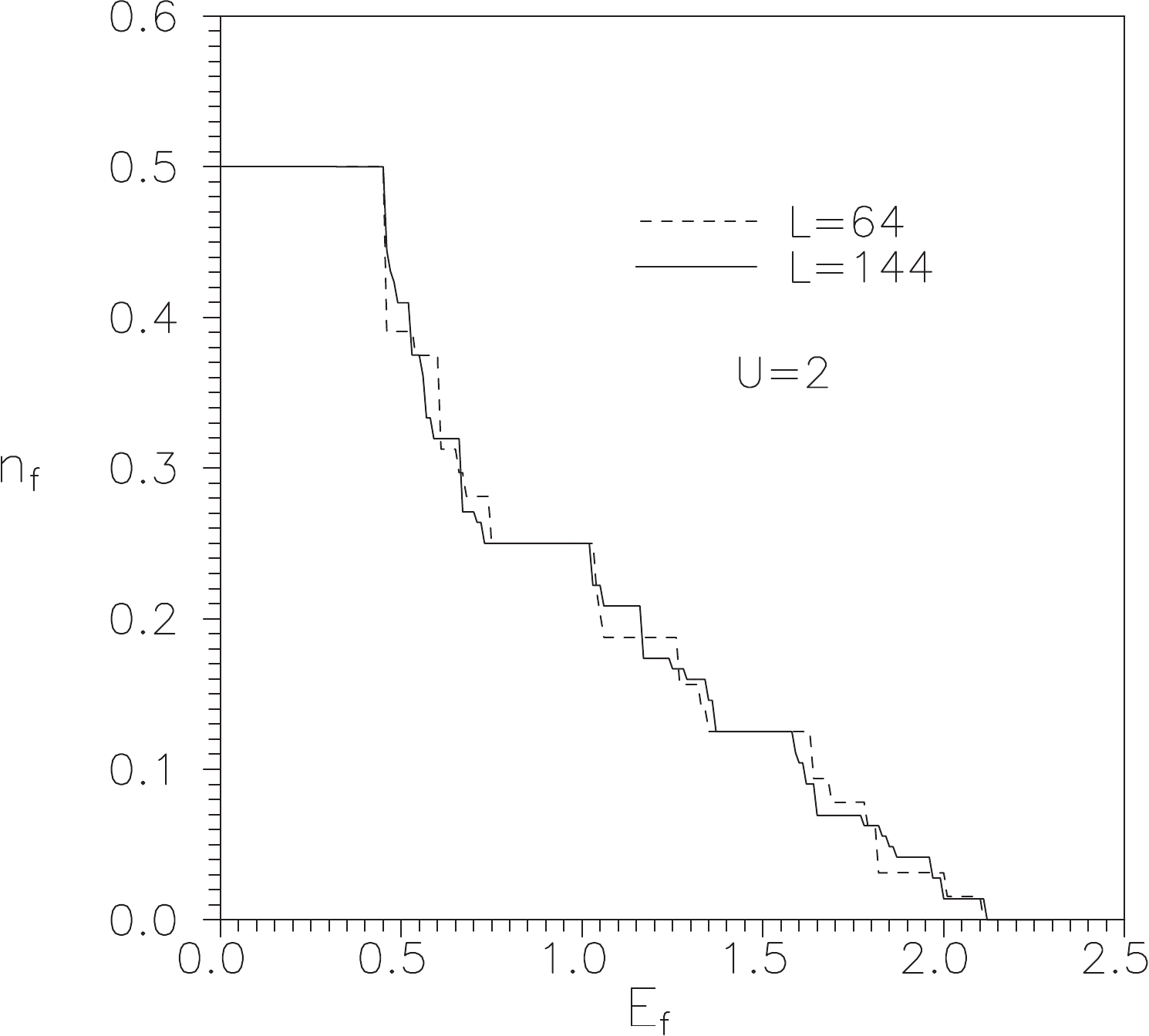}}\hspace*{1cm}
\mbox{\includegraphics[width=7.5cm,angle=0]{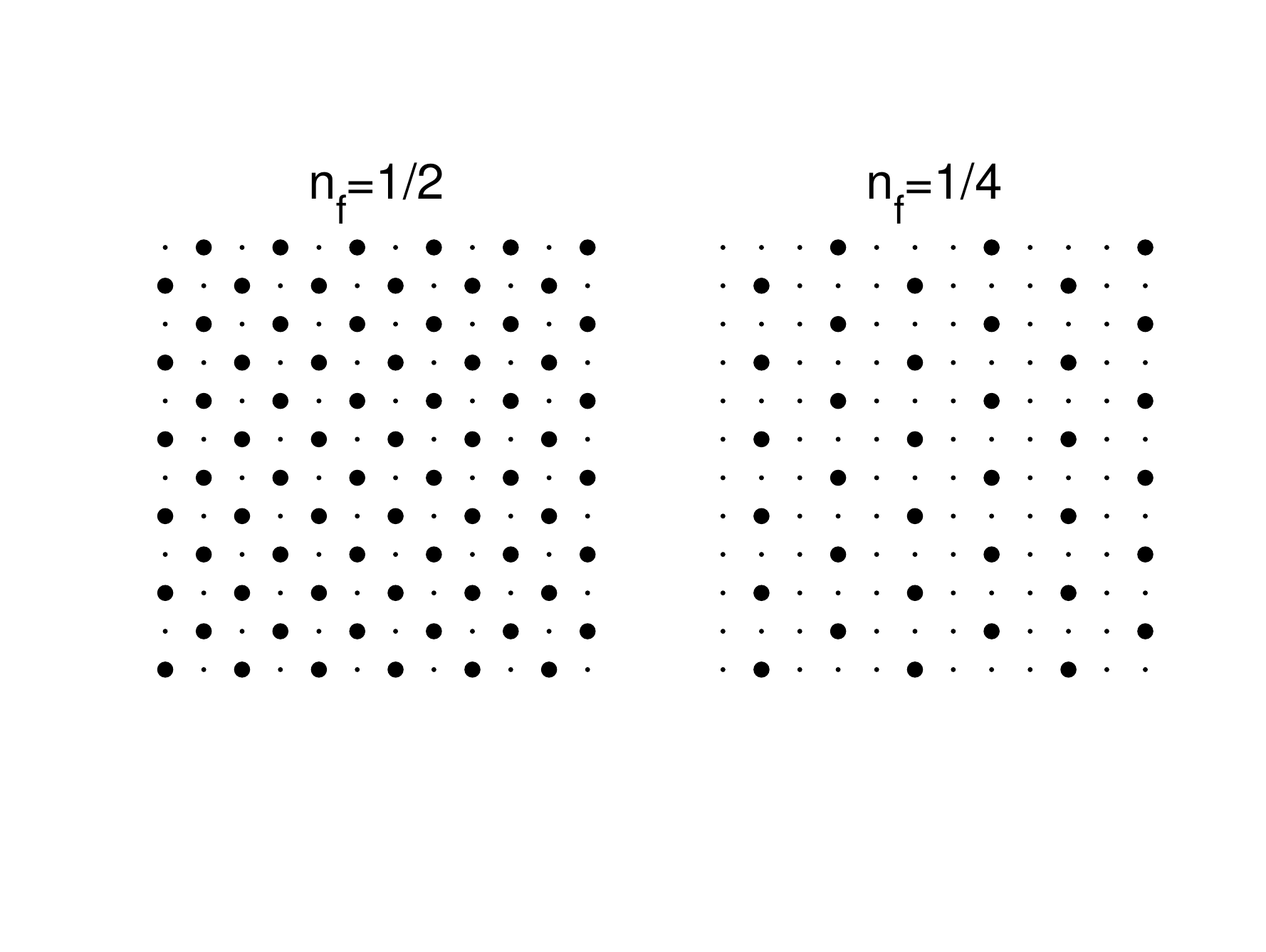}}
\end{center}
\vspace*{-0.5cm}
\caption{ {\it Left}: Dependence of the $f$-electron occupation number $n_f$ on the $f$-level position $E_f$ calculated for $U=2$ and two
different clusters ($8 \times 8$ and $12 \times 12$). {\it Right}: The ground-state configurations of the two-dimensional spinless Falicov-Kimball
model for $U=2$ and two different $f$-electron densities: $n_f=1/2$ and $n_f=1/4$~\cite{Farky17}.}
\label{epj1_4}
\end{figure*}
Figure~\ref{epj1_4} represents a cut of the phase diagram in the  $n_f-E_f$ plane (the valence transition) for the  intermediate value of the Coulomb interaction $U=2$ and $L=8\times 8$ and $L=12\times 12$.  Similarly to the  one-dimensional case, the largest regions of stability  again correspond to configurations with the rational $f$-electron concentrations  ($n_f$ = 1/2, 1/3, 1/4, $\dots $) and similar are  also  the charge distributions (see figure~\ref{epj1_4}) with the difference that the periodic one-dimensional distributions are reflected now in the diagonal charge stripes regularly repeated with the same periodicity like in the one-dimensional case  ($n_f=1/2$ and $n_f=1/4$)\footnote{For $n_f=1/2$ and $n_f=1/4$ our results are fully consistent with the previous results of Watson and Lemanski obtained within the method of restricted phase diagrams~\cite{Watson_Lemanski}.}. Below a certain critical value $n_f^{\rm c}$  the ground states are  phase separated. This means that $f$ electrons occupy only one part of the lattice, while the remaining part is empty. Typical examples of such phase-separated configurations are shown in figure~\ref{epj1_7} (left hand panel).
\begin{figure*}[!t]
\begin{center}
\mbox{\includegraphics[width=6.5cm,angle=0]{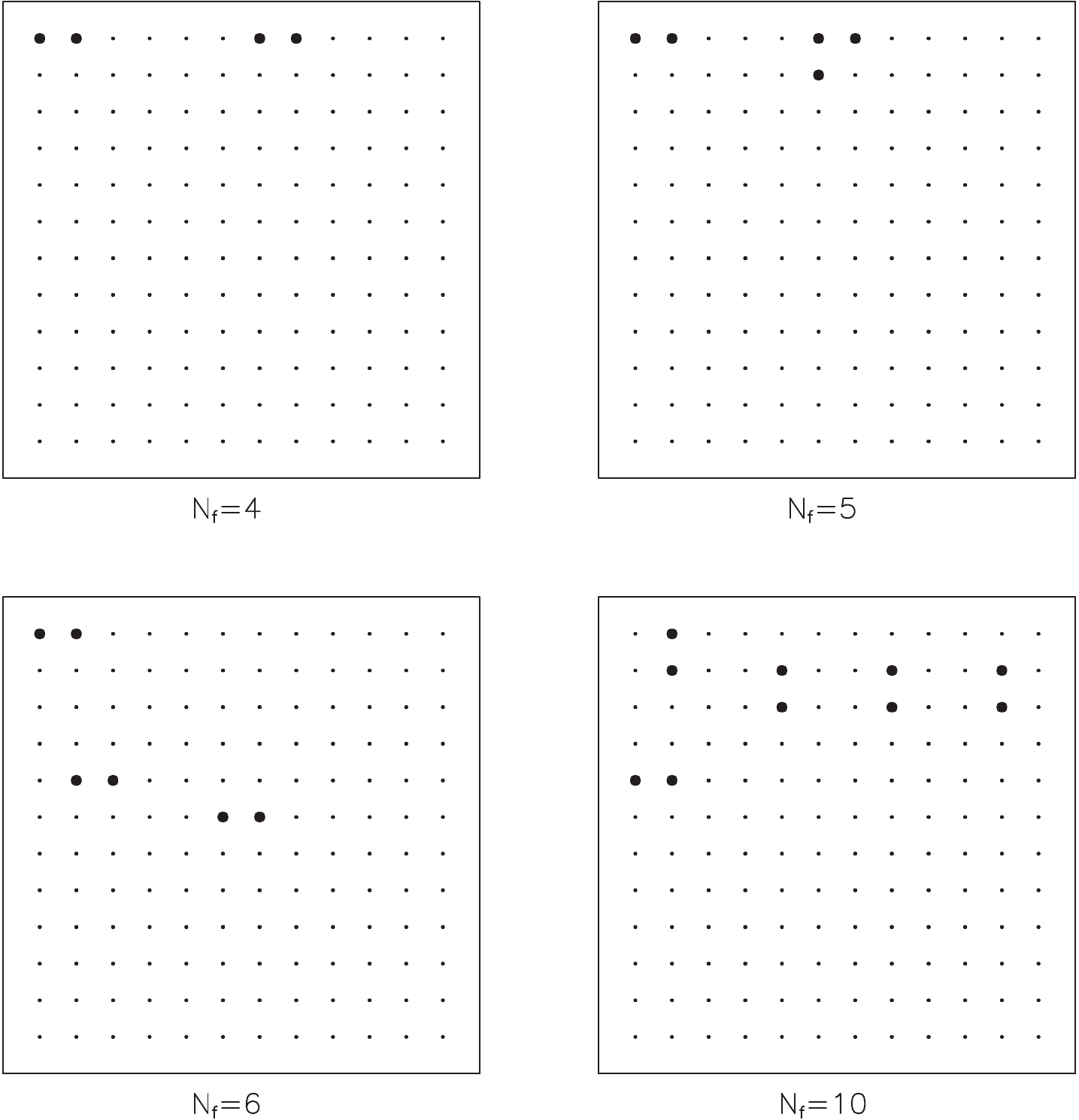}}\hspace*{0.7cm}
\mbox{\includegraphics[width=6.5cm,angle=0]{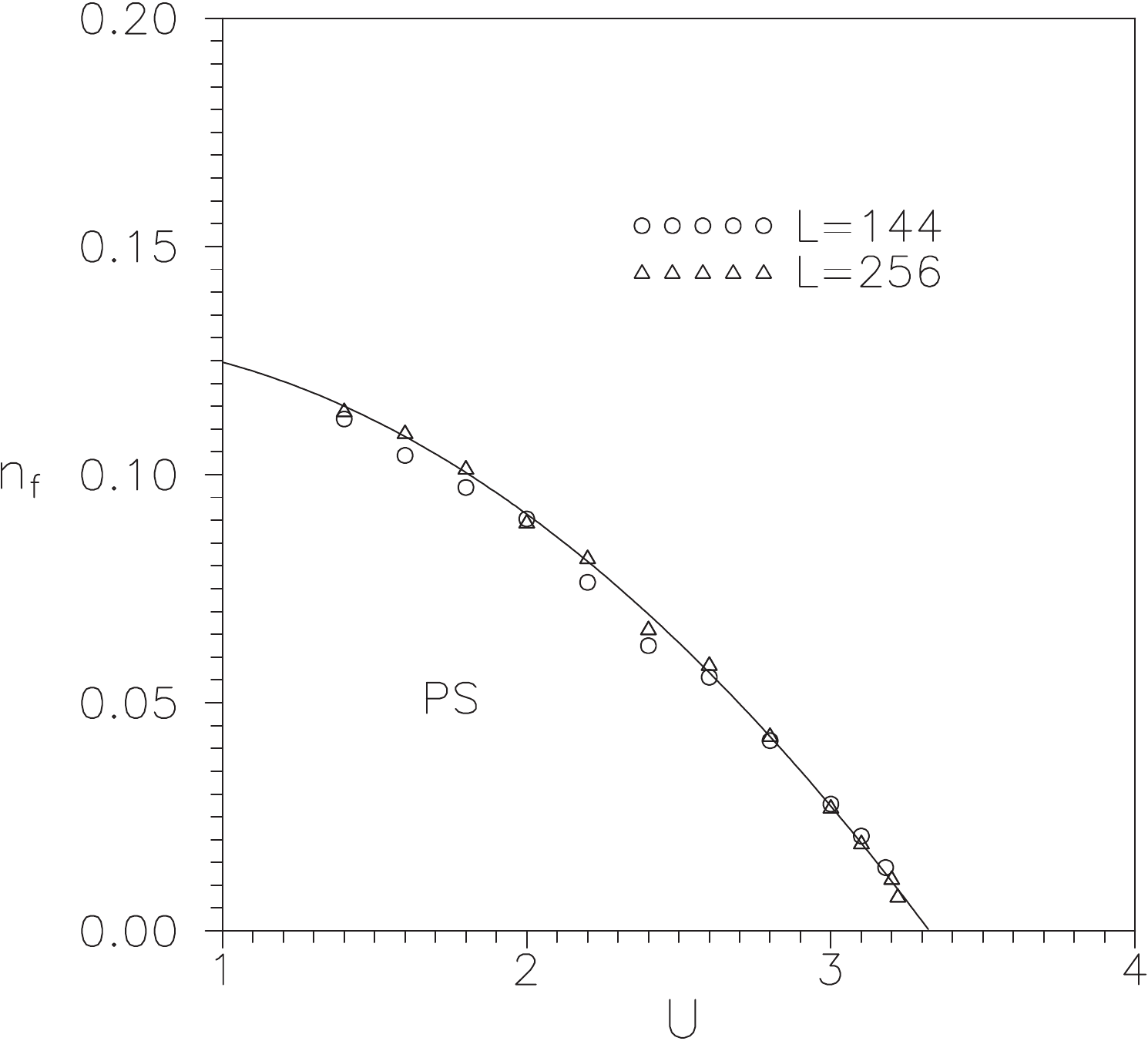}}
\end{center}
\vspace*{-0.5cm}
\caption{ {\it Left}: The ground-state configurations of the two-dimensional Falicov-Kimball model for $U=2$ and several $f$-electron densities $n_f < n^{\rm c}_{f}$. {\it Right}: The region of phase separation (PS) of the two-dimensional Falicov-Kimball model calculated for two different clusters ($12 \times 12$
and $16 \times 16$)~\cite{Farky17}.}
\label{epj1_7}
\end{figure*}
\begin{figure*}[b!]
\begin{center}
\hspace*{-1.cm}
\mbox{\includegraphics[width=16.5cm,angle=0]{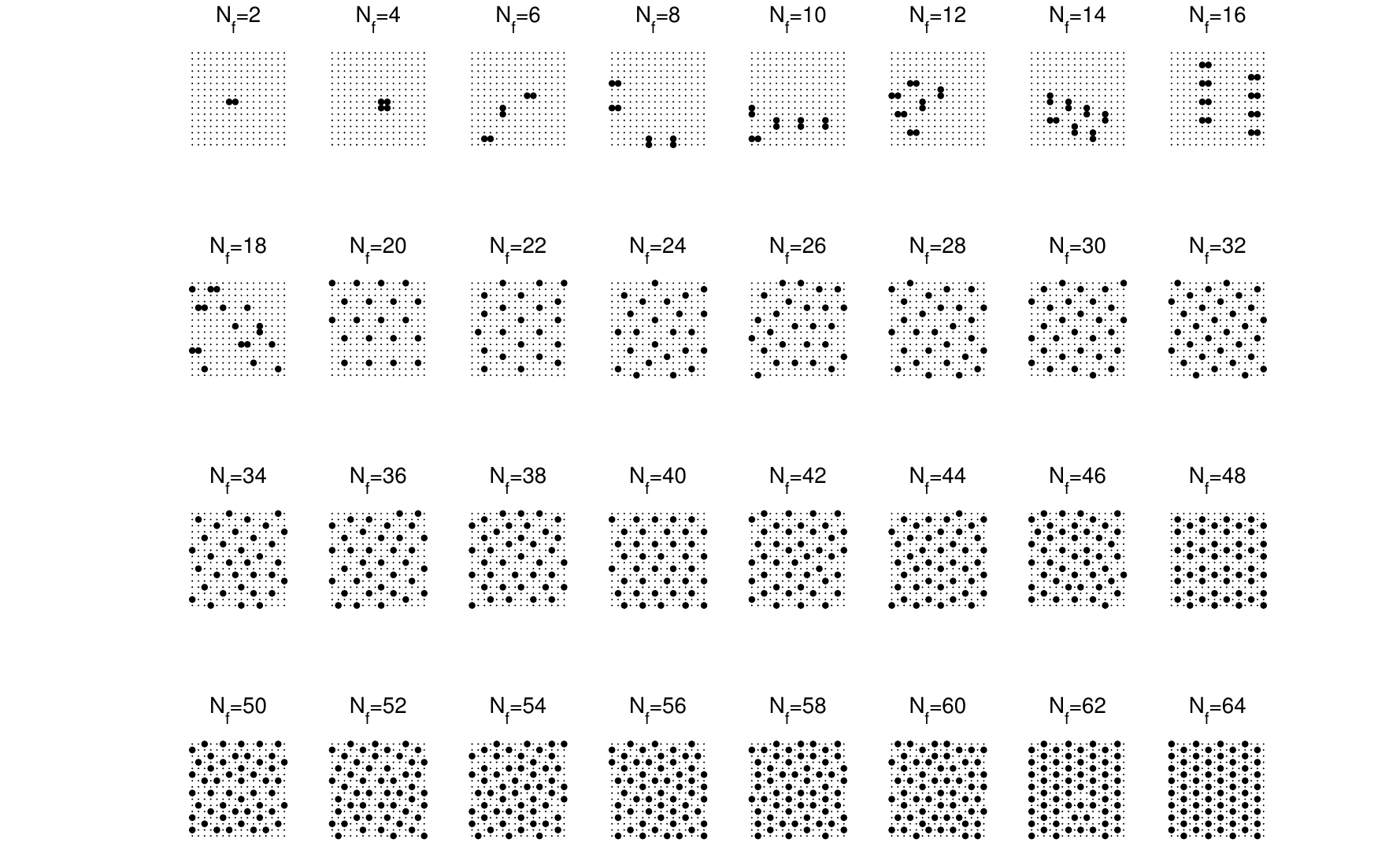}}
\end{center}
\vspace*{-0.3cm}
\caption{ The ground-state configurations of the two-dimensional Falicov-Kimball model obtained for $L=16\times 16$ and $U=2$ ($N_f\leqslant L/4$).}
\label{fig_konf_2d_d}
\end{figure*}

Similarly to  the one-dimensional case, these configurations are metallic, and thus the boundary between  the phase-separated domain and the rest of the phase diagram is the boundary of metal-insulator transitions. For $U\geqslant 1$, where the lattice effects are negligible, we have specified  the region of stability of this domain very precisely (figure~\ref{epj1_7}). A surprising finding was that in the two-dimensional case the stability region of phase-separated domain shifts very significantly to higher values of $U$ ($U\sim 3.3$). This result is very important from the point of view  of possible applications of the model for a  description of metal-insulator transitions in  rare-earth and transition-metal compounds. It is generally assumed that in these materials  the values of the local Coulomb interaction $U$ are much larger than the hopping integrals $t_{ij}$, and, therefore, for a  correct description of metal-insulator transitions in real materials one should consider the limit of $U>t$ rather than $U<t$.

\begin{figure*}[h!]
\begin{center}
\vspace*{0.3cm}
\hspace*{-1.cm}
\mbox{\includegraphics[width=16.5cm,angle=0]{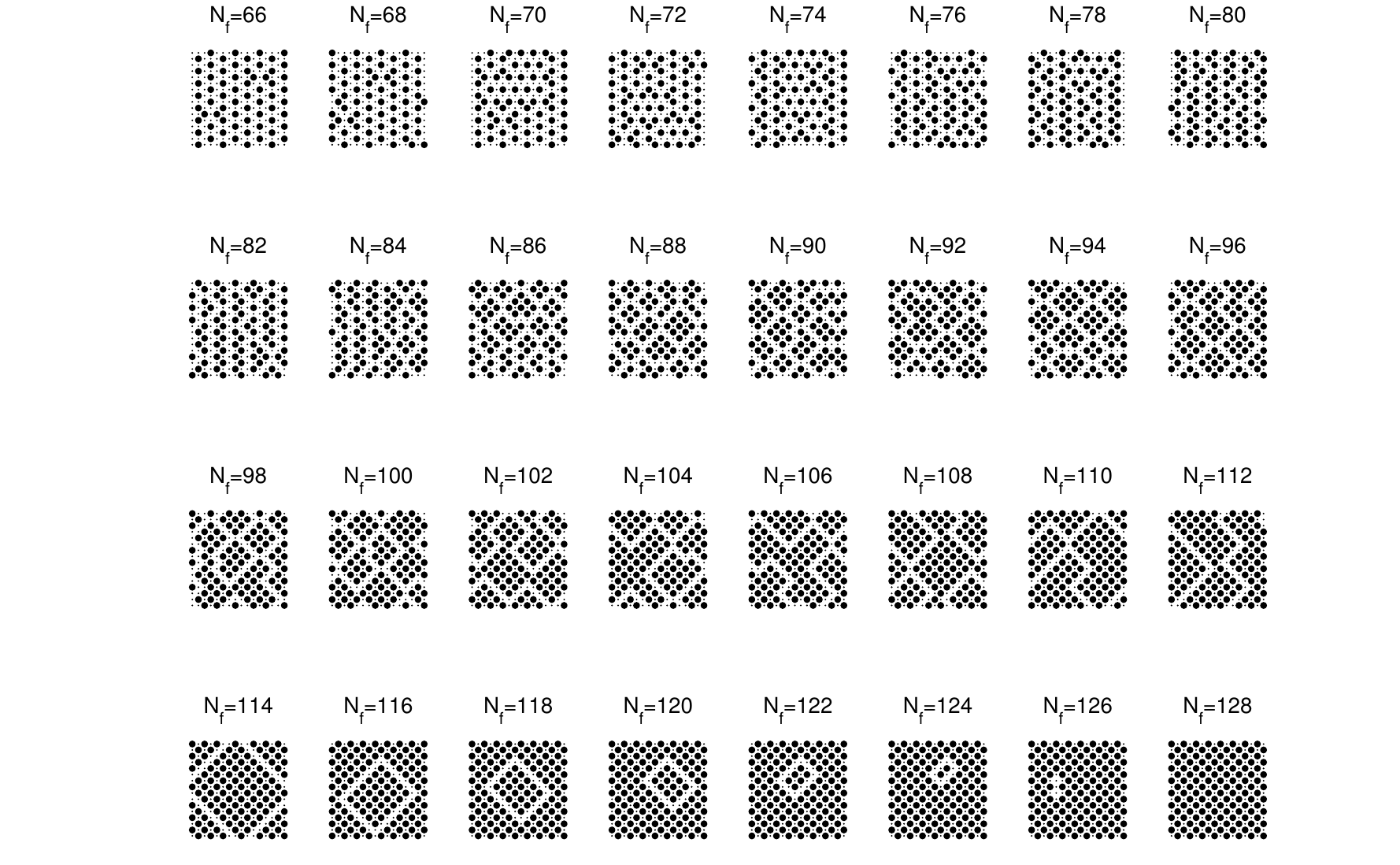}}
\end{center}
\vspace*{-0.3cm}
\caption{ The ground-state configurations of the two-dimensional Falicov-Kimball model obtained for $L=16\times 16$ and $U=2$ ($L/4<N_f\leqslant L/2$).}
\label{fig_konf_2d_u}
\end{figure*}

For this reason we have performed additional numerical studies of the model in the intermediate ($U=2$) and strong ($U=4$) coupling limit on the larger cluster of $L=16\times 16$ sites.
The complete list of ground-state configurations obtained in the intermediate coupling limit for even values of $N_f$ are displayed in figure~\ref{fig_konf_2d_d} ($N_f\leqslant L/4$) and figure~\ref{fig_konf_2d_u} ($N_f>L/4$).
Going with $N_f$ from zero to $L/2$ we have observed the following configuration types. For small $f$-electron concentrations ($N_f\leqslant 4$) the ground states are phase segregated. Then there follows the region of phase separation ($N_f\lesssim 18$) when two-molecule clusters of $f$ electrons are distributed only over one part of lattice leaving another part of lattice free of $f$ electrons. The two-molecule distributions  disappear at $N_f=20$. This is also the point of phase transition from the phase separated to the homogeneous/quasi-homogeneous phase, where the single $f$ electrons are distributed regularly/quasi-regularly over the whole lattice. This region ends at $N_f\sim 64$ where the axial stripes of empty and alternating configurations are observed. Above $N_f=64$ the axial bands of width $w_D=2$ are formed with the chessboard distribution of $f$ electrons separated by empty lines.
At $N_f\sim 84$ the chessboard structure starts to develop, first in the form of small clusters of four molecules and then in the form of larger and larger clusters with fully developed chessboard ordering separated by  diagonal lines  of empty sites.

We have observed the same picture  for larger values of Coulomb interactions. The larger values of $U$ only  modify the stability regions of some phases, but no new configuration types appear. In particular, for $U=4$, the region of phase segregation/separation is fully suppressed  and  the region of regular/quasi-regular distributions extends up to $n_f\to 0$.\\

{\bf Three-dimensional case}

\begin{figure*}[b!]
\begin{center}
%\vspace{-0.8cm}
\includegraphics[width=6.5cm]{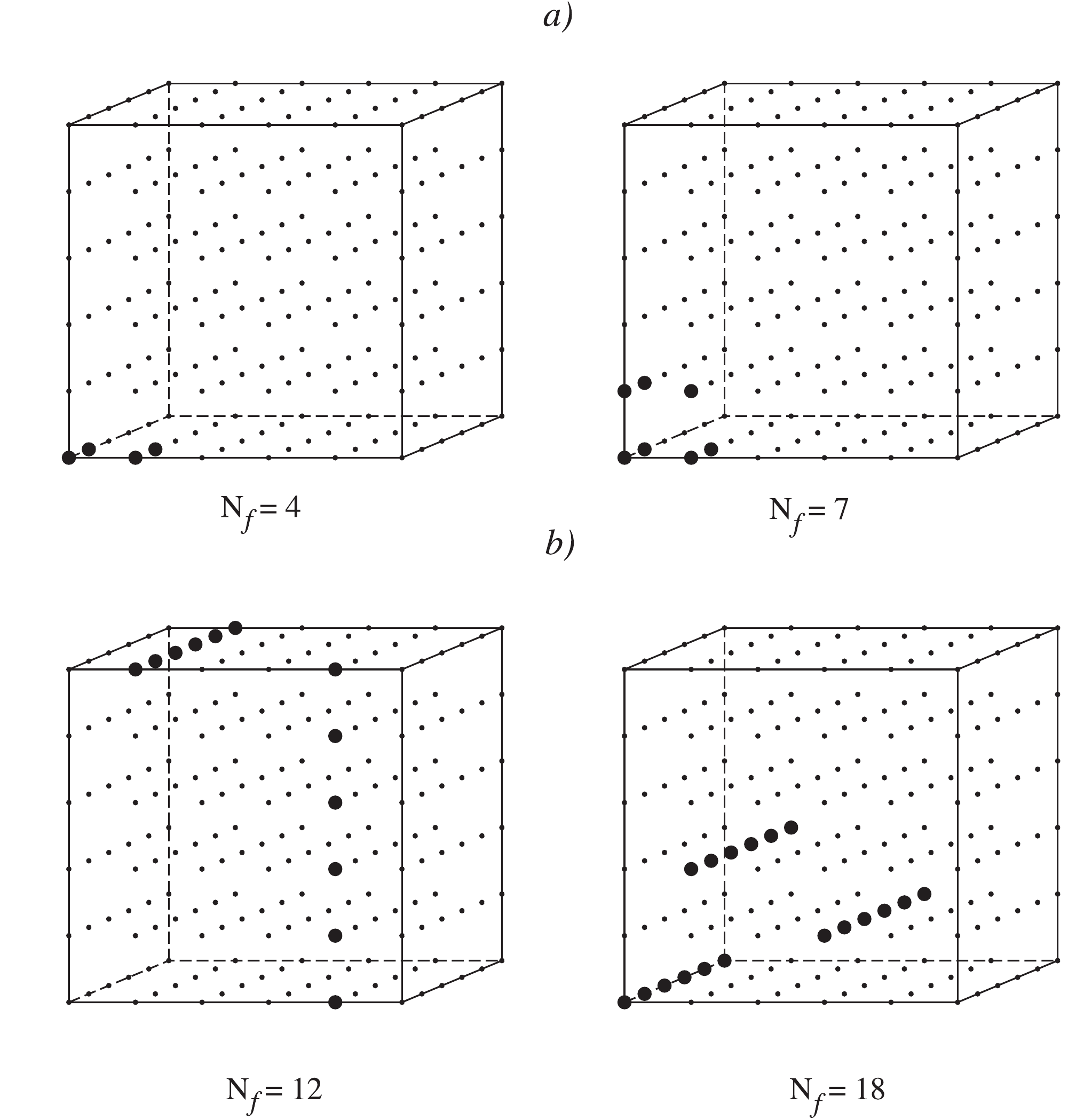}
\includegraphics[width=6.5cm]{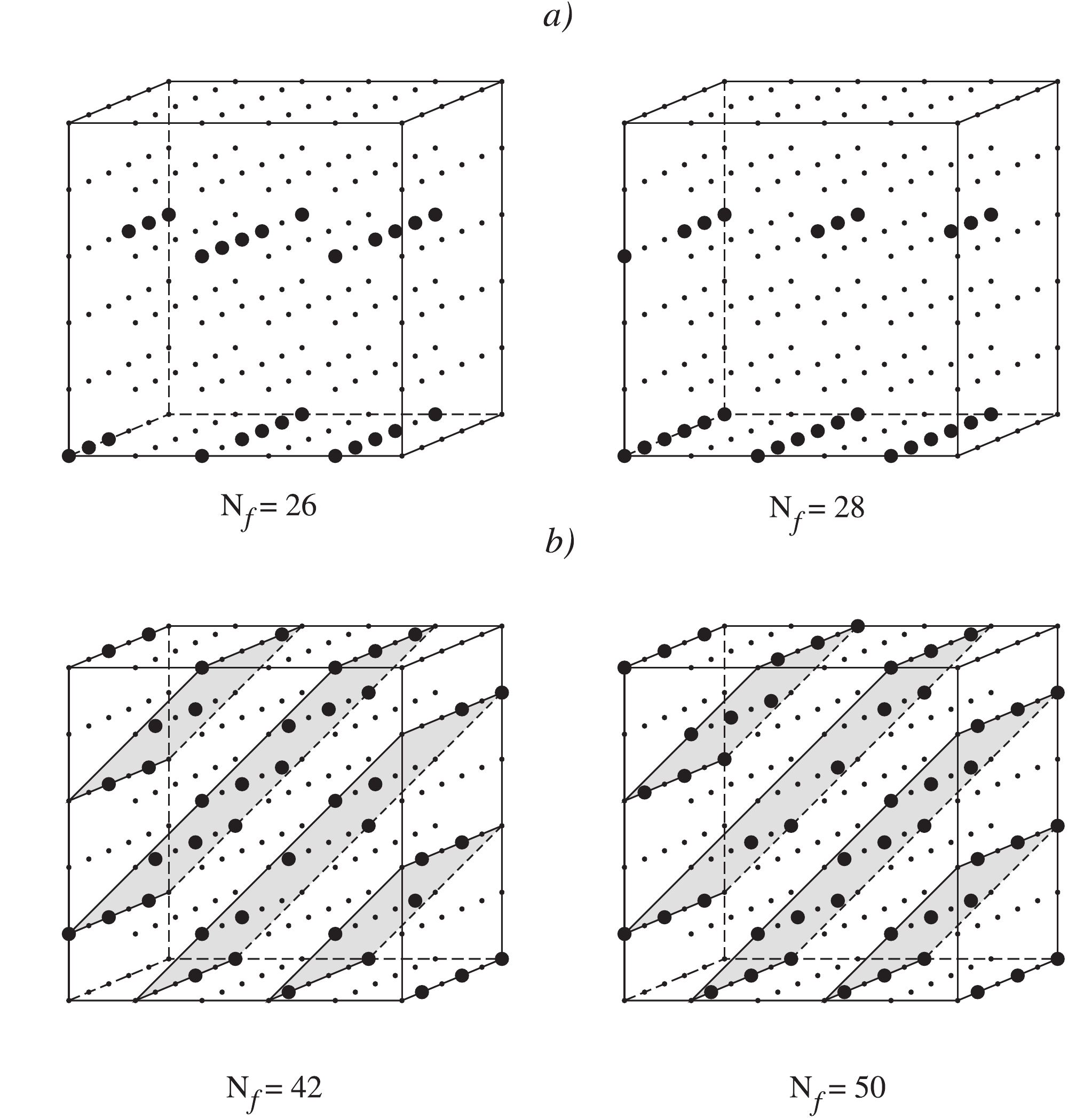}
\end{center}
%\vspace*{-1.2cm}
\caption{ {\it Left}:  Typical examples of phase-segregated (a) and striped (b) configurations obtained for $U=1$ and $L=6\times 6\times 6$.
 {\it Right}: Typical examples of striped configurations with regular distribution (a) and diagonal charge planes with an incomplete chessboard structure
(b)  obtained for $U=1$ and $L=6\times 6\times 6$~\cite{Farky32}.}
\label{3d_1}
\end{figure*}

\begin{figure*}[t!]
\begin{center}
%\vspace{-0.5cm}
\includegraphics[width=6.5cm]{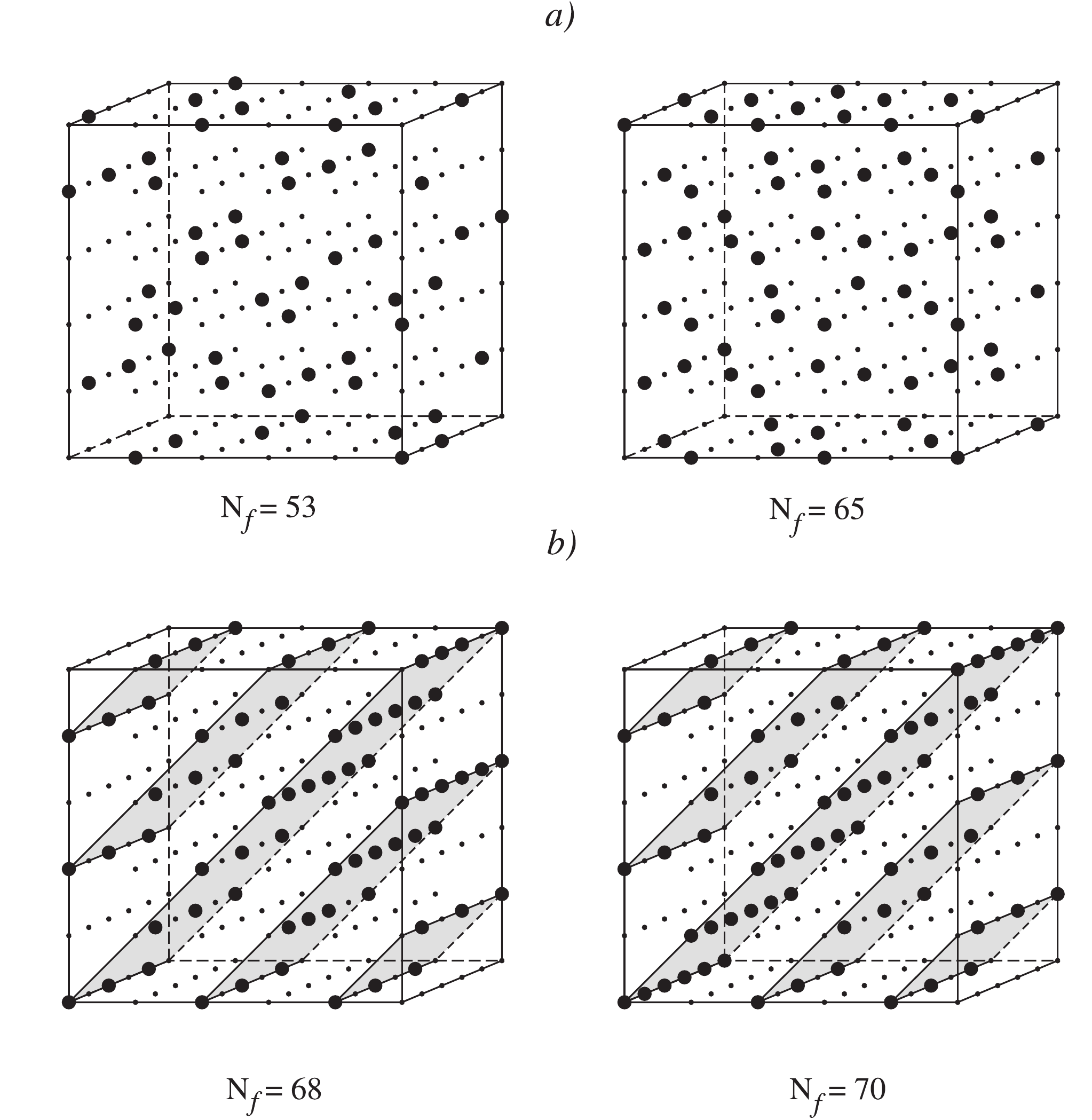}
\includegraphics[width=6.5cm]{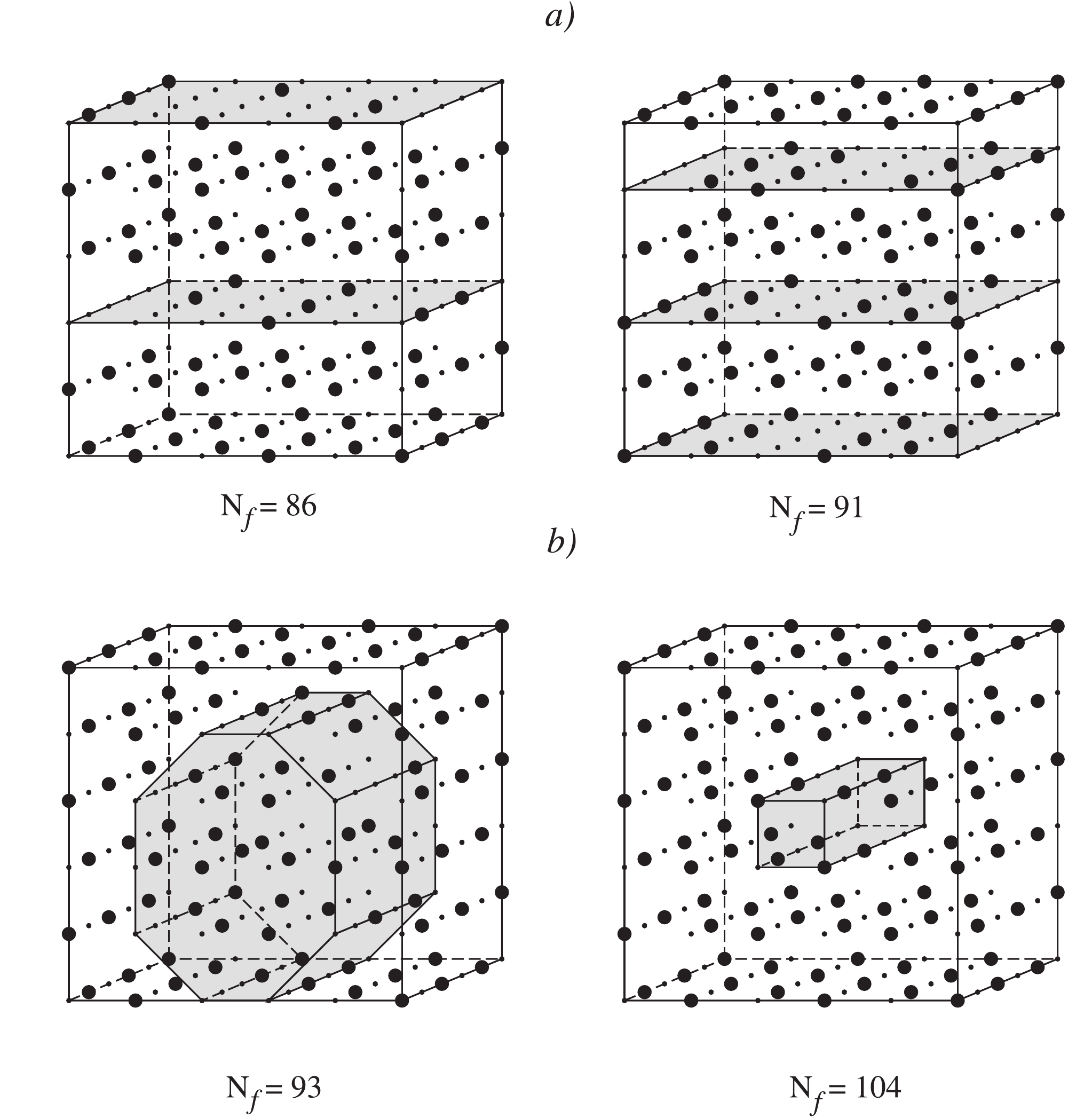}
\end{center}
%\vspace*{-1.2cm}
\caption{
{\it Left}:  Ground-state configurations for intermediate $f$-electron concentrations. (a) The formation of a chessboard structure. (b) The examples of ground-state configurations that can be considered as mixtures of configuration types with smaller $n_f$ ($U=1$, $L=6\times 6\times 6$).  {\it Right}: Examples of an incomplete chessboard structure obtained for $U=1$ and $L=6\times 6\times 6$.  (a) The chessboard structure is fully developed in some regions (planes) that are separated by planes with incompletely developed chessboard structure. (b) Both regions with complete and incomplete chessboard structure have a three-dimensional character~\cite{Farky32}. }
\label{3d_3}
\end{figure*}

In principle, the same procedure as the one used in $D=1$ and $D=2$ can be also used  in $D=3$. Due to the numerical complexity of the problem in $D=3$ we have performed numerical calculations only for selected values of the  Coulomb interaction $U$ representing the typical behaviour  of the model at the weak ($U=1$), intermediate ($U=2$) and strong ($U=8$) interactions~\cite{Farky32}. In order to reveal the finite-size effects, numerical calculations were made on two different clusters of $4\times 4 \times 4$ and $6\times 6\times 6$ sites. A direct comparison of numerical results obtained on $4\times 4 \times 4$ and $6\times 6\times 6$ clusters showed that the ground-state configurations fall into several different categories whose stability regions are practically independent of $L$.  For this reason we present here only the results obtained on  $L=6\times 6\times 6$ (figure~\ref{3d_1} and figure~\ref{3d_3}).
The largest number of configuration types is observed in the weak-coupling limit. Going with $N_f$ from zero to half-filling ($N_f=L/2$) we have observed the following configuration types for $U=1$. At low $f$-electron concentrations, the ground-states are the phase-segregated configurations
($f$ electrons clump together while the remaining part of lattice is free of $f$ electrons).
Typical examples of ground-state configurations from this region are depicted in figure~\ref{3d_1} (left hand panel).

Above the region of phase segregation, we have observed the region of stripe formation ($N_f=10,\ldots,20$). In this region the $f$ electrons form the one-dimensional charge lines (stripes) that can be  perpendicular or parallel. This result shows that the crucial mechanism leading to the formation of stripes in strongly correlated systems should be a competition between the kinetic and short-range Coulomb interaction.

Going with $N_f$ to higher values, the stripes vanish and again appear at $N_f=26$, though in a fully different distribution. While at smaller values of $N_f$ the stripes have been distributed inhomogeneously (only over one half of the lattice), the stripes in the region $N_f=26,\ldots,31$ are distributed regularly.

Above this region a new type of configurations  starts to develop. We call them diagonal charge planes with an  incomplete chessboard structure, since the $f$ electrons prefer to occupy the diagonal planes with slope~1, and within these planes they form a chessboard structure. This region is relatively broad and extends up to $N_f \sim 50$. Then there follows the region in which the chessboard structure starts to develop. As illustrated in figure~\ref{3d_3}, the $f$ electrons begin to  preferably occupy the sites of sublattice A, leaving the sublattice B free of $f$ electrons. Furthermore, the configurations that can be considered as mixtures of previous configuration types are also observed in this region. However, with increasing $N_f$ the configurations of chessboard type become dominant. Analysing these configurations we have found that the transition to the purely chessboard configuration is realized through several steps. The first step, the formation of the chessboard structure has been illustrated in figure~\ref{3d_1}. The second step is shown in figure~\ref{3d_3}. It is seen that the chessboard structure is fully developed in some regions (planes) that are separated by planes with an incomplete developed chessboard structure. Such type of distribution is replaced for larger values of $N_f$ by a new type of distributions (step three), where both regions with complete and incomplete chessboard structure have a three-dimensional character.

 \begin{figure}[b!]
\begin{center}
\includegraphics[width=6cm]{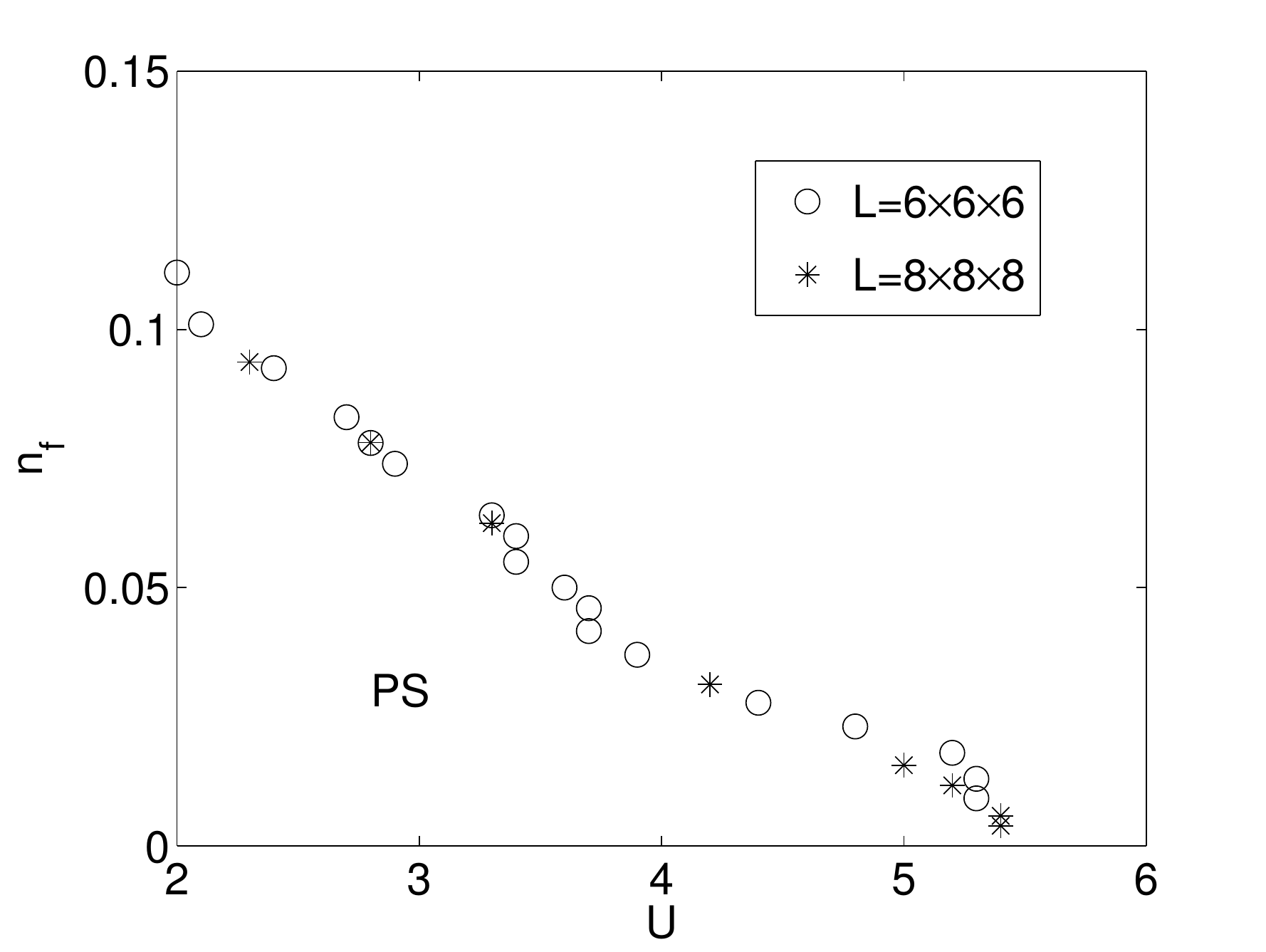}
\end{center}
\vspace*{-0.3cm}
\caption{ The region of phase separation (PS) of the three-dimensional Falicov-Kimball model calculated for two different clusters ($6\times6\times6$ and $8\times8\times8$)~\cite{Farky56}.}
\label{3d_fd}
\end{figure}

We have observed the same picture for intermediate values of Coulomb interactions \mbox{($U=2$)}. Larger values of $U$ only slightly modify the stability regions of some phases, but no new configuration types appear. In particular, the domain of phase segregation, as well as the domain of stripe formation are reduced while the domain of diagonal planes with chessboard structure increases. This trend is also observed for larger values of $U$. In the strong coupling limit ($U=8$) the phase segregated and striped phases are absent and the region of stability of the diagonal planes extends to relatively small values of $N_f \sim 20$. Below this value a homogeneous distribution of $f$ electrons is observed.

For the same reasons as discussed above for the two-dimensional case we have investigated the stability of  phase-separated (metallic) domain in the  three-dimensional case and found (see figure~\ref{3d_fd})
that the phase-separated region (and thus the metal-insulator transitions) extends up to $U\sim 5.5$,  which  is a much larger value than in the two-dimensional case~\cite{Farky56}.

\subsection{The effect of long-range hopping}
\label{Influence of long-range hopping}

\begin{figure*}[h!]
\begin{center}
%\hspace*{-0.3cm}
\mbox{\includegraphics[width=0.5\textwidth,angle=0]{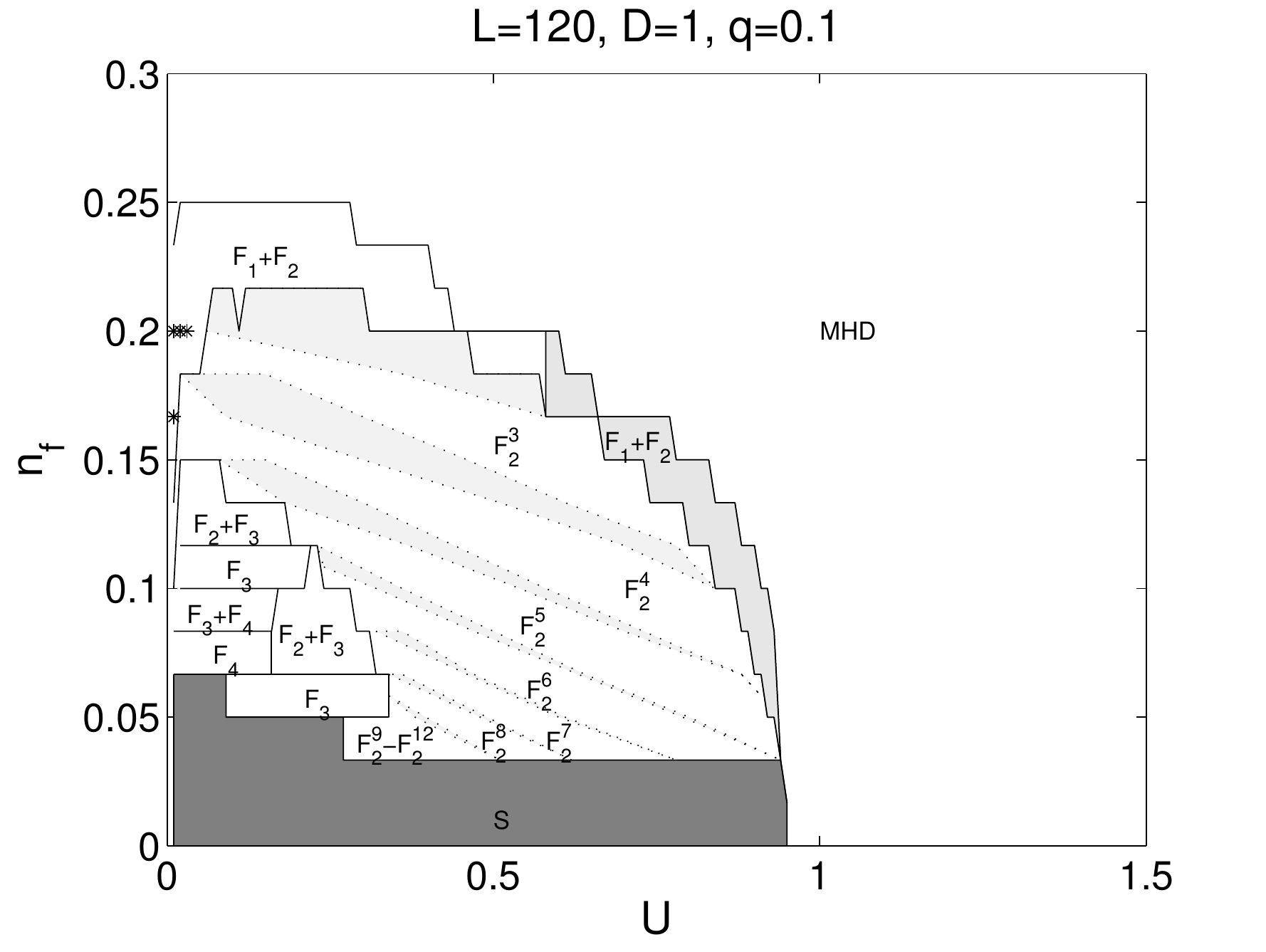}}\hspace*{-0.5cm}
\mbox{\includegraphics[width=0.5\textwidth,angle=0]{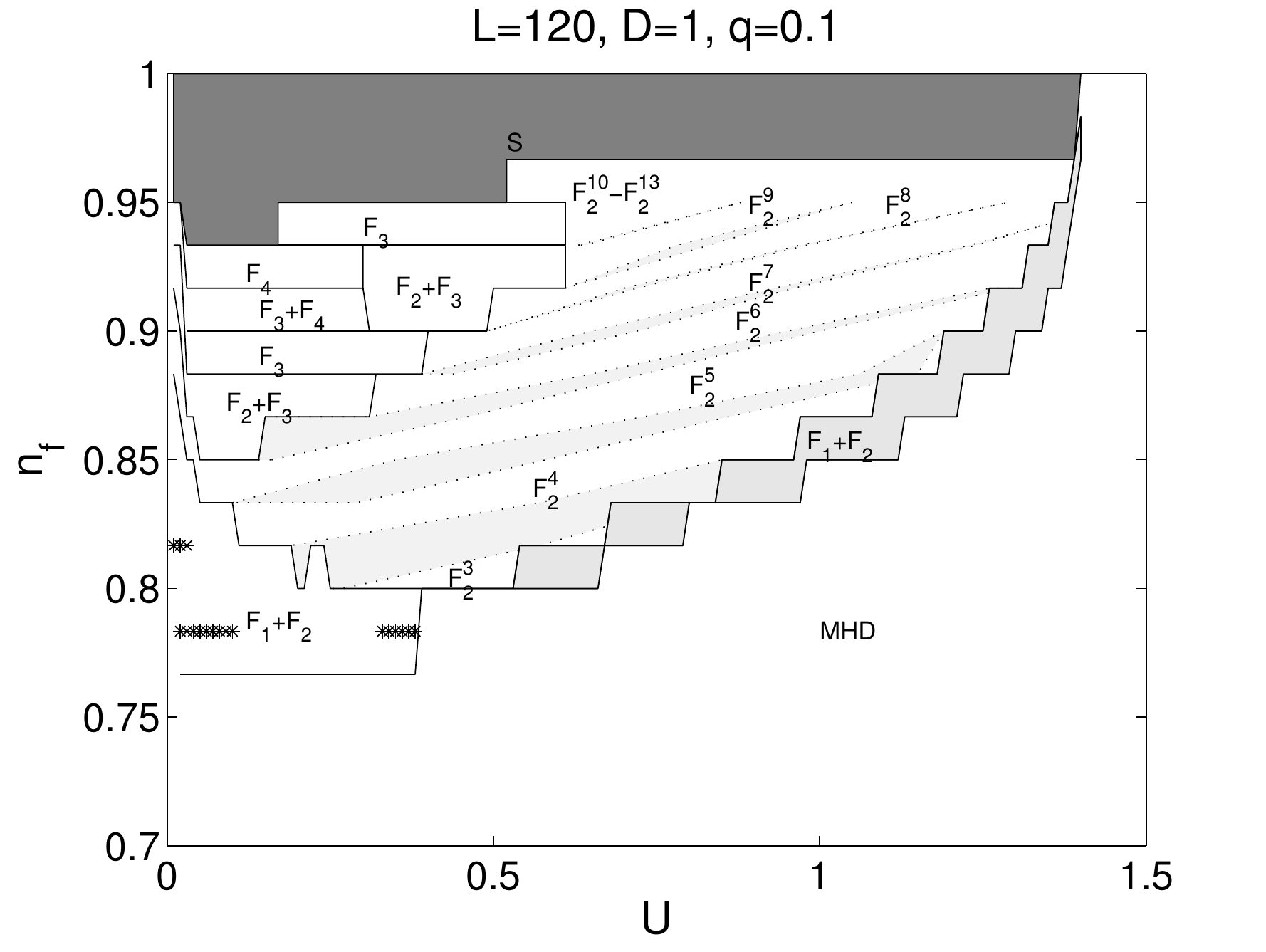}}
%\vspace*{-0.3cm}
\caption{  The ground-state phase diagram of the one-dimensional spinless Falicov-Kimball model with long-range hopping ($q=0.1$) for $n_f<0.3$ (the left hand panel) and $n_f>0.7$ (the right hand panel).  }
\label{q01}
\end{center}
\end{figure*}

Since the model including the electron hopping solely  to the nearest neighbours may seem at first glance a very crude approximation, in order to have a more realistic description of electron processes in rare-earth compounds, we have  generalized this model  by taking into account the transitions to other neighbours~\cite{Farky3,Farky4}. Basically, there were two possible ways of performing such a generalization. The first way was to assign independent transition amplitudes for the first ($t_1$), second ($t_2$), third ($t_3$), \ldots \ nearest neighbours, while the  second way was to describe the electron hopping by a one-parametric formula with power decaying transition amplitudes $t_ {ij}\sim q^{|i-j|}$, where $q\leqslant 1$. From the practical point of view, the second method is more suitable because it does not expand the model parameter space and has a clearer physical meaning, since the  atomic wave functions have also the power law decay with an increasing distance. For this reason, for a  description of electron hopping in the generalized model we have chosen the long-range hopping with power decreasing amplitudes. Explicit expressions of matrix elements  $t_{ij}$ for the case of periodic boundary conditions in the one-dimensional case have the form:
\begin{eqnarray}
t_{ij}=-tq^{L/2-||i-j|-L/2|-1}(1-\delta_{ij})\,,
\label{eq4.1.2.1}
\end{eqnarray}
from which it follows that the case $q\to 0$ approximates the nearest neighbours hopping ($t_{ij}=-t$), while the  $q=1$  case corresponds to an unconstrained hopping, when  the model is solvable exactly~\cite{Farky55}.
The results of our numerical simulations obtained within the AM are summarized in figure~\ref{q01} in the form of $n_f-U$ diagrams.
Figure~\ref{q01} presents the one-dimensional ground-state phase diagram of the model for $q=0.1$ obtained on the finite cluster of $L=120$ sites. Comparing this phase diagram with its $q=0$ counterpart one can find obvious similarities. The largest part of the $n_f-U$ phase diagram  corresponds to the most homogeneous domain, while the phase-separated domains appear only for $U$ small and $n_f\lesssim 0.25$ or $n_f\gtrsim 0.75$. Practically the same is  the internal structure of the phase-separated domains corresponding to ground-state phases of the model for $q=0$ and $q=0.1$. The term of long-range hopping only renormalizes the size of the phase-separated domains (for $n_f\lesssim 0.25$ the phase-separated domain is shifted to smaller $U$, while for $n_f\gtrsim 0.75$ it is shifted to higher $U$ and higher $n_f$), but practically no new phases are generated for $q=0.1$. However, for $q=0.3$ the situation changes dramatically.
\begin{figure*}[t!]
\begin{center}
\hspace*{-0.6cm}
\mbox{\includegraphics[width=0.6\textwidth,angle=0]{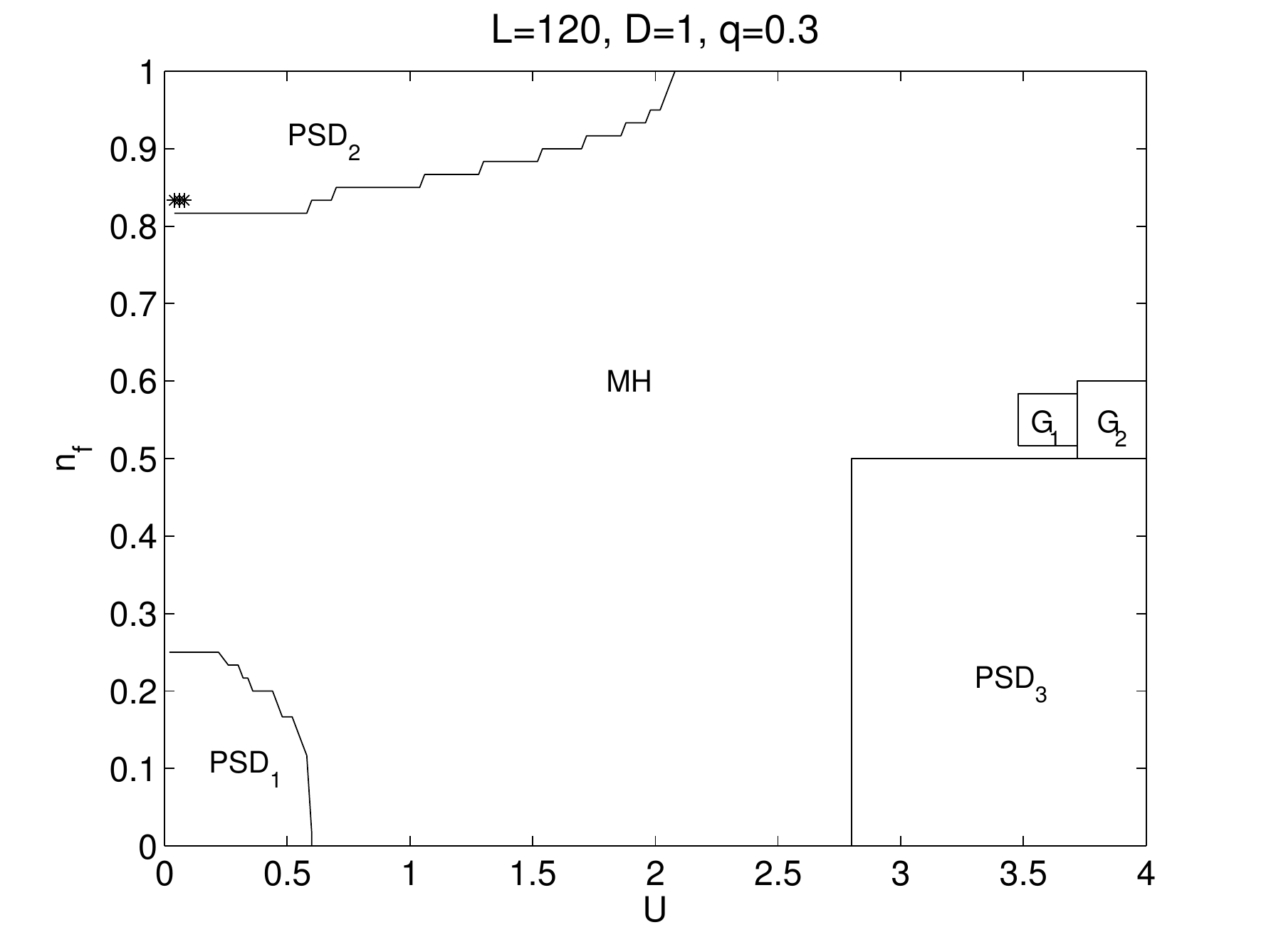}}
\vspace*{-0.2cm}
\caption{ The ground-state phase diagram of the one-dimensional spinless Falicov-Kimball model with long-range hopping calculated for $q=0.3$ and $L=120$.  }
\label{q03}
\end{center}
\end{figure*}
\begin{figure*}[b!]
\begin{center}
\hspace*{-1cm}
\centerline{\hspace{1cm}
\mbox{\includegraphics[width=0.48\textwidth,angle=0]{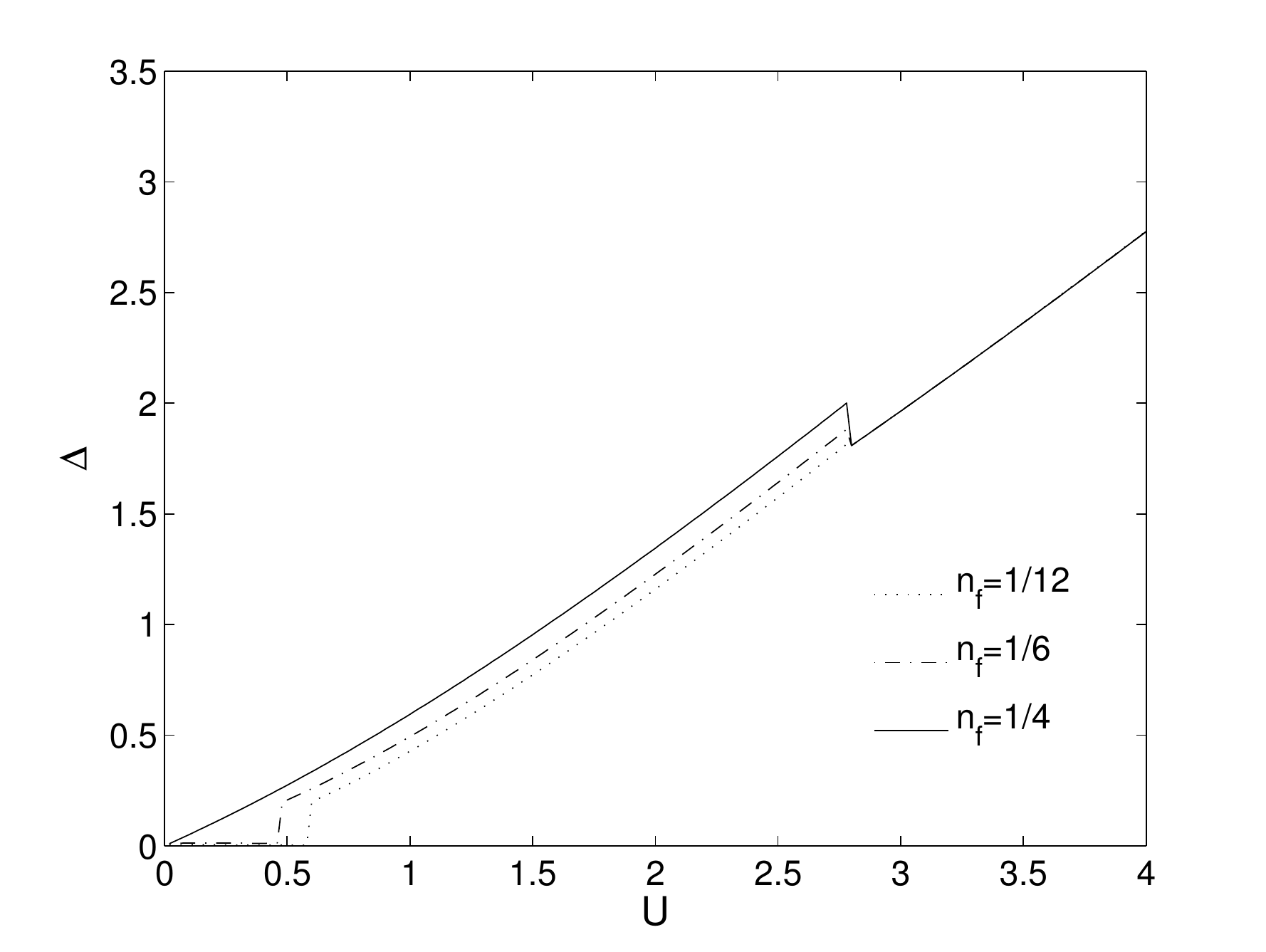}}
\hfill%\hspace*{-0.3cm}
\mbox{\includegraphics[width=0.48\textwidth,angle=0]{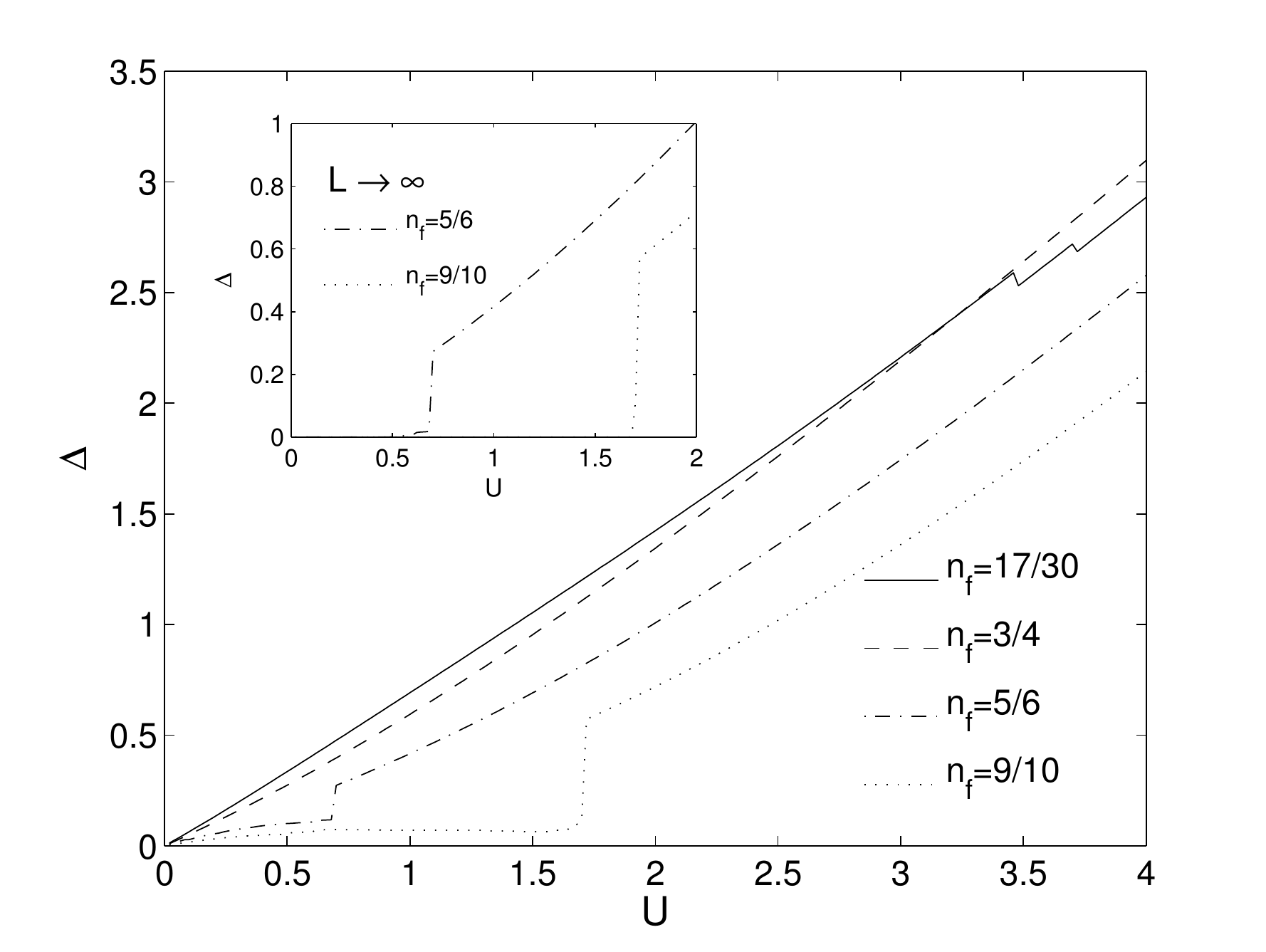}}}
\vspace*{-0.2cm}
\caption{ The $U$-dependence of the energy gap $\Delta$ for the one-dimensional spinless Falicov-Kimball model with long-range hopping ($q=0.3$) calculated for $L=120$ and $n_f<0.5$ (left  panel) and $n_f>0.5$ (right hand panel). The inset shows the extrapolated data ($L\to\infty$) for $n_f=5/6$ and $n_f=9/10$.}
\label{gap}
\end{center}
\end{figure*}
One can see (figure~\ref{q03}) that the largest region of stability still corresponds to the most homogeneous domain. However, in addition to this domain and two small phase-separated domains near $n_f=0$ (PSD$_1$) and $n_f=1$ (PSD$_2$), there appear two new large domains in the limit of intermediate and strong interactions. In the first region (PSD$_3$) located at $U\geqslant U_{\rm c}\approx 2.783$ and $n_f\leqslant 0.5$ the ground states are configurations that can be considered as mixtures of the empty configuration and the alternating configuration $w_a(N_f)=\{1010\dots 10\}$ with $N_f=1, 2, \dots, L/2$. The second domain located above $U_{\rm c}$ and $n_f>0.5$ consists of two subdomains $G_1$ and $G_2$, where the ground states are configurations of the following types:
\begin{eqnarray}
\mbox{$G_1$:\;\;} w&=&110[1010110]_k[10]_{\frac{L-7k-3}{2}}\,,\hspace*{0.55cm} \mbox{with k=3, 7, 11, 15}\;,
\\
\mbox{$G_2$:\;\;} w&=&110[10110]_k[10]_{\frac{L-5k-3}{2}}\,,\hspace*{0.9cm} \mbox{with k=7, 11, 15, 19}\;.
\label{eq:wt}
\end{eqnarray}
Analysing the energy gaps of ground-state configurations in all the above mentioned domains we have found (see figure~\ref{gap}) that only domains PSD$_1$ and PSD$_2$ are metallic, while all the remaining domains, including PSD$_3$ are insulating.
Thus, in accordance with the $q=0$ case, only the phase boundary between the small phase-separated domains PSD$_1$ and PSD$_2$ and the most homogeneous domain is a boundary of the metal-insulator transitions induced by Coulomb interaction (the $f$-electron concentration).

%\pagebreak

\subsection{The effect of nonlocal Coulomb interactions}
\label{Influence of nonlocal Coulomb interactions}

One of the shortcomings of the basic variant of the spinless Falicov-Kimball model is that it neglects all nonlocal interactions between electrons, which immediately evokes the question of possible instability of numerical solutions discussed above with respect to the case when some of these nonlocal interactions are turned on. To answer this question we have examined the effects of two nonlocal interactions, and namely, the correlated hopping
\begin{eqnarray}
H_{t'}=t'\sum_{\langle ij\rangle}(f^+_if_i + f^+_jf_j)d^+_id_j\,,
\label{eq4.1.2.2}
\end{eqnarray}
and the nearest-neighbour Coulomb interaction between $d$ a $f$ electrons
\begin{eqnarray}
H_{\rm non}=U_{\rm non} \sum_{\langle i,j\rangle}f^+_if_id^+_jd_j\,.
\end{eqnarray}

\subsubsection{The effect of correlated hopping}
\label{Influence of correlated hopping}

Let us first describe the effect of the first term. This term is in the literature usually referred to as a term of correlated hopping, since it can be interpreted as a single-particle Hamiltonian describing the hopping of $d$ electrons between the neighbouring $d$ orbitals with an amplitude that explicitly depends on the occupancy of $f$ orbitals. The selection of this term was motivated by earlier works~\cite{Michielsen, Wojtkiewicz, Gajek}, which showed its importance in describing the properties of strongly correlated systems, e.g., the superconducting state~\cite{Hirsch}. We have focused our attention on examining the effects of this term on charge ordering and valence and  metal-insulator transitions~\cite{Farky16}.
\begin{figure*}[h!]
\begin{center}
\centerline{
\mbox{\includegraphics[width=0.47\textwidth,angle=0]{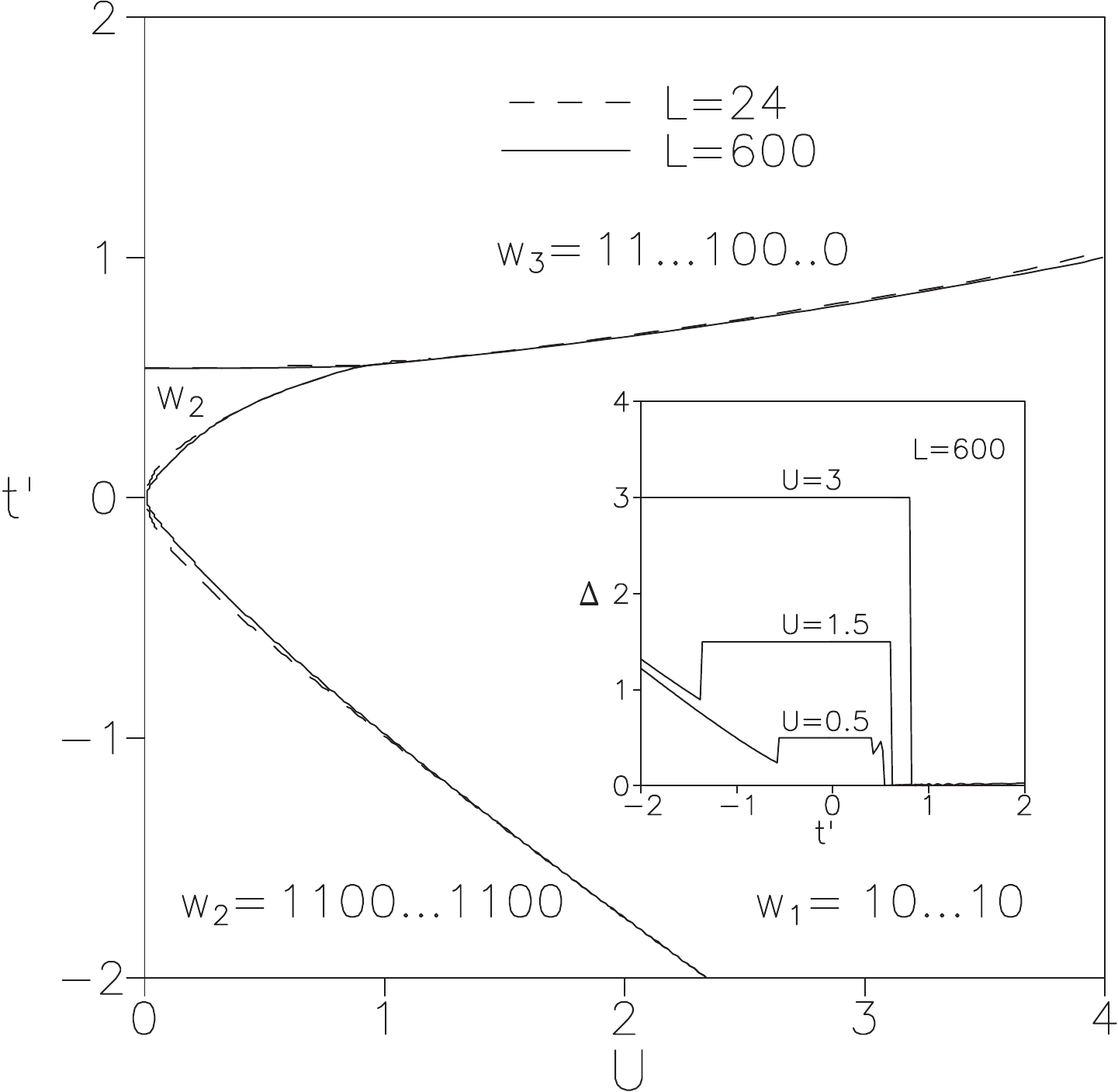}}%\hspace*{1cm}
\hfill
\mbox{\includegraphics[width=0.48\textwidth,angle=0]{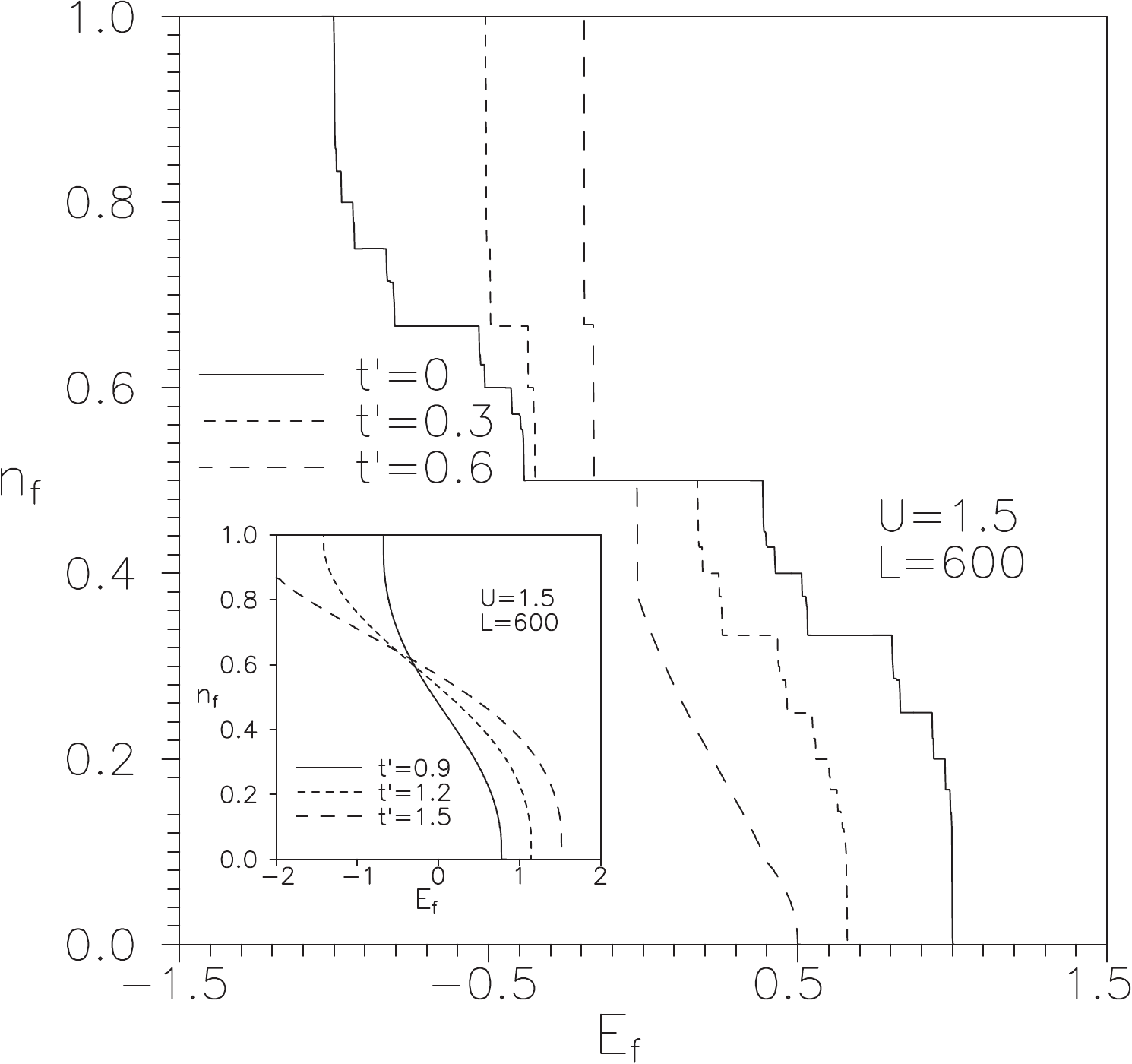}}}
\end{center}
\vspace*{-0.8cm}
\caption{ {\it Left}: $t'$-$U$ phase diagram of the one-dimensional  Falicov-Kimball model with correlated hopping at half-filling ($E_f=0$, $n_f=n_d=0.5$). Three different phases correspond to the alternating configuration $w_1=\{10 \dots 10\}$, the alternating configuration  $w_2=\{1100 \dots 1100\}$ with double period and the segregated configuration  $w_3=\{1 \dots 10\dots 0\}$. For  $L=24$ the phase diagram has been obtained over a full set of $f$-electron configurations, while for  $L=600$ only the restricted set of configurations   $w_1$, $w_2$ and $w_3$ has been used to determine the phase boundaries. The inset shows  $t'$-dependence of energy gaps corresponding to  the ground-state configurations for different  values of $U$.
{\it Right}: Dependence of  $f$-electron occupation number  $n_f$ on the $f$-level position $E_f$ for different values of correlated hopping  $t'$ at $U=1.5$~\cite{Farky16}.}
\label{pt1_1}
\end{figure*}
 A fundamental result of our numerical study is presented in figure~\ref{pt1_1},  where we have displayed the phase diagram of the generalized model in the $t'-U$ plane for the half-filled band case $n_f=n_d=0.5$.
One can see that already very small values of the correlated hopping term lead to the instability of alternating phase $w_1=\{1010 \dots 10\}$ (which is for $t'=0$ the ground state of the  model for all values of $U>0$). For $t'<0$, this phase transforms to the  alternating phase with a double period $w_2=\{1100 \dots 1100\}$ and for $t'>0$ on $w_2$ ($U<1$), respectively, the segregated configuration $w_3=\{11\dots 100\dots 0\}$ ($U>1$). Since the ground states corresponding to alternating phases  $w_1$ and $w_2$ are insulating, while the ground state corresponding to $w_3$ is metallic, the phase boundary between $w_1$ and $w_3$ as well as between $w_2$ and $w_3$ is the boundary of metal-insulator transitions induced by the correlated hopping term. Similar instabilities were  observed outside the symmetric case  $n_f=n_d=0.5$, which ultimately led to a completely different picture of valence and metal-insulator transitions for $t'\neq 0$. Nonzero values of $t'$ reduce the total width of valence transitions (figure~\ref{pt1_1}) as well as the  width of stairs and above some critical value, the phase transition becomes continuous, initially only in certain areas (e.g.,  $n_f< 0.5$ for $t'= 0.6$), and finally, in the whole area ($t'\geqslant 0.9$). The parts of valence transition with staircase structure correspond to the most homogeneous (insulating) phases, while the continuous parts correspond to the segregated (metallic) phases. The border points between these two phases are, therefore, critical points of pressure ($p\sim E_f$) induced metal-insulator transitions.

\begin{figure*}[!h]
\begin{center}
\mbox{\includegraphics[width=6.8cm,angle=0]{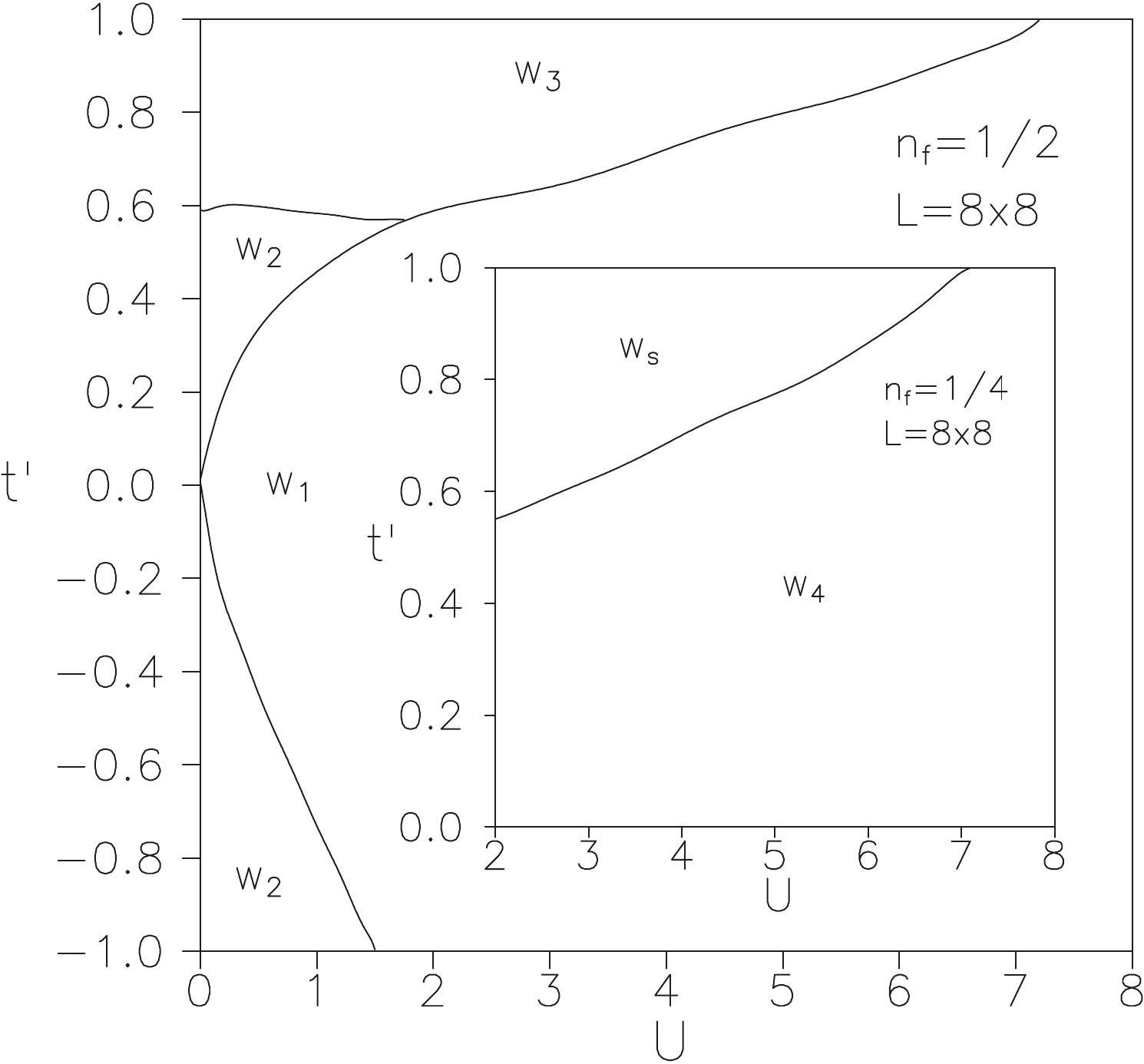}}\hspace*{1cm}
\mbox{\includegraphics[width=6cm,angle=0]{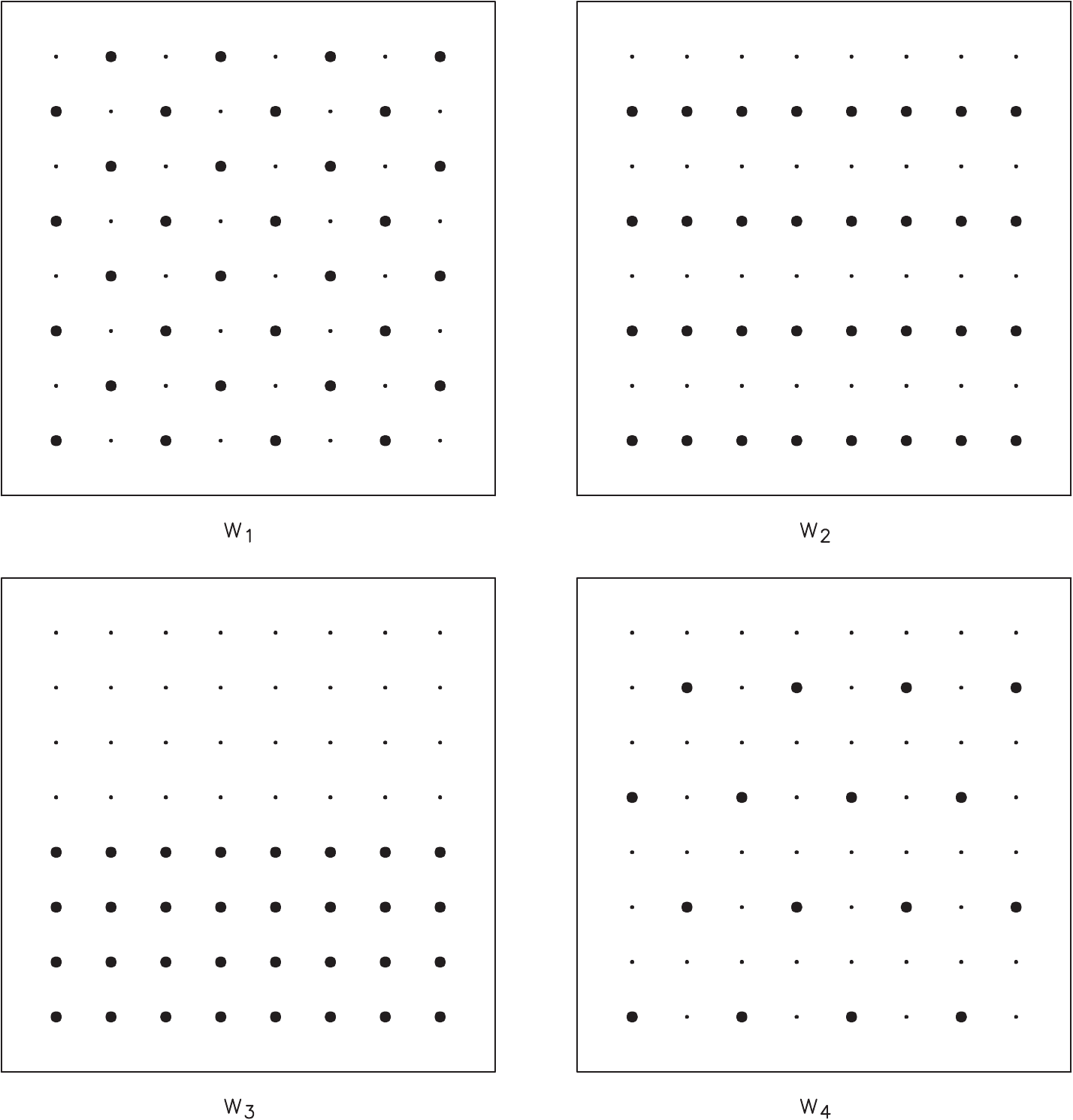}}
\end{center}
\vspace*{-0.5cm}
\caption{ {\it Left}: The $t'$-$U$ phase diagram of the two-dimensional  Falicov-Kimball model with correlated hopping at half-filling  ($E_f=0,n_f=n_d=0.5$). The inset shows the  $t'$-$U$ phase diagram for  $n_f=$1/4 and $U>2$. {\it Right}: The ground-state configurations for $n_f=$1/2 ($w_1$, $w_2$ and $w_3$) and $n_f=$1/4 ($w_4$)~\cite{Farky21}. }
\label{jp3_1}
\end{figure*}

We have performed the same study in the two-dimensional case, where we have used our AM. We have found~\cite{Farky21} that all characteristics discussed above, including the phase diagram for the half-filled band (figure~\ref{jp3_1})
as well as the picture of metal-insulator transitions remain unchanged in the two-dimensional case. A new and very interesting result is the observation of the charge striped ordering  of the type $w_2$ in the half-filled band case. Since the mechanism of formation of  inhomogeneous striped ordering in strongly correlated electron systems  is a very intensively discussed topic in recent years, this result is also valuable from the cognitive perspective, because it opens up a new way to the study of this certainly interesting phenomenon.

\begin{figure}[!h]
\begin{center}
\includegraphics[width=8.0cm]{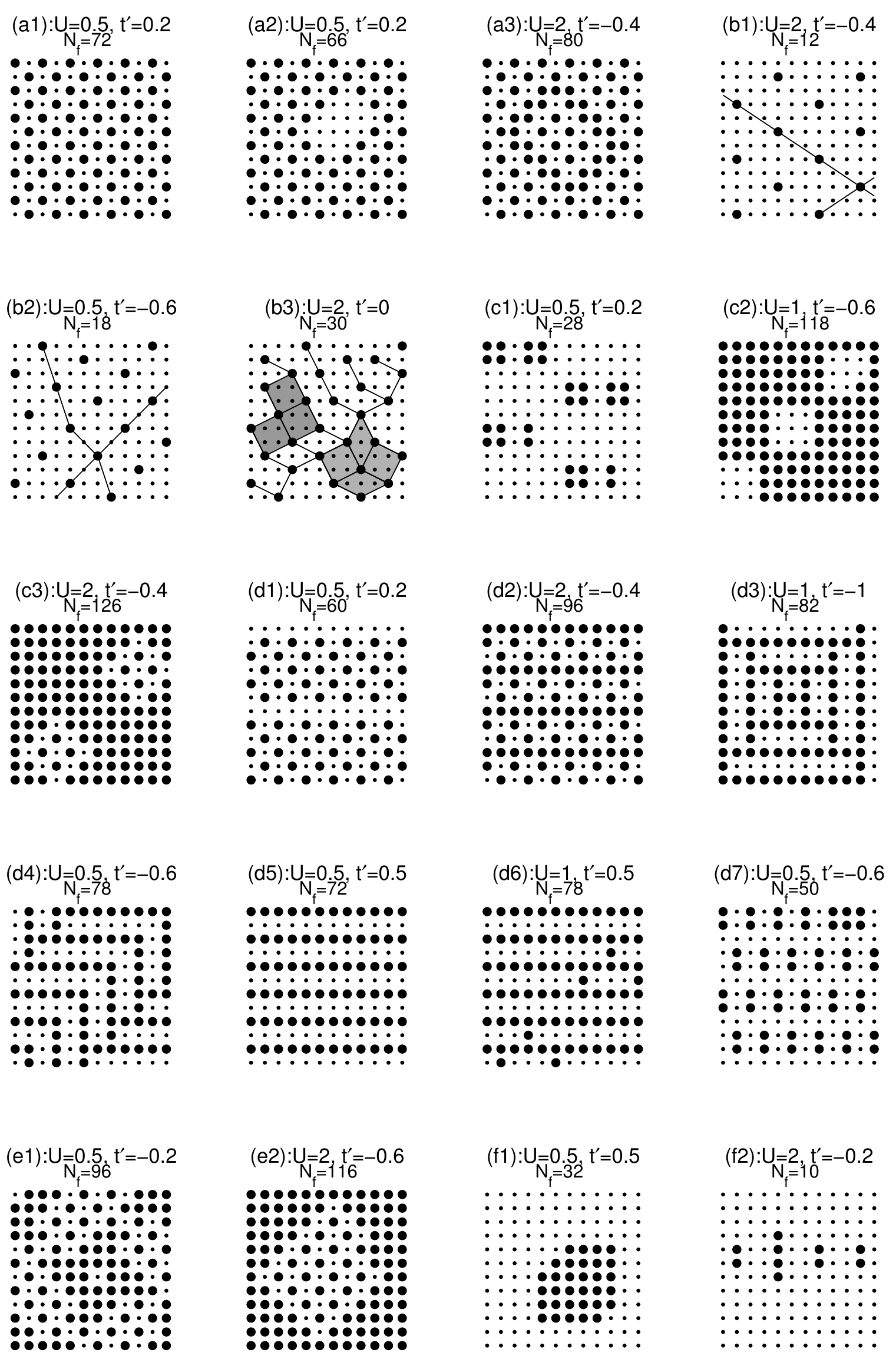}
\end{center}
%\vspace*{-0.5cm}
\caption{ Representative ground-state configurations that form the basic structure of the phase diagrams in the $n_f-t'$ plane: (a1) the chessboard phase, (a2, a3) the perturbed chessboard phases, (b1--b3) the diagonal stripes and perturbed diagonal stripes, (c1--c3) n-molecular phases, (d1--d7) the axial stripes and perturbed axial stripes, (e1, e2) the mixed phases and (f1, f2) the segregated and phase-separated configurations. The large dots correspond to occupied sites and the small dots correspond to vacant sites~\cite{Farky31}. }
\label{corr1}
\end{figure}

For this reason we have performed the exhaustive numerical studies of the model outside the half-filled band case. The primary goal of these studies was to identify all possible types of charge ordering induced by the term of correlated hopping.
To fulfill this goal we have performed an exhaustive numerical study of the two-dimensional spinless Falicov-Kimball model for a wide range of Coulomb interactions $U$ and correlated hopping $t'$~\cite{Farky31}. For each selected $U$ and $t'$  the ground-state configurations for $N_f=0,1,\ldots,L$ are calculated using our AM.  To minimize the finite-size effects the same procedure is repeated on several different clusters. Of course, such a procedure demands in practice a considerable amount of CPU time, which imposes severe restrictions on the size of clusters that can be studied using this method ($L=4\times 4, 6\times 6, 8\times 8,10 \times 10, 12\times 12$). Fortunately, we have found that the main features of the phase diagrams hold on all the examined lattices and thus can be used  satisfactorily to represent the behaviour of macroscopic systems. In particular, we have found that for each $L$ there is a finite number of basic types of ground-state configurations that form the basic structure of the phase diagram. This structure depends only very weakly on the size of clusters and covers practically the whole area of the phase diagram in the $n_f-t'$ plane. Let us start a discussion of these phase diagrams (for $U=0.5,1$ and 2) with a description of configuration types that form their basic structure (see figure~\ref{corr1}).
{\it (a)  The chessboard phase} ($n_f=1/2$). The $f$ electrons occupy the A sublattice of the bipartite lattice and the B sublattice is empty (a1). {\it The perturbed chessboard phases} ($n_f$ close 1/2), denote the chessboard structure decorated by two-dimensional patterns of occupied or empty sites (e.g., a2, a3). {\it(b) The diagonal stripes and perturbed diagonal stripes}. The mentioned phases could be divided into three principal categories that are represented by examples b1, b2 and b3. {\it (c) n-molecular phases} (c1--c3). {\it (d) The axial stripes and  perturbed axial stri\-pes}. This group consists of several subgroups. In the first subgroup, ground states are configurations that can be constructed by spaced lines of occupations (vacancies) aligned with the lattice axes, into the chessboard structure (d1, d2). The second subgroup includes perpendicular axial stripes (d3, d4). The third subgroup is formed by simple axial and perturbed axial stripes (d5, d6). The last subgroup consists of n-molecular axial stripes (d7). {\it (e) The mixed phases}. They can be  considered as mixtures of chessboard and fully occupied (empty) lattice (e1, e2). {\it (f)  The segregated and phase-separated phases} (f1, f2). The $f$ electrons clump together, or they are distributed only over  one half of the lattice, leaving another part of the lattice free of $f$ electrons. {\it (g) Unspecified phases} (various mixtures of previous configuration types).
The stability regions of all the above described phases are displayed in
\begin{figure}[h!]
\begin{center}
\vspace{3mm}
\includegraphics[width=9.5cm,angle=-90]{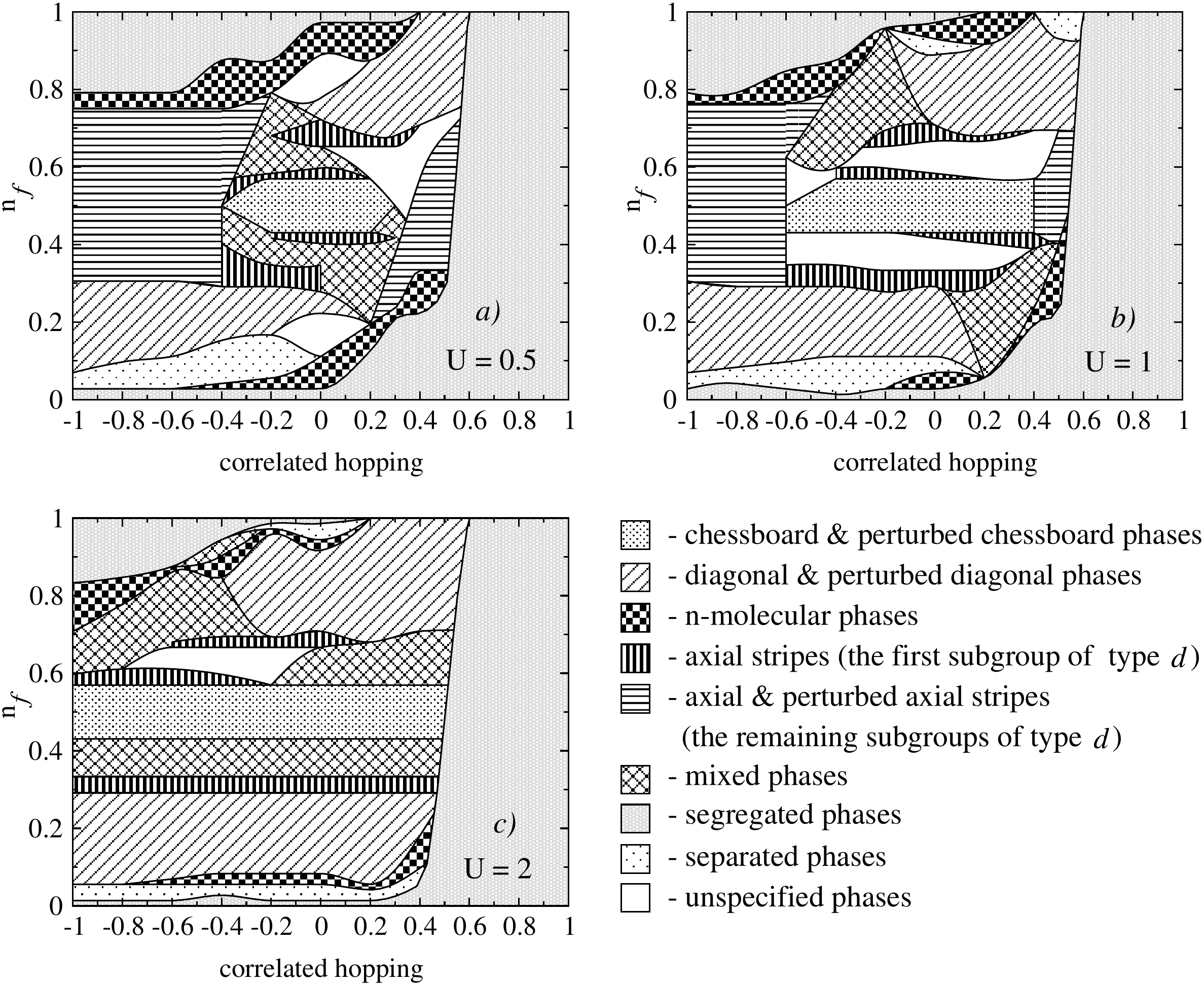}
\end{center}
%\vspace*{-0.5cm}
\caption{ $n_f-t'$ phase diagrams of the two-dimensional Falicov-Kimball model with correlated hopping for $U=0.5$ (a), $U=1$ (b) and $U=2$ (c)~\cite{Farky31}.}
\label{corr2}
\end{figure}
figure~\ref{corr2}, where the comprehensive phase diagrams of the Falicov-Kimball model with correlated hopping are presented for weak, intermediate and strong interactions. A direct comparison of these results reveals one general trend, and namely that the structure of phase diagrams is gradually simplified with increasing $U$ and becomes very simple in the strong-coupling limit. In this case the basic structure of the phase diagram (in the $n_f-t'$ plane) is formed by large segregated domains and several horizontal (band) domains corresponding to the separated phases, n-molecular phases, diagonal and perturbed diagonal stripes, axial stripes (the first subgroup discussed above), mixed phases and finally  perturbed chessboard configurations. Small deviations from the horizontal structure are observed for n-molecular phases for small ($n_f \leqslant 0.2$) and large ($n_f \geqslant 0.7$) $f$-electron concentrations. Comparing the phase diagram with conventional Falicov-Kimball model ($t'=0$) it is seen that the positive values of $t'$ do not essentially effect the ground states up to some critical value $t'_{\rm c}(n_f)$. However, at $t'_{\rm c}(n_f)$ the system exhibits a steep transition to the segregated phase that is stable for all $n_f$ and $t'>t'_{\rm c}$. A slightly different picture occurs for negative values of $t'$. In this case the correlated hopping term induces new regions of axial stripes (the type d1, d2). The strong effect of negative $t'$ is apparent for $f$-electron concentrations close to 1, where the diagonal configuration type (the type~${\it b}$) gradually disappears, while the segregated region is stabilized.

As was discussed above, the phase diagrams become more complicated when the Coulomb interaction decreases. The simple band structure observed in the strong-coupling limit persists only for $f$-electron concentrations close to half-filling and  it is suppressed gradually by axial-stripe configurations (the type d3--d7)  for both positive and negative values of $t'$. It should be noted that these axial stripe configurations have an arrangement  principally different from the axial stripes observed in the conventional Falicov-Kimball model (d1, d2). The appearance of new types of axial stripes is one of the most interesting effects of correlated hopping on the ground-state  properties of the two-dimensional Falicov-Kimball model. At the same time, this result positively answers the question whether the correlated hopping term can or cannot stabilize the stripe phases. Our results show that already relatively small values of $t'$ (positive as well as negative) stabilize this inhomogeneous charge ordering. Moreover, it was found (see figure~\ref{corr2}) that the capability of correlated hopping to generate stripe ordering increases with a decreasing Coulomb interaction between localized and itinerant electrons. This opens up a new route towards the understanding of the nature of stripe formations in strongly correlated electron systems.

\subsubsection{The effect of nearest-neighbour Coulomb interaction}
\label{Influence of nearest-neighbour  Coulomb interaction}

Let us now discuss the effects of another nonlocal interaction term, i.e., the nearest-neighbour Coulomb interaction between the localized $f$ and itinerant $d$ electrons, that is of the same order as the term of correlated hopping.
To study the effect of nonlocal Coulomb interaction $U_{\rm non}$ on ground-state properties of the one- and two-dimensional  Falicov-Kimball model, we have performed an exhaustive  numerical study of the model for weak ($U=0.5$), intermediate ($U=2$) and strong  ($U=8$) on-site Coulomb interactions and for a wide range of nonlocal Coulomb  interaction $U_{\rm non}$~\cite{Farky41}. To determine the ground states
of the model we used the EDM (up to $L=36$ lattice sites) in combination with AM (up to $L=120$ sites). We have started our study with the one dimensional case and $U$ large ($U=8$), which are relatively simple for a description.

\begin{figure}[!h]
\begin{center}
\includegraphics[width=5.250cm]{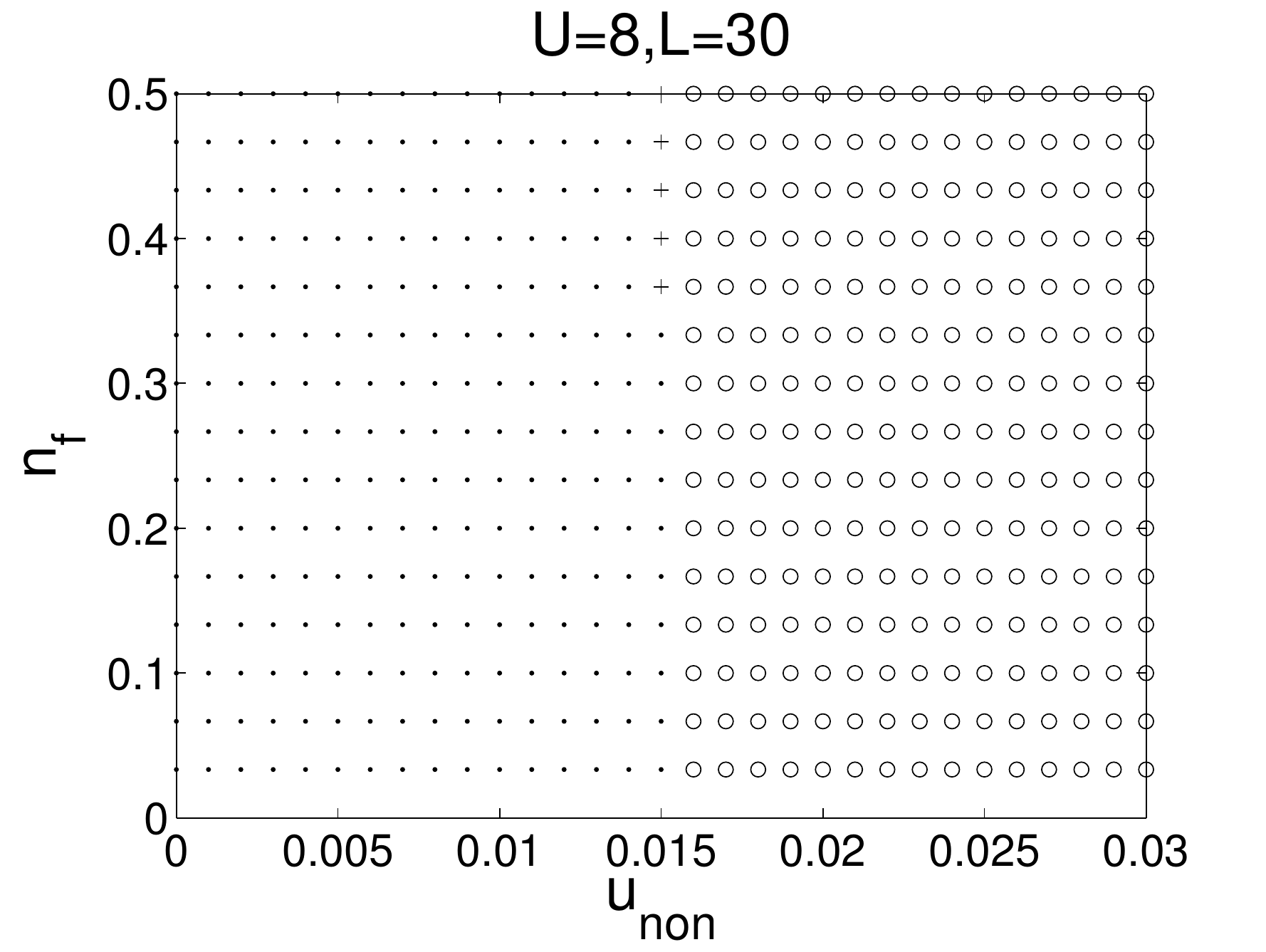}\hspace*{-0.3cm}
\includegraphics[width=5.250cm]{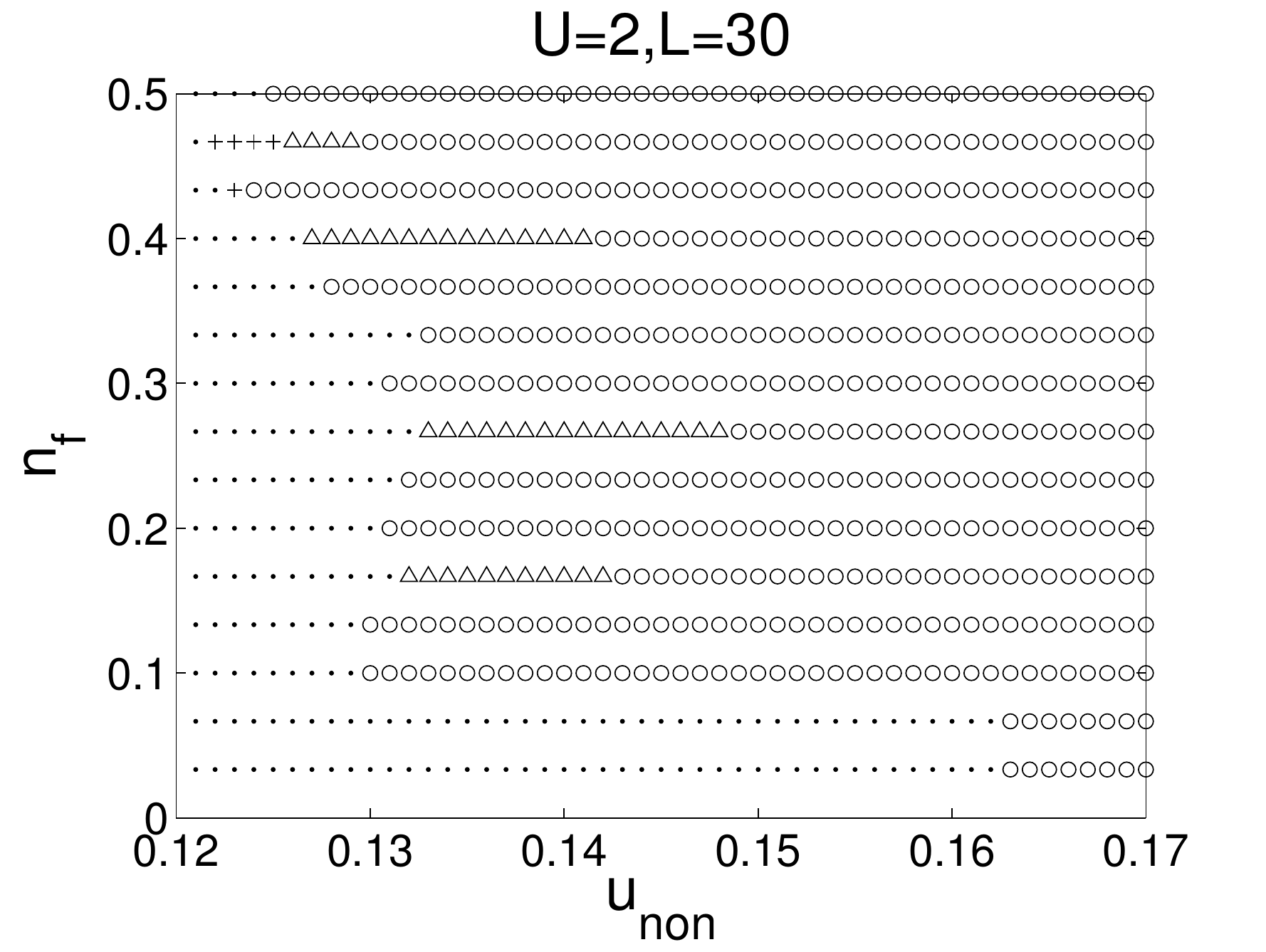}\hspace*{-0.3cm}
\includegraphics[width=5.250cm]{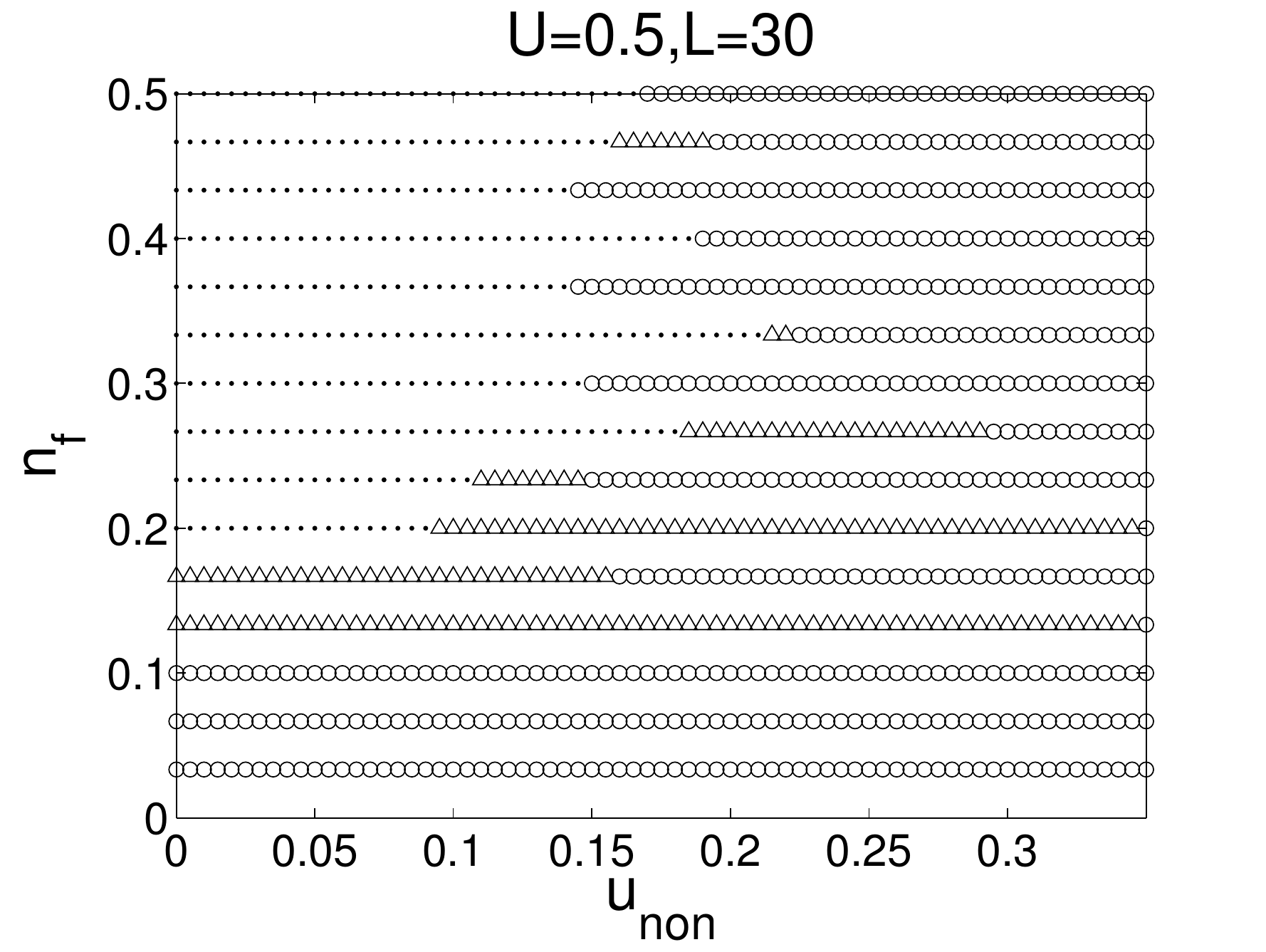}\\
\includegraphics[width=4.60cm]{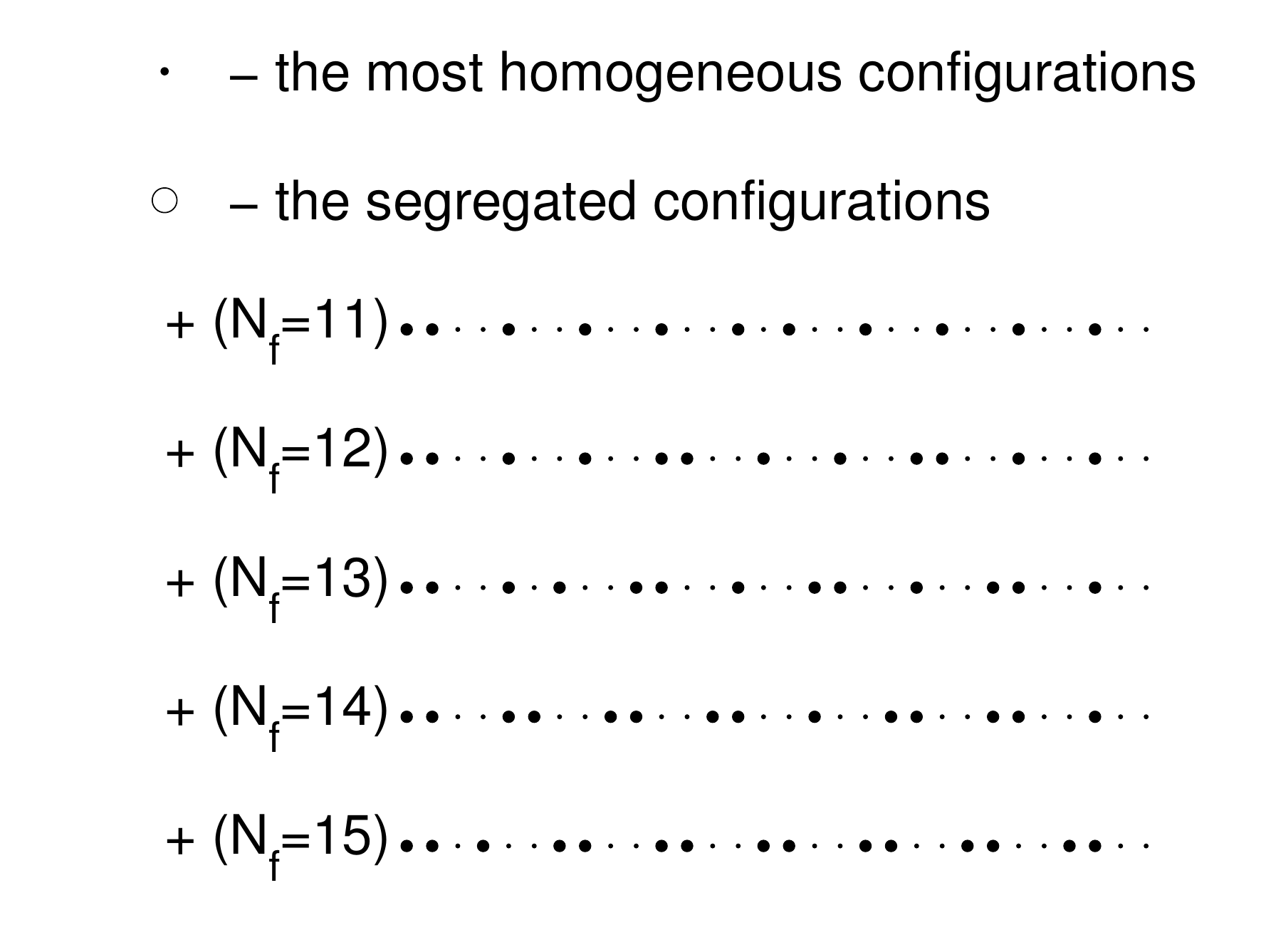}\hspace*{0.5cm}
\includegraphics[width=4.60cm]{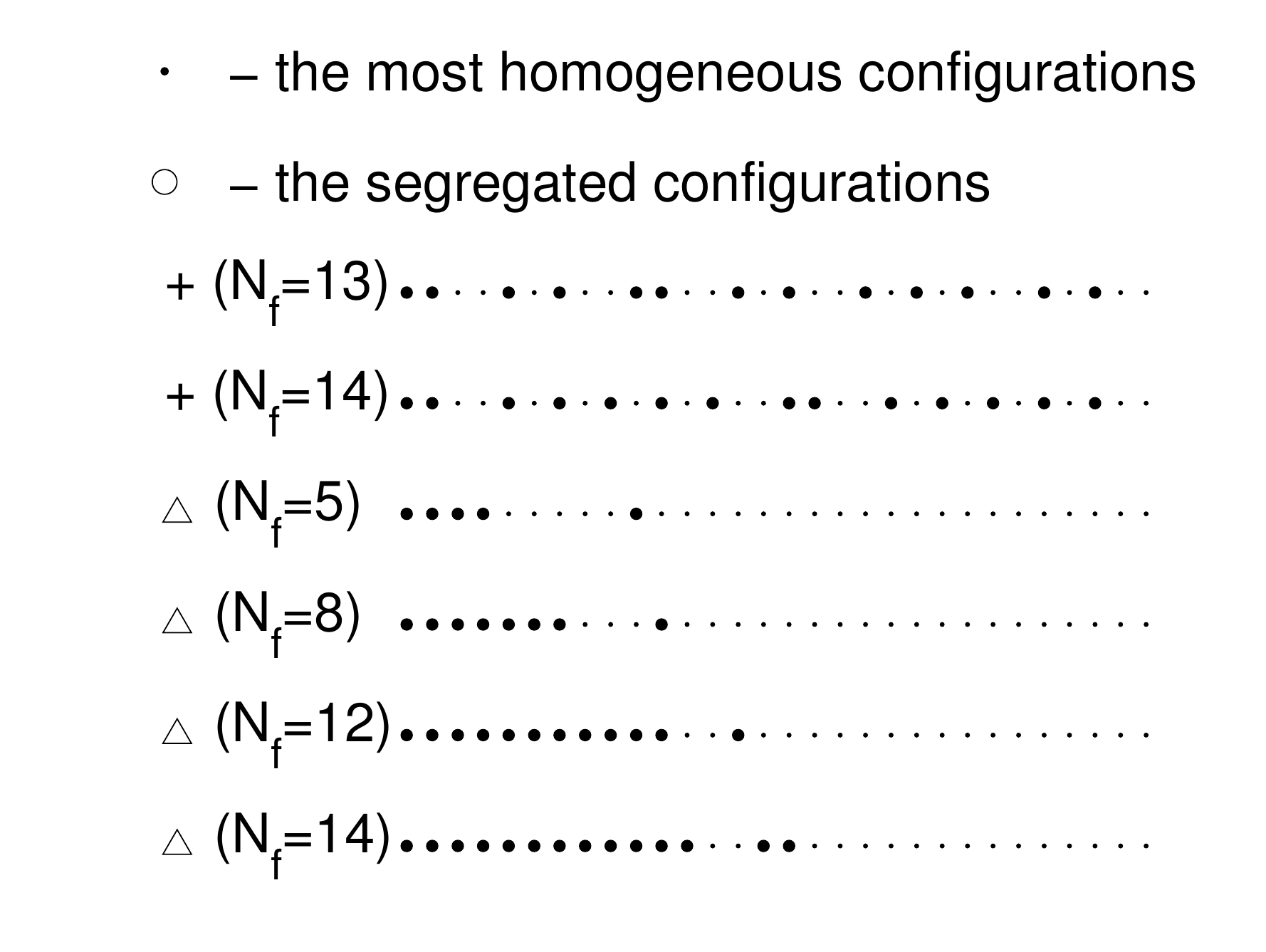}\hspace*{0.5cm}
\includegraphics[width=4.60cm]{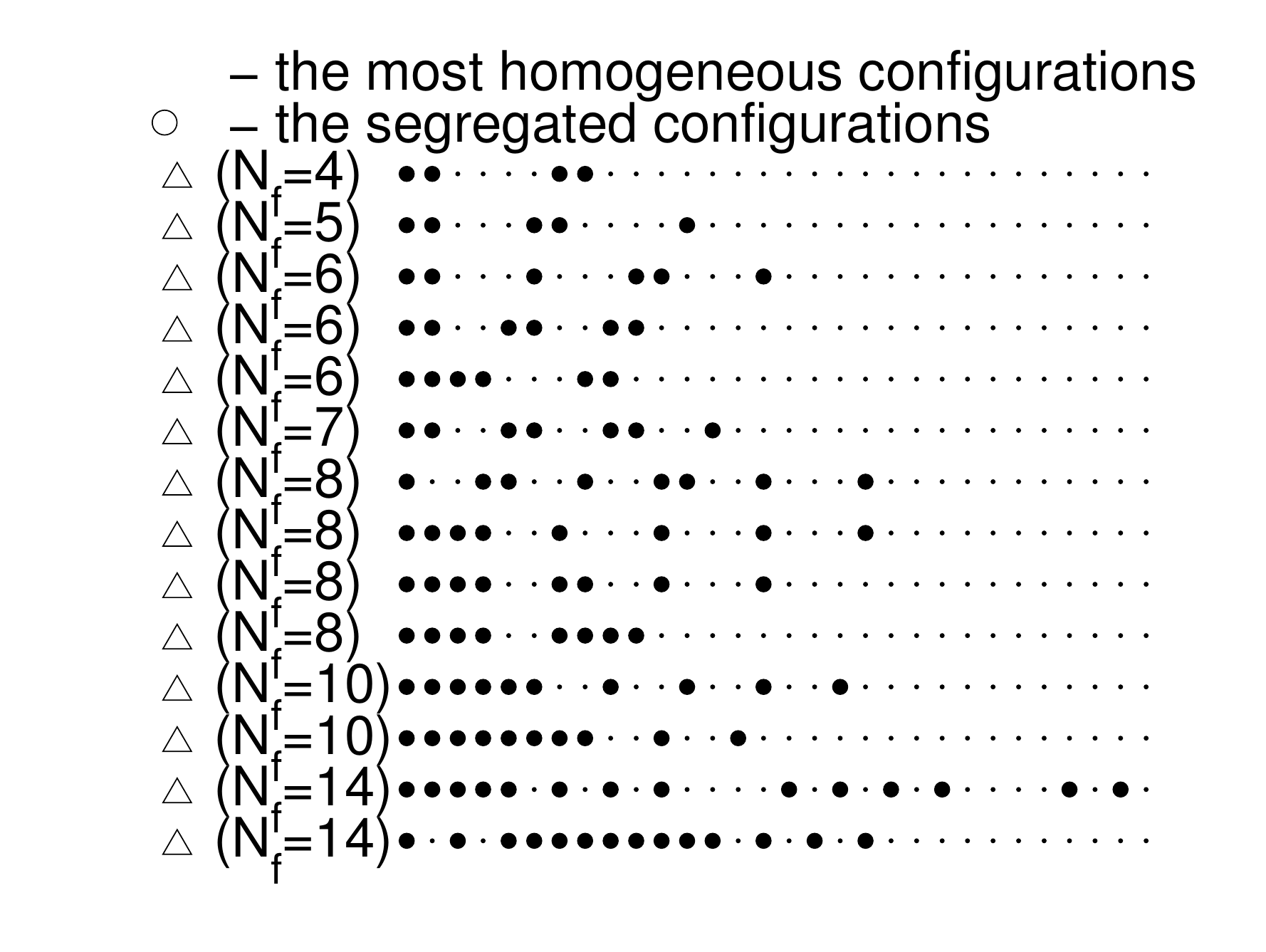}
\caption{ {\it Up:} $n_f-u_{\rm non}$ ground-state phase diagrams of the spinless Falicov-Kimball model extended by nonlocal Coulomb interaction for $L=30$ in the strong ($U=8$), intermediate ($U=2$) and weak ($U=0.5$) interaction limit, calculated for $n_f+n_d=1$. The one-dimensional exact-diagonalization results. {\it Down:} The complete lists of ground-state configurations corresponding to different interaction limits. The large (small) dots  correspond to  occupied (vacant) sites~\cite{Farky41}.}
\label{o_int22_1}
\end{center}
\end{figure}

Firstly, we have studied the ground-state phase diagram of the model in the $n_f-u_{\rm non}$ plane  ($u_{\rm non}=U_{\rm non}/U$ changes from  0 to 1 with step 0.001). To reveal the finite-size effects on the ground states of the model, we have performed numerical calculations  for three different clusters of $L=$ 12, 24 and 30 sites. The numerical results obtained for $L=30$  are displayed in figure~\ref{o_int22_1}\footnote{We have found that the ground-state phase  diagrams depend on $L$ only very weakly and thus already the results obtained for $L=30$ can be used
satisfactorily to represent the behaviour of macroscopic systems.}.
Our numerical results clearly demonstrate that already relatively small changes of $u_{\rm non}$ can produce large changes in the ground-state $f$-electron distributions. Indeed, we have found that already values of $U_{\rm non}$ around 60 times smaller than $U$ ($u_{\rm non}^{\rm c}=0.015$) are capable of fully destroying the most homogeneous distributions of $f$-electrons (that are the ground states of the conventional Falicov-Kimball model in the strong coupling limit for all $n_f$). It is interesting that only two configuration types are stabilized above the critical value of nonlocal interaction $u^{\rm c}_{\rm non}$. The first configuration type ($w_{\rm h}$)  is formed by the homogeneous distribution of $[1100]$ and $[10]$ clusters. A  complete set of these configurations that are stable only in a very narrow region of $u_{\rm non}$ (for $1/3<n_f<2/3$) are listed in figure~\ref{o_int22_1} (the first panel below). The second type of configurations determined  in the phase diagram of the Falicov-Kimball model with nonlocal Coulomb interaction in the strong coupling are the segregated configurations that are preferred as a ground state for all $n_f$ started at $u_{\rm non}=0.016$.

In figure~\ref{o_int22_1}  we present the ground-state phase diagrams of the extended Falicov-Kimball model for intermediate ($U=2$) and weak ($U=0.5$) interactions. One can see that the main feature of the phase diagram found for $U=8$, the nonlocal Coulomb  interaction induced transition from the regular $f$-electron distributions to the phase-segregated distributions, also holds  for smaller values of $U$, although the phase boundaries of different ground-state configuration types are now not so obvious due to the finite-size effects. A detailed analysis of the phase diagram for intermediate couplings showed that besides the most homogeneous configurations  only three other configuration types enter the ground-state phase diagram, and namely: the segregated configurations ($\circ$), the weakly perturbed segregated configurations    ($\bigtriangleup$) and $w_{\rm h}(N_f)$ distributions (+). Contrary to $U=8$, $w_{\rm h}(N_f)$ configurations occur rarely at $n_f=1/2$, while weakly perturbed segregated  configurations ($\bigtriangleup$) are observed on relatively large $u_{\rm non}$ intervals. As  shown in figure~\ref{o_int22_1}, in the weak interaction limit, the configurations $w_{\rm h}$ (+) fully vanish and the set of ground-state  configurations is much richer. Apart from  the most homogeneous configurations ($\cdot$) and the segregated configurations ($\circ$) we have determined a number of phase-separated configurations (denoted by $\bigtriangleup$). A complete list of these configurations is given in figure~\ref{o_int22_1}.

We have performed the same calculations in two dimensions.  To minimize the finite-size effects, the numerical calculations have been performed on three different clusters of $4\times 4$, $6\times 6$ and $8\times 8$ sites. On the $4\times 4$ cluster, the calculations have been performed by the EDM, and on larger clusters our AM was used.

\begin{figure}[!t]
\begin{center}
\includegraphics[width=5.250cm]{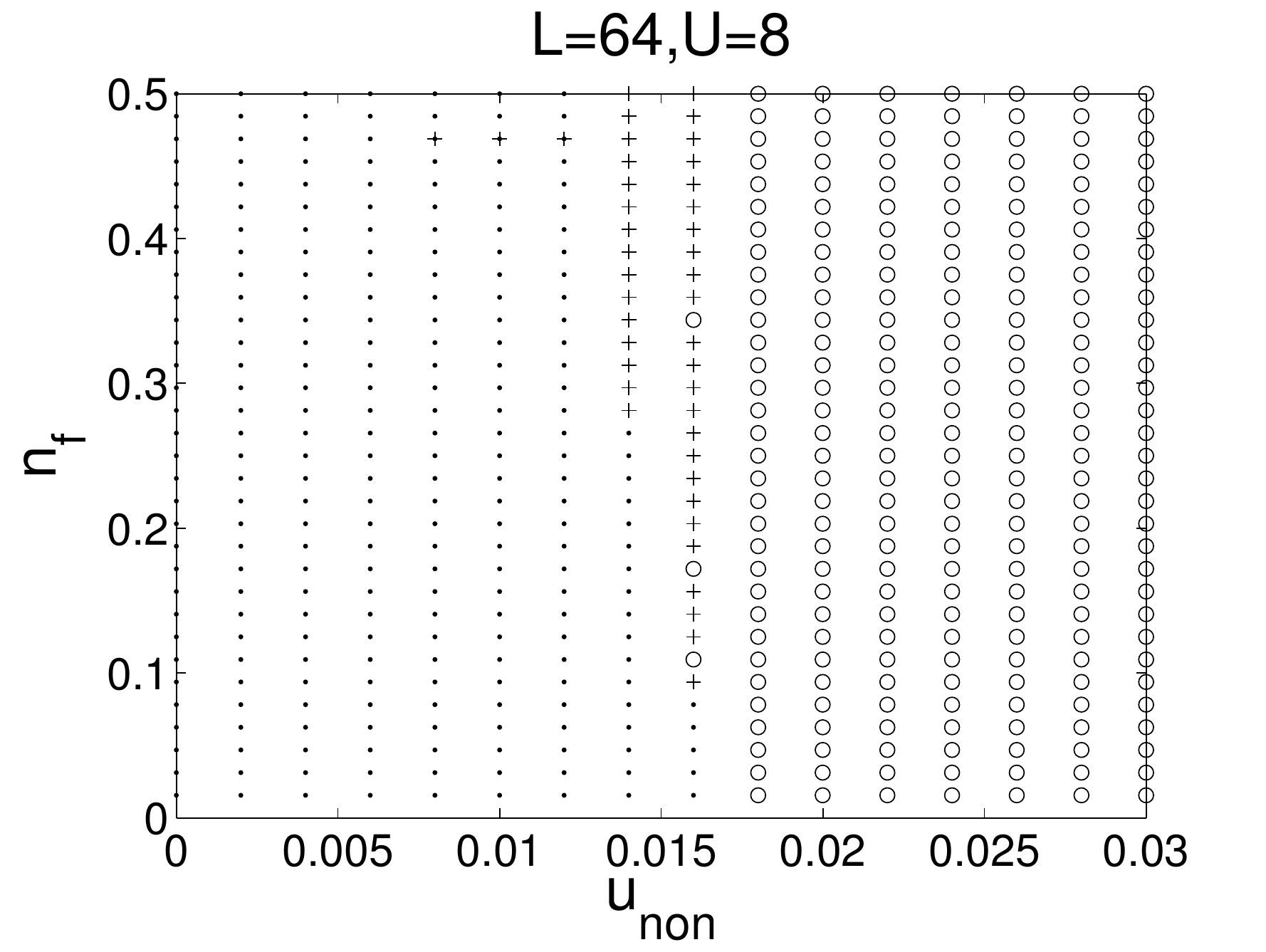}\hspace*{-0.3cm}
\includegraphics[width=5.250cm]{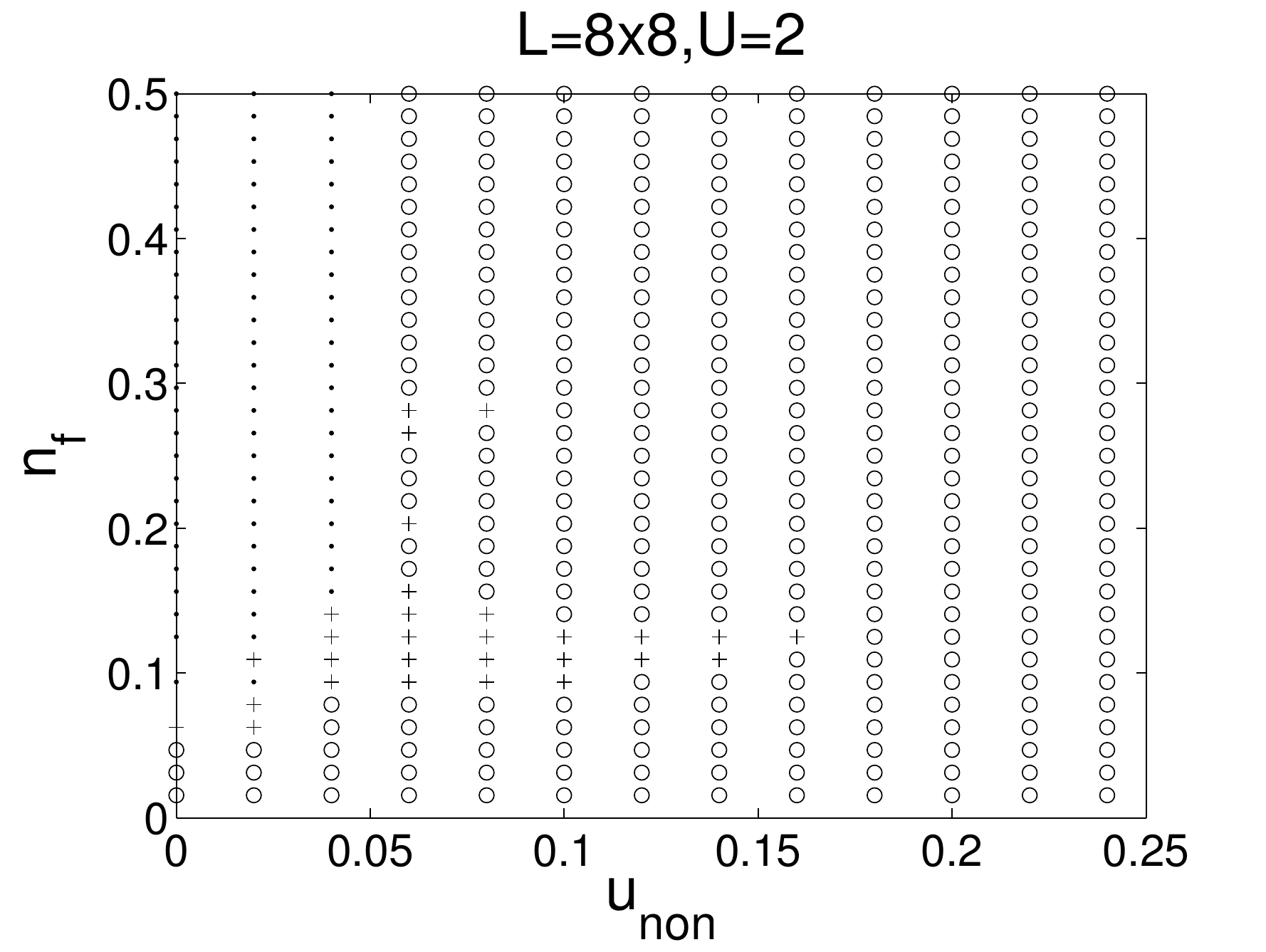}\hspace*{-0.3cm}
\includegraphics[width=5.250cm]{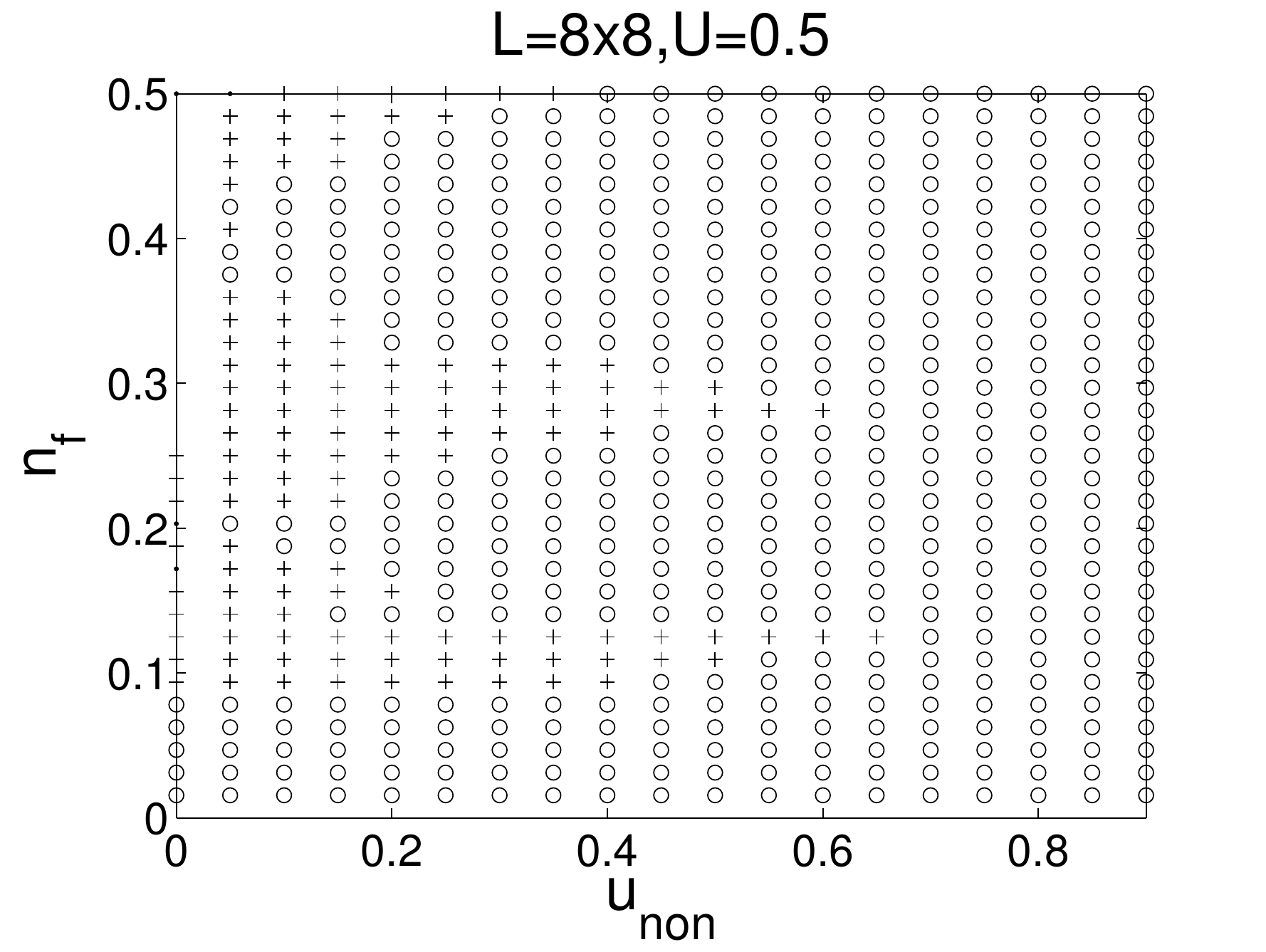}
\caption{ $n_f-u_{\rm non}$ ground-state phase diagrams of the spinless Falicov-Kimball model extended by nonlocal Coulomb interaction for  $L=64$ in the strong ($U=8$), intermediate ($U=2$) and weak ($U=0.5$) interaction limit, calculated for $n_f+n_d=1$. The two-dimensional approximative     results. Three different regions of stability corresponding to regular distributions, phase-segregated distributions and ``other phases'' are denoted as $\cdot$, $\circ$ and +~\cite{Farky41}.}
\label{o_int22_2}
\end{center}
\end{figure}

\begin{figure}[b!]
\begin{center}
\includegraphics[width=13.50cm]{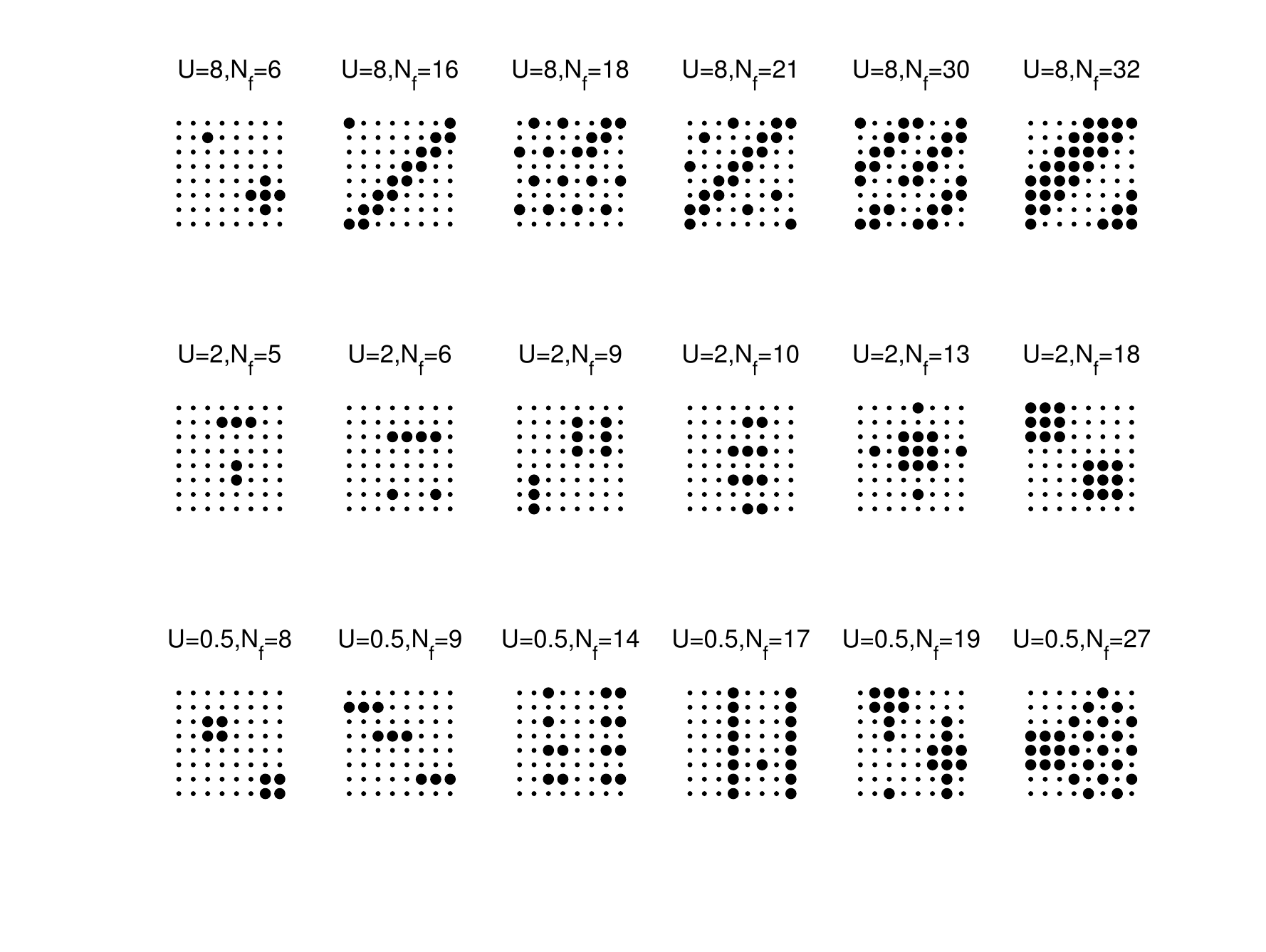}
\vspace*{-1cm}
\caption{ Typical examples of ground states of the spinless Falicov-Kimball model extended by nonlocal
Coulomb interaction from the area depicted by symbol (+) for strong ($U=8$), intermediate ($U=2$) and weak ($U=0.5$) interaction limit. The large (small)  dots correspond to occupied (vacant) sites~\cite{Farky41}.}
\label{o_int22_3}
\end{center}
\end{figure}

Similarly to the one-dimension we have started our two-dimensional studies with $U=8$. The ground-state phase diagram  (calculated for $8\times 8$) is shown in figure~\ref{o_int22_2} (the first panel).
Comparing this phase diagram with its one-dimensional counterpart one can find obvious similarities. In both cases the basic structure of the phase diagram only very weakly depends on $L$ and consists of only three   configuration types. Again one can see that the relatively small values of  nonlocal
interaction lead to changes of ground-state configurations from the regular distributions ($\cdot$) to the segregated  arrangements ($\circ$) straightforwardly, or through  some $n$-molecular distributions (+) (usually diagonal $n$-molecules but also the mixture of regular and diagonal 2-molecular distributions).  Typical examples of $n$-molecular distributions (for $L=8\times 8$) are displayed in figure~\ref{o_int22_3} (the first row).
While in the one-dimension the area of these configurations is stable only in isolated points of $u_{\rm non}$ ($u_{\rm non}=0.015$) and for $2/3>n_f>1/3$, in two dimensions these phases also persist for smaller $n_f$ and on the wider $u_{\rm non}$ interval. On the other hand, it is interesting to note that the critical value of  $u_{\rm non}^{\rm c}=0.018$ above which all ground-state configurations are only the segregated phases is almost identical to 1D case ($u_{\rm non}^{\rm c}=0.016$).   Similarities between phase diagrams of 1D and 2D cases, can be also observed  for intermediate and weak interactions. For both cases one can see a transition from ground states corresponding to  $u_{\rm non}$=0 to phase-segregated distributions, due to the nonlocal Coulomb interaction. For $U=2$, the obtained results show that the ground-state phase diagram consists of three different configuration types, denoted as regular distributions ($\cdot$), phase-segregated distributions ($\circ$) and other types (+) including many $n$-molecular distributions  (usually arranged to the ``ladders'' or to the blocks). Typical examples of these  distributions (for $L=8\times 8$) are displayed in figure~\ref{o_int22_3} (the second row). Similarly, in the weak interaction limit the nonlocal Coulomb interaction $u_{\rm non}$ prefers only a few types of ground-state configurations. We again observed regular distributions ($\cdot$), phase  segregated distributions ($\circ$) and some specific arrangements (+) discussed below. As was shown in figure~\ref{o_int22_2}  the critical value of $u_{\rm non}$, above which phase-segregated configurations  are ground-states (for all $n_f$), shifts to higher values of $u_{\rm non}$. On the other hand, the critical value of $u_{\rm non}$, where the ground states of conventional Falicov-Kimball model are changed into the other ones, are observed already for $u_{\rm non}\sim 0.05$ (for $L=8\times 8$). Between these two boundaries one can find  various $f$-electron distributions. In particular, there exist regular $n$-molecules, $n$-molecular ``ladders'', mixtures of chessboard structure and phase-segregated distributions, some periodic structures as well as the stripe formations (see figure~\ref{o_int22_3}, the third row). This  clearly shows that nonlocal interaction can stabilize various types of inhomogeneous charge ordering in strongly correlated electron systems.

\subsection{The effect of lattice geometry}
\label{Influence of lattice geometry}

\begin{figure*}[h!]
\begin{center}
\includegraphics[width=6.cm,angle=0]{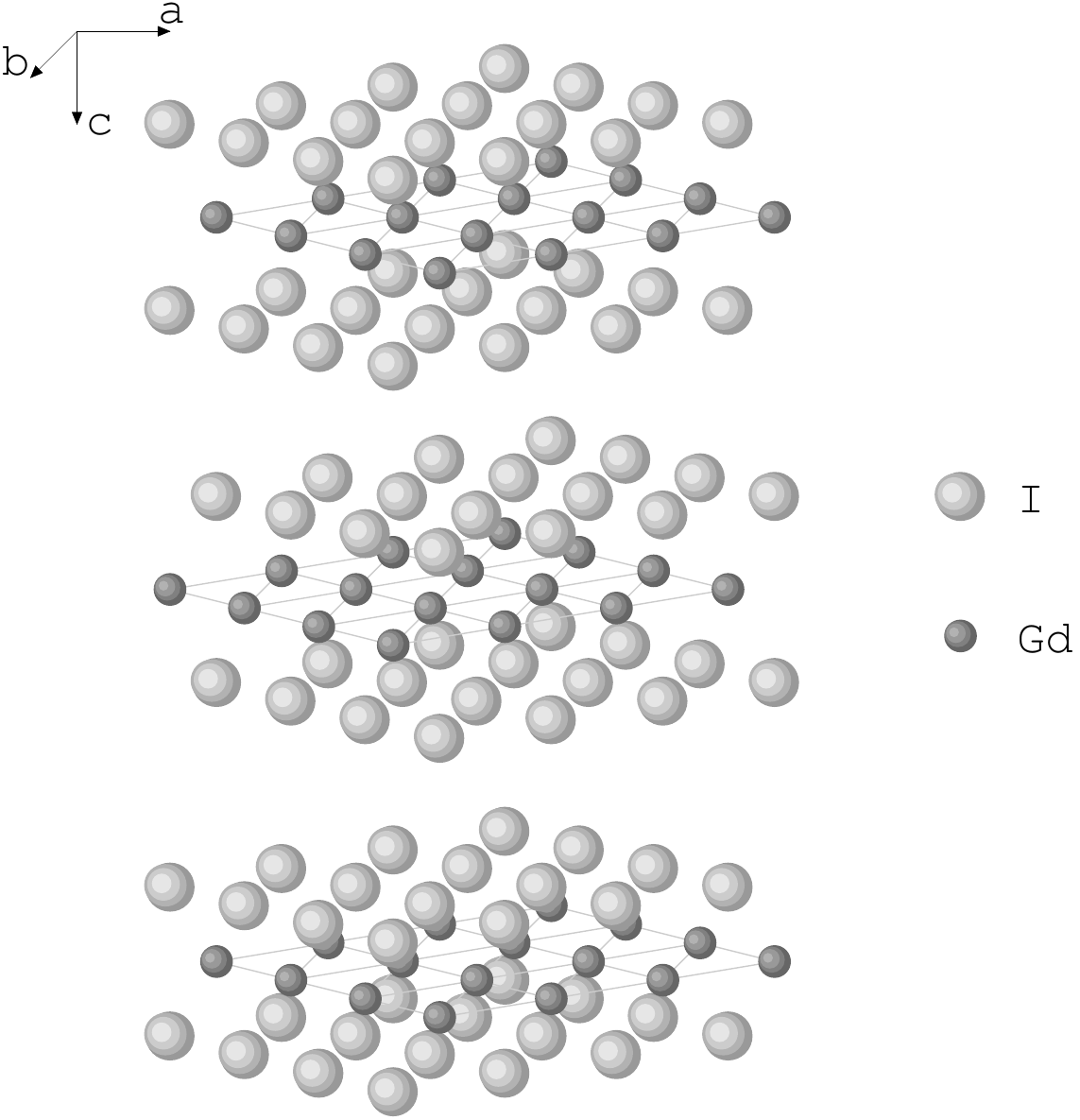}\hspace*{1.5cm}
\includegraphics[width=3cm,angle=0]{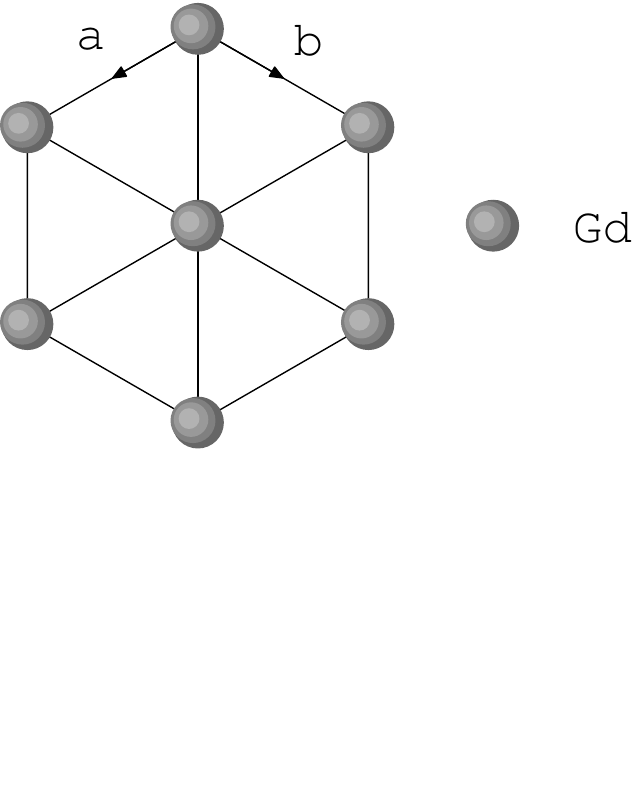}
\end{center}
\vspace*{-0.3cm}
\caption{ Crystallographic structure of GdI$_2$ ($P6/mmc$ group).}
\label{o_gdi}
\end{figure*}

There exists a large group of rare-earth and transition-metal compounds (e.g., GdI$_2$, Na$_x$CoO$_2$, etc.) in which atoms instead of a square or cubic lattice decorate a triangular lattice (see figure~\ref{o_gdi}) and they exhibit a number of anomalous physical characteristics~\cite{Ni1,Mn,Co1,Cu1,GdI1,GdI2}. From this point of view it is interesting to perform the same numerical study for a two-dimensional spinless Falicov-Kimball model on a triangular lattice.
To reveal the effects of the lattice geometry on the ground-state properties of the Falicov-Kimball model we have started with the half-filled band case  for which the nature of the ground state, its structural and energetic  properties are quite understandable on the square lattice~\cite{Farky39}.
In this case the localized $f$ electrons fill up one of two sublattices of the square lattice (the checkerboard structure) and the corresponding ground state is insulating for all $U>0$. Thus, for finite interaction    strength $U$ there is no correlation-induced phase or metal-insulator    transition at half-filling.

\begin{figure}[!t]
\begin{center}
\includegraphics[width=8.5cm]{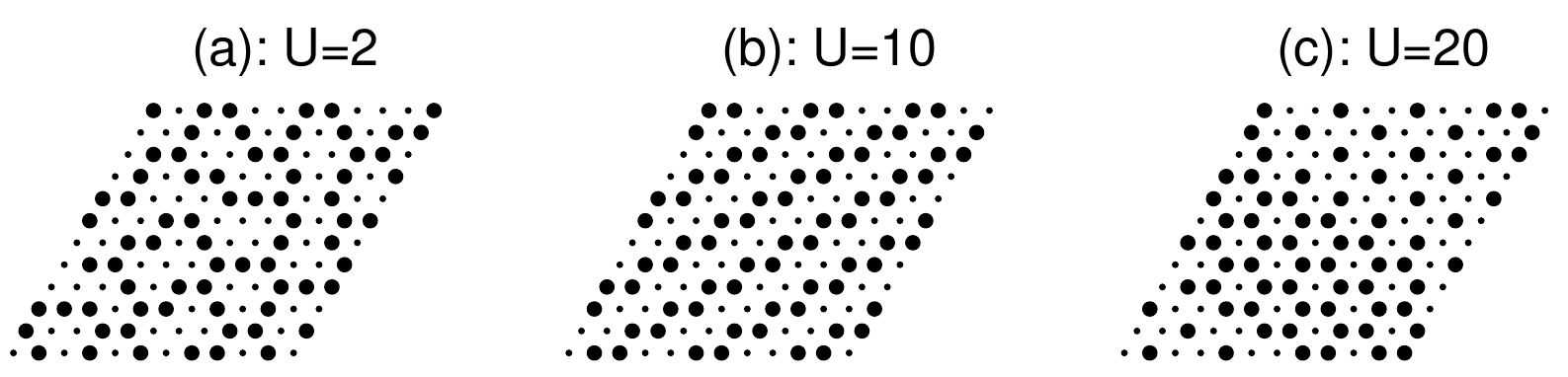}
\caption{ The ground-state configurations of the triangular Falicov-Kimball model in the half-filled band case~\cite{Farky39}.}
\label{pss244_1}
\end{center}
\end{figure}

The numerical calculations that we have performed in the half-filled band case on the triangular lattice revealed a completely different behaviour of the model for  nonzero $U$ in comparison with the square variant.  Indeed, with increasing $U$ we have observed a sequence of two correlation-induced phase transitions, indicating strong effects of the  lattice geometry on the ground-state characteristics of correlated electron systems. In the weak and intermediate coupling, the $f$ electrons  preferably form the closed lines that surround clusters of empty sites, without apparent long-range order (see figure~\ref{pss244_1}~(a)).
At $U \sim 4$, the system undergoes a correlation induced phase transition to the ordered phase (figure~\ref{pss244_1}~(b))     characterized  by a diagonal distribution of $f$-electron pairs (stripes). This phase persists up to relatively large values of $U$ $(U\sim 12)$ where   the system undergoes the second-phase transition into the phase-separated phase (figure~\ref{pss244_1}~(c)) formed by a mixture of $n_f=1/3$ and $n_f=2/3$ phase.  The dramatic effects of the lattice geometry on the ground-state characteristics of the Falicov-Kimball model at half-filling indicate that the picture of valence and metal-insulator transitions on the triangular lattice should be significantly changed in comparison with the square variant. To verify this conjecture we have performed an exhaustive numerical study of the model outside the half-filled case for a wide range of the Coulomb interaction $U=1,2,\ldots,9,10,12,16$ and 20 and several different cluster sizes ($L=6\times 6, 8\times 8, 10\times 10, 12\times 12$). For each selected $U$ the ground-state configurations for all $N_f=0,1,\ldots,L$ have been calculated using our AM.
\begin{figure}[b!]
\begin{center}
\includegraphics[width=8.5cm]{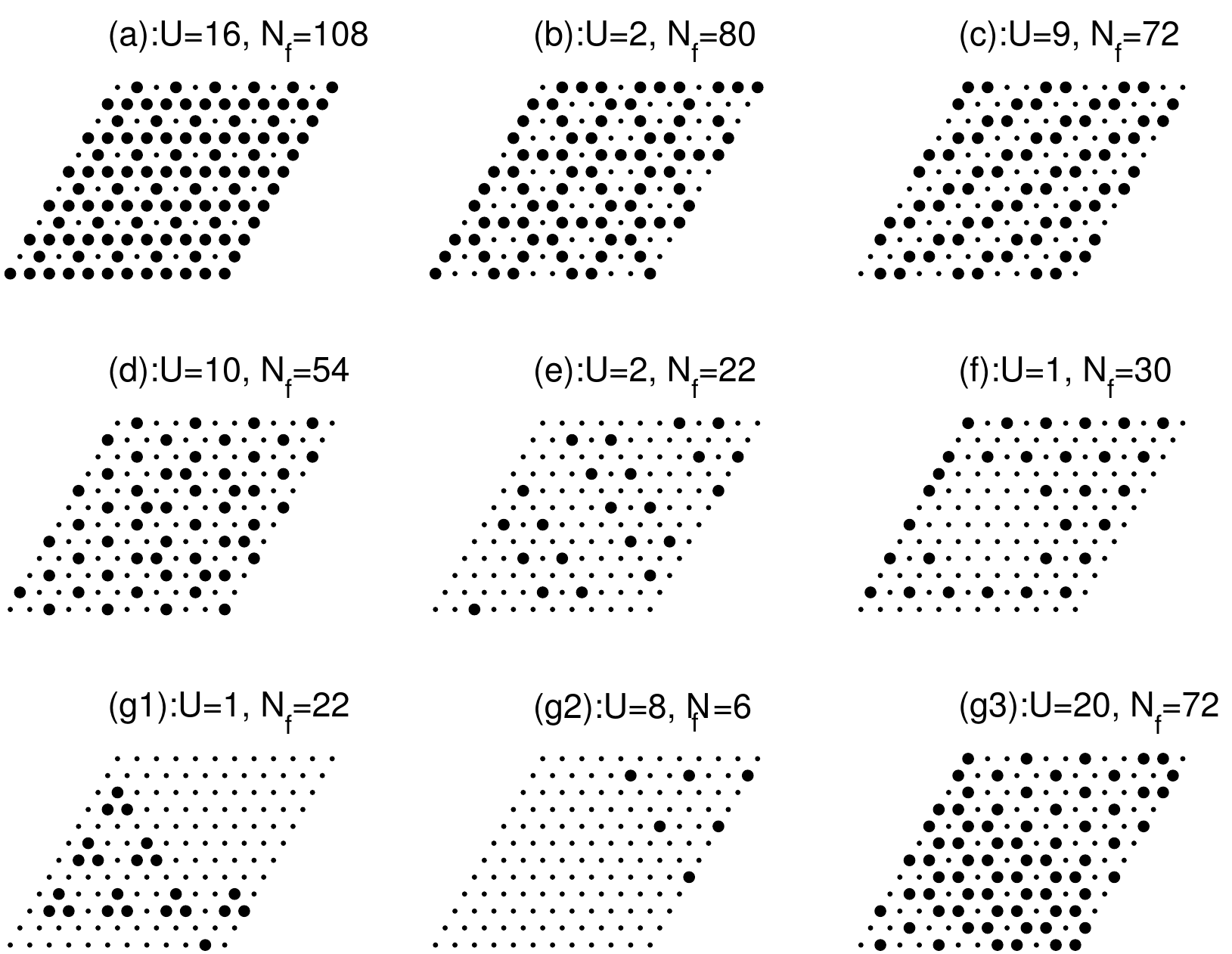}
\vspace{0.3cm}
\caption{ Representative ground-state configurations that form the basic structure of the phase diagram of the triangular Falicov-Kimball model in the $n_f-U$ plane. Large dots correspond to occupied sites and small dots correspond to vacant sites~\cite{Farky39}.}
\label{pss244_2}
\end{center}
\end{figure}
Analysing the ground-state configurations we have found that for each $L$ there is a finite number of basic types of distributions that     form the basic structure of a phase diagram in the $n_f-U$ plane.   Including the above described configurations  we have identified the  following basic configuration types (depicted in figure~\ref{pss244_2}): {\it (a) The regular or quasi-regular distributions.} This type of distributions is dominant in the phase diagram and can be found for all the investigated
$U$.  {\it (b) The bounded phases}, where small regions of empty sites are encircled by occupied sites. {\it (c) The diagonal striped configurations}  and {\it (d) the mixtures of 4-molecular distributions} and the most homogeneous phase for $n_f=1/3$.
{\it (e) The double axial or diagonal  stripes}, observed only for small $n_f$ similarly to {\it (f) mixtures of the most homogeneous  distribution} with $n_f=1/3$ and the empty lattice. {\it (g) The phase-separated configurations}. This group consists of several subgroups. In the first subgroup (g1) the $f$-electrons clump to n-molecules (usually 3-molecules) and they are distributed only over one half of the lattice, leaving another part of the lattice free of $f$-electrons. This configuration type is observed  only for $n_f\rightarrow 0$ and $U<5$. The second type (g2), observed for $n_f\rightarrow 0$/$n_f\rightarrow 1$, is  a typical phase-separated  distribution leaving at least one half of the lattice free of $f$-electrons while the single $f$ electrons are distributed over the remaining part of the lattice. The last subgroup (g3) is formed by mixtures of  $n_f=1/3$ and $n_f=2/3$ phases.

\begin{figure}[!t]
\begin{center}
\includegraphics[width=8.cm]{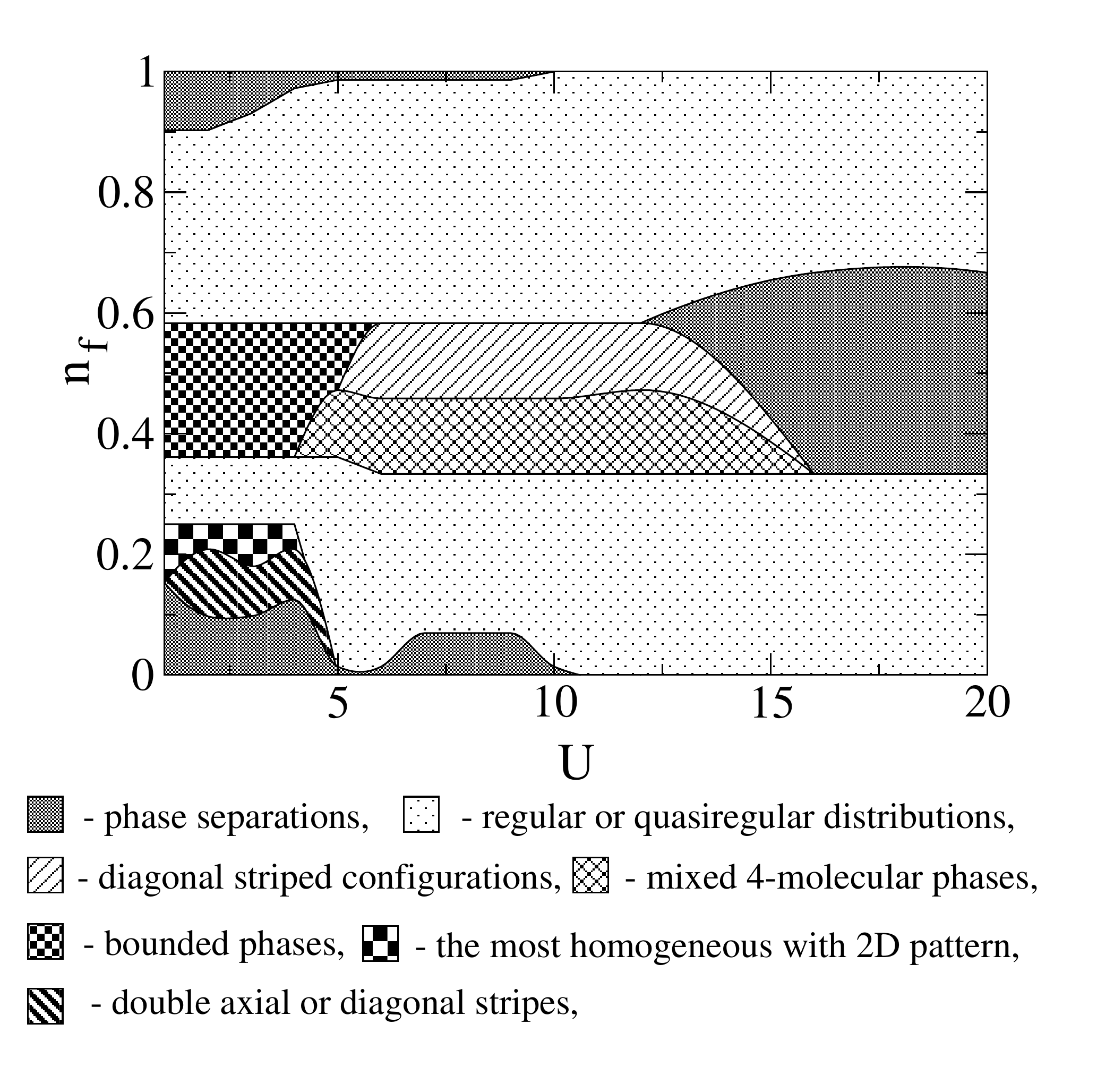}
\vspace*{-0.2cm}
\caption{ The $n_f-U$ phase diagram of the two-dimensional Falicov-Kimball model with triangular lattice~\cite{Farky39}.}
\label{pss244_3}
\end{center}
\end{figure}
\begin{figure}[!b]
\begin{center}
\includegraphics[width=7cm]{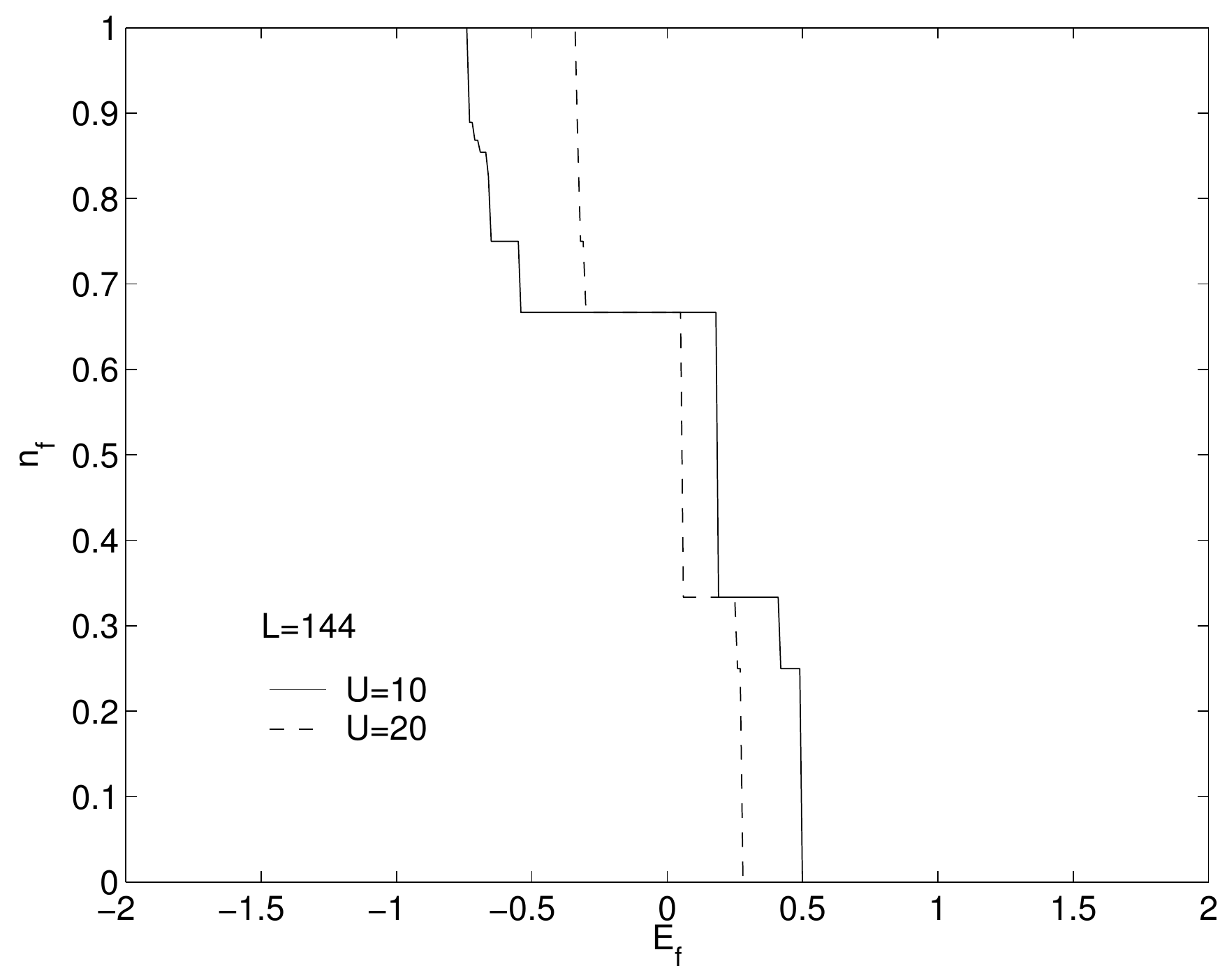}
\end{center}
\vspace{-0.3cm}
\caption{ The valence transition $n_f-E_f$ depicted for $L=144$ and two different values of Coulomb interaction $U=10$ and $U=20$~\cite{Farky39}.}
\label{pss244_4}
\end{figure}

The stability regions of all the above described phases are displayed in figure~\ref{pss244_3}.
It is seen that the phase diagram of the triangular Falicov-Kimball model still keeps  the band structure  similarly to its square equivalent~\cite{Farky31}, with dominant area corresponding to the regular and quasi-regular distributions.
The central band consists of the bounded phases, that are replaced with increasing $U$ by other new phases, namely the diagonal stripes, the mixtures of 4-molecular configurations, the phase-separated configurations (around $n_f \sim 1/2$), while the central band of the phase  diagram of the square Falicov-Kimball model consists of only the most homogeneous configurations  decorated by 2$D$ pattern~\cite{Farky31}.  Moreover, with decreasing $n_f$ and decreasing Coulomb interaction, the new types of $f$-electron distributions (missing in the square Falicov-Kimball model) have occurred (the type ${\it e}$ and ${\it f}$).
\begin{figure}[!t]
\begin{center}
\includegraphics[width=8cm]{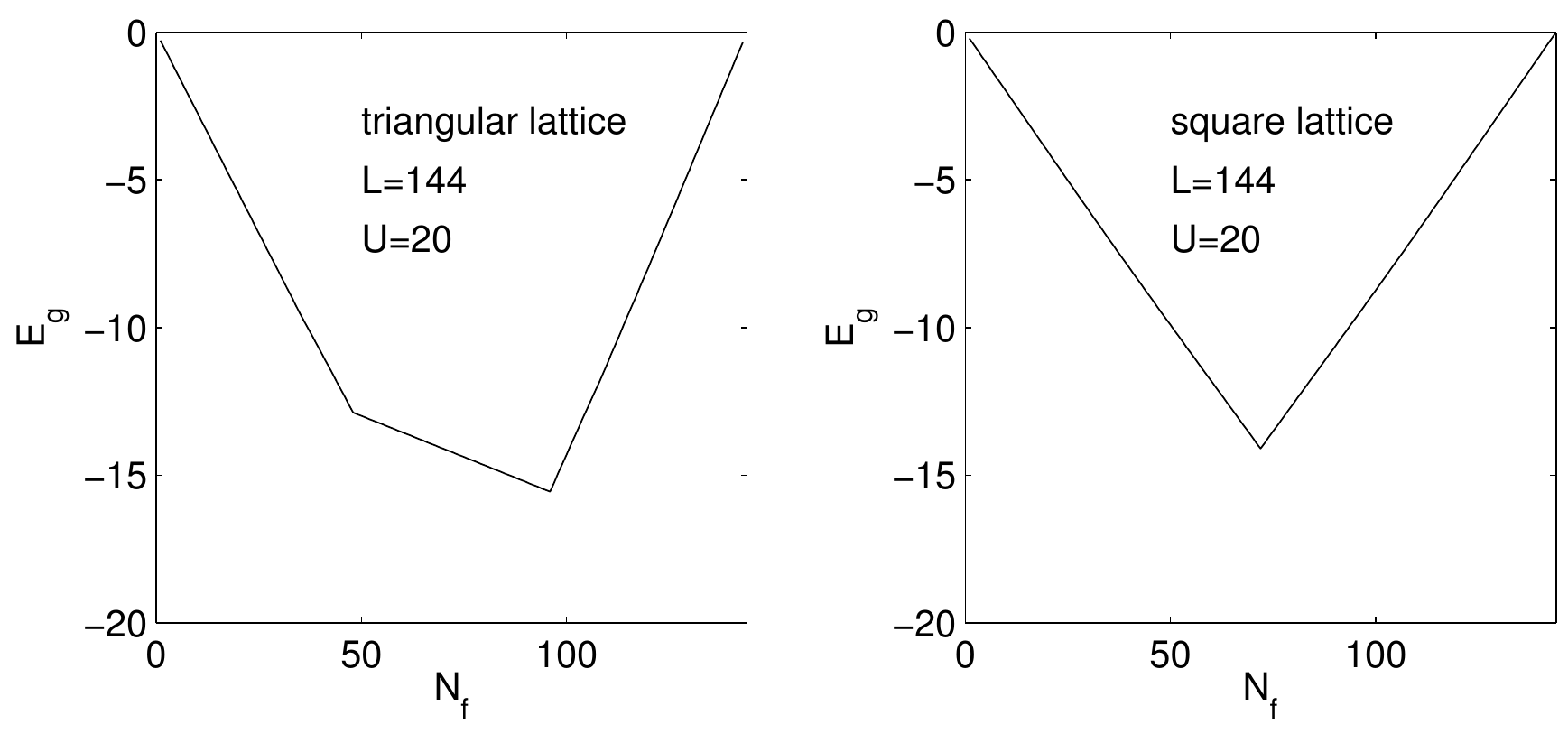}
\end{center}
\caption{ Dependences of the total energy $E_g$ on the number of localized electrons for triangular as well as square lattices with $L=144$ lattice sites in the strong coupling limit ($U=20$)~\cite{Farky39}.}
\label{pss244_5}
\end{figure}
\begin{figure}[!b]
\begin{center}
\includegraphics[width=7.5cm]{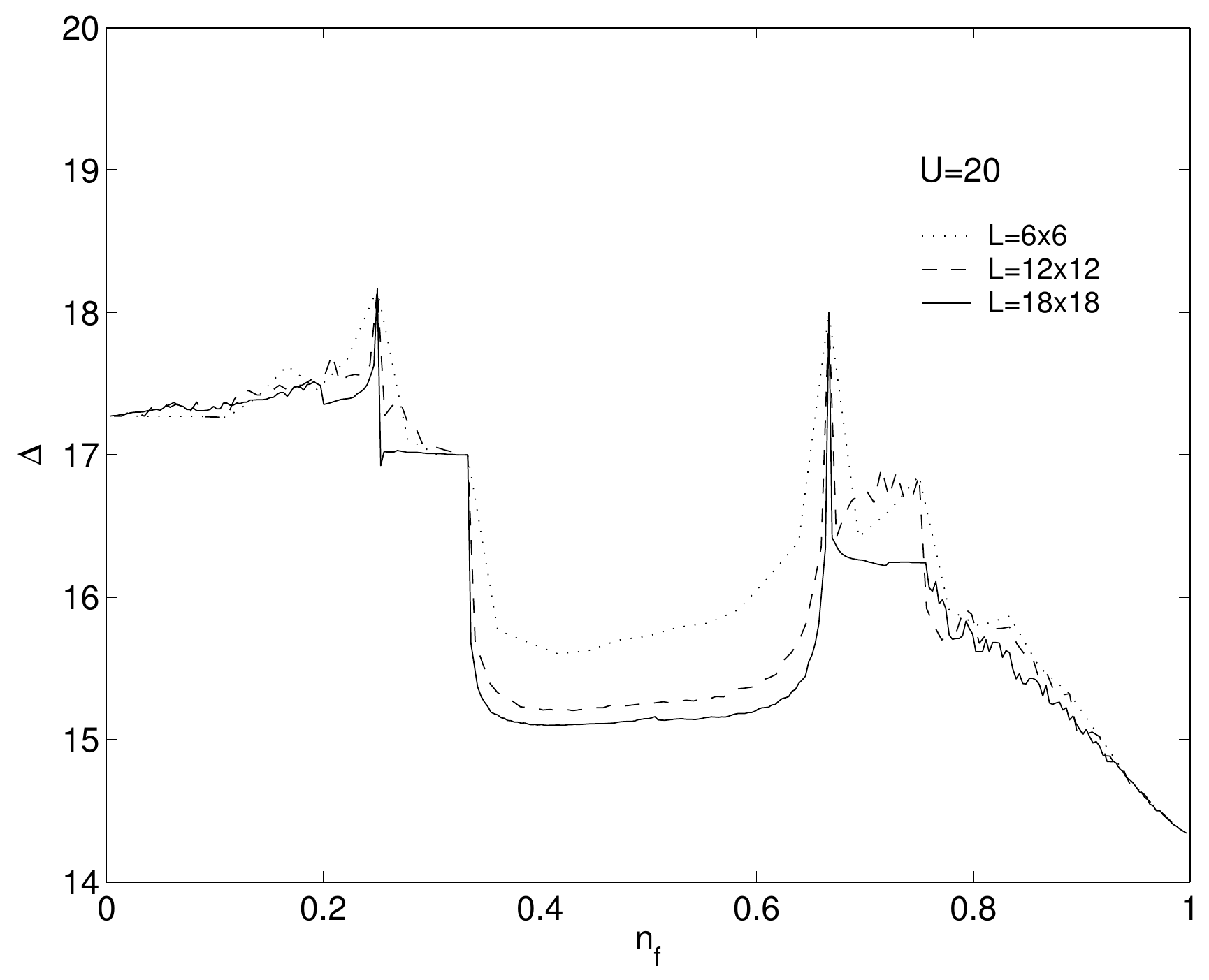}
\end{center}
\vspace{-0.3cm}
\caption{ Dependence of the energy gap $\Delta$ at the Fermi level on the $f$-electron density $n_f$ for $U=20$ and different clusters~\cite{Farky39}.}
\label{pss244_6}
\end{figure}
\begin{figure}[!h]
\begin{center}
\includegraphics[width=8.5cm]{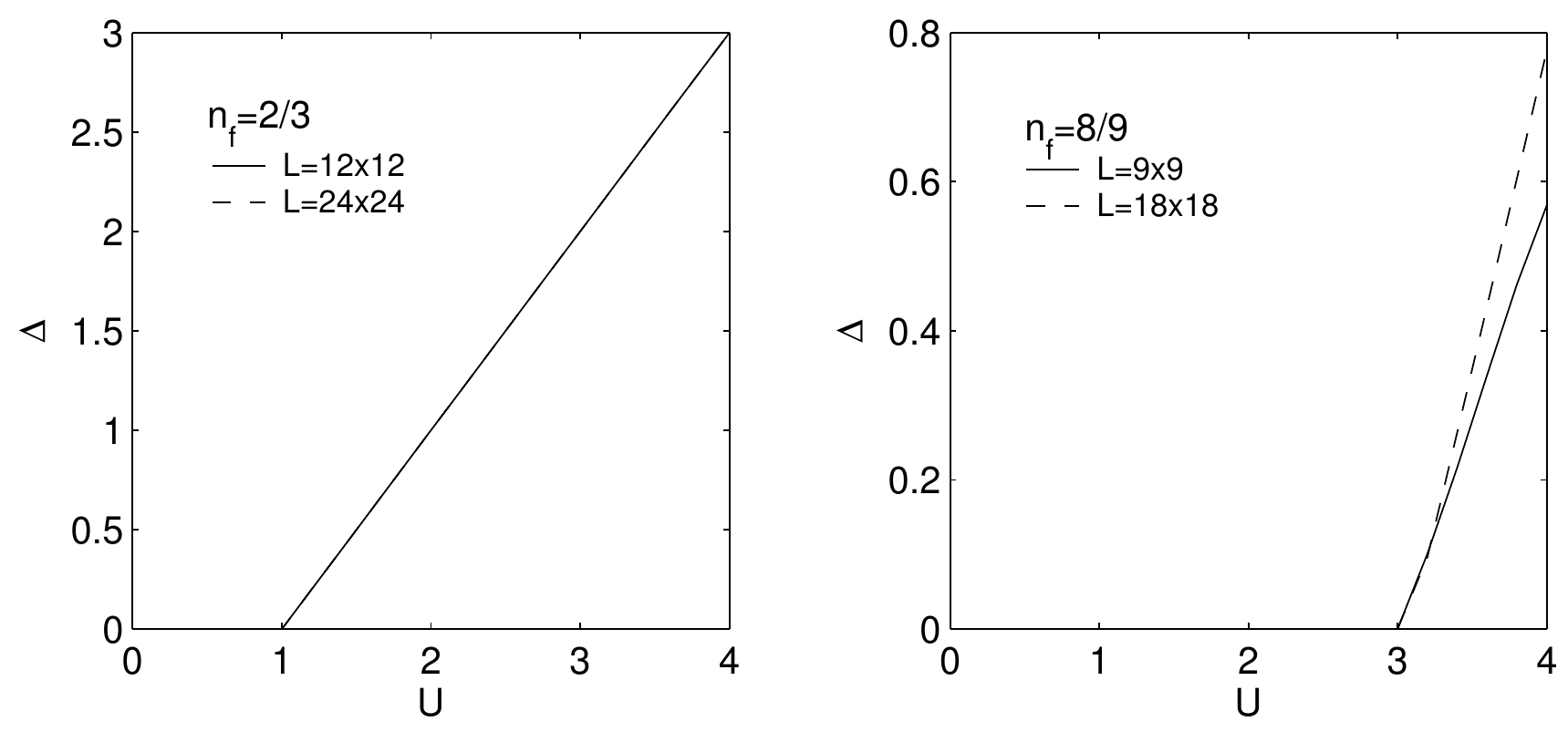}
\end{center}
\vspace{-0.3cm}
\caption{ Dependences of the energy gap $\Delta$ at the Fermi level on the Coulomb interaction $U$ for $n_f=2/3$ ($L=12\times 12$, $L=24\times 24$) and $n_f=8/9$ ($L=9\times 9$, $L=18\times 18$)~\cite{Farky39}.}
\label{pss244_7}
\end{figure}

Having a complete set of ground-state configurations we have tried to construct the picture of valence transitions in different interaction limits. The resultant behaviours obtained for intermediate and strong values of Coulomb interactions are depicted in figure~\ref{pss244_4}.
Comparing these behaviours with their square counterparts~\cite{Farky31} one can find significant differences. Indeed, while the valence transitions  for square Falicov-Kimball model is symmetric with the largest step at $n_f=1/2$, the valence transitions on the triangular lattice are    asymmetric and without any step at $n_f=1/2$. Instead of a significant step  at $n_f=1/2$ there are now two large steps at $n_f=1/3$ and $n_f=2/3$. The origin of this different behaviour relates with the  behaviour of the total  energy $E_{\rm g}$ (see figure~\ref{pss244_5}),
which instead of the total minimum at $n_f=1/2$ (the square Falicov-Kimball model) exhibits the total minimum at $n_f=2/3$ with the linear behaviour between $n_f=1/3$ and $n_f=2/3$. For this reason the valence transition for all finite clusters has a staircase structure, that follows the sequence $n_f=1 \rightarrow n_f=2/3\rightarrow n_f=1/3$ to $n_f=0$ and it is practically independent of $L$ for intermediate and strong interactions. For $U\rightarrow \infty$, the total energy $E_{\rm g}=0$ (for all $f$-electron concentrations), the steps at $n_f=2/3$ and $n_f=1/3$ vanish  and the valence transition is discontinuous from $n_f=1$ to $n_f=0$.

Although in the strong coupling limit there exists a special type of phase separation, the dependence of the energy gap $\Delta$ at the Fermi level on the $f$-electron concentration exhibits an insulating behaviour in the whole region.
One can see  (figure~\ref{pss244_6}) that outside the interval $n_f\in (1/3,2/3)$ the energy gap $\Delta$    depends on the  cluster size only very weakly and always has a finite value indicating an insulating behaviour. Inside the mentioned interval the energy gap slightly falls down and even depends on $L$, but the  insulating behaviour is evident. The detailed analysis performed in the weak coupling limit at selected values of $n_f$ ($n_f=2/3$ and 8/9) showed (see figure~\ref{pss244_7}),
that there exists a critical value of $U$ ($U_{\rm c}\sim 1$ for $n_f=2/3$ and $U_{\rm c}\sim 3$ for $n_f=8/9$) below which the Falicov-Kimball model on the triangular lattice also exhibits  a metallic behaviour. From this point of view, the principal question occurring  in the literature, namely, whether the systems with triangular lattice are capable of exhibiting the Mott-Hubbard transition, has been answered positively.

\section{Charge and spin ordering in the spin-1/2 Falicov-Kimball model}
\label{Charge and spin ordering in the spin-1/2 Falicov-Kimball model}

\subsection{Spin-1/2 Falicov-Kimball model without the Ising interaction}
\label{Spin-1/2 Falicov-Kimball model without Ising interaction}

The transition from spinless to spin version of the Falicov-Kimball model is formally trivial, and it is sufficient to add the spin variable to the creation and annihilation operators of itinerant and localized electrons:
\begin{eqnarray}
H=\sum_{ij\sigma}t_{ij}d^+_{i\sigma}d_{j\sigma}
+ \sum_{i\sigma\sigma '}d^+_{i\sigma}d_{i\sigma}f^+_{i\sigma '}f_{i\sigma'}
+ E_f\sum_{i\sigma}f^+_{i\sigma}f_{i\sigma}\,.
\label{eq4.1.7.1}
\end{eqnarray}
It should be noted, however, that from the physical point of view the spin Falicov-Kimball model describes a fully different physical reality. While in the spinless  model, all states with double occupancy are projected out (the Coulomb interaction between $d$ electrons with opposite  spins $U_{dd}$ as well as between  $f$ electrons with opposite spins $U_{ff}$ are infinitely large) in the spin model, such states are permitted ($U_{dd}=0$, $U_{ff}=0$). The total omission of Coulomb interactions between the itinerant and localized electrons with opposite spins is, however, a too crude simplification of physical reality in real systems, and therefore as the first step of our study we have generalized the model Hamiltonian (\ref{eq4.1.7.1}) by the term:

\begin{eqnarray}
U_{ff}\sum_{i}f^+_{i\uparrow}f_{i\uparrow}f^+_{i\downarrow}f_{i\downarrow}
\label{eq4.1.7.2}
\end{eqnarray}
describing the Coulomb repulsion of two $f$ electrons with oppositely oriented spins localized at the same position, and studied its effect on the valence and metal-insulator transitions~\cite{Farky6}.

Since (\ref{eq4.1.7.2}) does not violate the commutativity of  $f^+_{i\sigma}f_{i\sigma}$ with the total Hamiltonian of the system, one can again replace $f^+_{i\sigma}f_{i\sigma}$ by the classical variable $w_{i\sigma}=0,1$ and use for the study of the  generalized spin-1/2 Falicov-Kimball model the same procedures and methods as for the spinless model (the exact diagonalization on finite clusters followed by extrapolation of the results to the thermodynamic limit).
First we have investigated, the spin-1/2 Falicov-Kimball model for small finite clusters (up to 24 sites) and for all possible configurations of the localized $f$ electrons. The small-cluster exact-diagonalization calculations have been performed for the following set of $U_{ff}$ and $U$ values: $U_{ff}=0, 0.1, 0.2,\dots, 1.5$,  $U=1.5, 2, 3, 4, 5, 10$. We summarize our results with some observations. (i) For $U_{ff}<2$, the results do not sensitively depend upon $d$-$f$ interaction strength $U$. (Next the value $U=3$ is chosen to represent the typical behaviour of the model in a strong coupling limit.) (ii) The ground state for  $U_{ff}=0$ is a segregated configuration of the local $f$  pairs ($w_{\mathrm p}=\{22\dots200\dots0\}$ for $N_f$ even and $w_{\mathrm p}=\{22\dots2100\dots0\}$ for $N_f$ odd). (iii) For a given $N_f$ the segregated configuration $w_{\mathrm p}$ persists as the ground state for $U_{ff}<U_{ff_1}$. (iv) For $U_{ff_1}<U_{ff}<U_{ff_2}$, the number of local $f$ pairs is reduced with increasing $U_{ff}$. For $U_{ff}>U_{ff_2}$, the ground state is the segregated configuration with singly occupied sites ($w_{\mathrm s}=\{11\dots100\dots0\}$).

Furthermore, we have found that the transition from $w_{\mathrm p}$ to $w_{\mathrm s}$ is realized through the following steps
\begin{eqnarray}
w_{\mathrm p}&=&\{2\dots20\dots0\}\rightarrow\{12\dots210\dots0\}
\nonumber\\
&\rightarrow & \{112\dots2110\dots0\}\rightarrow\dots\to\{1\dots10\dots0\}=w_{\mathrm s}
\label{wp1}
\end{eqnarray}

or
\begin{eqnarray}
w_{\mathrm p}&=&\{2\dots210\dots0\}\rightarrow\{12\dots2110\dots0\}
\nonumber\\
&\rightarrow & \{112\dots21110\dots0\}\rightarrow\dots\to\{1\dots10\dots0\}=w_{\mathrm s}\, .
\label{wp1a}
\end{eqnarray}

The last observation is very important for the extrapolation of small-cluster exact-diagonali\-zation calculations since it allows us to avoid technical difficulties associated with a large number of configurations and consequently to study much larger systems.
\begin{figure}[h!]
\begin{center}
\mbox{\includegraphics[width=7cm,angle=0]{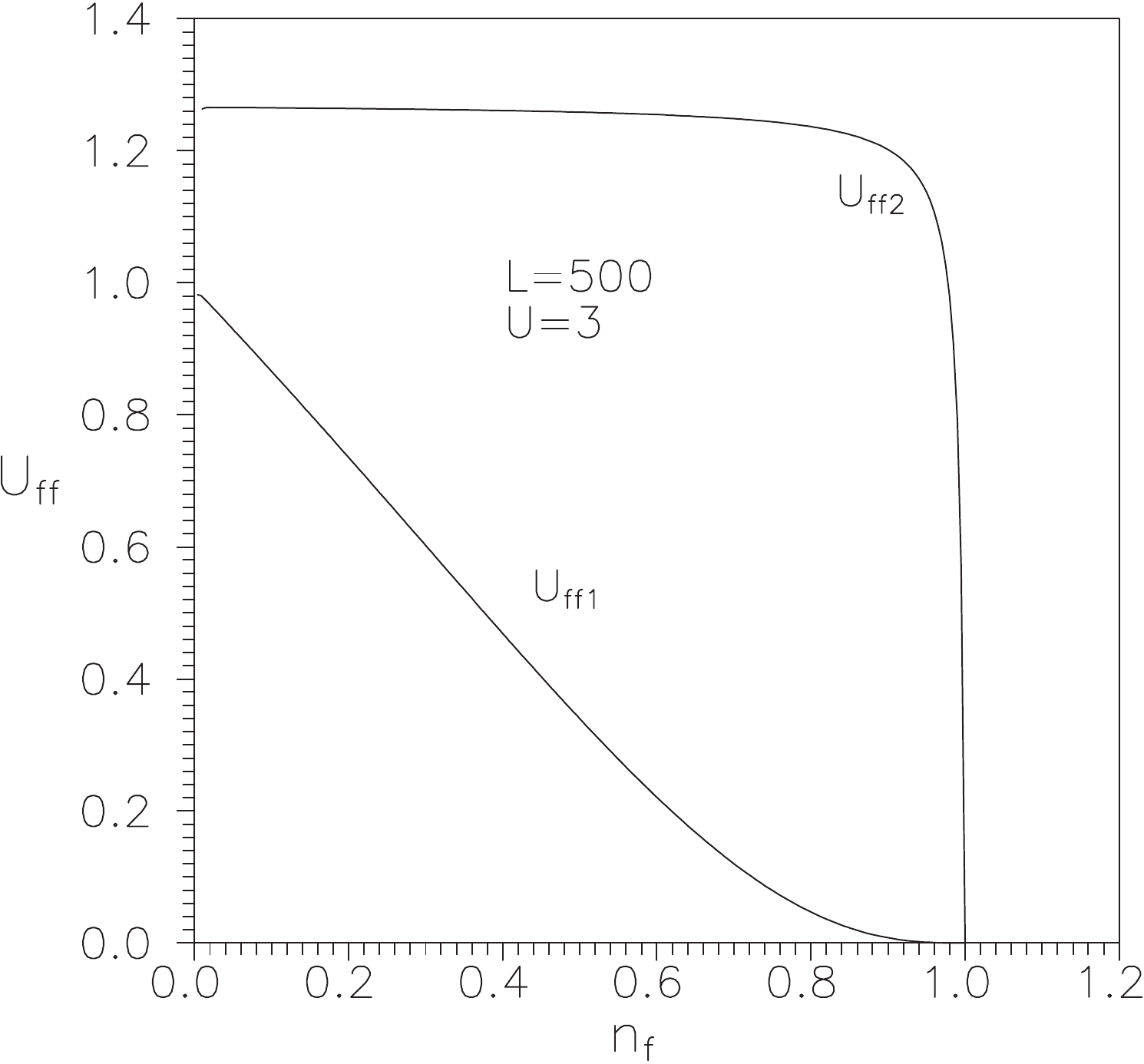}}
\end{center}
\vspace*{-0.5cm}
\caption{ Critical interaction strengths $U_{ff_1}$ and $U_{ff_2}$ as a function of $n_f$ calculated for $L=500$ and $U=3$~\cite{Farky6}. }
\label{prb5_1}
\end{figure}
Figure~\ref{prb5_1} presents numerical results for critical interaction strengths $U_{ff_1}$ and $U_{ff_2}$ as functions of the $f$-electron occupation number $n_f$ obtained for $U=3$ and $L=500$. It is seen that there is a relative large region of $U_{ff}$ values where the configurations with a nonzero number of local $f$ pairs are the ground states. The fact that the $f$ electrons form the local $f$ pairs, in spite of a relatively large repulsive interaction $U_{ff}$, indicates that there is an attractive interaction that is capable of overcoming this direct repulsion. One of the most important results for the spin-1/2 Falicov-Kimball model is that the interaction of the localized $f$ electrons with the itinerant $d$-band electrons leads to an effective on-site attraction between the localized $f$ electrons. It is interesting to study whether this feature changes the picture of valence and metal-insulator transitions found for the spinless version of the model. The numerical results for $E_f$ dependence of $n_f$ (calculated for configurations of type (\ref{wp1}) or (\ref{wp1a})) are plotted in figure~\ref{prb5_2} for $U=3$ and different values of $U_{ff}$.
\begin{figure}[!t]
\begin{center}
\mbox{\includegraphics[width=6cm,angle=0]{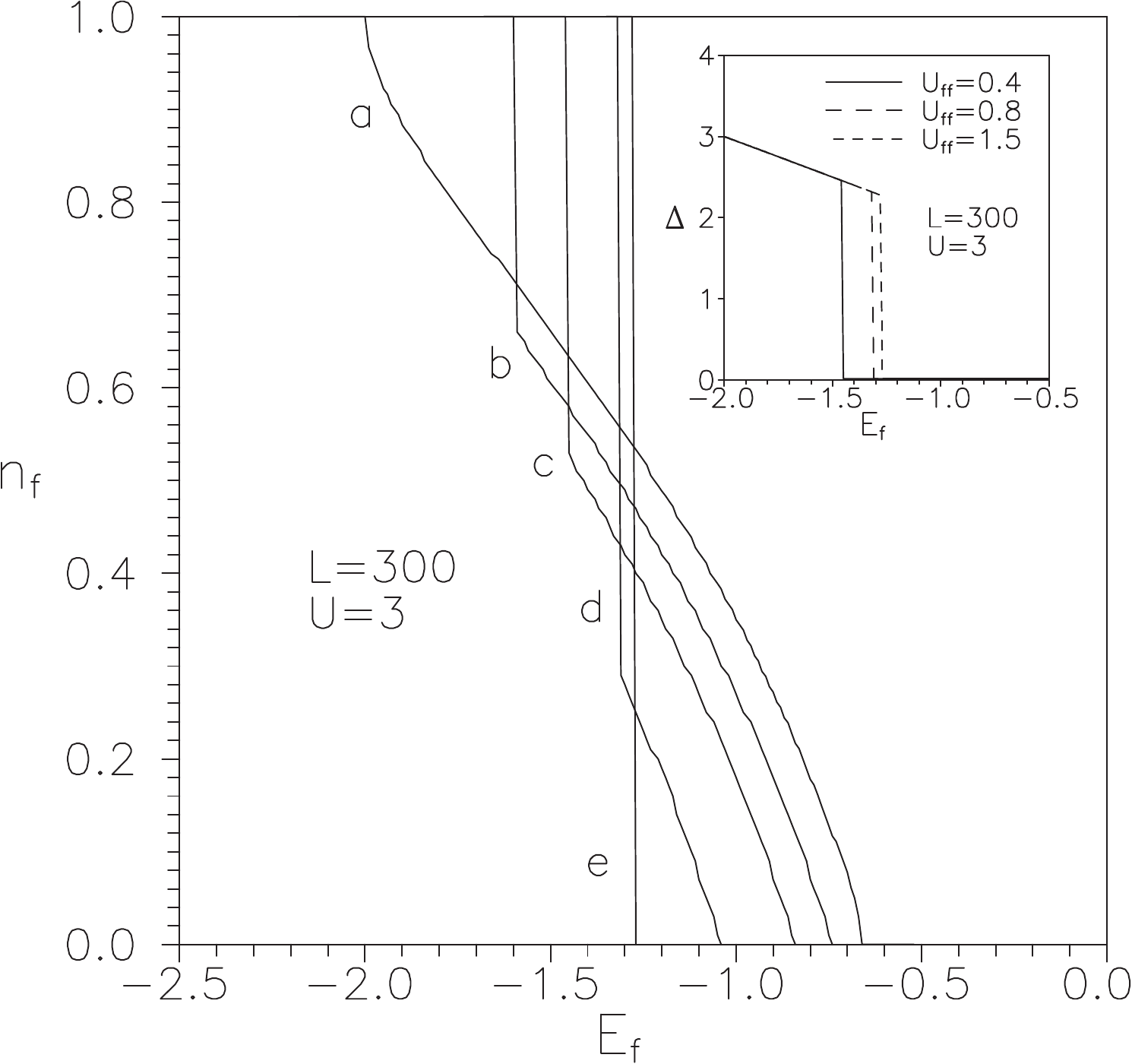}}\hspace*{0.5cm}
\mbox{\includegraphics[width=6cm,angle=0]{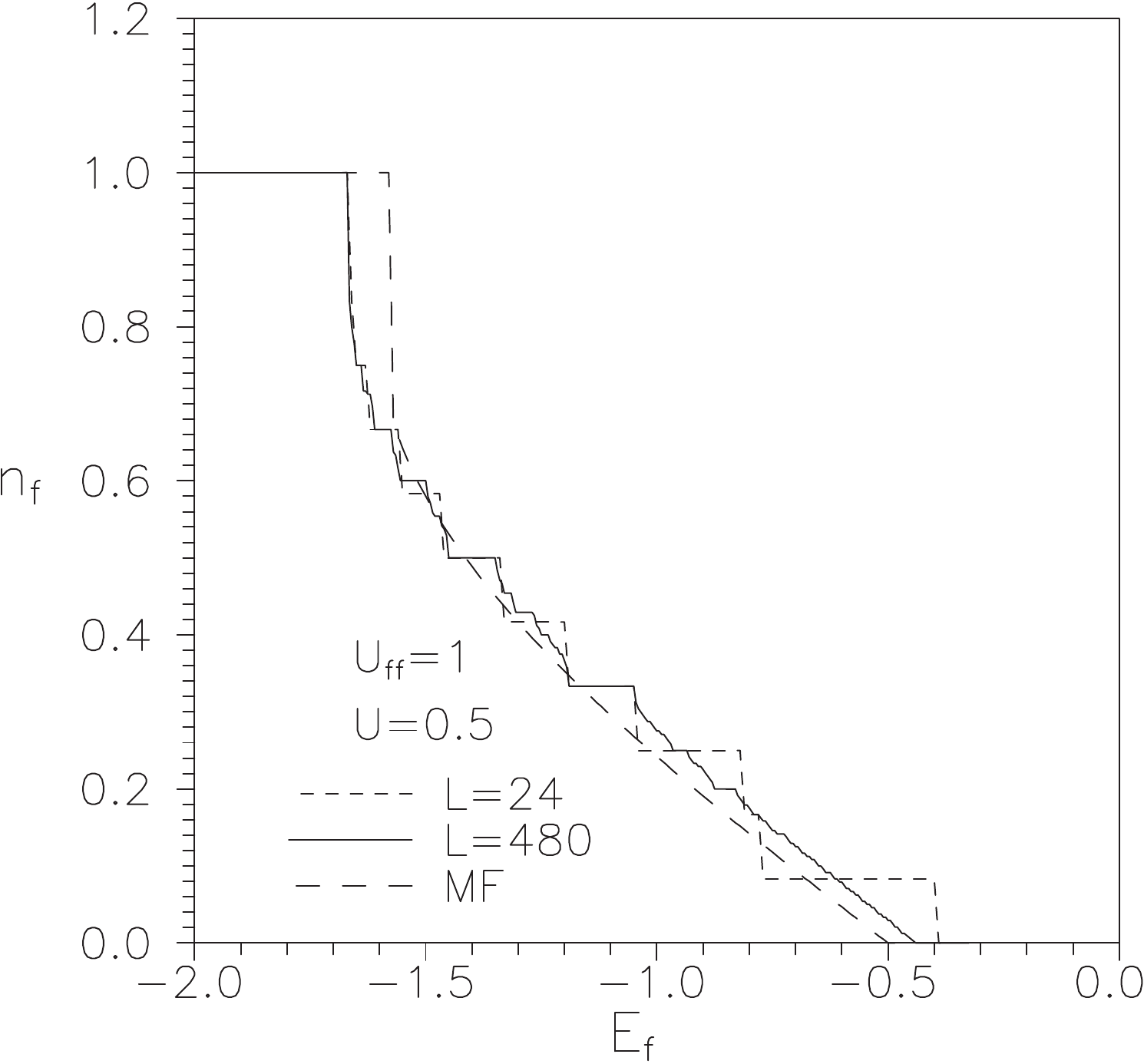}}
\end{center}
\vspace*{-0.5cm}
\caption{ {\it Left}: Dependence of the $f$-electron occupation number  $n_f$ on the  $f$-level position $E_f$ for
$L=300$, $U=3$ and five different values of  $U_{ff}$: (a) $U_{ff}=0$, (b) $U_{ff}=0.2$, (c) $U_{ff}=0.4$, (d) $U_{ff}=0.8$, (e) $U_{ff}=1.5$. Inset: dependence of energy gap $\Delta$ on the  $f$-level position $E_f$ calculated for  $L=300, U=3$ and three different values of  $U_{ff}$. {\it Right}: Dependence of the $f$-electron occupation number  $n_f$ on the  $f$-level position $E_f$ for   $U=0.5$  and $U_{ff}=1$. The behaviour for $L=24$ has been calculated for all possible $f$-electron configurations, for  $L=480$ only the most homogeneous configurations have been considered, and the MF curve represents the mean-field result obtained using the exact density of states~\cite{Farky6}. }
\label{prb5_2}
\end{figure}
\begin{figure}[b!]
\begin{center}
\vspace{-2mm}
\mbox{\includegraphics[width=8.3cm,angle=0]{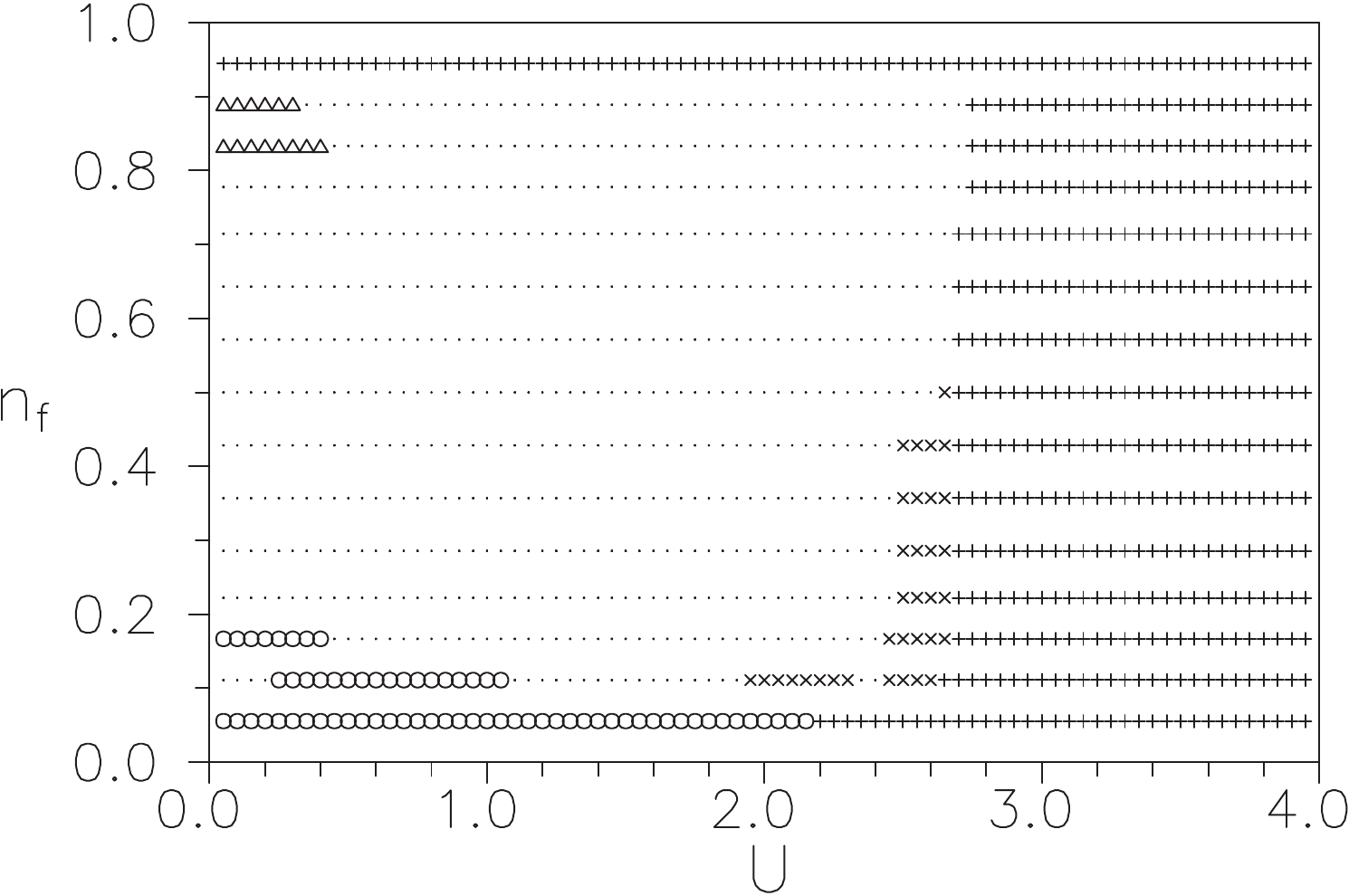}}
\vspace*{0.1cm}
\caption{ The ground-state phase diagram of the spin-1/2 Falicov-Kimball model
obtained over the full set of $f$-electron configurations.
For $1/4<n_f<3/4$ numerical calculations have been done
on the lattice with $L=28$ while for $n_f\leqslant 1/4$ and
$n_f\geqslant 3/4$ on the lattice with $L=36$.
Four different regions of stability corresponding to mixtures
$w_a\&w_e$, $w_b\&w_e$, $w_c\&w_e$ and $w_d\&w_f$  are denoted as
$\circ, \times,+$ and $\triangle$
~\cite{Farky15}. }
\label{prb6_1}
\end{center}
\end{figure}
They lead to the following conclusions. (i) The transition is continuous for $U_{ff}=0$. (ii) For $0<U_{ff}<U_{ff}^{\rm c}$ ($U_{ff}^{\rm c}\sim 1.273$) there are discontinuous transitions from an integer-valence state $n_f=1$ into an inhomogeneous intermediate-valence state $n_f\neq1$ at $E_f=E_f^{\rm c}(U_{ff})$. (iii) For $U_{ff}>U_{ff}^{\rm c}$ the transitions are discontinuous from $n_f=1$ to $n_f=0$. They take place at $E_{\rm c}=-U_{ff}^{\rm c}$ independently of $U_{ff}$.

In the weak coupling limit, we have found strong finite-size effects and, therefore, we have turned our attention to the case $U_{ff}=\infty$ that permits to reduce the total number of the investigated  $f$-electron configurations from $4^L$ to $2^L$, and thus to greatly increase the size of the clusters studied, which is very important for the correct analysis of structural properties of the model~\cite{Farky15}.

To reveal the basic structure of the phase diagram in the $n_f$-$U$
plane ($E_f=0$) we have performed an exhaustive study of the model on finite
(even) clusters up to 36 sites. For fixed $L$, the numerical calculations
have been  performed along the lines discussed above with a step $\Delta N_f=2$
and $\Delta U =0.05$. The results of numerical computations
are summarized in figure~\ref{prb6_1}.
These results show that the phase diagram of the
spin-1/2 Falicov-Kimball model consists of three main domains:
the most homogeneous domain MHD~(in figure~\ref{prb6_1} denoted as $\cdot$) and two phase
separation domains PSD$_1$~(denoted as $\circ, \times$ and $+$)
and PSD$_2$~(denoted as $\triangle$).
In the MHD the ground states are configurations in which the atomic or
$n$-molecule clusters of $f$ electrons are distributed in such a manner that
the distances between two consecutive clusters are either $d$ or
$d+2$. Furthermore, distribution of the distances of $d$ and $d+2$ has
to be the most homogeneous. Two basic types of the ground state configurations
that fill up practically the whole MHD are displayed in table~\ref{tab5}.
In the PSD$_1$ the ground states are configurations in which
all $f$ electrons are distributed only in one part of the lattice
($w$) while another (connecting) part of the lattice ($w_e$) is
free of $f$ electrons (phase separation). In accordance with
Gruber et al.~\cite{Gruber_Ueltschi} we refer to such configurations as mixtures and denote
them by $w\&w_e$.
We have found three basic types of configurations $w$
which form these  mixtures: (i) aperiodic atomic configurations
$w_a=\{10_{k_1}10_{k_2}\dots10_{k_i}1\dots0_{k_2}10_{k_1}1\}$
(with $k_i>0$), (ii) aperiodic $n_i$-molecule configurations
$w_b=\{1_{n_1}0_{k_1}1_{n_2}0_{k_2}\dots1_{n_i}0_{k_i}1_{n_i}\dots
0_{k_2}1_{n_2}0_{k_1}1_{n_1}\}$ (with $1\leqslant n_i\leqslant N_f/2$ and $k_i>0$),
and (iii) $N_f$-molecule (segregated) configurations $w_c=\{11\dots 1\}$.
Three different regions of stability corresponding to mixtures
$w_a\&w_e$, $w_b\&w_e$ and $w_c\&w_e$ are denoted in figure~\ref{prb6_1} as
$\circ, \times$ and $+$.
It is seen that the mixtures of the atomic configuration $w_a$
and the empty configuration are stable only at low $f$-electron
concentrations and the Coulomb interactions $U<2.2$.
\begin{table}[h]
\caption{ Two basic types of the most homogeneous configurations
that fill up practically the whole MHD for  $L=24$~\cite{Farky15}.}
\vspace{2ex}
\begin{center}
\begin{tabular}{|l|l|l|}
\hline
$N_f$            & $w^a_{\rm h}$& $w^b_{\rm h}$\\
\hline\hline
4  & 100001000000100001000000 &110000000000110000000000\\
6  & 100100001001000010010000 &110000001100000011000000\\
8  & 100100100100100100100100 &110000110000110000110000\\
10 & 110010011001001001100100 &110011000011001100110000\\
12 & 110011001100110011001100 &110011001100110011001100\\
14 & 111001110011100111001100 & \\
16 & 111100111100111100111100 & \\
18 & 111111001111110011111100 & \\
20 & 111111111100111111111100 & \\
\hline
\end{tabular}
\end{center}
\label{tab5}
\end{table}
A direct comparison of the results obtained for the spin-1/2 and spinless
Falicov-Kimball model~\cite{Gruber_Ueltschi} shows that this region roughly corresponds to a region of
phase separation in the spinless Falicov-Kimball model. Outside these regions, the phase diagrams
of the spin-1/2 and spinless Falicov-Kimball model are, however, strongly different.
While the phase separation in the spinless Falicov-Kimball model takes place only for
weak interactions ($U<1.2$) the spin-1/2 Falicov-Kimball model exhibits the phase
separation for all Coulomb interactions. Even with an increasing $U$,
the phase separation shifts to higher $f$-electron concentrations.
Particularly, in the region $2.5<U<2.7$ where the ground states are
the mixtures of $n$-molecule configurations $w_b$
with the empty configuration, the phase separation takes place for
all $N_f<L/2$ and in the region $U>2.7$ where the ground states are
the segregated configurations $w_{\rm S}=w_c\&w_e$ even for all $N_f<L$.
At large $f$-electron concentrations $n_f$
but in the opposite limit $(U<0.4)$ there exists another small domain
of a phase separation PSD$_2$ (denoted by $\triangle$).
The numerical results on finite lattices up to 36 sites
revealed only one type of configurations that can be the ground
state configurations in this domain, and namely, the mixtures of the periodic
$n$-molecule configurations $w_d$ with the fully occupied lattice
$w_f=\{11\dots 1\}$ (the length of a connected cluster of the occupied sites
in these mixtures is at least $L/2$).

The second step in our numerical studies has been the extrapolation
of small-cluster exact diagonalization results on large lattices.
\begin{figure}[!t]
\begin{center}
\mbox{\includegraphics[width=6.5cm,angle=0]{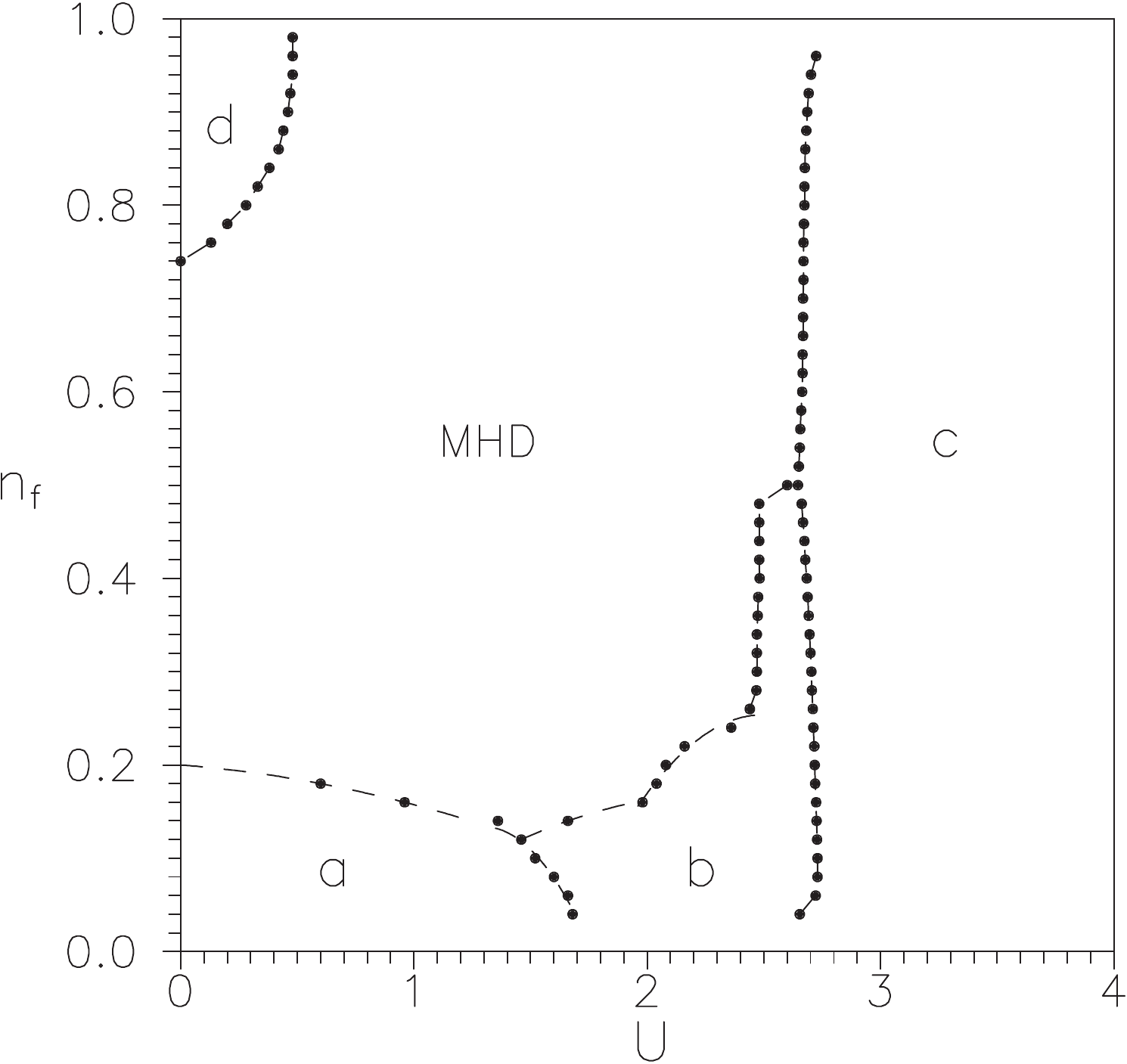}}
\end{center}
\vspace*{-0.5cm}
\caption{ Ground-state phase diagram of the spin-1/2 Falicov-Kimball model obtained on the extrapolated set of $f$-electron configurations for  $L=100$. Four different regions of stability correspond to mixtures
$w_a\&w_e$ ($a$), $w_b\&w_e$ ($b$), $w_c\&w_e$ ($c$) and $w_d\&w_e$ ($d$) and MHD denotes the region of the most homogeneous configurations~\cite{Farky15}. }
\label{prb6_2}
\end{figure}
In figure~\ref{prb6_2} we present the ground state phase diagram of the spin-1/2
Falicov-Kimball model obtained for $L=100$ on the extrapolated set of configurations that
includes practically all possible types of the ground-state configurations
found on finite lattices up to 36 sites. In particular,
we have considered (i) the most homogeneous configurations $w_{\rm h}$ of the
type $a$ and $b$ (see table~\ref{tab5}), (ii) all mixtures $w_a\&w_e$ with $k_i$
smaller than 6, (iii) all mixtures $w_b\&w_e$ with $n_i$ and $k_i$ smaller
than 6, (iv) all mixtures $w_d\&w_f$ with periods smaller than 12,
and (v) all segregated configurations.
One can see that all fundamental features of the phase diagram found
on small lattices hold on much larger lattices too.
Of course, the phase boundaries  of different regions corresponding
to mixtures $w_a\&w_e$ $(a)$, $w_b\&w_e$ $(b)$, $w_c\&w_e$ $(c)$ and
$w_d\&w_f$ $(d)$ are now more obvious.
Since the mixtures $w_a\&w_e$, $w_b\&w_e$, $w_c\&w_e$ and $w_d\&w_f$ are
metallic~\cite{Gruber_Ueltschi} one can expect that the phase boundary between
the MHD and PSD is also the boundary  of the correlation induced
metal-insulator transition.
\begin{figure*}[!b]
\begin{center}
\mbox{\includegraphics[width=5cm,angle=0]{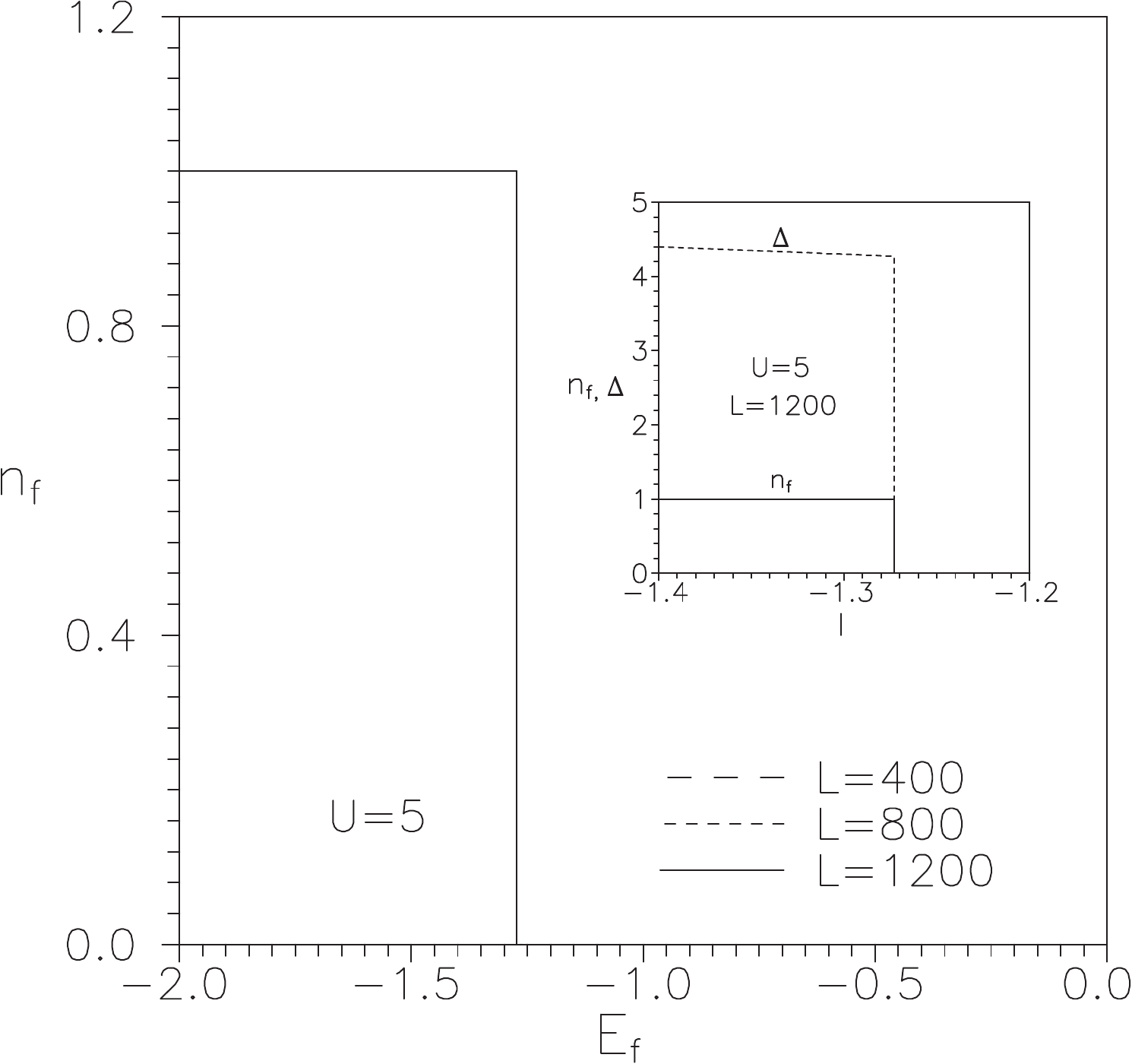}}%\hspace{-0.2cm}
\mbox{\includegraphics[width=5cm,angle=0]{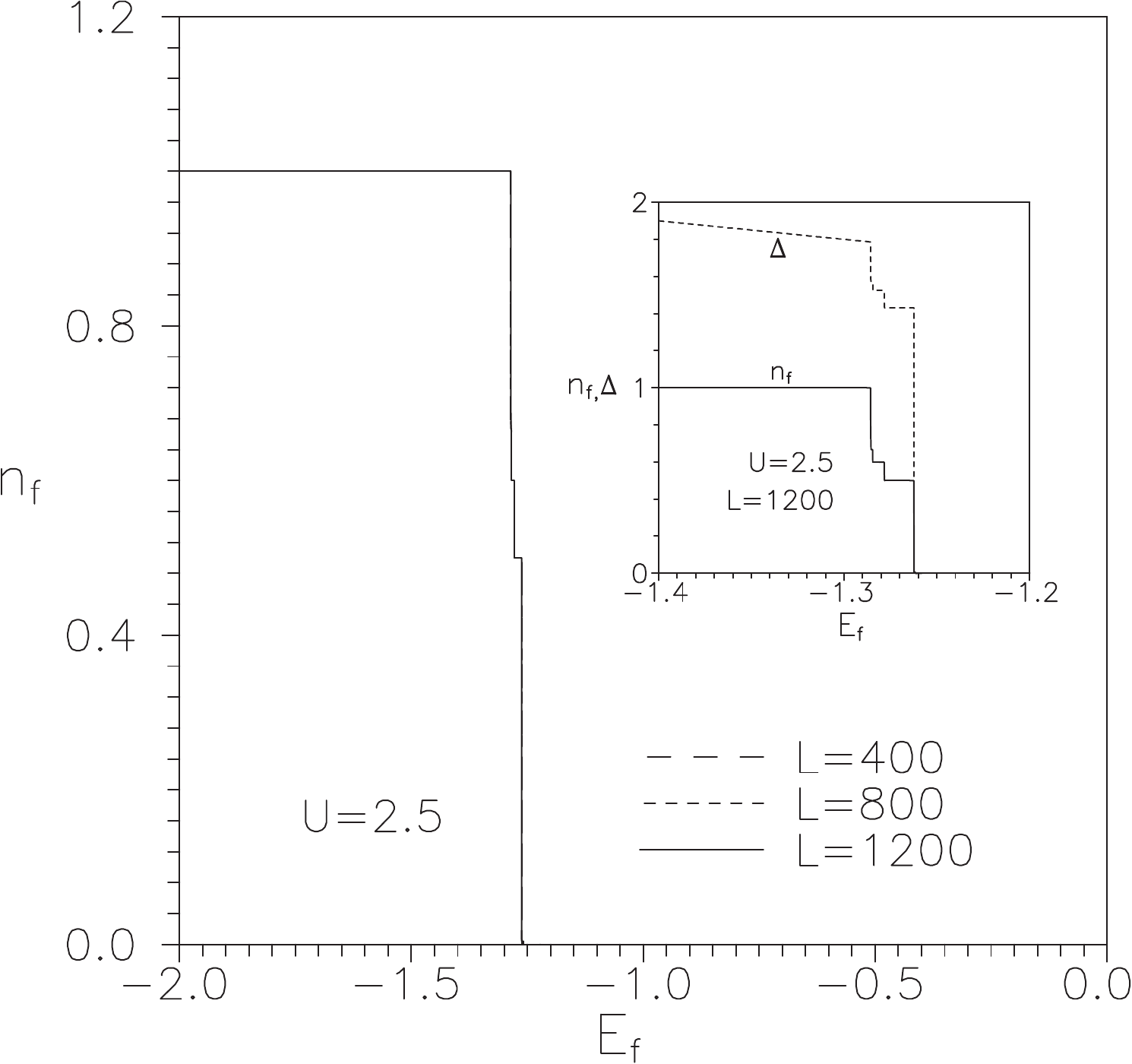}}%\hspace{-0.2cm}
\mbox{\includegraphics[width=5cm,angle=0]{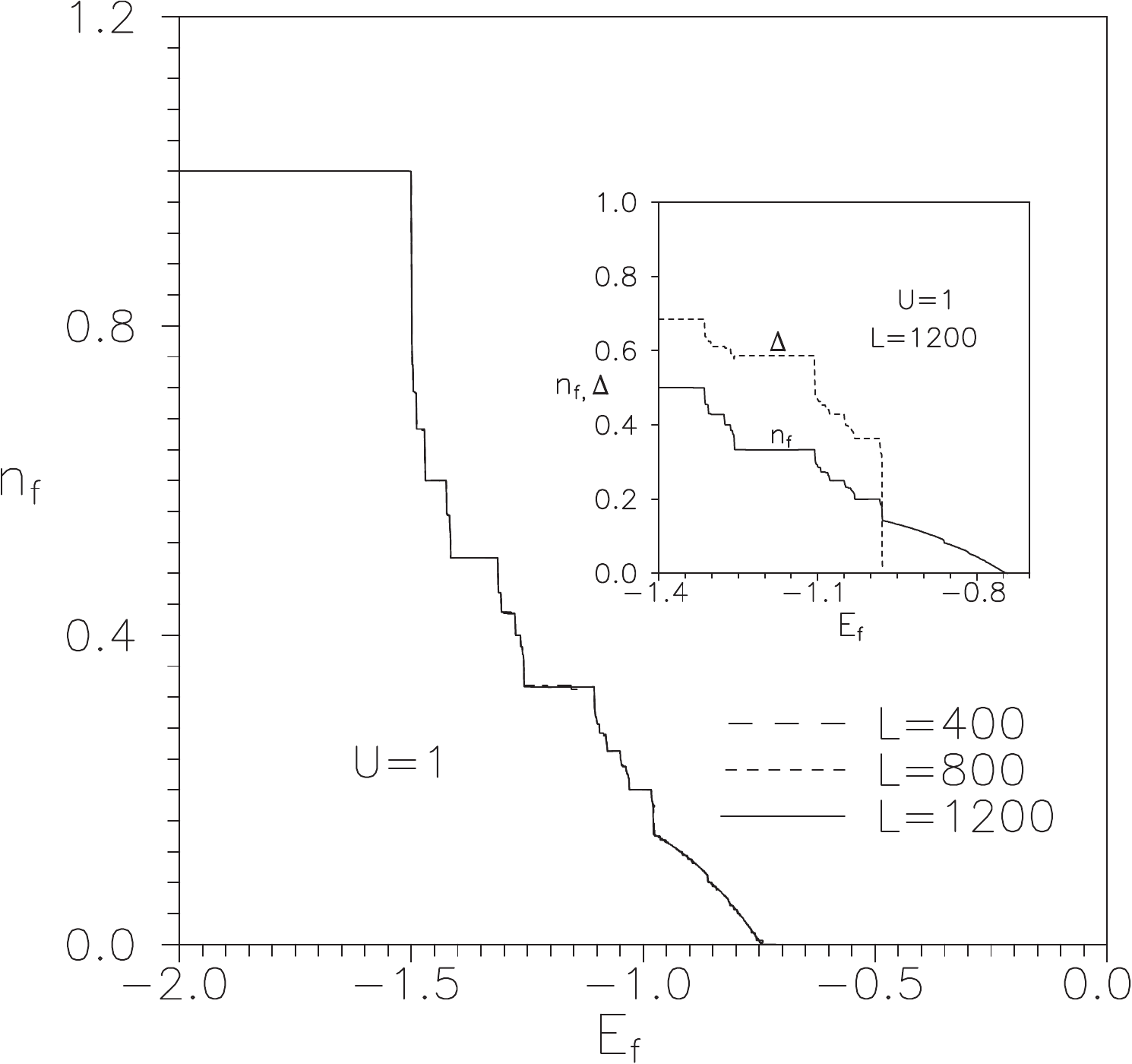}}
\end{center}
\caption{ Dependence of the  $f$-electron occupation number   $n_f$  on the $f$-level position $E_f$ calculated for $U=5$ (the first panel), $U=2.5$ (the second panel) and $U=1$ (the third panel) and  $L=400$, 800 a 1200. Insets show the behaviour of $n_f$ and $\Delta$ close to the insulator-metal transition point~\cite{Farky15}.}
\label{prb6_4}
\end{figure*}
To confirm this conjecture it is necessary
to show that the most homogeneous configurations from the MHD are insulating,
or, in other words, that there is a finite energy gap $\Delta$ at the Fermi
energy in the spectra of these configurations\footnote{Since we consider the case $N_f+N_d=L$, the Fermi level $E_{\rm F}$ and the energy gap $\Delta$ of some configuration $w$ are presented by $E_{\rm F}=\lambda_{L-N_f}$ and $\Delta=\lambda_{L-N_f+1}-\lambda_{L-N_f}$, respectively.}. The numerical
results for the most-homogeneous configurations of the type $a$ and $b$ that
fill up practically the whole MHD are displayed in figure~\ref{prb6_4}.
The insulating
character of these configurations is clearly demonstrated by the finite
$\Delta$ that exists for all nonzero $f$-electron concentrations $n_f$
and Coulomb interactions $U$. Thus we can conclude that
the spin-1/2 Falicov-Kimball model undergoes (on the phase boundary between the
MHD and PSD) the correlation induced metal-insulator transition that
is accompanied by a discontinuous change of the energy gap $\Delta$.
This result is similar to what was found for the spinless
Falicov-Kimball model by numerical~\cite{Gruber_Ueltschi} and analytical
calculations in the strong~\cite{Lemberger92} and weak-coupling
limit~\cite{Freericks_Gruber}.

Of course, this fact has to lead to a different picture of valence and metal-insulator transitions  (figure~\ref{prb6_4}) induced by pressure (increasing $E_f$). We have found that: (i) In the strong-coupling limit ($U>4$), the model exhibits a pressure induced discontinuous insulator-metal transition from an integer-valence state ($n_f=1$) into another integer-valence state ($n_f=0$). (ii) For intermediate values of $U$ ($U\sim 2.5$) the Falicov-Kimball model undergoes a few discontinuous intermediate-valence transitions. There are several discontinuous insulator-insulator transitions from $n_f=1$ to $n_f=1/2$ and discontinuous insulator-metal transition from $n_f=1/2$ to $n_f=0$. (iii) In the weak-coupling limit  ($U<2$), the model undergoes a few consecutive discontinuous and continuous intermediate-valence transitions as well as a discontinuous metal-insulator transition.

\subsection{Spin-1/2 Falicov-Kimball model with the Ising interaction}
\label{Spin-1/2 Falicov-Kimball model with Ising interaction}

Despite the unquestionable success of the spinless  Falicov-Kimball model in describing  the  charge ordering in strongly correlated systems, this relatively simple model was not capable of accounting for all aspects of real experiments, and namely,  that the charge superstructure is often accompanied by a magnetic superstructure.  In order to more realistically describe  electronic and spin processes in real materials,  the original spin-1/2 Falicov-Kimball model has been extended by the spin-dependent interaction (of the Ising type) between localized and itinerant electrons~\cite{Lemanski05,Farky29, Lemanski_Wrzodak}
\begin{eqnarray}
H=\sum_{ij\sigma}t_{ij}d^+_{i\sigma}d_{j\sigma}
+ U\sum_{i\sigma\sigma '}f^+_{i\sigma}f_{i\sigma}d^+_{i\sigma '}d_{i\sigma '}
+ J\sum_{i\sigma}(f^+_{i-\sigma}f_{i-\sigma} - f^+_{i\sigma}f_{i\sigma})d^+_{i\sigma}d_{i\sigma}\;
\label{eq.4.3.2.1}
\end{eqnarray}
and, in addition, the local Coulomb interaction between the $f$~electrons in the limit of  $U_{ff}\rightarrow \infty$ has been included.  Thus, from the major interaction terms that come into account for the interacting $d$ and $f$-electron subsystems, only the Hubbard type interaction
\begin{eqnarray}
H_{dd}=U_{dd}\sum_{i}d^+_{i\uparrow}d_{i\uparrow}d^+_{i\downarrow}d_{i\downarrow}
\label{eq.4.3.2.1a}
\end{eqnarray}
 between the spin-up and spin-down $d$ electrons has been omitted in the Hamiltonian~(\ref{eq.4.3.2.1}).

In the work~\cite{Lemanski05} Lemanski presented a simple justification for the omission of this term based on an intuitive argument: the longer time electrons occupy the same site, the more important becomes the interaction between them. According to this rule the interaction between the itinerant $d$ electrons $(U_{dd})$ is smaller than the interaction between the localized $f$ electrons $(U_{ff})$ as well as smaller than the spin-independent interaction between the localized and itinerant electrons $U$.

\begin{figure*}[!t]
\begin{center}
\includegraphics[angle=0,width=7cm,height=5.5cm]{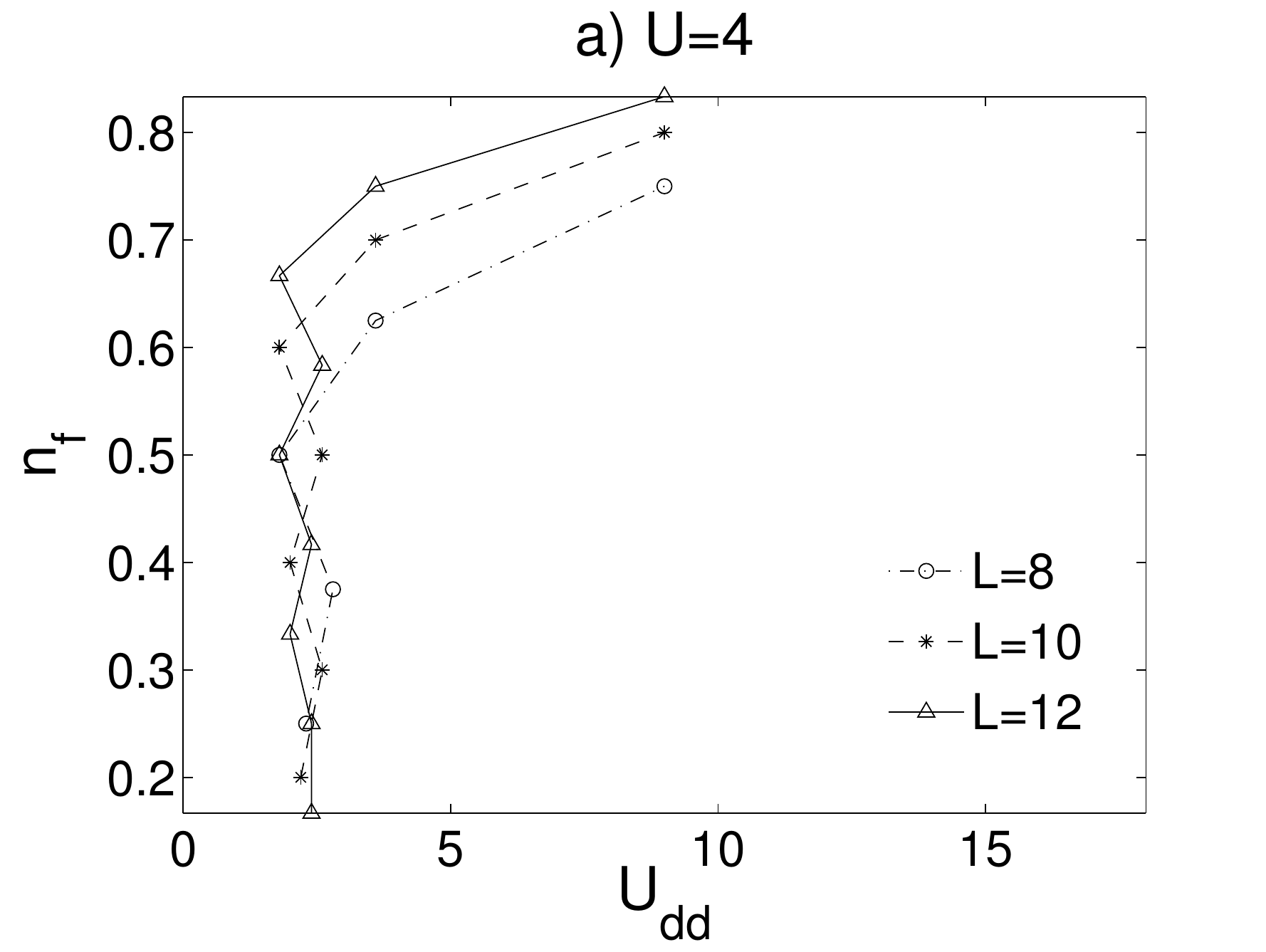}\hspace*{0.7cm}
\includegraphics[angle=0,width=7cm,height=5.5cm]{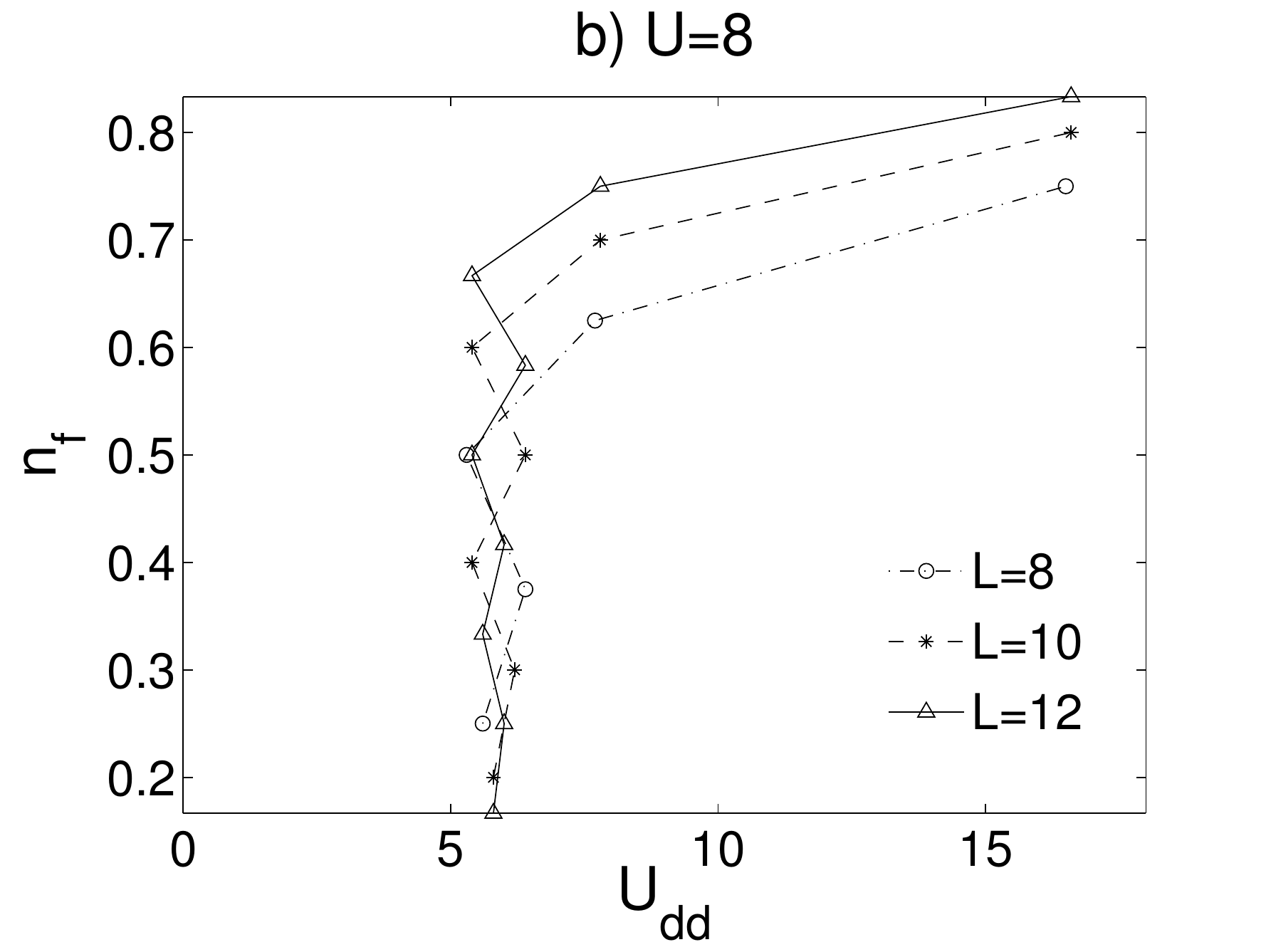}
\end{center}
\vspace*{-0.5cm}
\caption{
The ground-state phase diagrams of the spin-1/2 Falicov-Kimball model extended by the Hubbard interaction between the itinerant electrons calculated for two different values of $U$ on small finite clusters of $L=8,10$ and 12 sites. Below $U^{\rm c}_{dd}$ the ground states are the ground-state configurations of the conventional spin-1/2 Falicov-Kimball model ($U_{dd}=0$). Above $U^{\rm c}_{dd}$ these ground states become unstable. The one-dimensional exact-diagonalization results~\cite{Farky29}.}
\label{ising_1}
\end{figure*}

We have tried to verify the validity of these intuitive statements numerically and we have found that their validity is restricted only to the region of intermediate and strong Coulomb interactions $U$ between the localized and itinerant electrons. Our numerical proof  is based on the exact diagonalization of the  model Hamiltonian (\ref{eq.4.3.2.1}) extended by  $H_{dd}$ on small  finite clusters. Such a study is numerically very exhaustive, since it is necessary for each configuration of localized electrons/spins to perform the  Lanczos diagonalization~\cite{Dagotto2} of  the Hamiltonian over the full Hilbert space of the model and this process should be repeated $\displaystyle\left(\begin{smallmatrix} L\\N_f\end{smallmatrix}\right)$ times. Of course, such a procedure demands in practice a considerable amount of CPU time, which imposes severe restrictions on the size of clusters that can be studied within the exact diagonalization calculations. For this reason we were able to  exactly investigate only the clusters up to $L=12$. Fortunately, it was found that in some parameter regimes the ground-state characteristics of the model are practically independent of $L$ and thus already such small clusters can be  satisfactorily used to represent the behaviour of macroscopic systems. In particular, we have studied the stability of the ground-state configuration $w^0(N_f)$ (obtained for $U_{dd}=0$ and fixed $N_f$) at finite values of $U_{dd}$. The results of numerical calculations obtained for $U=4$ and $U=8$ are summarized in figure~\ref{ising_1}
\begin{figure*}[!b]
\begin{center}
\hspace*{-0.35cm}
\includegraphics[angle=0,width=7.50cm,scale=1]{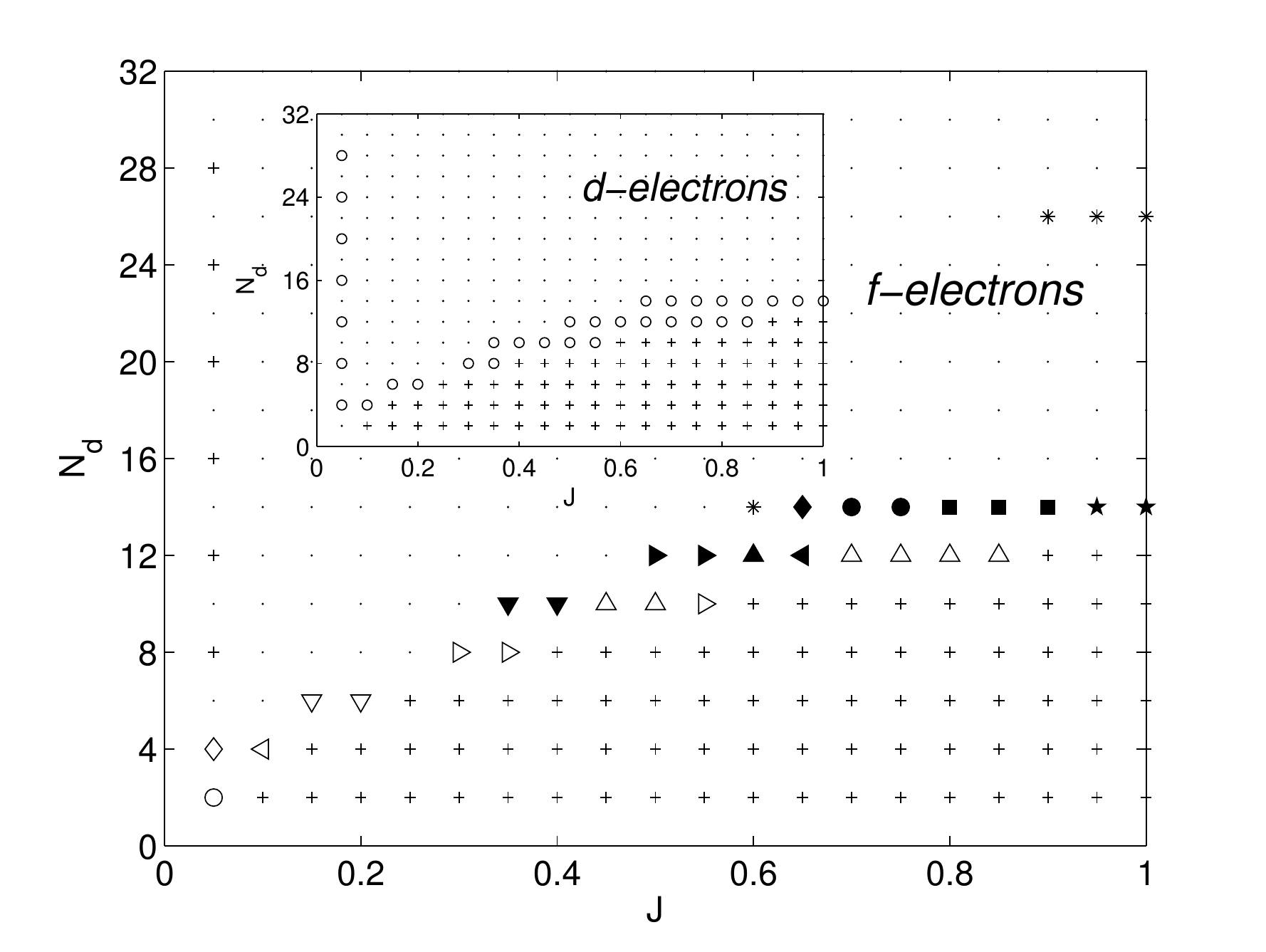}\hspace*{-0.7cm}
\includegraphics[angle=0,width=8.5cm,scale=1]{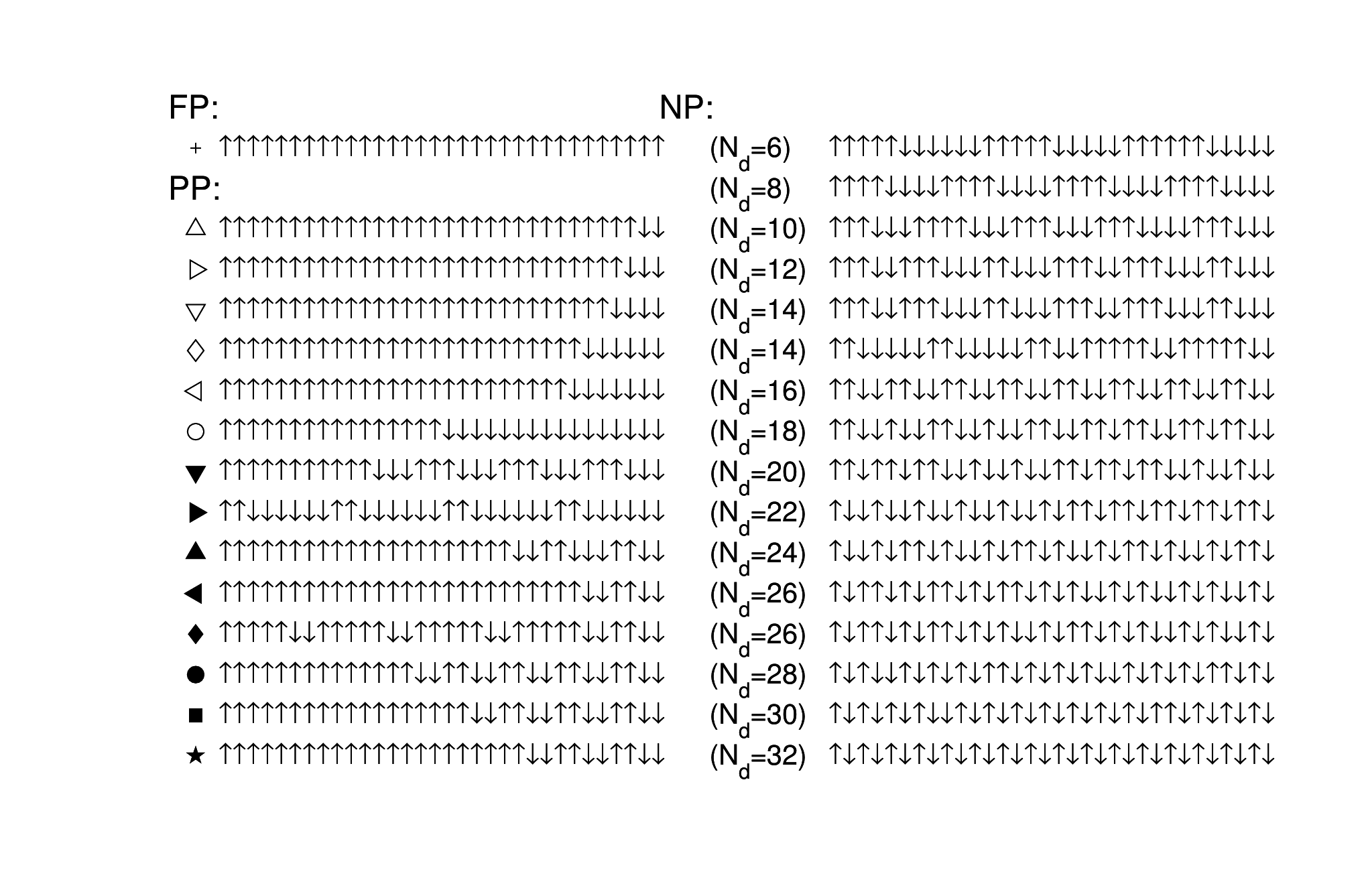}
\end{center}
\vspace*{-0.5cm}
\caption{ The $f$-electron ground-state phase diagram of the spin-1/2 Falicov-Kimball model
extended by the spin-dependent interaction  calculated for $N_f=L$ ($L=32$). The one-dimensional exact-diagonalization results. Inset: The $d$-electron ground-state phase diagram of the model calculated at the same values of model parameters. ($\cdot$): the NP phase, (+): the FP phase, ($\circ$): the PP phase~\cite{Farky29}.}
\label{ising_2}
\end{figure*}
in the form of $n_f-U_{dd}$ phase diagrams (the half-filled band case $n_f+n_d=1$ is considered). One can see that the ground-state configuration $w^0(N_f)$ found for $U_{dd}=0$ persists as a ground state up to relatively large values of $U_{dd}$ ($U^{\rm c}_{dd}\sim 2$, for $U=4$ and $U^{\rm c}_{dd}\sim 6$, for $U=8$), revealing small effects of the $U_{dd}$ term on the ground state of the model in the strong $U$ interaction limit. Contrary to the strong coupling case, for small ($U=1$) and intermediate ($U=2$) values of the Coulomb interaction between the localized and itinerant electrons, very strong effects of $U_{dd}$ term on the ground states of the model have been observed.  In these cases the typical values of $U^{\rm c}_{dd}$ are of the order of 0.5 but for some $N_f$ even much smaller values were found. Thus, we can conclude that the Hubbard type interaction between the spin-up and spin-down $d$ electrons can be neglected in the strong interaction limit between the localized and itinerant electrons ($U\geqslant 4$). We have obtained the same conclusion  for the spin-dependent interaction $J>0$ \cite{Farky40}. For this reason, all subsequent calculations on the spin-1/2 Falicov-Kimball model with spin-dependent Coulomb interaction $J$ between the $f$ and $d$ electrons have  been done at $U=4$.

\begin{figure}[!t]
\begin{center}
\includegraphics[width=9cm,scale=1]{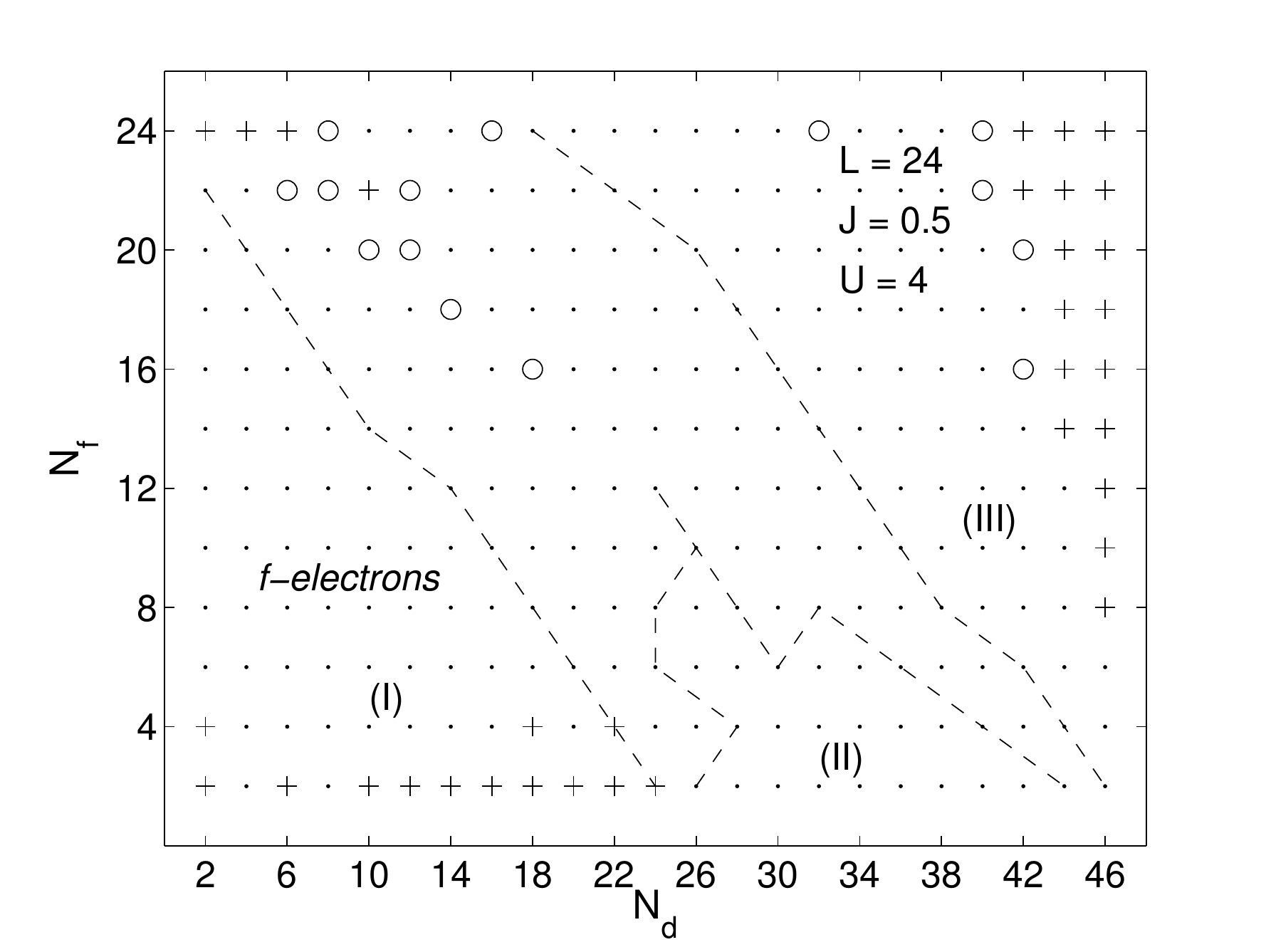}
\end{center}
\vspace*{-0.5cm}
\caption{ The $f$-electron skeleton phase diagram of the spin-1/2 Falicov-Kimball model extended by the spin-dependent interaction  calculated for $N_f\leqslant L, J=0.5$ and $L=24$. ($\cdot$): the NP phase, (+): the FP phase, ($\circ$): the PP phase. See the text for a definition of configuration types that are ground states in the areas I, II and III. The one-dimensional exact-diagonalization results~\cite{Farky29}.}
\label{ising_3}
\end{figure}
\begin{figure*}[!b]
\begin{center}
\includegraphics[angle=0,width=7.5cm,scale=1]{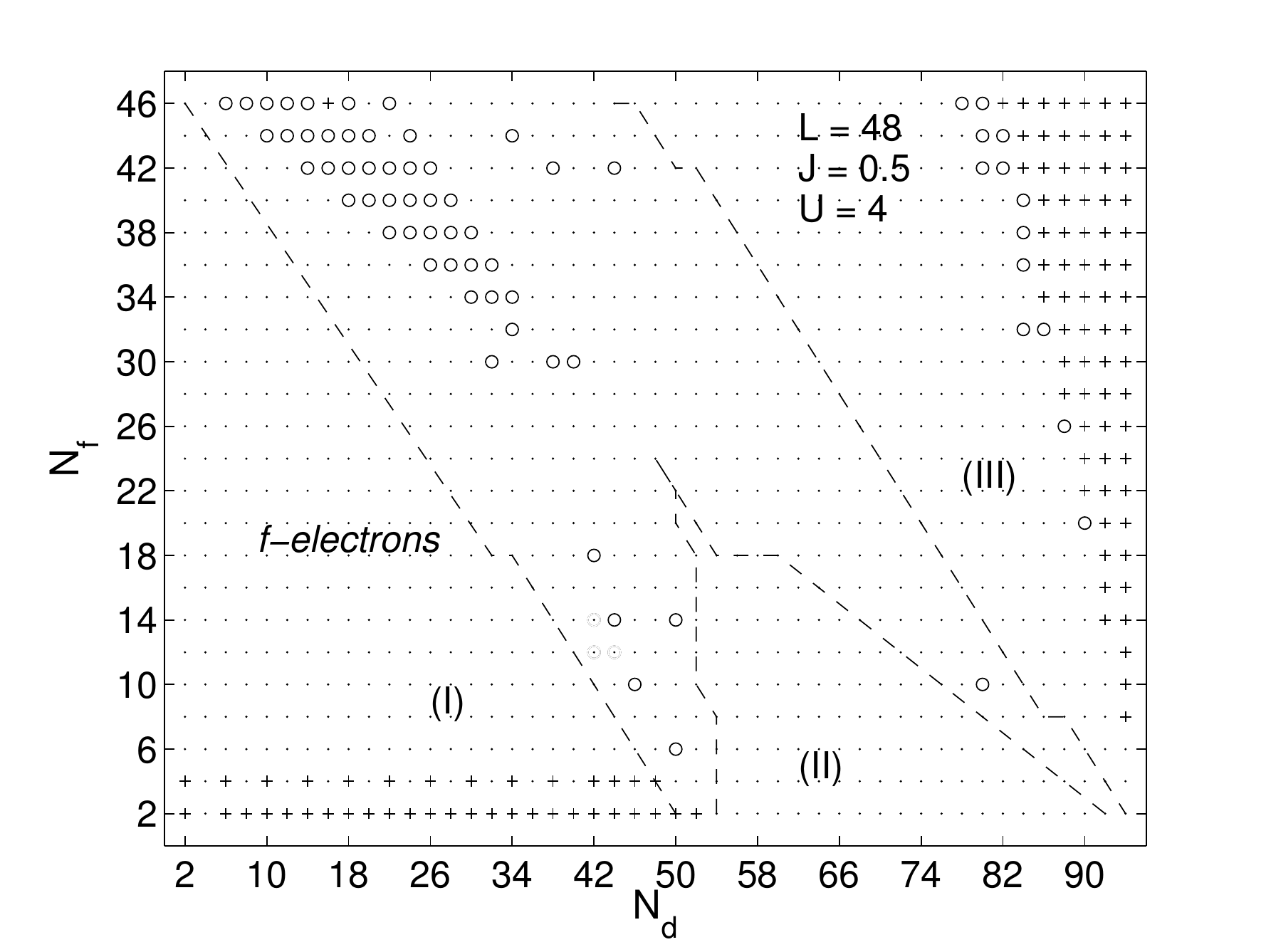}\hspace*{0.5cm}
\includegraphics[angle=0,width=7.5cm,scale=1]{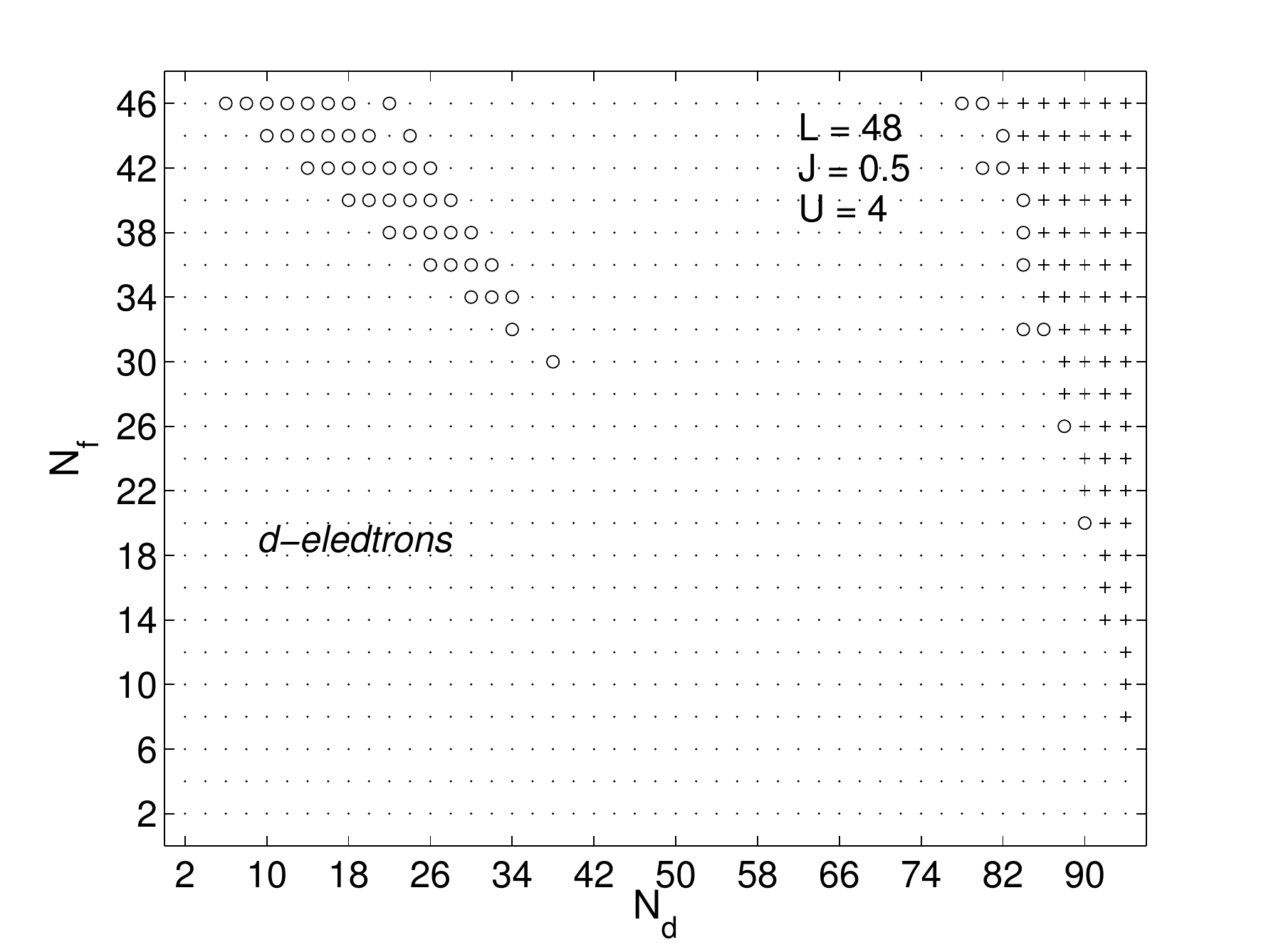}
\end{center}
\vspace*{-0.5cm}
\caption{ The $f$ and $d$-electron phase diagrams of the spin-1/2 Falicov-Kimball model
extended by the spin-dependent interaction  calculated for $N_f < L, J=0.5$ and $L=48$. ($\cdot$): the NP phase, (+): the FP phase, ($\circ$): the PP phase. The one-dimensional approximative results~\cite{Farky29}.}
\label{ising_4}
\end{figure*}

We have started the study of the effect of anisotropic, spin-dependent interaction between localized and itinerant electrons on
the ground-state properties of the model in the simplest case $N_f=L$ ($D=1$), which is accessible to  exact numerical calculations on relatively large clusters ($L=32$). The results of our numerical study are summarized in figure~\ref{ising_2} in the form of $N_d-J$  phase diagram.
Various    phases that enter  the phase diagram are classified according to $S^z_f=\sum_i(w^0_{i\uparrow}-w^0_{i\downarrow})$ and $S^z_d=N_{d\uparrow}-N_{d\downarrow}$: the fully polarized (FP) phase is characterized by $|S^z_f|=N_f$, $|S^z_d|=N_d$, the partially polarized (PP) phases are characterized by $0<|S^z_f|<N_f$, $0<|S^z_d|<N_d$ and the non-polarized (NP) phases  are characterized by $|S^z_f|=0$, $|S^z_d|=0$.

\begin{figure*}[!b]
\begin{center}
\includegraphics[angle=0,width=7.5cm,scale=1]{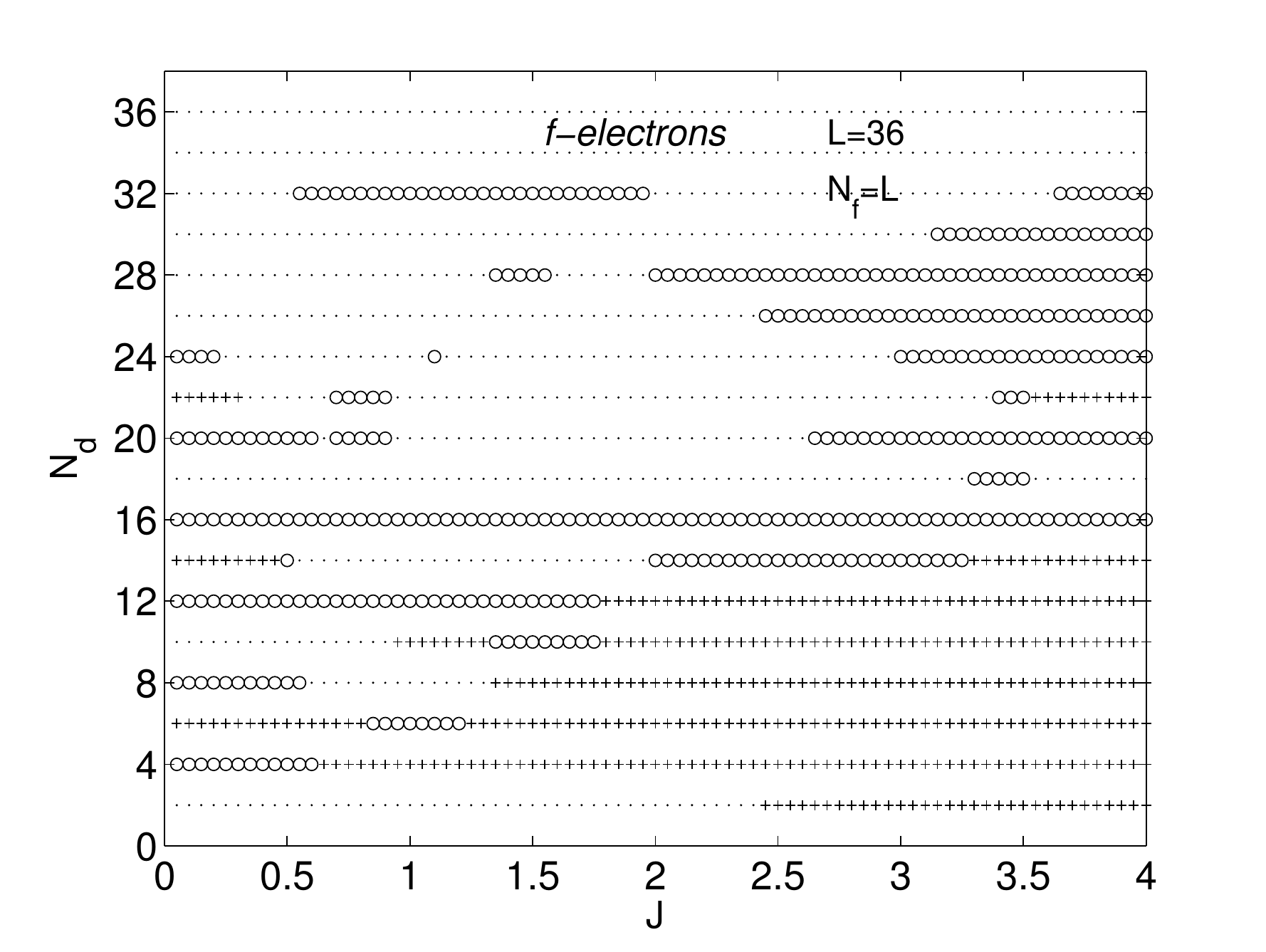}
\includegraphics[angle=0,width=7.5cm,scale=1]{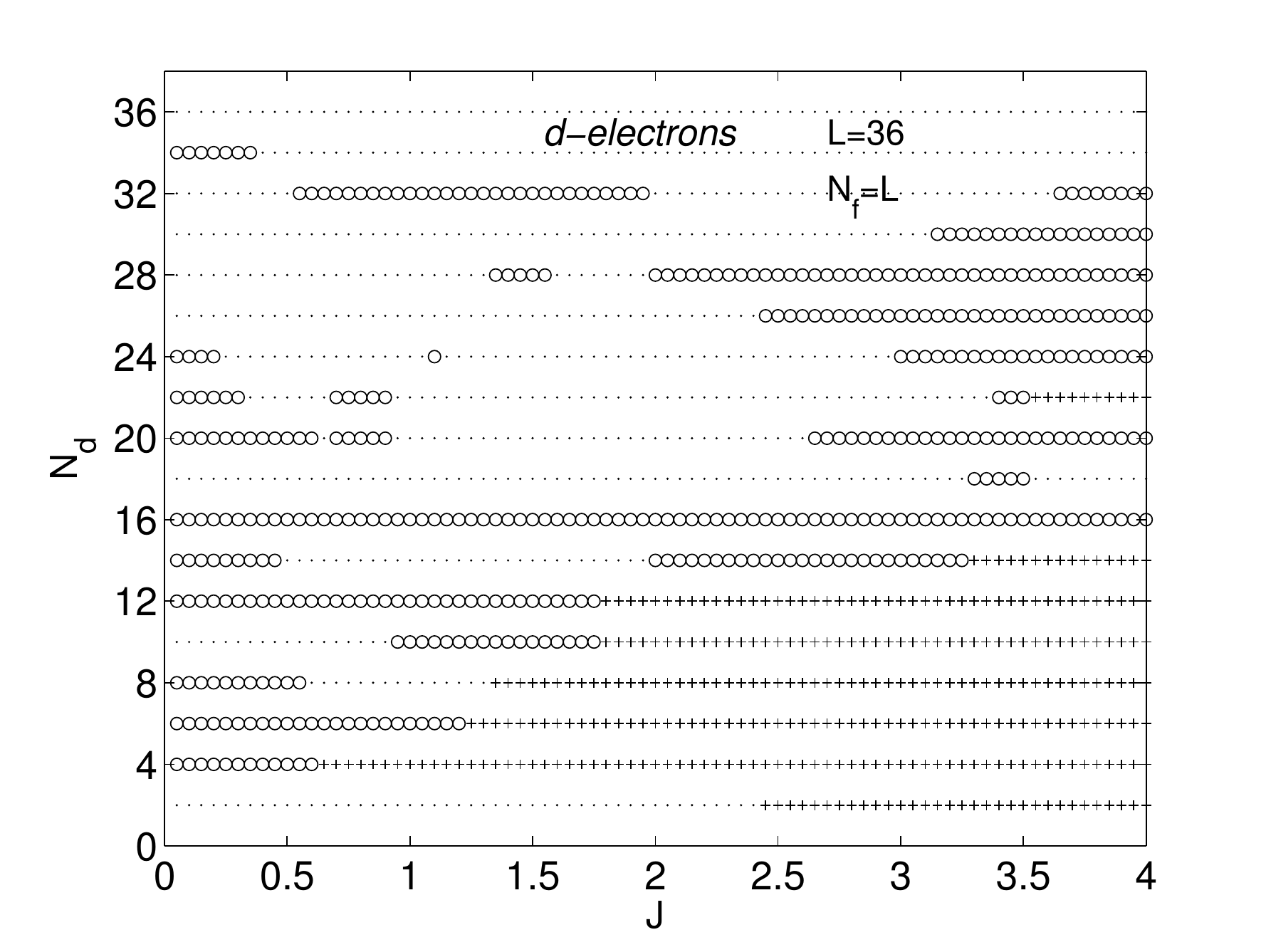}\\
\includegraphics[angle=0,width=2cm,scale=1]{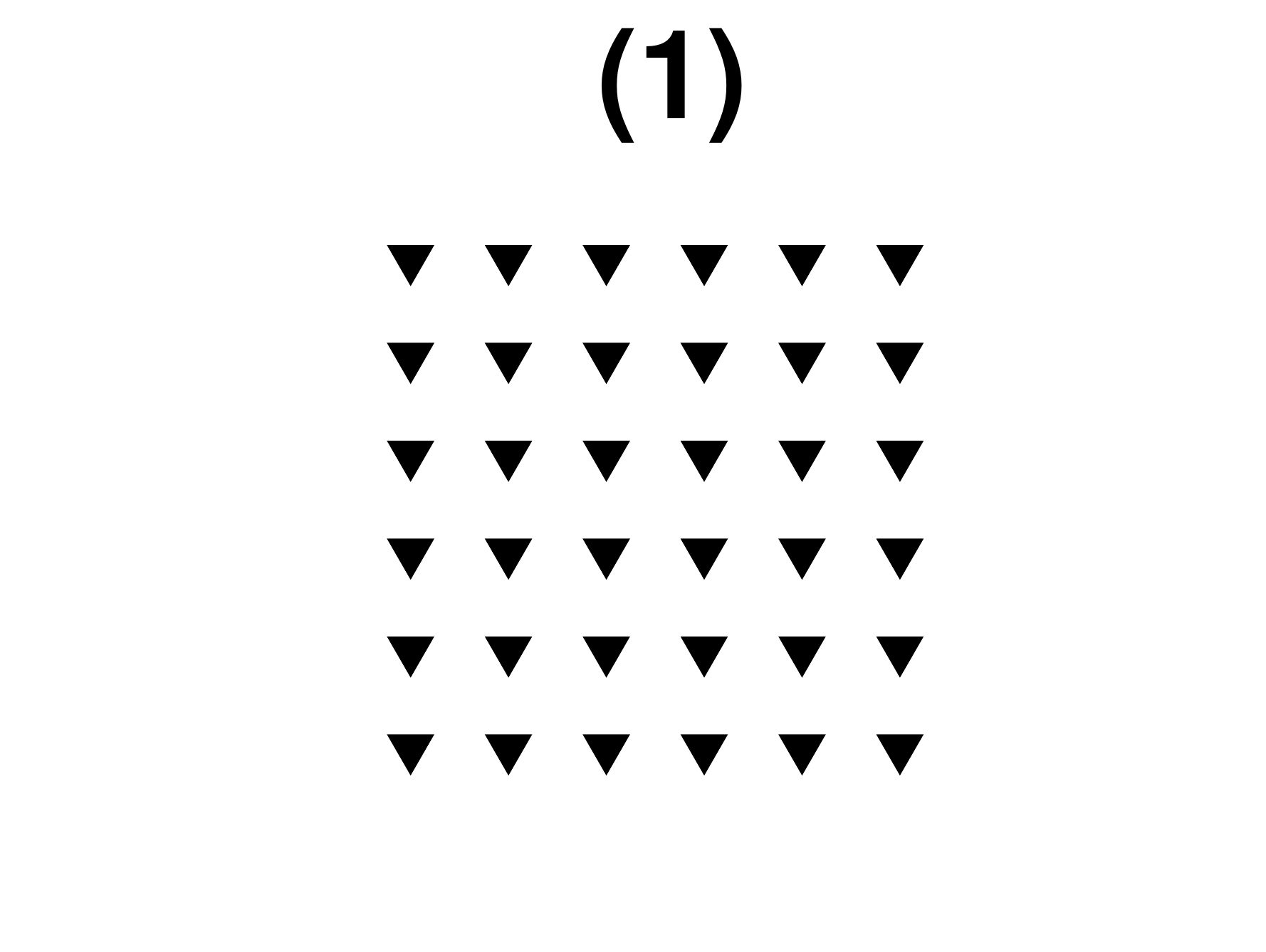}\hspace*{-0.5cm}
\includegraphics[angle=0,width=2cm,scale=1]{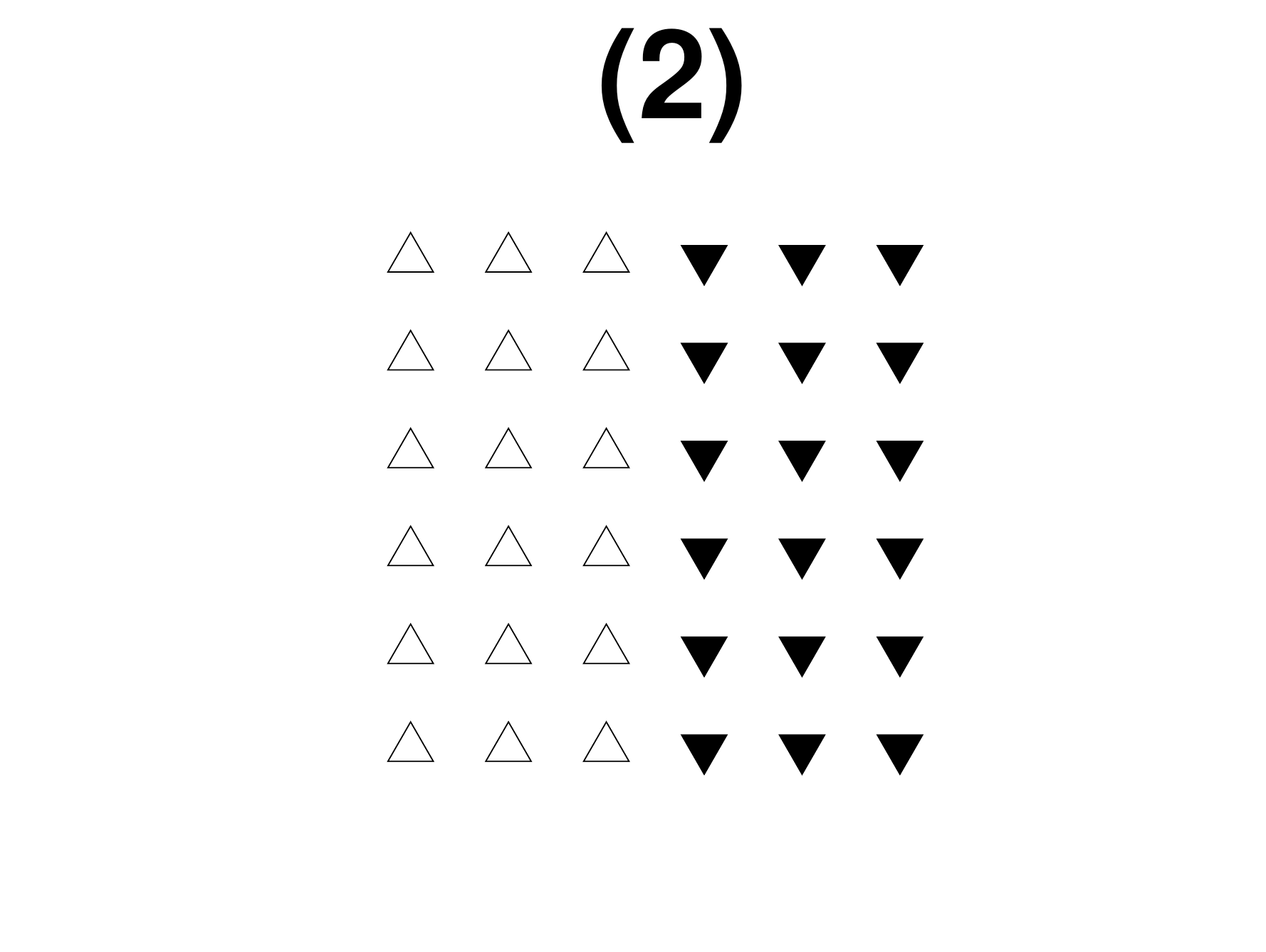}\hspace*{-0.5cm}
\includegraphics[angle=0,width=2cm,scale=1]{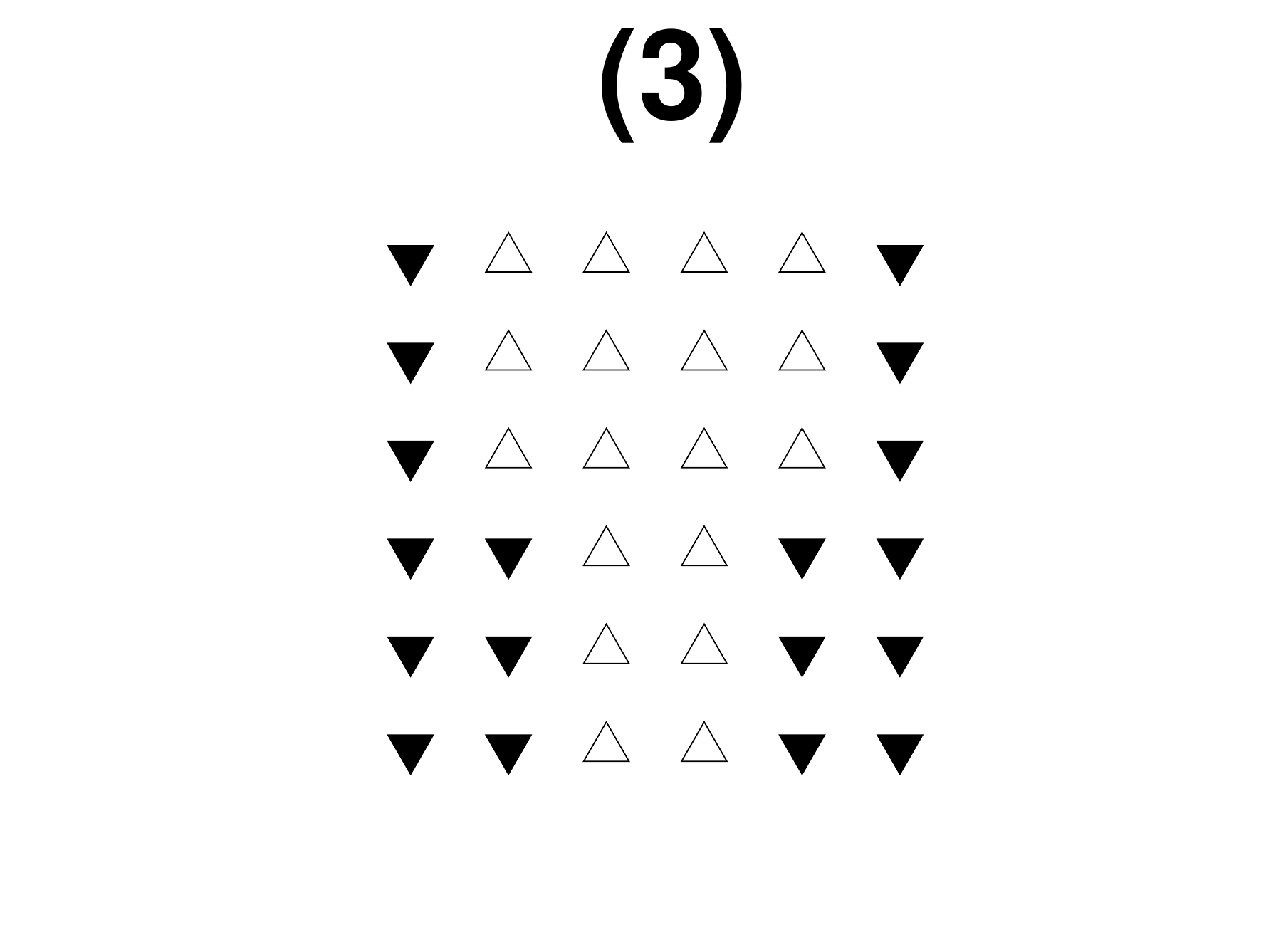}\hspace*{-0.5cm}
\includegraphics[angle=0,width=2cm,scale=1]{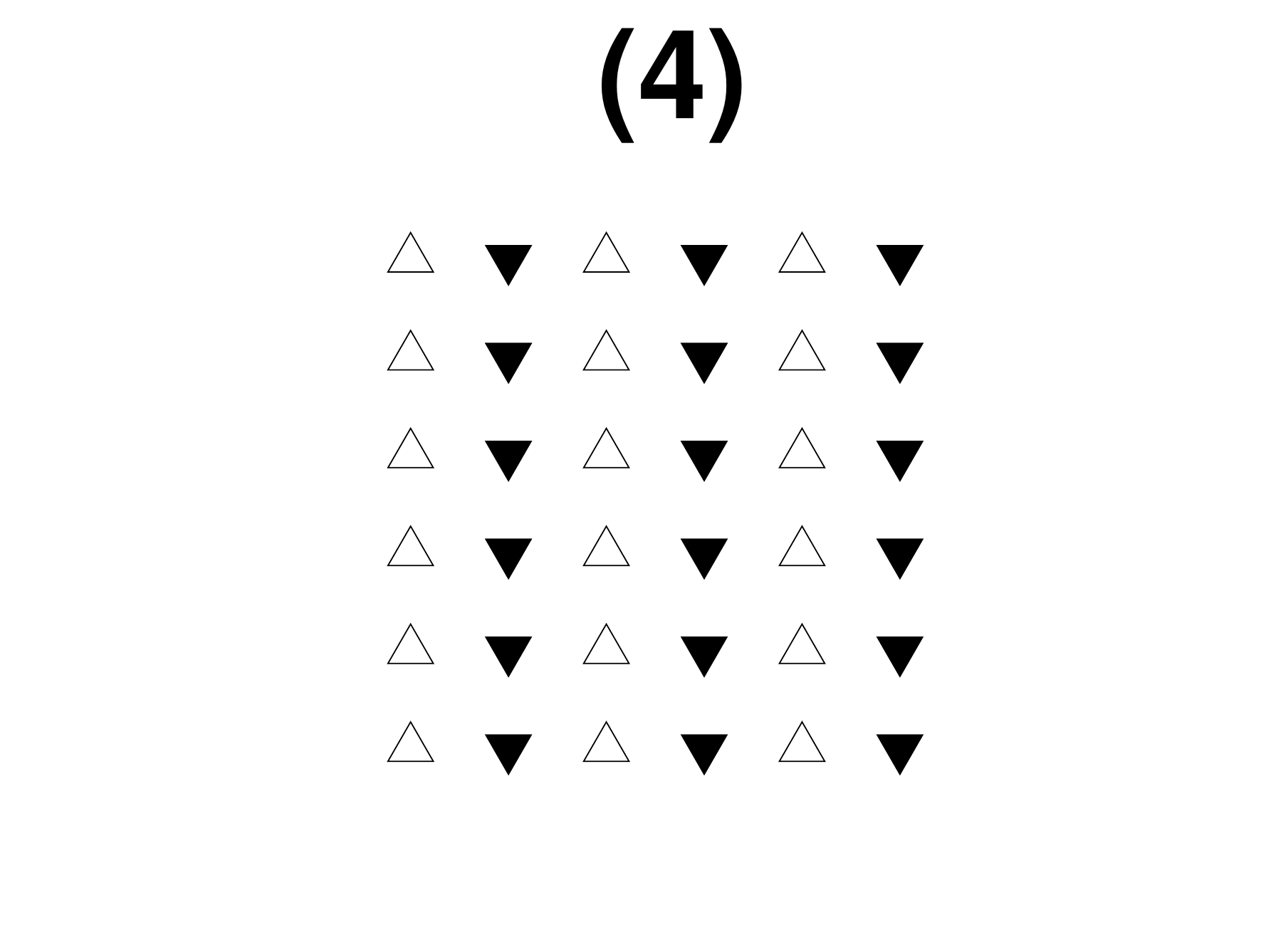}\hspace*{-0.5cm}
\includegraphics[angle=0,width=2cm,scale=1]{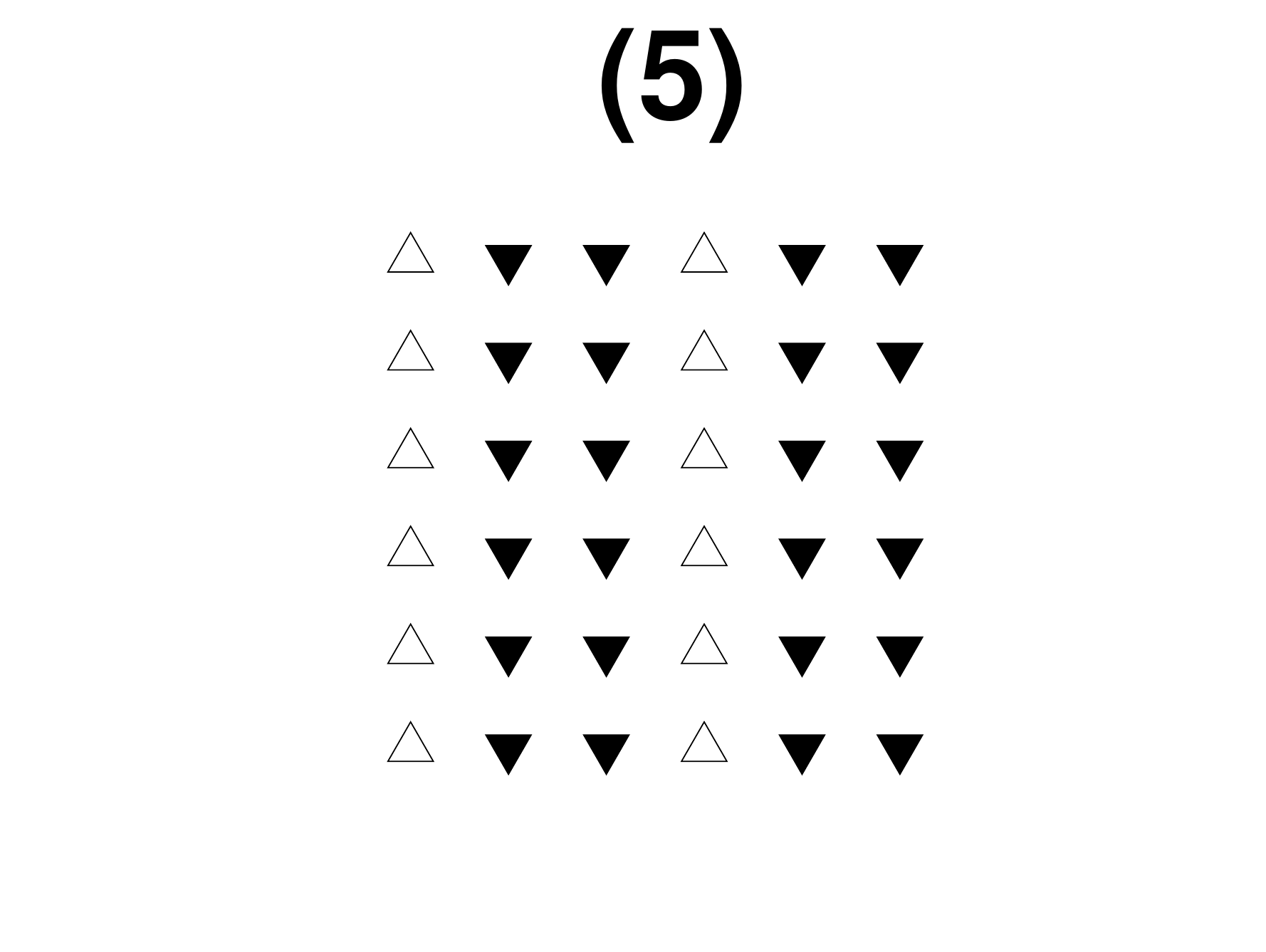}\hspace*{-0.5cm}
\includegraphics[angle=0,width=2cm,scale=1]{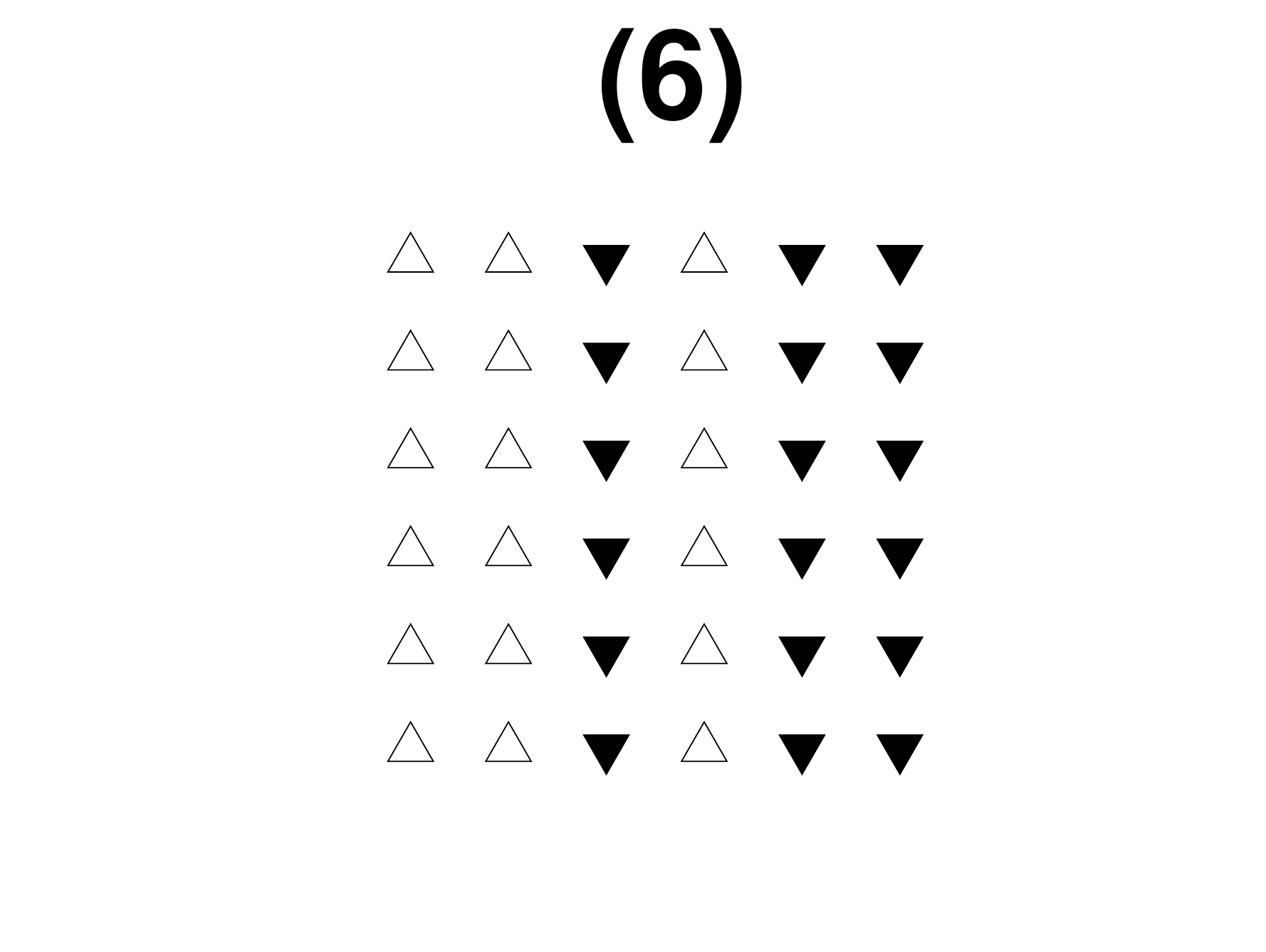}\hspace*{-0.5cm}
\includegraphics[angle=0,width=2cm,scale=1]{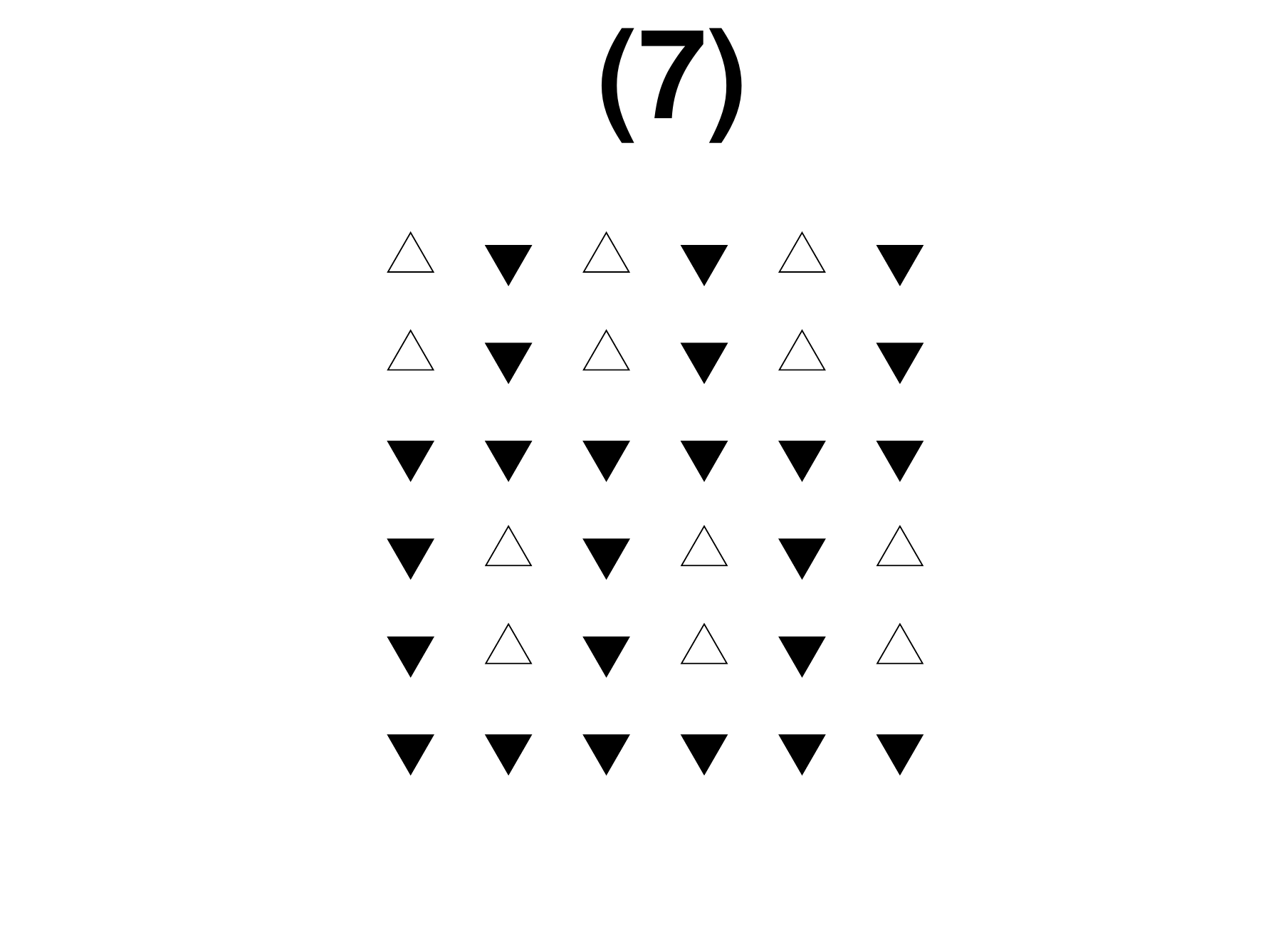}\hspace*{-0.5cm}
\includegraphics[angle=0,width=2cm,scale=1]{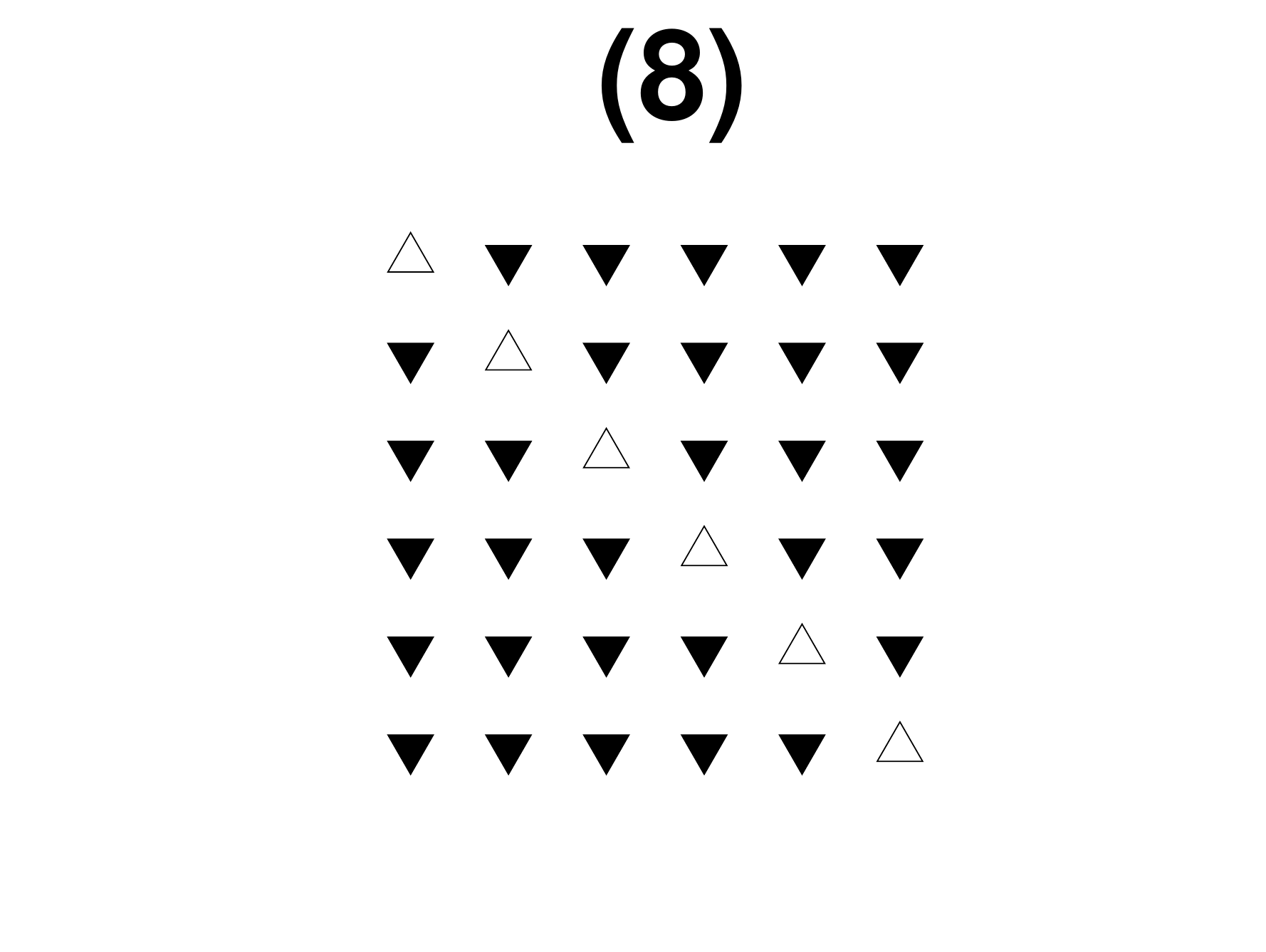}\\
\includegraphics[angle=0,width=2cm,scale=1]{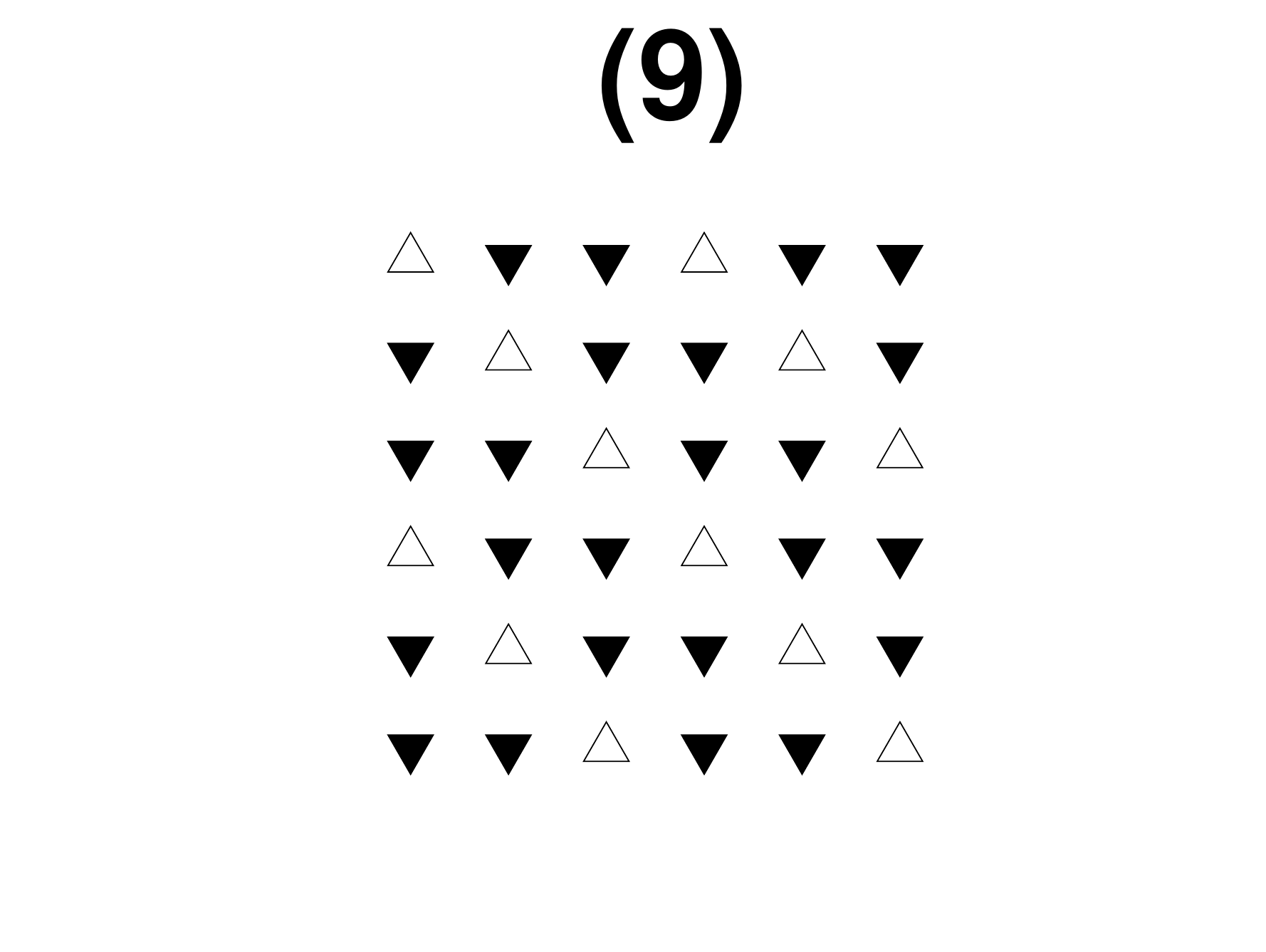}\hspace*{-0.5cm}
\includegraphics[angle=0,width=2cm,scale=1]{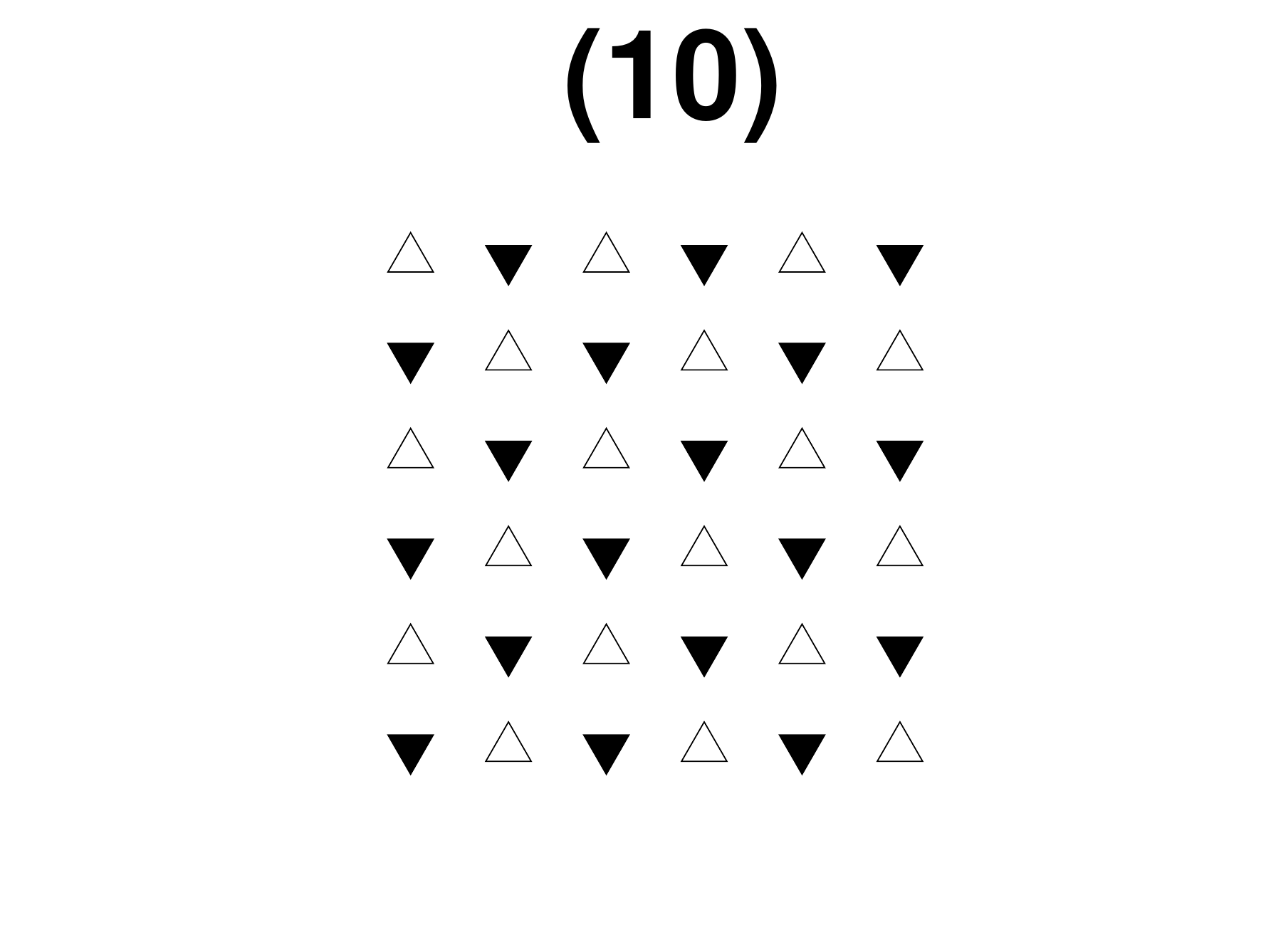}\hspace*{-0.5cm}
\includegraphics[angle=0,width=2cm,scale=1]{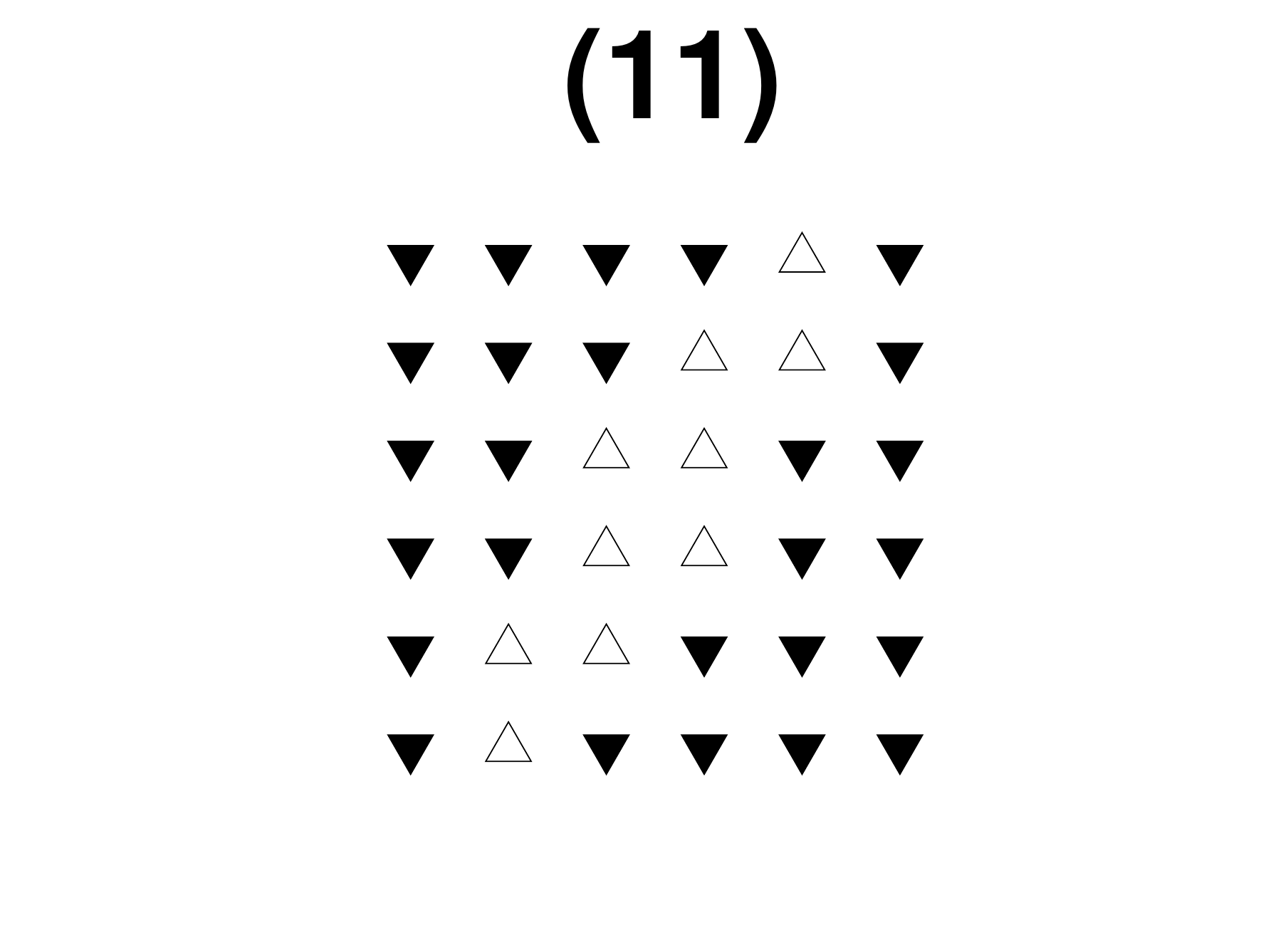}\hspace*{-0.5cm}
\includegraphics[angle=0,width=2cm,scale=1]{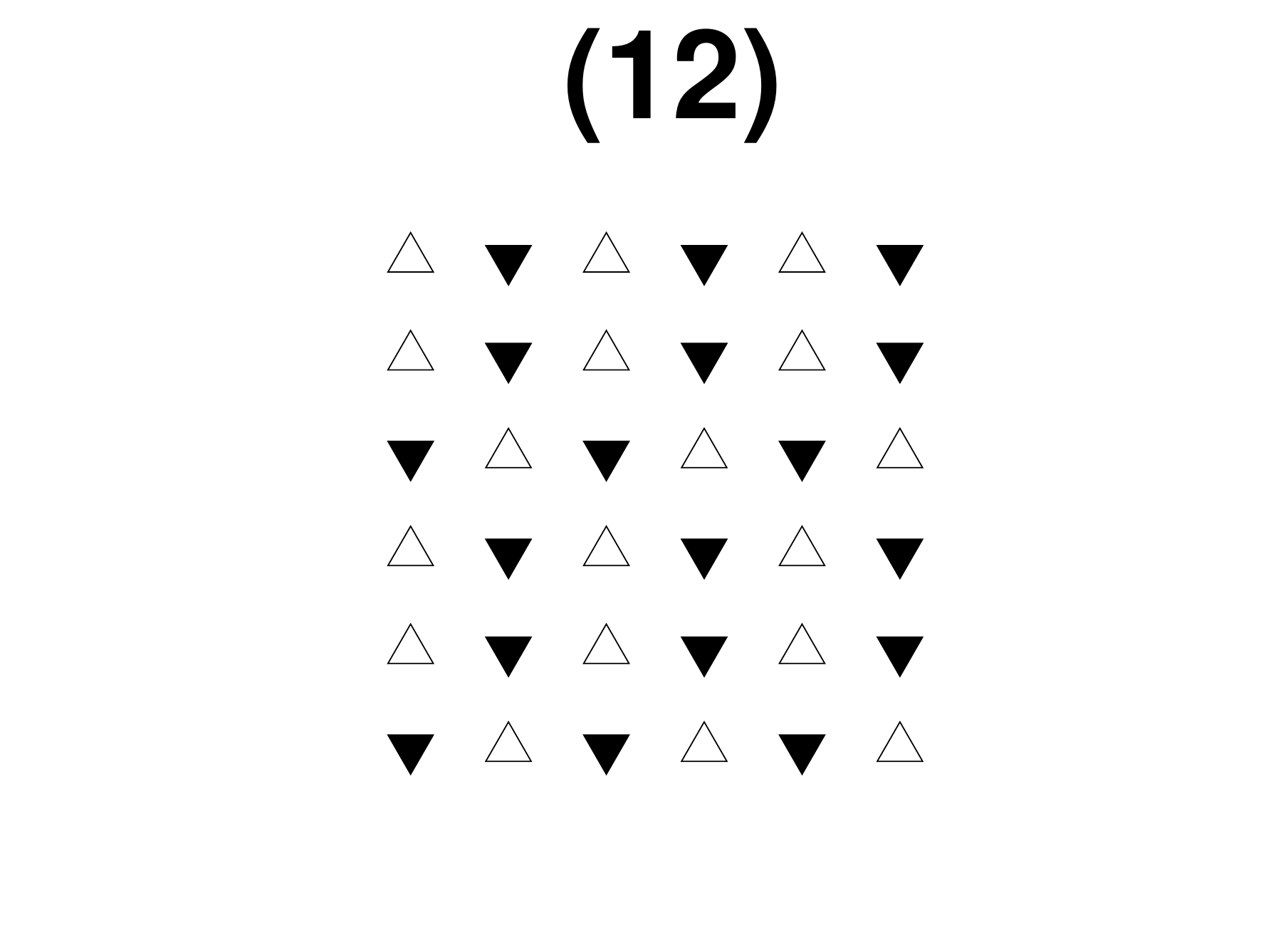}\hspace*{-0.5cm}
\includegraphics[angle=0,width=2cm,scale=1]{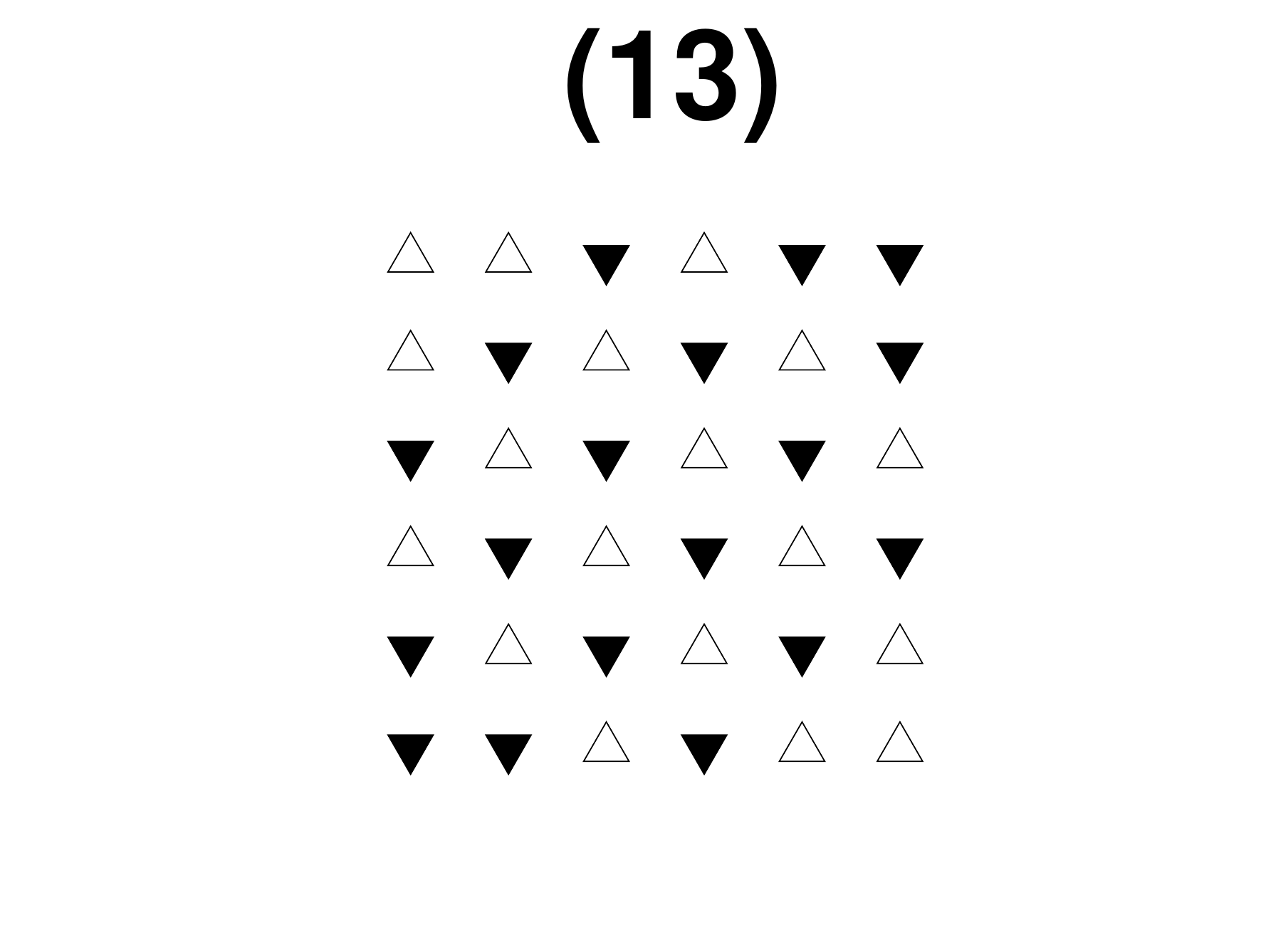}\hspace*{-0.5cm}
\includegraphics[angle=0,width=2cm,scale=1]{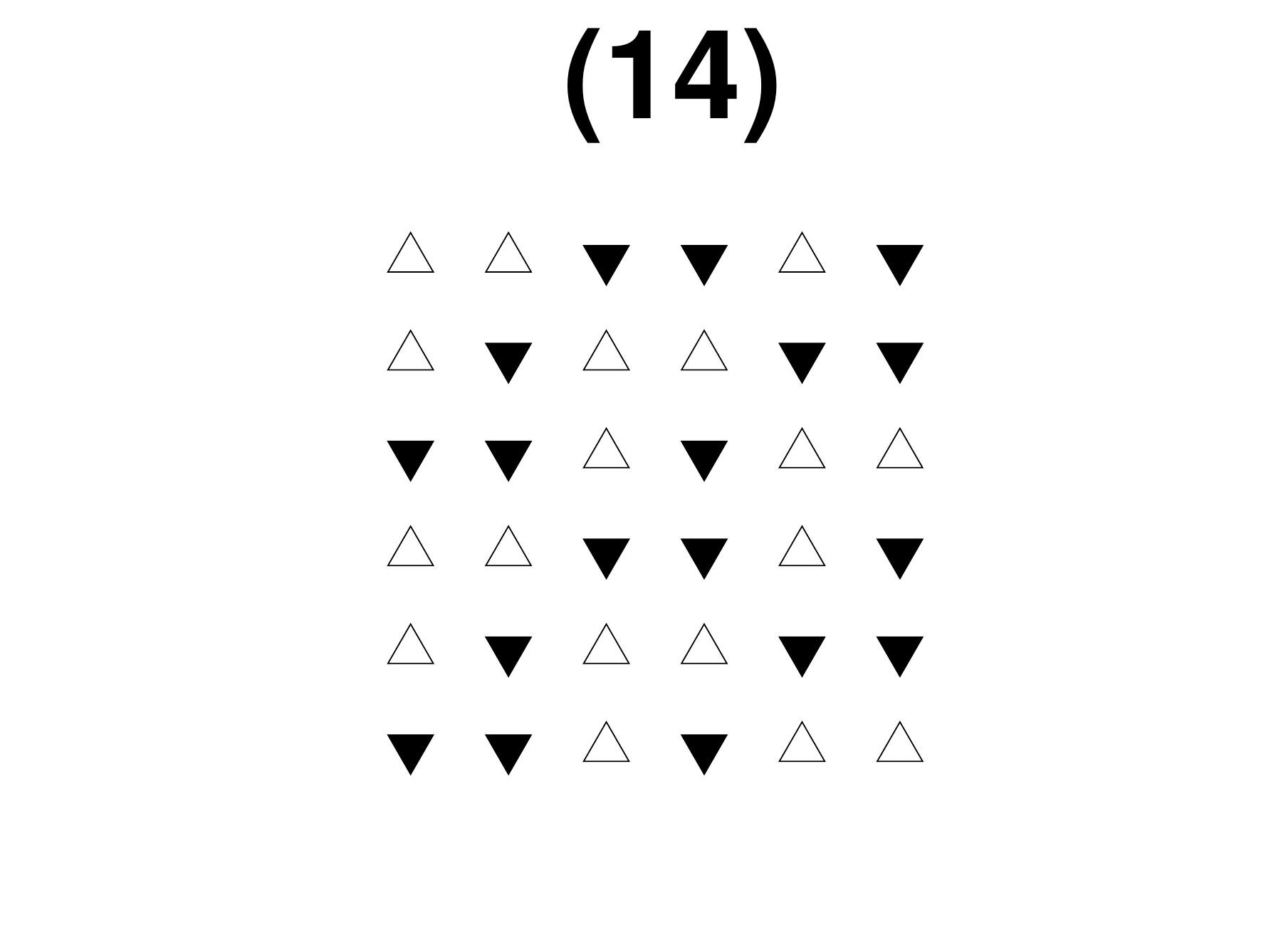}\hspace*{-0.5cm}
\includegraphics[angle=0,width=2cm,scale=1]{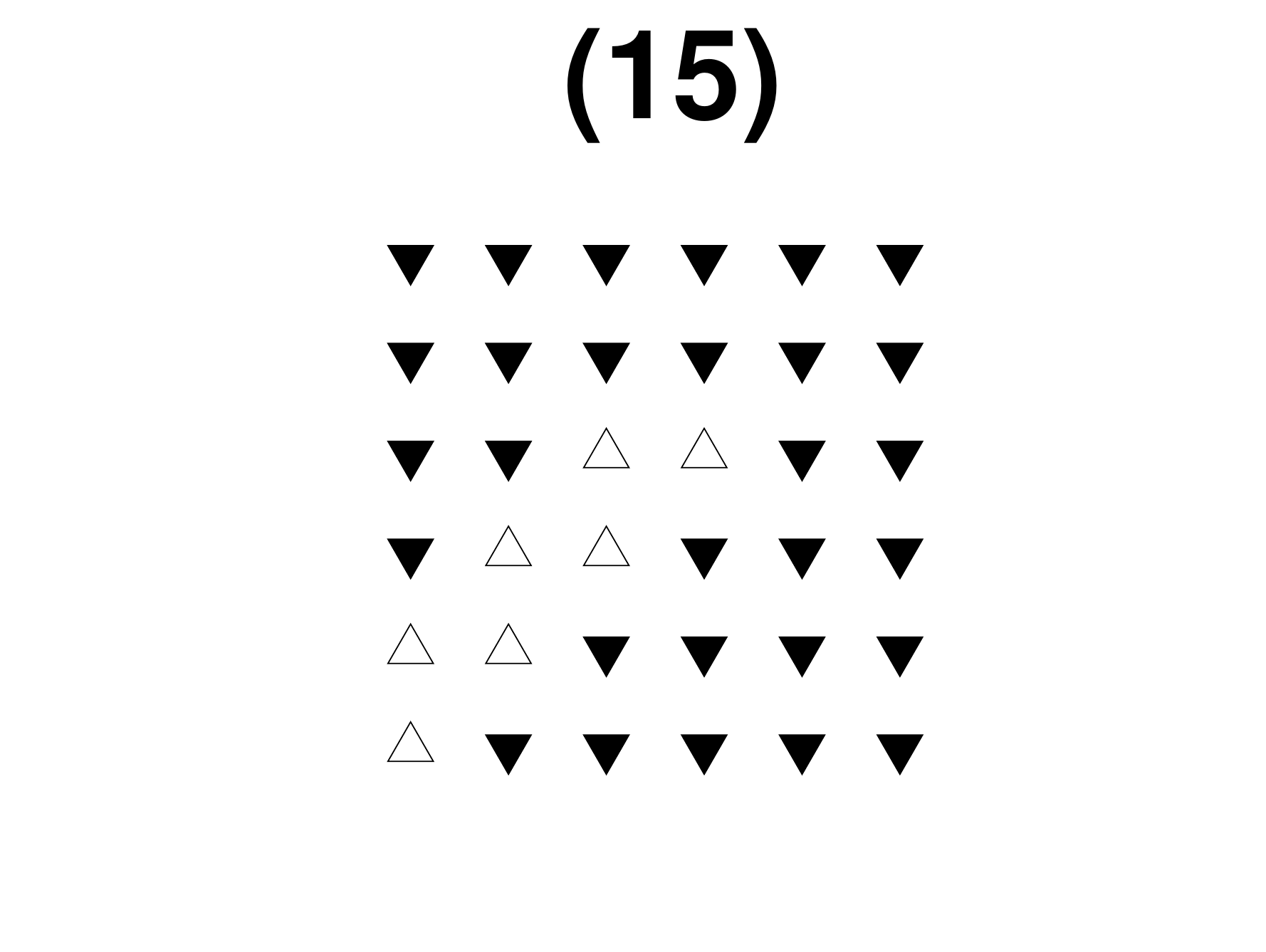}
\end{center}
\vspace*{-0.5cm}
\caption{ {\it Up}: The $f$ and $d$-electron ground-state phase diagrams of the spin-1/2 Falicov-Kimball model extended by the spin-dependent interaction  calculated for $N_f=L$ ($L=36$). The two-dimensional approximative results. {\it Down}: Typical examples of ground states of the spin-1/2 Falicov-Kimball model extended by the spin-dependent interaction  obtained for $N_f=L$ ($L=36$). To visualize the spin distributions we use $\bigtriangleup$ for the up spin electrons and $\blacktriangledown$ for the down spin electrons. The two-dimensional approximative results~\cite{Farky29}. }
\label{ising_5}
\end{figure*}

Comparing numerical results obtained for $|S^z_f|$ and $|S^z_d|$ one can find a nice correspondence between the magnetic phase diagrams of localized ($f$) and itinerant ($d$) subsystems. Indeed, with the exception of several isolated points at $J=0.05$, the corresponding FP, PP and NP phases perfectly coincide over the remaining part of diagrams showing the strong coupling between the magnetic subsystems of localized and itinerant electrons for nonzero values of $J$.

\begin{figure*}[!b]
\begin{center}
\hspace*{-0.35cm}
\includegraphics[trim = 0.5cm 1cm 0.5cm 0cm, clip,width=13cm]{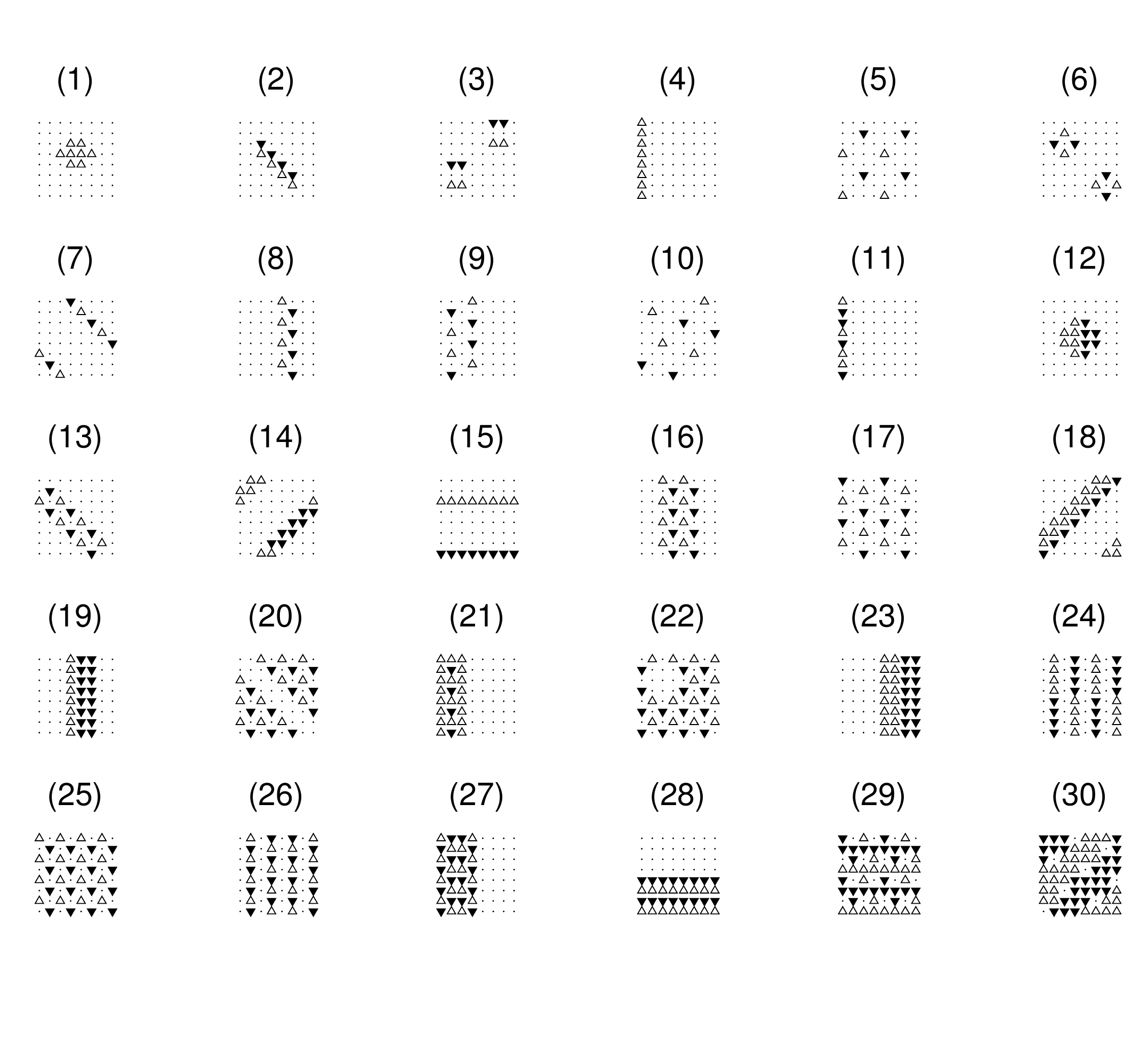}
\end{center}
\vspace*{-0.8cm}
\caption{ Typical examples of ground states of the spin-1/2 Falicov-Kimball model extended by the spin-dependent interaction  obtained for $N_f < L,J=0.5$ and $L=64$. The two-dimensional approximative results~\cite{Farky29}.}
\label{ising_7}
\end{figure*}

In general, the spin-dependent interaction $J$ stabilizes the FP and PP phases, while the NP phase is gradually suppressed with increasing $J$. In figure~\ref{ising_2} we present a complete set of ground-state  configurations  from the NP region. Among them one can find different types of periodic and nonperiodic configurations, but the most interesting examples represent configurations formed by antiparallel ferromagnetic domains, that  convincingly illustrate the cooperative effects of  spin-dependent interaction $J$ between the localized and itinerant electrons.

\begin{figure}[!b]
\begin{center}
\includegraphics[width=10.50cm]{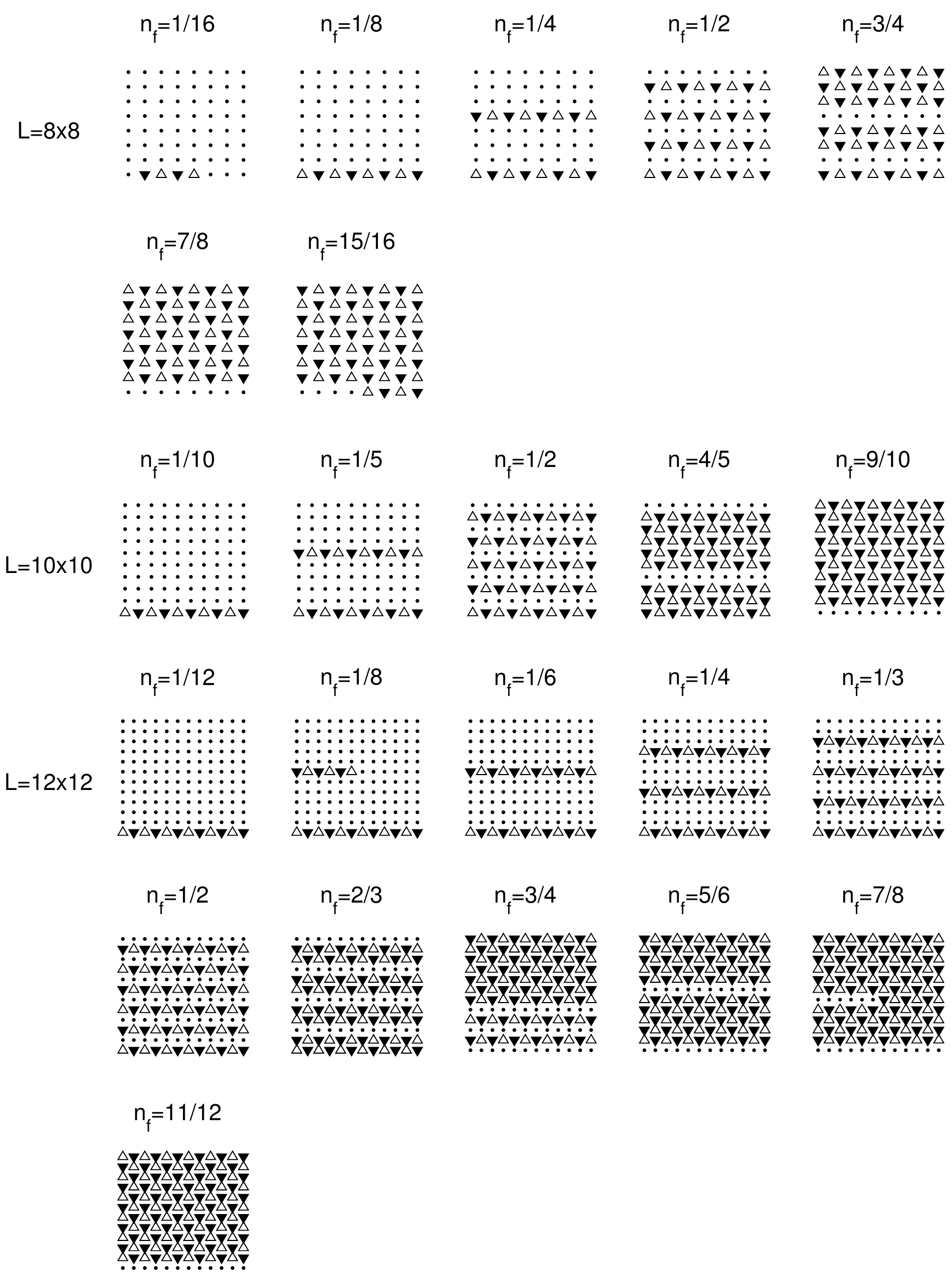}
\end{center}
\caption{ Typical ground-state configurations of the two-dimensional generalized
Falicov-Kimball model obtained for selected values of $n_f$ on finite clusters of $L=8\times 8$,
$L=10\times 10$ and  $L=12\times 12$ sites at
$U=4$, $J=0.5$ and $n_f+n_d=2$~\cite{Farky40}.}
\label{pss245_3}
\end{figure}
As the next step, we have performed numerical studies of the model for the case   $N_f\neq L$.  From the numerical point of view this case is considerably exact, since now we have to minimize the ground-state energy not only over all different spin configurations but also over all different $f$-electron distributions.
This takes a considerable amount of CPU time and for this reason we were able to investigate exactly only the clusters up to $L=24$. We have found that  the spectrum of magnetic solutions that yields the model for the NP and PP phases is very rich.
Indeed, for $L=24$ and $J=0.5$ we have found 140 different NP phases and 20 different PP phases that enter the $N_f-N_d$ phase diagram. The main configuration types, with the largest stability regions are presented in figure~\ref{ising_3} in the form of a skeleton phase diagram.
From the NP phases, the largest stability region (denoted by I) corresponds to configurations of the type $\uparrow_n\downarrow_n0_{L-2n}$. The second largest region (denoted by II) corresponds to NP configurations of the type $[\uparrow0_n\downarrow0_n]_k0_{L-2k(n+1)}$. Typical examples of the NP ground states from the central region of the phase diagram represent periodic configurations of the type $\uparrow_n0_{m}\downarrow_n0_{m}$ (below the main diagonal) and configurations of the type $\uparrow_2[\uparrow\downarrow]_{k_1}\downarrow_20_m[\downarrow 0_p\uparrow 0_p]_{k_2}$, or $\uparrow_2[\uparrow\downarrow]_{k_1}\downarrow_20_m[\downarrow 0_p\uparrow 0_{p-1}]_{k_2}$, above the main diagonal. Between these configurations and the FP region, the ground states are the segregated configurations of the type $\uparrow_2[\downarrow\uparrow]_k\downarrow_20_{m}$, or $[\uparrow\downarrow]_{k_1}\uparrow_2[\downarrow\uparrow]_{k_2}\downarrow_2 [\uparrow\downarrow]_{k_3}0_{m}$, or their modifications (the region denoted by III). In the PP region, the typical examples of ground states represent configurations of the type $\uparrow_n[0\downarrow0\uparrow]_k0\downarrow0$.

In order to minimize the finite-size effects we have performed the same study using our AM on two times larger clusters ($L=48$). Our numerical results showed that all the main results obtained on small clusters  also hold on larger clusters. Again we have observed a strong coupling between two magnetic subsystems and a coincidence of corresponding magnetic phases, that stability regions are practically unchanged with increasing $L$ (figure~\ref{ising_4}).

\begin{figure}[!b]
\begin{center}
\includegraphics[width=7.0cm]{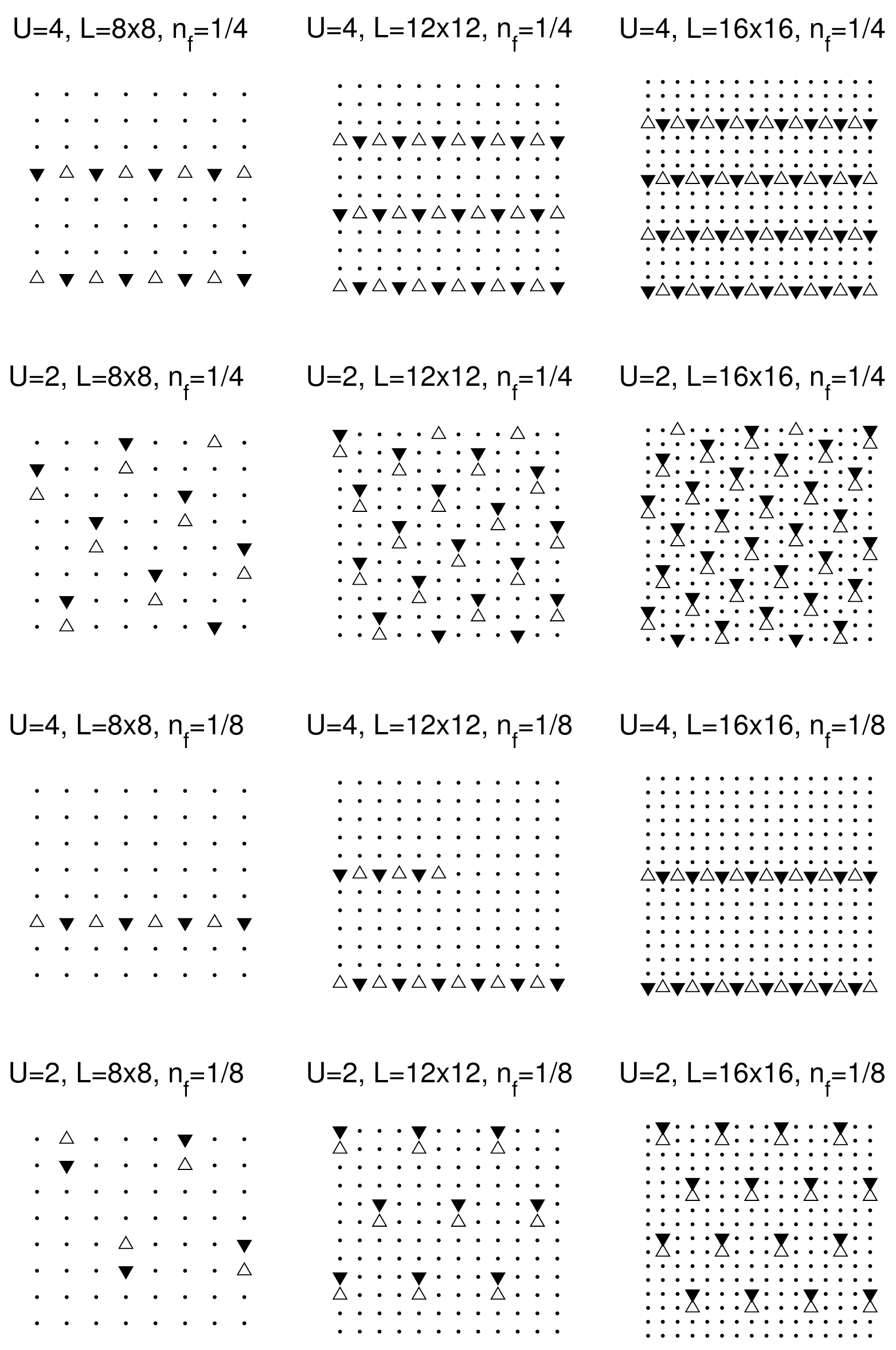}
\caption{ Ground-state configurations of the two-dimensional generalized Falicov-Kimball model
obtained for different $U$ ($U=4$ and $U=2$) and $n_f$ ($n_f=1/4$, $n_f=1/8$
at $n_f+n_d=2$)~\cite{Farky40}.}
\label{pss245_4}
\end{center}
\end{figure}

However, from the experimental point of view, the two-dimensional case is much more  interesting. Numerical results obtained for this case and $N_f=L$ are displayed in figure~\ref{ising_5}.
Similarly to the one-dimensional case, the basic structure of the phase
diagram in $D=2$ is formed by three large  FP, PP and NP domains that are accompanied by secondary phases (points in which the $d$ and $f$-electron phase diagrams do not coincide). However, while in the one dimension the secondary phases are stable only in isolated points at very small values of $J$, in two dimensions these secondary phases  also persist for large $J$. Calculations that we have performed on different clusters ($4 \times 4, 6 \times 6$, and $8 \times 8$) showed that the secondary structure depends very strongly on the cluster size and with increasing $L$ it is gradually suppressed.

The typical examples of ground-state configurations (that represent the most frequently appearing types of ground states in the $N_d-J$ phase diagram) are displayed in figure~\ref{ising_5}.  Again one can see that the spectrum of magnetic solutions that yields the Falicov-Kimball model extended by spin-dependent interaction is very rich.
In addition to the FP phase (that the stability region shifts to higher $d$-electron concentrations when $J$ increases) there are various types of NP and PP structures like the antiparallel ferromagnetic domains (2--3), the axial magnetic stripes (4--7), the diagonal magnetic stripes (8--11) and the perturbed diagonal magnetic stripes (12--15). This again demonstrates strong effects of the spin-dependent interaction on the formation of magnetic superstructures in the extended Falicov-Kimball model and its importance for a correct description of correlated electron systems.

Finally, we have performed the same numerical study for the case $N_f<L$. The typical examples of the resulting charge and spin ordering are displayed in figure~\ref{ising_7}.
Among them one can find various types of phase-segregated (1), phase-separated (16) and n-molecular (3) configurations with FP, PP and NP ground states as well as various types of axial (9) and diagonal (5) magnetic/charge stripes. In general, we have observed that the system shows tendency towards phase segregation for small and large $d$-electron concentrations, while near the $n_d=1$ point the system prefers to form various types of axial and diagonal stripes.

For  the most physically interesting cases  $N_f+N_d=L$ and $N_f+N_d=2L$ we have also performed  a detailed analysis of charge and magnetic ordering.
We have started our  study with the
case $N_f+N_d=2L$, which is slightly simpler for a description. As shown in figure~\ref{pss245_3} the ground states for $N_f+N_d=2L$ are
antiferromagnetic (AF) with alternating pattern, where the electrons
(for $n_f<1/2$) or holes (for $n_f>1/2$) form axial distributions.
Our calculations showed that these inhomogeneous stripe distributions are
stable for large Coulomb interactions, while decreasing $U$ leads to their
destruction and prefers a homogeneous electron arrangement (see figure~\ref{pss245_4}).

%\pagebreak

\begin{figure}[!h]
\begin{center}
\includegraphics[width=12.0cm]{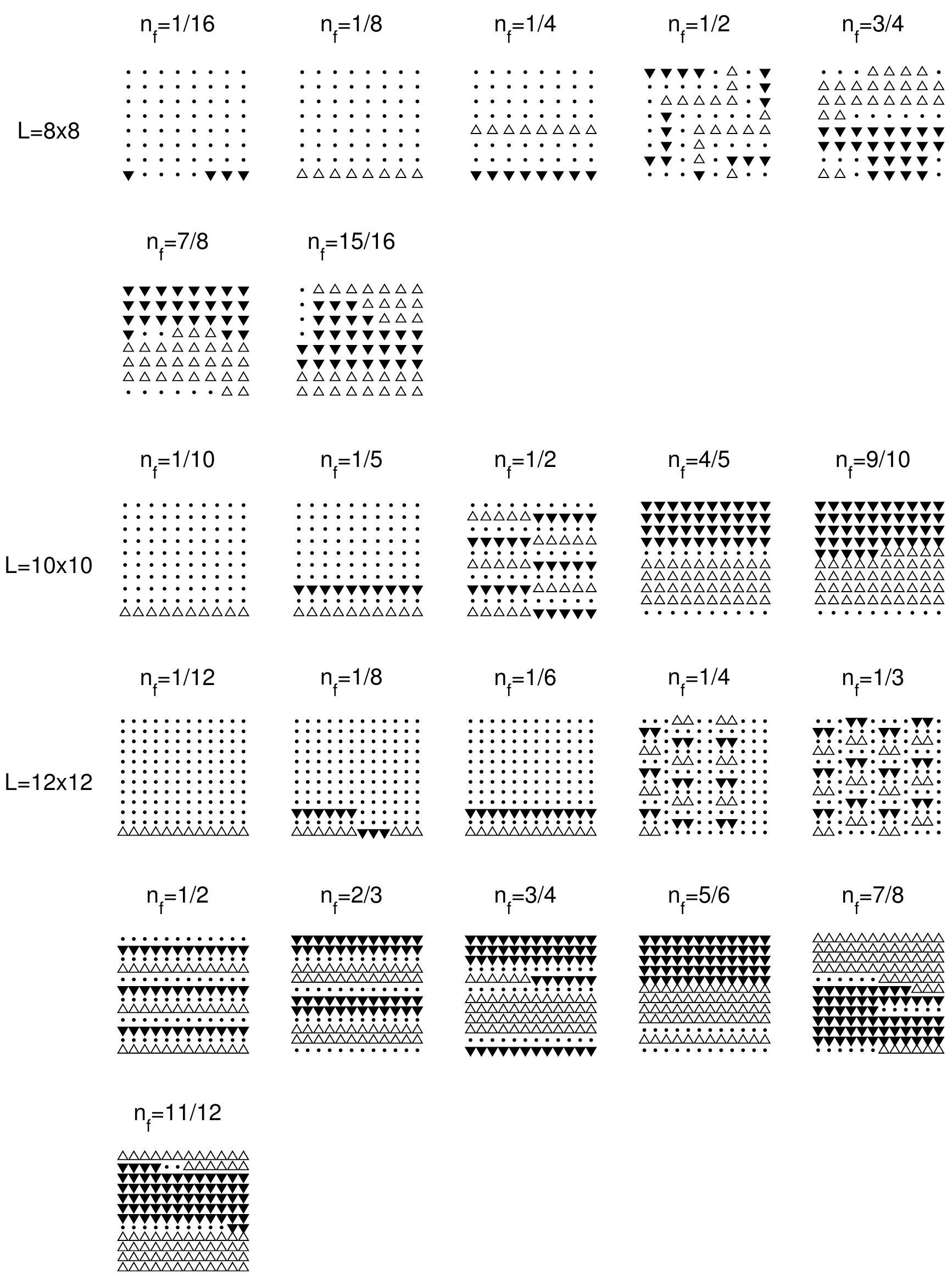}
\caption{
Typical ground-state configurations of the two-dimensional generalized
Falicov-Kimball model obtained for selected values of $n_f$ on finite clusters of $L=8\times 8$,
$L=10\times 10$ and  $L=12\times 12$ sites at
$U=4$, $J=0.5$ and $n_f+n_d=1$~\cite{Farky40}.}
\label{pss245_5}
\end{center}
\end{figure}

For $N_f+N_d=L$ the situation is fully different. The first fundamental difference is that
for sufficiently small $f$-electron concentrations $n_f$ the ground state
could be ferromagnetically (F) ordered (see figure~\ref{pss245_5}).

The second difference is that
although with increasing $f$-electron concentration the ground states are
the AF, these AF arrangements are formed by F ordered clusters (domains).
In  addition, for $n_f=1/4$ and $n_f=1/3$ ($L\geqslant 144$) a new type of stripes
(known as the ladders) occurs. And finally, a detailed analyses showed that
there exists a critical $f$-electron concentration $n_f^{\rm c}\sim 1/4$ below
which the ground-states are phase separated.

%\pagebreak

\section{Applications of the Falicov-Kimball model to a description of real\\ materials}
\label{Applications of the Falicov-Kimball model for a description of charge and spin ordering in real systems}

\subsection{Ground-state properties of Na$_x$CoO$_2$}
\label{Ground-state properties of $Na_xCoO_2$}

\begin{figure*}[h!]
\vspace*{0.3cm}
\begin{center}
\includegraphics[width=6.5cm,angle=0]{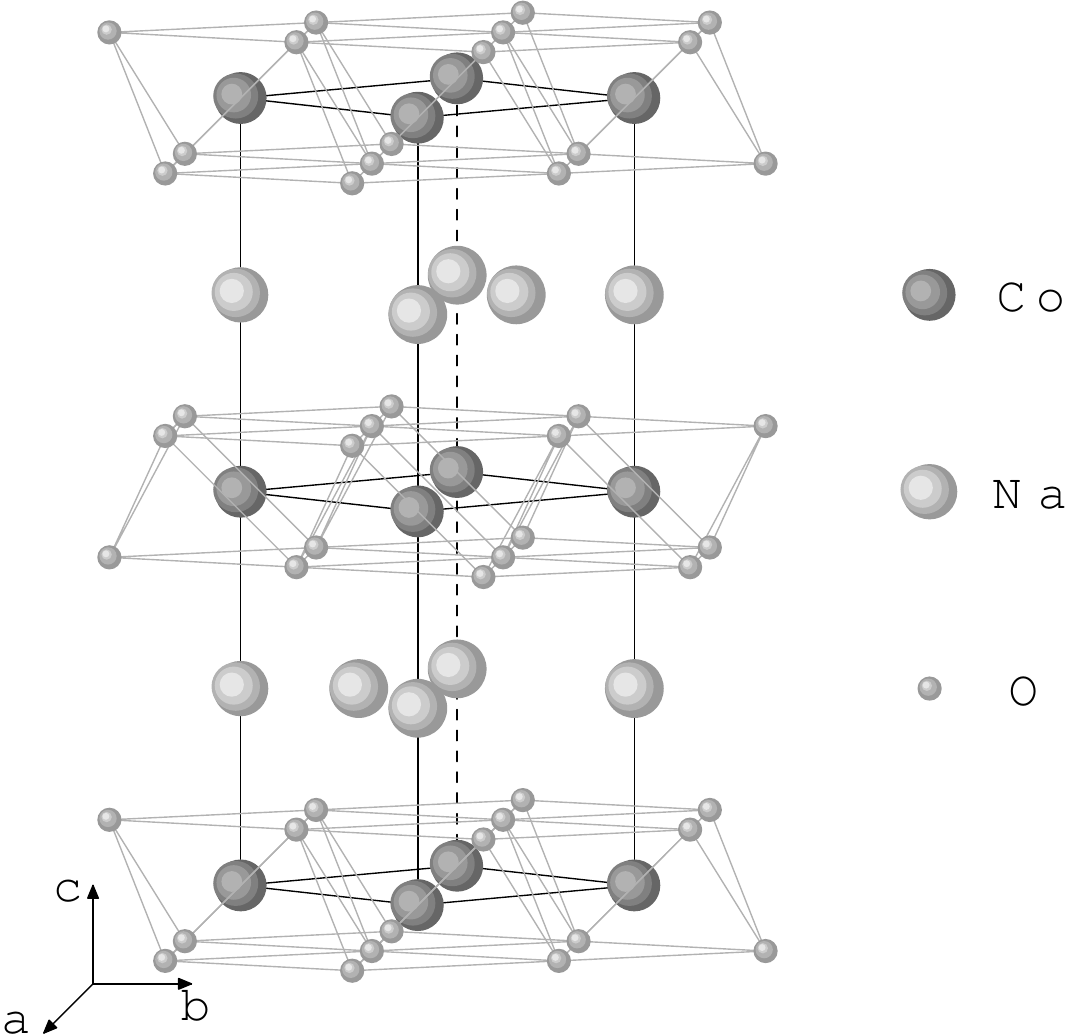}\hspace*{2cm}
\includegraphics[width=5.cm,angle=0]{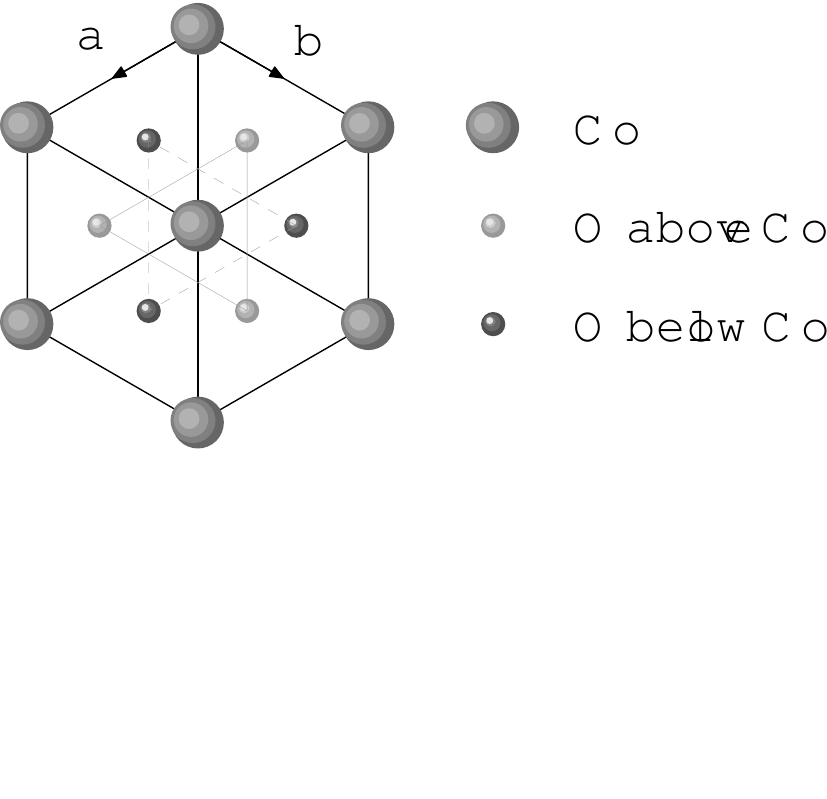}
\end{center}
\vspace*{-0.3cm}
\caption{ Crystal structure of Na$_x$CoO$_2$ (left hand panel). The (a)--(b) planes, showing the orientation of the oxygen tetrahedron around a central Co ion (right hand panel).}
\label{o_nax}
\end{figure*}

The Na$_x$CoO$_2$ system, which forms the basis  for a quasi-two-dimensional transition metal oxide superconductor ($T_{\rm c}=4.5$~K) when hydrated~\cite{Takada} shows a wide variety of unexplained behaviours in the accessible range $0<x<1$.  It consists of alternate stacks of electronically active triangular CoO$_2$ layers (with edge sharing CoO$_6$ octahedra) separated by Na layers that act not only as spacers, leading to electronic two-dimensionality (figure~\ref{o_nax}), but also as charge reservoirs~\cite{Foo1, Lynn, Foo2}.
With increasing Na content $x$, the phase diagram nonhydrated Na$_x$CoO$_2$ system exhibits a succession of ground states~\cite{Foo2} changing  from a paramagnetic metal at $x=0.3$ through a charge-ordered insulator at $x=0.5$, a ``Curie-Weiss metal'' for $x=0.75$, and finally a magnetically ordered state for $x>0.75$. At $x=0.5$, detailed electron and neutron diffraction measurements~\cite{Zandbergen, Huang}  revealed an ordering of the Na ions along one crystallographic direction which decorates the chains of Co ions with different amounts of charge~\cite{Huang}.

It is suggested that there are strong correlations between the Na ions and electrons (holes) from the CoO$_2$ layers, even though they occupy separate layers, and that a charge ordering in the CoO$_2$ layers is induced by sodium ion ordering. Unfortunately, numerical calculations performed at the Density Function Theory (DFT) and DFT+U level by Li et al.~\cite{Li} did not confirm this conjecture. In the spirit of these results we have tried to describe the conducting properties and a formation of charge ordering in the CoO$_2$ layers within a relatively simple model that takes into account both the interplane interactions between the ordered Na ions and electrons (holes) form the CoO$_2$ planes as well as the intraplane interaction between electrons within CoO$_2$ layers~\cite{Farky36}. The system is modelled by the spinless Falicov-Kimball Hamiltonian that automatically projects out the states with two holes on the $3d$ ($d_{z^2}$) orbitals. Thus, our starting Hamiltonian can be written as a sum of three terms:
\begin{eqnarray}
H=\sum_{\langle ij\rangle}t_{ij}d^+_{i}d_{j} - \sum_i\omega_i d^+_id_i+ V\sum_{\langle ij\rangle}n_in_j\,.
\label{omega}
\end{eqnarray}
The first term of (\ref{omega}) describes hopping of $3d$-electrons on the triangle network of Co$^{4+}$ ions. These intersite hopping transitions are described by the matrix elements $t_{ij}$ which are $-t$ if $i$ and $j$ are the nearest neighbours and zero otherwise. The second term  describes the interplane interaction between the Na ions and the $d$ electrons from the CoO$_2$ layers, where $\omega_i$ represents the static potential of the first, second and third nearest Na ions to site $i$. The strengths of the Coulomb interactions corresponding to the first ($U_1$), second ($U_2$) and third ($U_3$) nearest neighbours are considered as the model parameters. The third term represents the nearest-neighbour Coulomb repulsion between two $d$ electrons on the Co triangle network.

\begin{figure}[!t]
\begin{center}
\includegraphics[width=10cm]{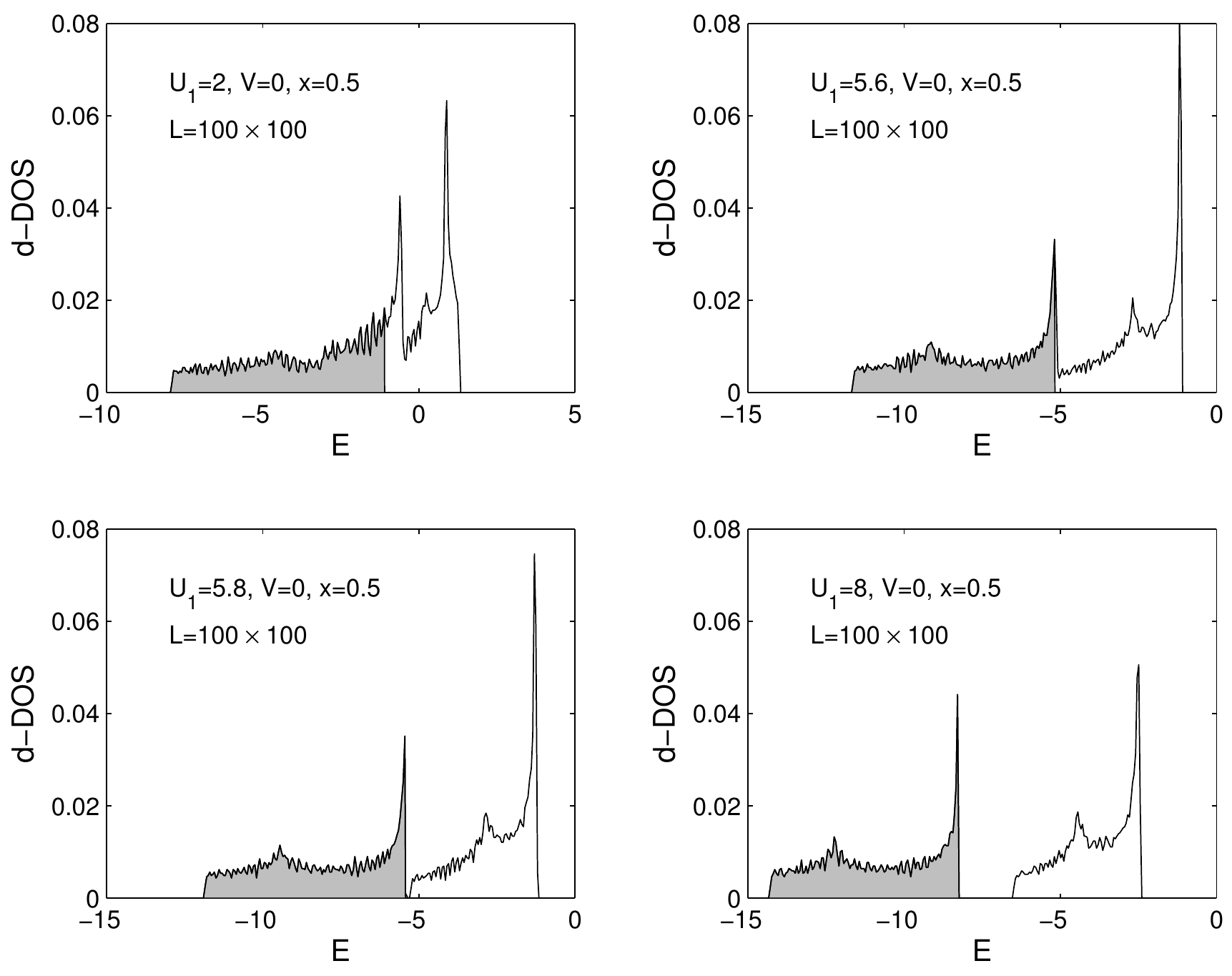}
\end{center}
\vspace{-0.5cm}
\caption{ The $d$-electron density of states of the model (\ref{omega}) calculated for $x=0.5$, $V=0$, $L=100\times 100$ and four different values of the nearest neighbour interplane interaction $U_1$. The second ($U_2$) and third ($U_3$) nearest neighbour interplane interactions are modelled by $U_2=0.2U_1$ and $U_3=0.2U_2$~\cite{Farky36}.}
\label{nax_1}
\end{figure}
\begin{figure}[!b]
\begin{center}
\includegraphics[width=7.5cm]{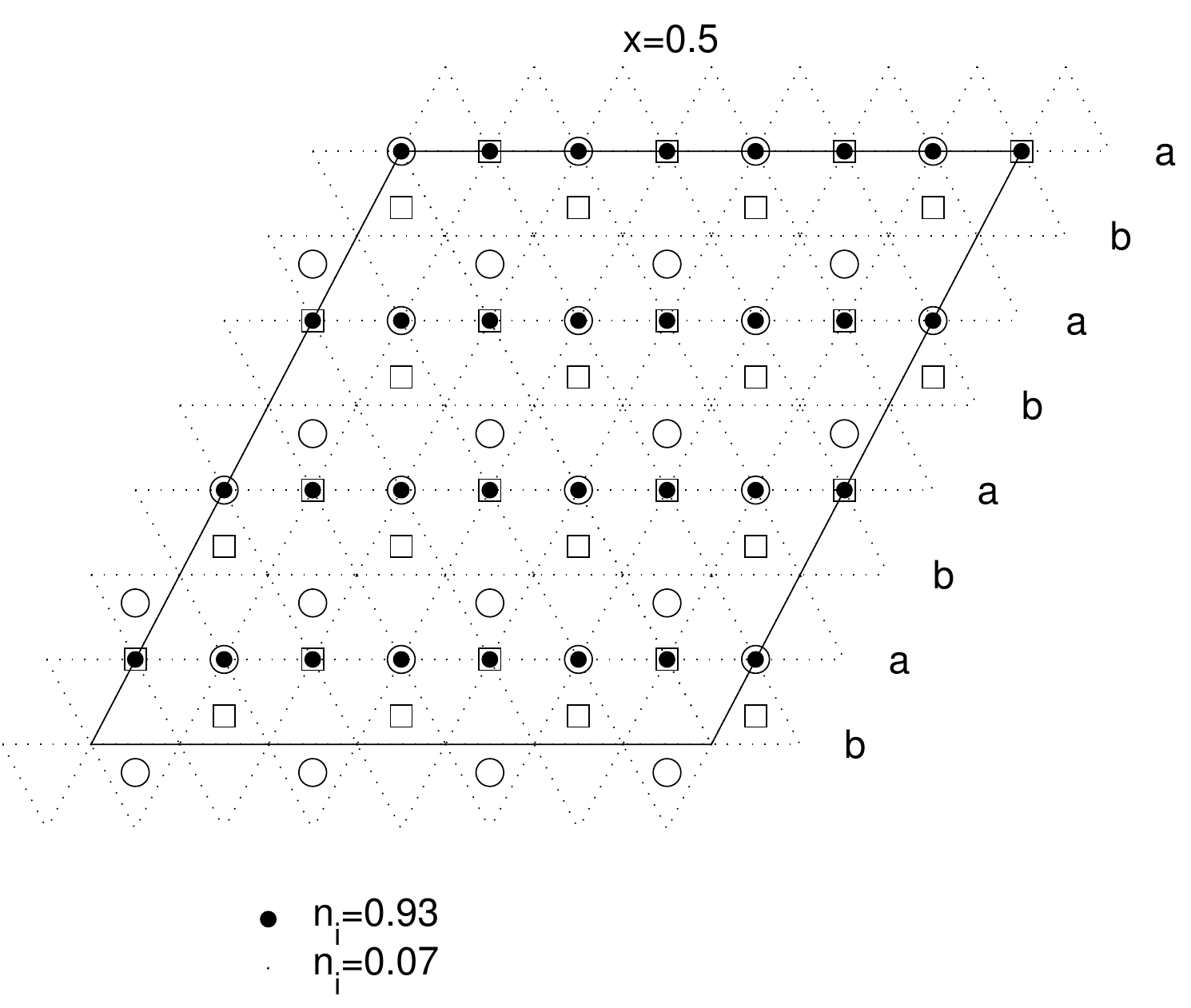}
\end{center}
\vspace{-0.5cm}
\caption{ The $d$-electron on-site occupation $n_i=\langle d^+_id_i \rangle$ calculated for $x=0.5$, $V=0$, $U_1=8$ $(U_2=0.2U_1$, $U_3=0.2U_2)$ on $L=8\times 8$ cluster.  The on-site occupation $n_i$ is represented by filled-circles ($\bullet$), where the radius of the circle on site $i$ is proportional to $n_i$. The open symbols represent the Na positions above (circles) and below (squares) the CoO$_2$ layer~\cite{Farky36}.}
\label{nax_2}
\end{figure}
\begin{table}[!t]
\begin{center}
\caption{ The $d$-electron on-site occupation $n_i$ on stripes $a$ and $b$ calculated for $x=0.5$, $U_1=8$ $(U_2=0.2U_1$, $U_3=0.2U_2)$ and four different values of intraplane interaction $V$~\cite{Farky36}.}
\label{tab7}
\vspace*{2ex}
\begin{tabular}{|c|c|c|c|c|}
\hline
$n_i$     & $V=0$ & $V=2$ & $V=4$ & $V=8$ \\ \hline\hline
stripes $a$     & 0.92936153 & 0.9664959 & 0.9811269 & 0.99188500\\
stripes $b$     & 0.07063846 & 0.03350408 & 0.0188730 & 0.00811499\\ \hline
\end{tabular}
\end{center}
\end{table}
\begin{figure*}[!b]
\begin{center}
\includegraphics[width=10.0cm]{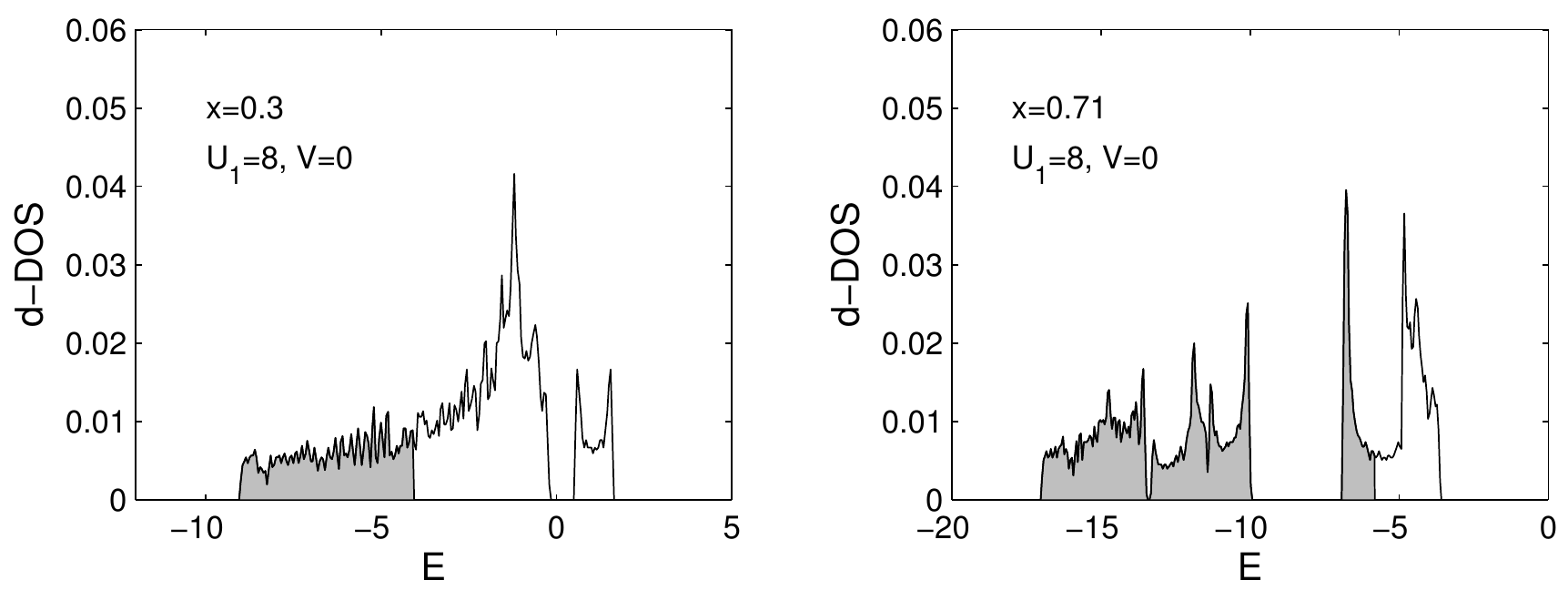}
\includegraphics[width=5.0cm]{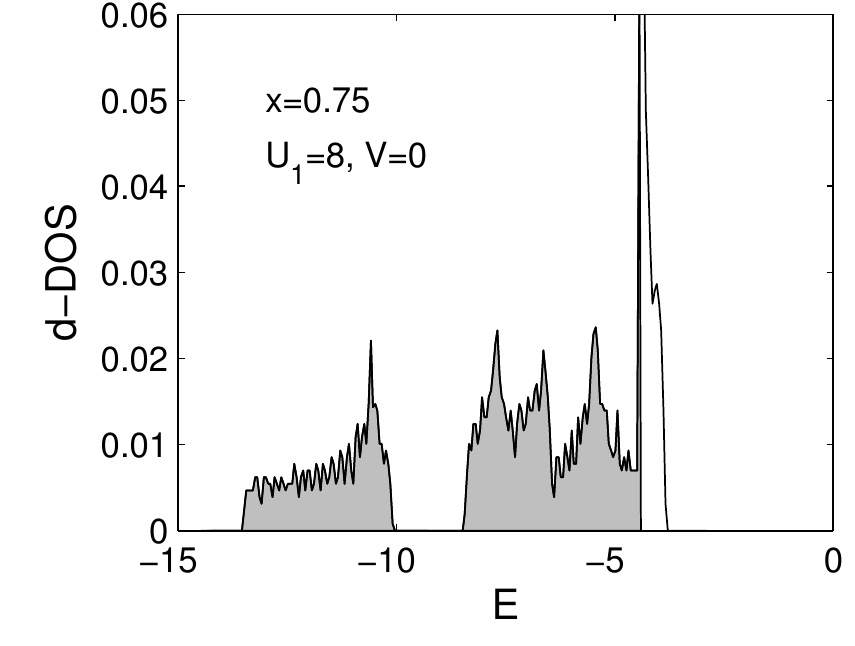}
\hspace*{-1.0cm}
\includegraphics[width=12cm]{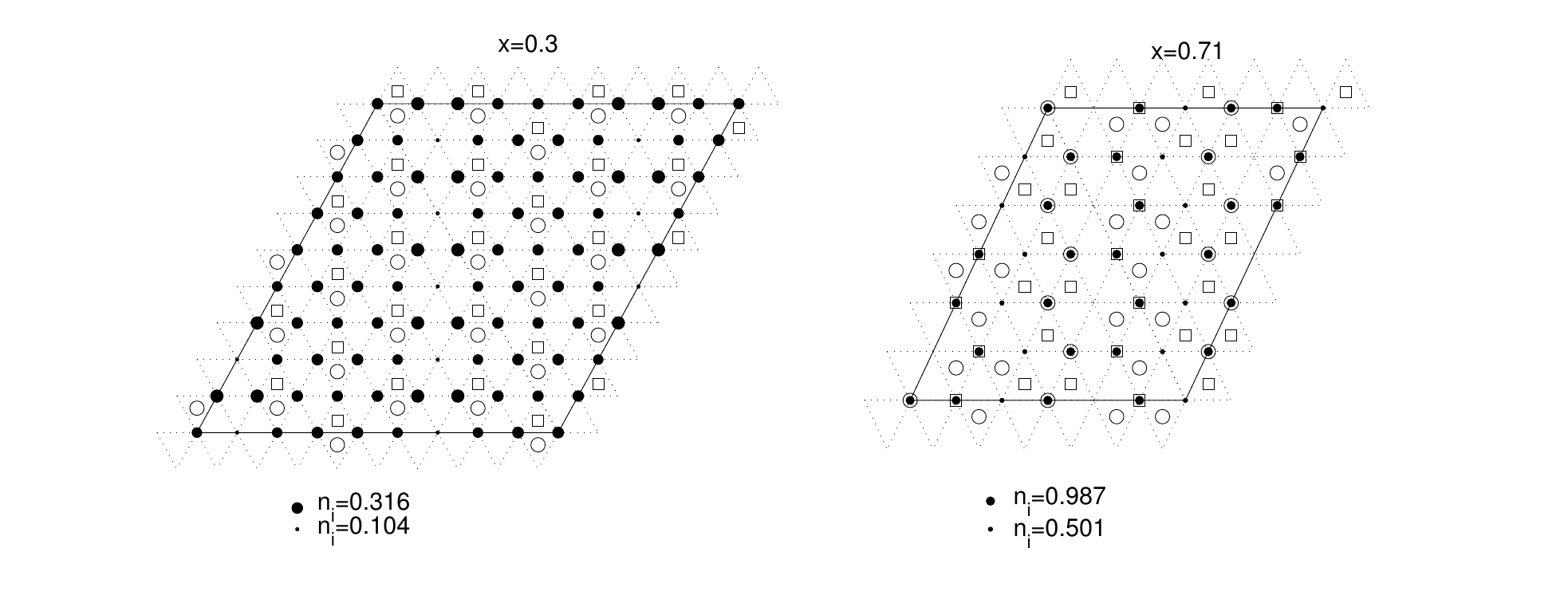}\hspace*{-2cm}
\includegraphics[width=6cm]{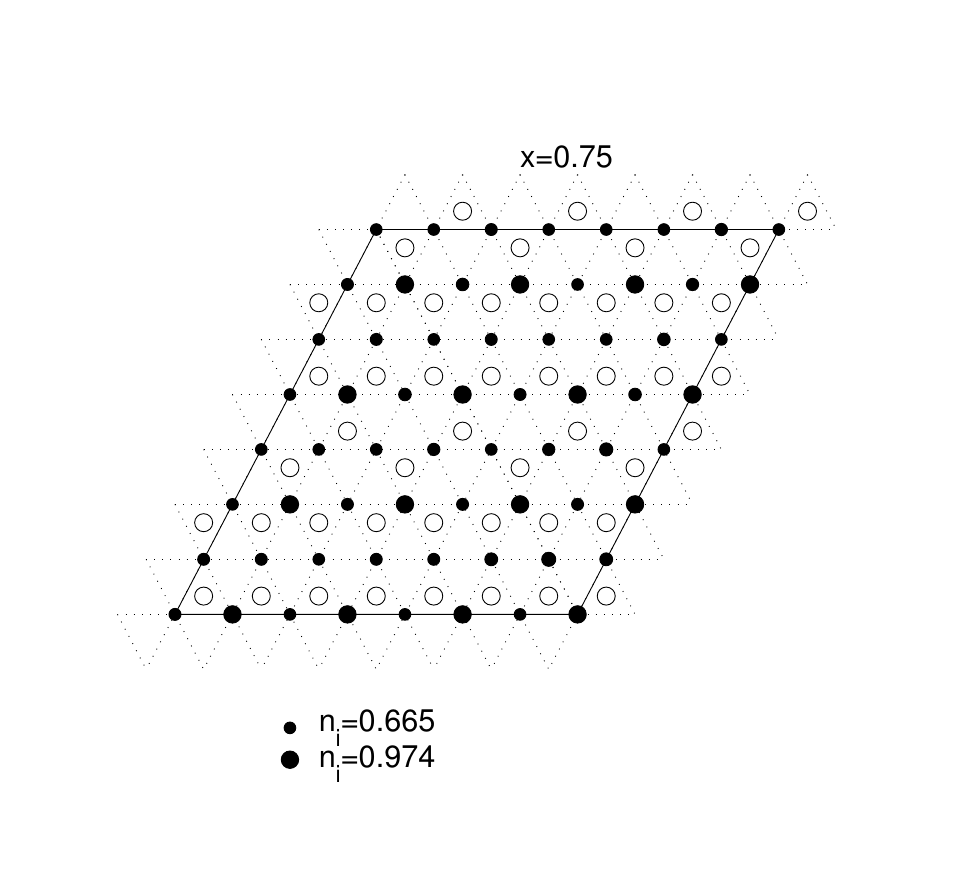}
\end{center}
\vspace{-0.5cm}
\caption{ {\it Up}: The $d$-electron density of states of the model  (\ref{omega}) calculated for $V=0$, $U_1=8$ $(U_2=0.2U_1$, $U_3=0.2U_2)$ and three different Na ion distributions taken from experiments. {\it Down}: The $d$-electron on-site occupation $n_i$ calculated for three different Na ion distributions  $x=0.3$, $x=0.71$ and $x=0.75$ taken from experiment. The on-site occupation $n_i$ is represented by filled-circles ($\bullet$) whose radius is proportional to $n_i$~\cite{Farky36}. }
\label{nax_3}
\end{figure*}

We have started our numerical study  with the case $x=0.5$ and $V=0$.   As mentioned above, at $x=0.5$, the ordered Na ions form one-dimensional zig-zag chains and so, one can easily reconstruct the static potential $\omega_i$ for any selected values of $U_1$, $U_2$ and $U_3$. Having $\omega_i$, the Hamiltonian~(\ref{omega}) can be diagonalized by standard numerical routines and used directly in the analysis of desired zero-temperature characteristics like the $d$-electron density of states, the on-site occupation $n_i=\langle d^+_id_i \rangle$, etc. The numerical results for the $d$-electron density of states obtained for four selected values of the nearest interplane Coulomb interaction $U_1$ (in the rest of the paper the second and third nearest neighbour interplane interactions are modelled by $U_2=0.2U_1$, $U_3=0.2U_2$) are shown in figure~\ref{nax_1}.
One can see that for small and intermediate values of $U_1$ the system is metallic, but with increasing $U_1$ the charge gap at the Fermi level opens, and the system undergoes the metal-insulator transition at $U_1 \sim 5.8$. This confirms a supposition that the interplane interactions between the ordered Na ions and the $d$ electrons from Co layers play a crucial role in the stabilization of the insulating state in the Na$_{0.5}$CoO$_2$ material.

From this point of view it is interesting to ask if these interactions could also stabilize  the charge ordering within the Co layers. To answer this question we have calculated the on-site occupation $n_i=\langle \psi_{\rm G} |d^+_id_i| \psi_{\rm G} \rangle$, where $|\psi_{\rm G} \rangle$ is the ground state of Hamiltonian~(\ref{omega}) obtained directly from the exact numerical diagonalization solution. The typical results for $n_i$ within the insulating phase are shown in figure~\ref{nax_2} for a finite cluster of $L=8 \times 8$ sites.

Our results clearly demonstrate that an inhomogeneous charge ordering, of the stripe-like form, develops within the Co layers. For the sites below (above) the Na ions the occupation number $n_i$ is equal to 0.93 (the stripes $a$), while for the remaining sites $n_i=0.07$ (the stripes $b$).  Thus, we arrive at the  conclusion that interplane interactions between the ordered Na ions and the $d$ electrons from the CoO$_2$ layers not only stabilize the insulating ground state in Na$_{0.5}$CoO$_2$, but they are also capable of describing the formation of inhomogeneous charge ordering within the CoO$_2$ layers.

Since the Coulomb interactions between $d$ electrons inside CoO$_2$ planes are not negligible in comparison with interplane interactions, it was necessary to  examine what happens if these interactions are switched on ($V\ne 0$). We have solved this task in a manner analogous to the case $V=0$, i.e., by a direct calculation of the $d$-electron density of states and the  $d$-electron on-site occupation  $n_i$, with only one technical complication, namely,  that the diagonalization of the full Hamiltonian is performed by the Lanczos method, because for $V\neq 0$, the  Hamiltonian~(\ref{omega}) is no longer a single particle. The results of our numerical calculations for $n_i$ obtained for four different values of the intraplane interaction $V$ are summarized in table~\ref{tab7}.
One can see that for nonzero interplane interactions, the nearest-neighbour intraplane interaction does not destroy the charge ordering induced by Na ions, but on the contrary, increasing $V$ further stabilizes the inhomogeneous charge ordering. This effect is very strong and for sufficiently large $V$, practically all $d$ electrons are localized on sites below (above) Na ions (the stripes $a$).
Taking into account the fact that $n_i=0$ (in our notation) corresponds to Co$^{4+}$ and $n_i=1$ to Co$^{3+}$, these results are consistent with the model proposed recently by Choy~\cite{Choy} for a description of the sign change of the Hall-coefficient (as a function of temperature), in which the rows of Co$^{4+}$ ions alternate with rows of Co$^{3+}$ ions. This indicates that our model, in spite of its relative simplicity, may still contain the relevant physics of cobaltate systems.

We have  used the same procedure for the case of $x\neq 0.5$. Our main motivation in this case was to answer the question whether our simple model is capable of  describing correctly the conducting properties of Na$_{x}$CoO$_2$  for $x<0.5$ as well as for $x>0.5$.  In
figure~\ref{nax_3} we present numerical results for $d$-electron density of states for three selected experimentally confirmed  Na structures with  $x=0.3$, $x=0.71$ a $x=0.75$, representing the  typical examples from the region of  $x<0.5$ and $x>0.5$ (the experimentally proposed Na ion distributions are  shown in figure~\ref{nax_3},  together with our results for $n_i$).

It is seen that in all three investigated cases the Fermi level lies inside the conduction band, and thus all three cases correspond to a metallic state which is perfectly consistent with experimental measurements~\cite{Foo2}.

\subsection{Magnetization processes in rare-earth tetraborides}
\label{Magnetization processes in rare-earth tetraborides}

\begin{figure*}[!t]
\begin{center}
\includegraphics*[width=4.3cm]{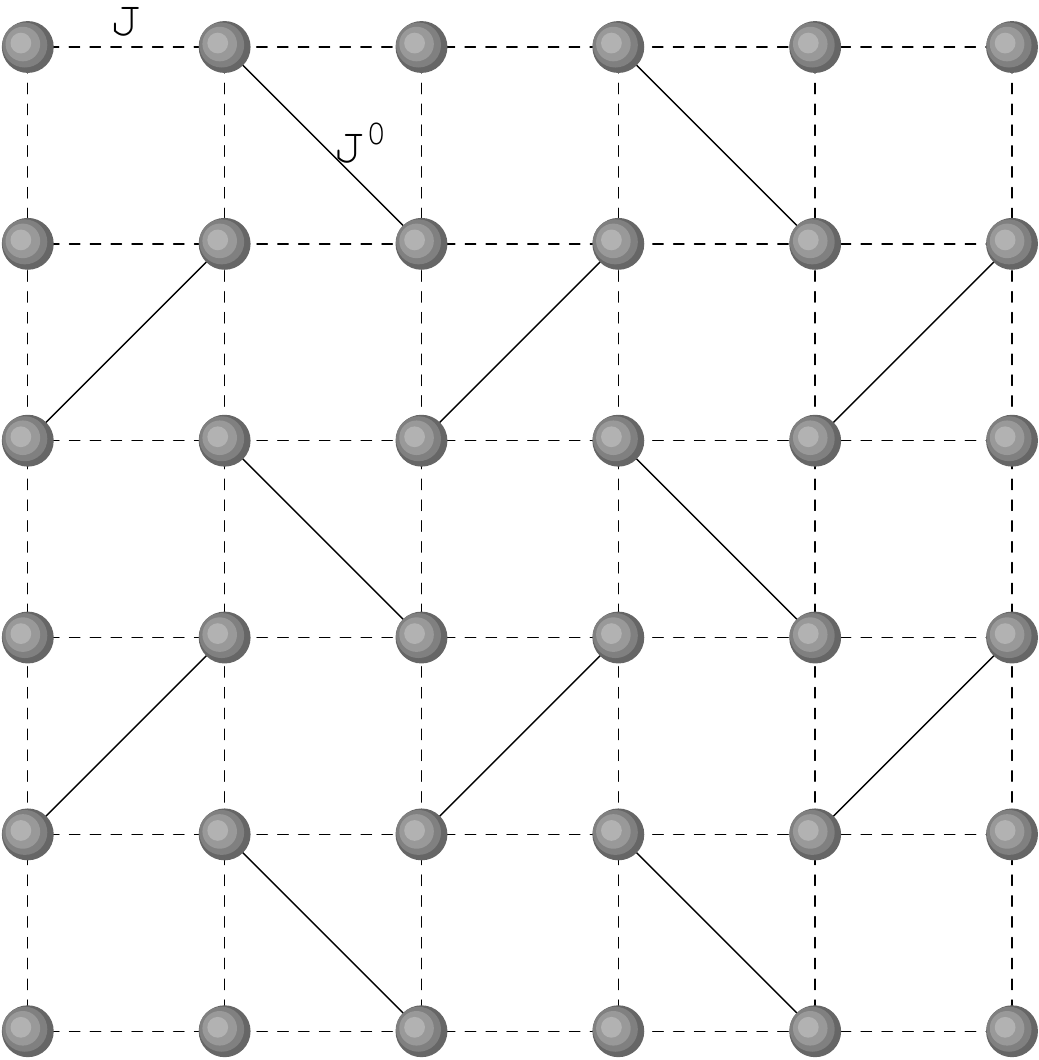}\hspace*{2.7cm}
\includegraphics*[width=4.6cm]{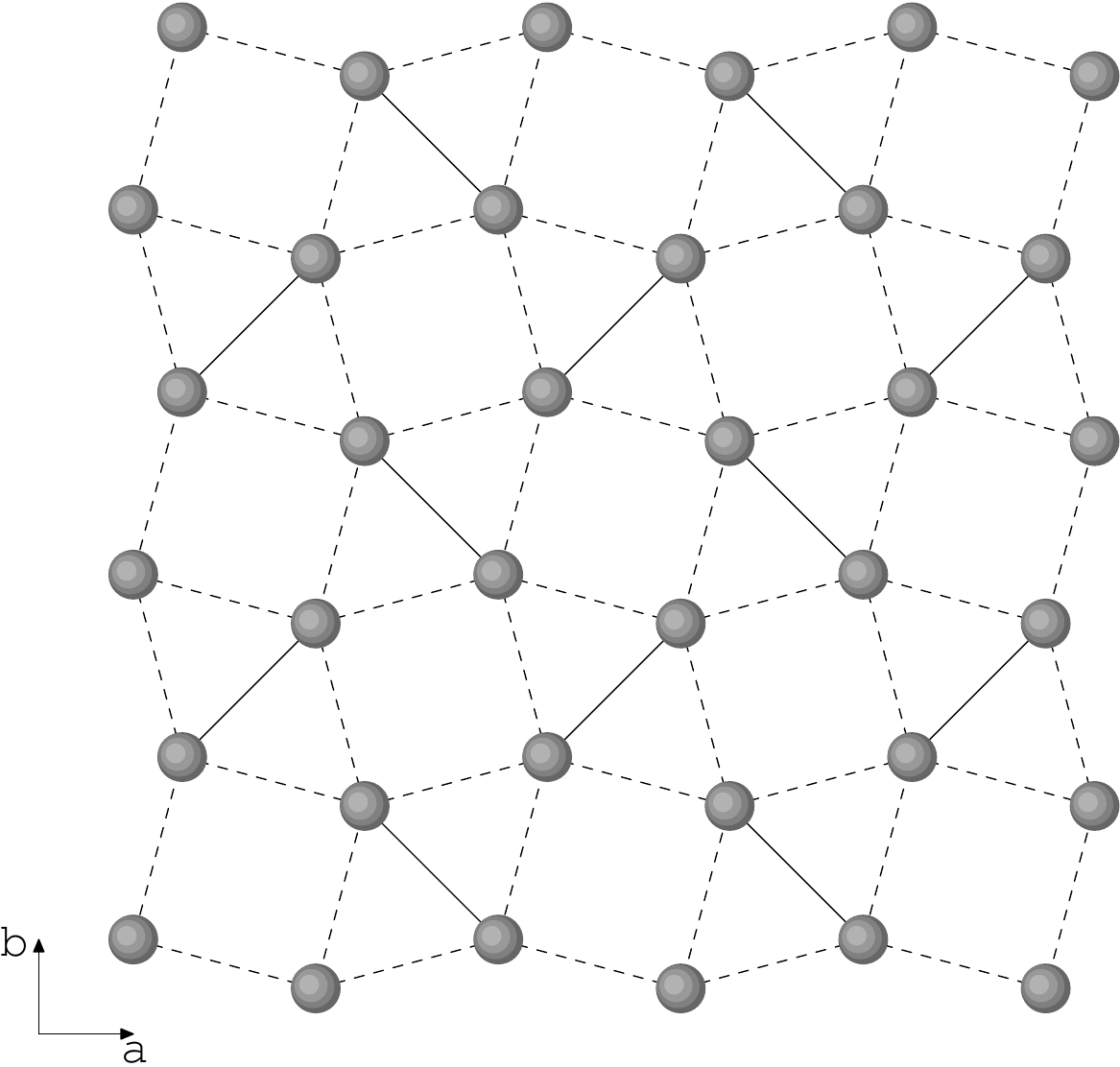}
\end{center}
\vspace*{-0.1cm}
\caption{ The Shastry-Sutherland lattice with magnetic couplings $J$ bonds along the edges of the squares and $J'$ along the diagonals (left hand panel), and the topologically identical structure realized in the (001) plane of SrCu$_2$(BO$_3$)$_2$ and rare-earth tetraborides (right hand panel).}
\label{fig1_f}
\end{figure*}

A spin system is frustrated when all local interactions between spin pairs cannot be satisfied at the same time. Frustration can arise from the competing interactions or/and from a particular geometry of the lattice, as seen in the triangular lattice. The Shastry-Sutherland lattice was considered more than 20 years ago  by Shastry and Sutherland~\cite{Shastry} as an interesting example of a frustrated quantum spin system with an exact ground state. It can be described as a square lattice with antiferromagnetic couplings $J$ between  nearest neighbours and additional antiferromagnetic couplings $J'$ between  next-nearest neighbours in every second square (see figure~\ref{fig1_f}). This lattice attracted much attention after its experimental realization in the SrCu$_2$(BO$_3$)$_2$ compound~\cite{SrCu}. The observation of a fascinating sequence of magnetization ($m/m_{\rm s}=$1/2, 1/3, 1/4 and 1/8 of the saturated magnetization $m_{\rm s}$) in this material~\cite{SrCu} stimulated further theoretical and experimental studies of the Shastry-Sutherland lattice~\cite{Sebastian,Dorier}.

Similar phenomena of magnetization plateaus is also observed in the rare-earth tetraborid TmB$_4$~\cite{Siem}. Since fully polarized state can be reached for experimentally accessible magnetic fields, this compound permits exploration of its complete magnetization process. It was found that the magnetization diagram of TmB$_4$ consists of magnetization plateaus located at small fractional values of  $m/m_{\rm s}=$1/9, 1/8, 1/7 of the saturated magnetization, followed by the major magnetization plateau located at $m/m_{\rm s}=1/2$. Due to strong  crystal field effects, the effective spin model for TmB$_4$ has been suggested to be described by the spin-1/2 Shastry-Sutherland model under strong Ising (or easy-axis) anisotropy in a magnetic field $h$~\cite{Siem}
\begin{equation}
H_{JJ'}=J\sum_{\langle i,j\rangle} S^z_iS^z_j + J'\sum_{\langle\langle i,j\rangle\rangle} S^z_iS^z_j - h\sum_i S^z_i\ ,
\label{eq1}
\end{equation}
where $S^z_i=\pm1/2$ denotes the $z$-component of a spin-1/2 degree of freedom on site $i$ of a square lattice and $J$, $J'$ are the antiferromagnetic exchange couplings between all nearest neighbour bonds ($J$) and next-nearest neighbour bonds in every second square ($J'$), as indicated in figure~\ref{fig1_f} (left hand panel).

Numerical simulations obtained within the Monte-Carlo~\cite{Meng} and tensor renormalization group methods~\cite{Chang} on large systems  showed that the Ising model on the Shastry-Sutherland lattice exhibits, in the presence of the magnetic field, a magnetization plateau only at 1/3 of the saturated magnetization. The existence of the magnetization plateau at only 1/3 of the saturated magnetization and its absence at 1/2 indicates that it is necessary to go beyond the classical Ising limit to reach the correct description of the magnetization process in TmB$_4$ and other rare-earth tetraborides. The first attempt of this kind has been made by Meng and Wessel~\cite{Meng} who studied the spin-1/2 easy-axis Heisenberg model on the Shastry-Sutherland lattice with ferromagnetic transverse spin exchange using quantum Monte-Carlo and degenerate perturbation theory. Besides the magnetization plateau at 1/3 of the saturated magnetization they found a further plateau at 1/2, which persists only in the quantum regime.

It should be noted that the behaviour similar to TmB$_4$ has been also observed for other rare-earth tetraborides. For example, for ErB$_4$ the magnetization plateau has been found at $m/m_{\rm s}=1/2$~\cite{Michi,Matas}, for TbB$_4$ at $m/m_{\rm s}=1/2,4/9,1/3,2/9$ and $7/9$~\cite{Yoshii} and for HoB$_4$ at $m/m_{\rm s}=1/3,4/9$ and $3/5$~\cite{Matas}.

Quite recently we have proposed an alternative model~\cite{Farky51} of stabilizing the magnetization plateaus in the rare-earth tetraborides based on the fact that these materials, in contrast to SrCu$_2$(BO$_3$)$_2$, are metallic. Thus, for a correct description of ground-state properties of rare-earth tetraborides one should take into account both spin and electron subsystems as well as the coupling between them. Supposing that electron and spin subsystems interact only via the spin dependent Ising interaction $J_z$, the Hamiltonian of the system can be written as
\begin{eqnarray}
H=\sum_{ij\sigma}t_{ij}d^+_{i\sigma}d_{j\sigma} + J_z\sum_{i}(n_{i\uparrow}-n_{i\downarrow})S^z_i
%\nonumber\\
- h\sum_i(n_{i\uparrow}-n_{i\downarrow}) + H_{JJ'}\, .
\label{eq2}
\end{eqnarray}
The model described by (\ref{eq2}) is a straightforward extension of the spin-1/2 Falicov-Kimball model with anisotropic spin-dependent interaction discussed in detail in~\cite{Lemanski05}. The only differences are that we consider here a direct spin interaction (of the Ising type) between the localized spins and that the underlying lattice is of the  Shastry-Sutherland type.

\begin{figure}[!h]
\begin{center}
\includegraphics[scale=0.6]{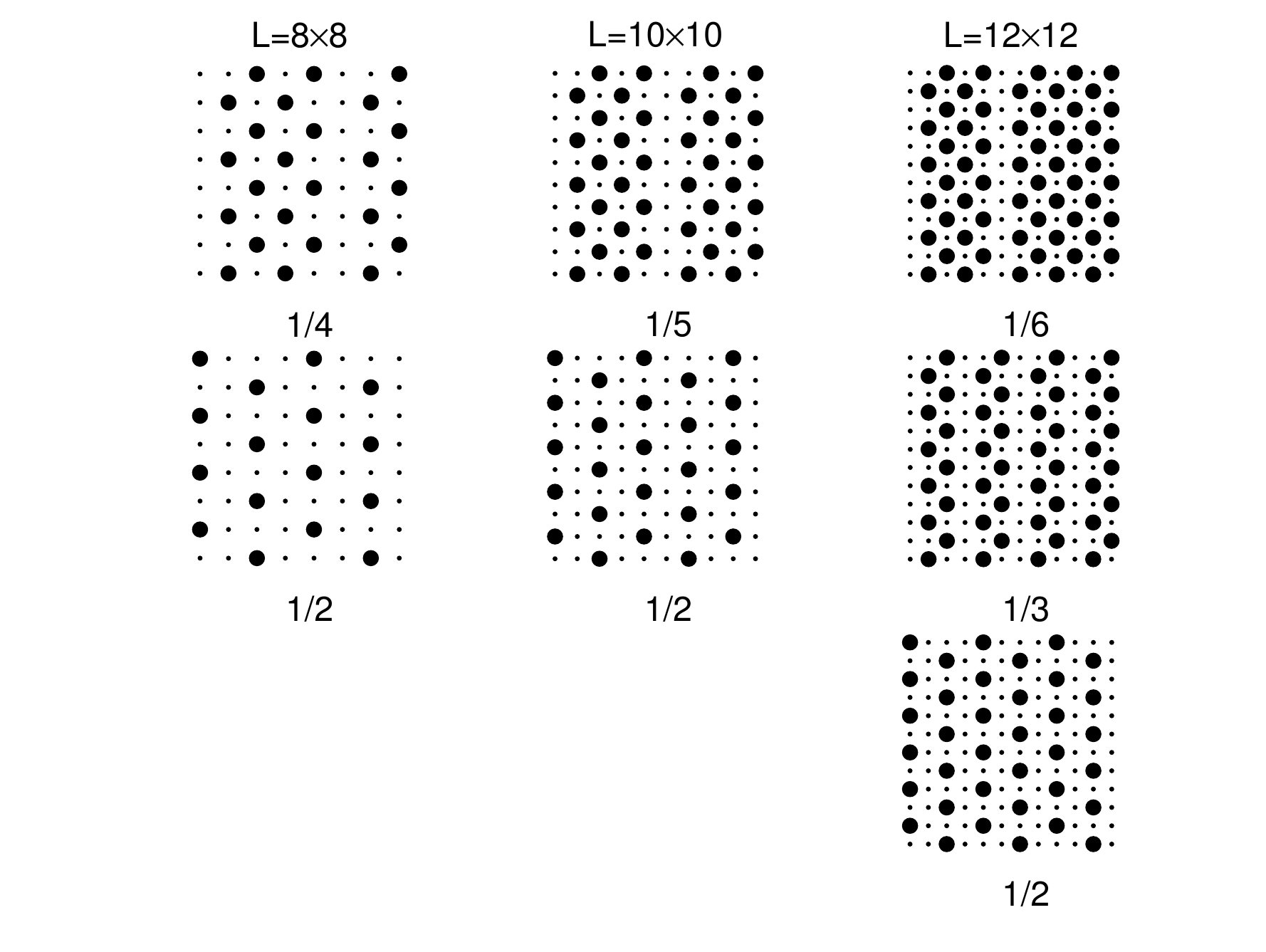}
\vspace*{-0.5cm}
\caption{  The complete list of the ground-state spin configurations that are stable on finite intervals of magnetic field for $L=8\times8$, $L=10\times10$ and $L=12\times12$. The big dots correspond to the up-spin orientation and the small dots correspond to the down-spin orientation~\cite{Farky51}.}
\label{prb82}
\end{center}
\end{figure}

\begin{figure*}[!t]
\begin{center}
\includegraphics[scale=0.4]{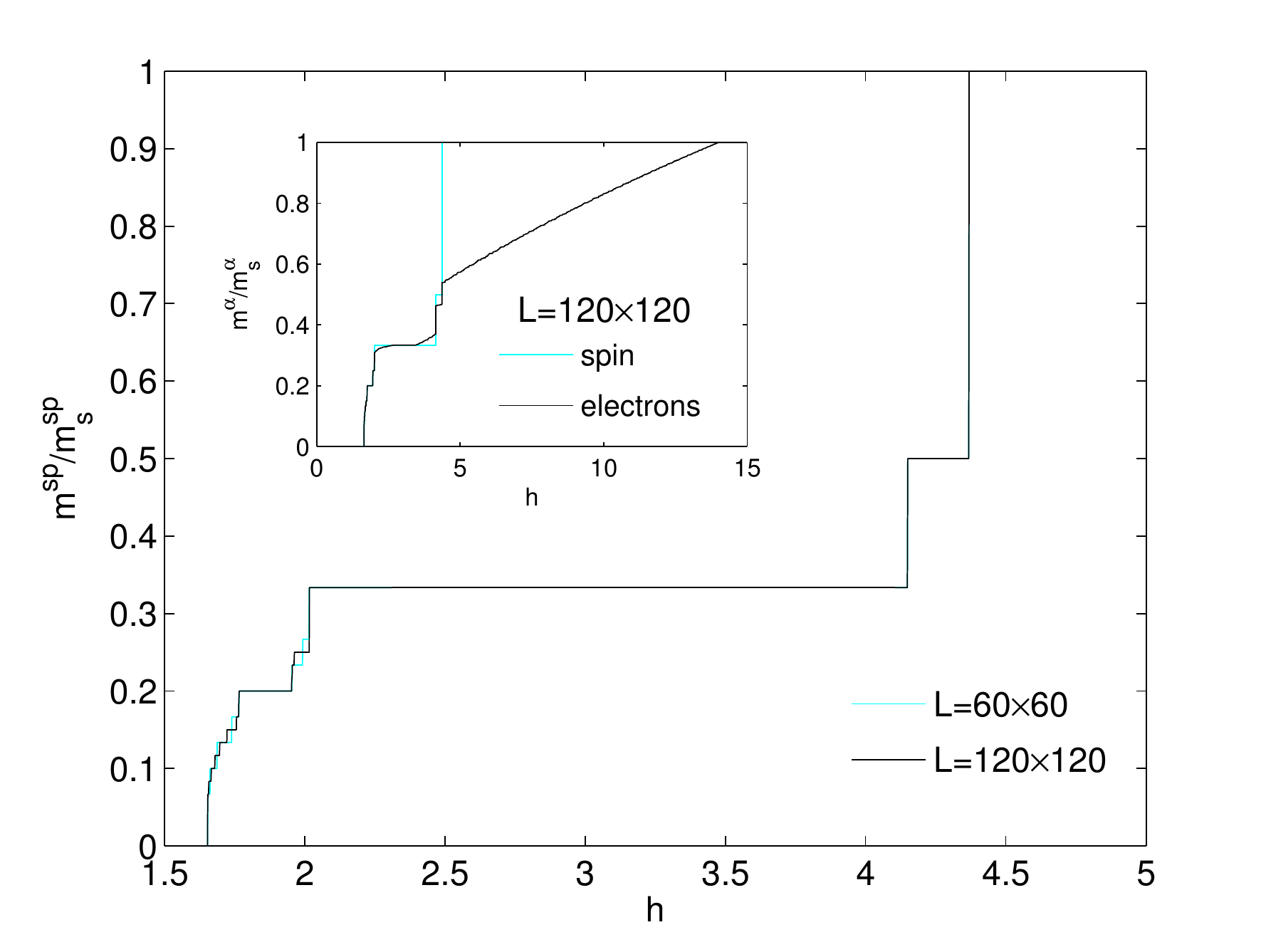}
\includegraphics[scale=0.4]{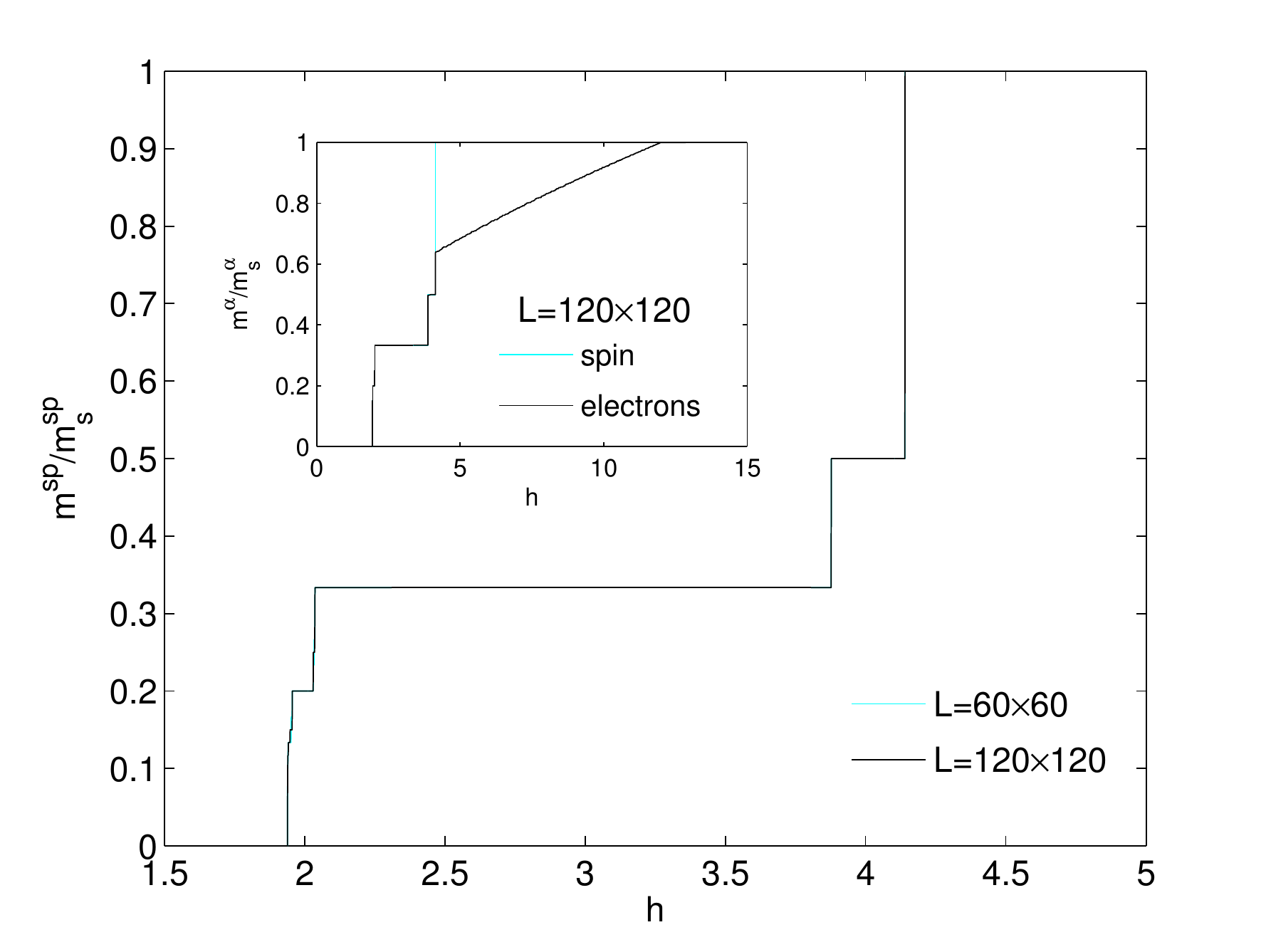}
\end{center}
\vspace*{-0.5cm}
\caption{ (Color online) The magnetization curves calculated for $J'/J=1$, $J_z=2$, $t=4$, $t'=0$ (left hand panel) and for $J'/J=1$, $J_z=4, t=4, t'=0$ (right hand panel). Insets present the behaviour of magnetization curves for the spin and the electron subsystem calculated on $L=120\times 120$ ($\alpha=$ el or sp)~\cite{Farky51}. }
\label{ssm_3}
\end{figure*}

To examine the magnetization curve corresponding to the model Hamiltonian we have used our AM. Using this method we have performed exhaustive numerical studies of the model (\ref{eq2}) for a wide range of model parameters $h$, $J_z$, $t$ (the hopping integral between the nearest-neighbours), $t'$ (the hopping integral between the next-nearest neighbours) and $J/J'=1$ selected based on the experimental measurements~\cite{Siem}. To exclude the finite-size effects, the numerical calculations have been performed for several different Shastry-Sutherland clusters consisting of $L=8\times 8$, $10\times 10$ and $12\times 12$ sites. The most important result obtained from these calculations is that the switching on of  $J_z$ and $t$ leads to a stabilization of the uniform ground-state spin arrangement consisting of parallel antiferromagnetic bands separated by ferromagnetic stripes. A complete list of the ground-state spin arrangements (for $0<m^{\rm sp}/m_{\rm s}^{\rm sp}<1$) that are stable on finite intervals of magnetic field values is depicted in figure~\ref{prb82}.
The second important observation is that the width $w$ of the antiferromagnetic bands cannot be arbitrary, but fulfills severe restrictions. Indeed, we have found that with the exception of the case $m^{\rm sp}/m^{\rm sp}_{\rm s}=1/2$, in all the remaining cases the permitted width of the antiferromagnetic band is only $w$ or $w+2$, where $w$ is an even number. This fact is very important from the numerical point of view, since it allows us to perform the numerical calculations on much larger clusters with the extrapolated set of configurations of the above described type.
The resulting magnetization curves obtained on the extrapolated set of ground-state spin configurations consisting of parallel antiferromagnetic bands of width $w$ ($w$ and $w+2$) separated by ferromagnetic stripes are shown in figure~\ref{ssm_3}  for the selected values of model parameters that represent the typical behaviour of the model.
One can see that the switching on of  the spin-dependent interaction $J_z$ and taking into account the electron hopping $t$ on the nearest lattice sites  of the Shastry-Sutherland lattice leads to a stabilization of new magnetization plateaus (the next-nearest hopping integrals $t'$ only renormalize the width of the magnetization plateaus). In addition to the Ising magnetization plateau at $m^{\rm sp}/m^{\rm sp}_{\rm s}=1/3$ we have found two new magnetization plateaus located at $m^{\rm sp}/m^{\rm sp}_{\rm s}=1/2$  and $m^{\rm sp}/m^{\rm sp}_{\rm s}=1/5$. The ground-state spin arrangements corresponding to these magnetization plateaus have the same structure consisting of parallel antiferromagnetic bands of a width $w$ (where $w=1$ for $m^{\rm sp}/m_{\rm s}^{\rm sp}=1/2$, $w=2$ for $m^{\rm sp}/m_{\rm s}^{\rm sp}=1/3$ and   $w=4$ for $m^{\rm sp}/m_{\rm s}^{\rm sp}=1/5$) separated by ferromagnetic stripes. Thus, our numerical results show that besides the pure spin mechanism (e.g., the easy-axis Heisenberg model on the Shastry-Sutherland lattice~\cite{Meng}) of stabilization the magnetization plateaus in rare-earth tetraborides, there also exists  an alternative mechanism based on the coexistence of electron and spin subsystems that are present in these materials. From this point of view it is interesting to compare in detail the ground states obtained within these two different approaches. For $m^{\rm sp}/m^{\rm sp}_{\rm s}=1/3$ our results are identical to the ones obtained within the Ising~\cite{Siem, Chang} as well as easy-axis Heisenberg~\cite{Meng,Liu} model on the Shastry-Sutherland lattice. The accordance between our and the easy-axis Heisenberg solution~\cite{Meng} is  surprisingly found for $m^{\rm sp}/m^{\rm sp}_{\rm s}=1/2$ as well. In this case both approaches predict the sequence of parallel antiferromagnetic and ferromagnetic stripes. For  $m^{\rm sp}/m^{\rm sp}_{\rm s}=1/5$ our results postulate a new type of spin ordering.

Finally, it should be noted that more exhaustive studies of the model performed on much larger lattices have revealed the existence of magnetization plateaus at $m^{\rm sp}/m^{\rm sp}_{\rm s}=1/7$, 1/9 and 1/11 (in accordance with  experimental measurements in TmB$_4$) but the stability regions of these phases are much narrower in comparison with the ones of 1/2, 1/3 and 1/5 plateau phases.

\subsection{Doping-induced valence changes in rare-earth compounds}
\label{Doping-induced valence changes in rare-earth compounds}

The substitution of one type of atoms/ions by another type changes electronic relations in a given material which permits to  directly study manifestations of electron correlations on physical properties of the system. From the perspective of the Falicov-Kimball model, which was originally introduced to describe valence and  metal-insulator transitions in rare-earth compounds, it is therefore particularly interesting to examine what predictions the model is capable to provide for the case when we replace the rare-earth ions by other ions. Since previous theoretical works have shown~\cite{Farky2,Farky9} that the Falicov-Kimball model can provide good qualitative predictions for the transport, magnetic and spectroscopic properties of some mixed valence compounds (e.g., SmB$_6$, SmS, etc.), we have focused on studying the effects of doping just for this type of materials. For SmB$_6$ it has been found, for example, that the substitution of Sm by nonmagnetic divalent ions (e.g., Sr$^{2+}$, Yb$^{2+}$) increases the average samarium valence (the average  occupancy of $f$ orbitals decreases), while the substitution of Sm by nonmagnetic trivalent ions (e.g., Y$^{3+}$, La$^{3+}$) produces the opposite effect~\cite{Tarascon, Gabani02}.

\begin{figure}[!t]
\vspace{-0.5cm}
\begin{center}
\includegraphics[angle=0,width=12cm,scale=1]{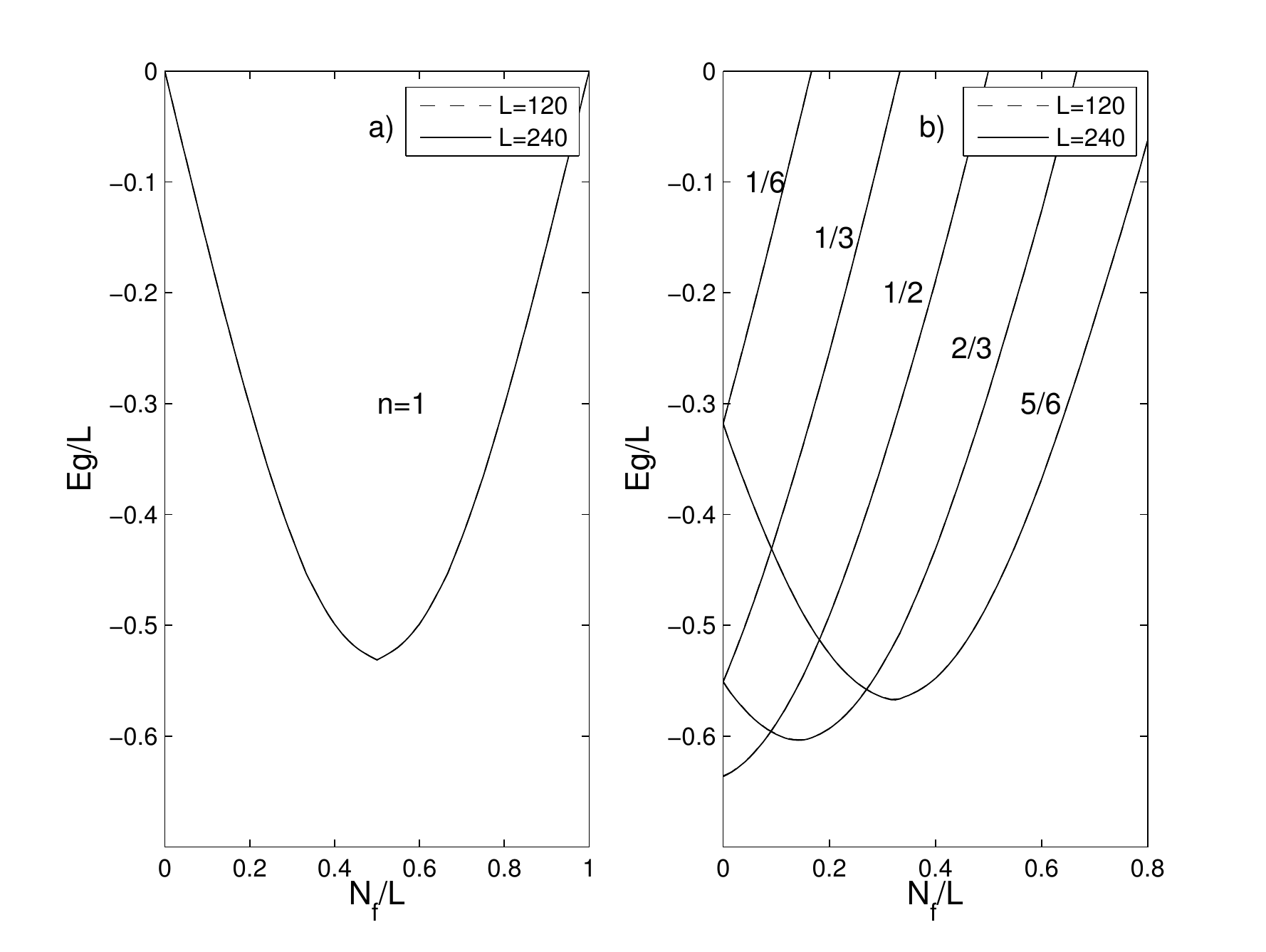}
\end{center}
\vspace{-0.6cm}
\caption{
(a) The ground-state energy $E_{\rm g}$ of the one dimensional Falicov-Kimball model as
a function of $N_f/L$ calculated
for two finite clusters of $L=120$ and L=240 sites at
$n=1$ $(n=N/L=(N_d+N_f)/L)$.
(b) The ground-state energy $E_{\rm g}$ as a function of $N_f/L$ calculated
for two finite clusters of $L=120$ and L=240 sites at
$n=1/6, 1/3, 1/2, 2/3, 5/6$. In both cases $E_f=0$ and $U=0.5$~\cite{Farky57}.}
\label{fig1_x}
\end{figure}
\begin{figure}[!b]
\vspace{-0.5cm}
\begin{center}
\includegraphics[angle=0,width=9.cm,scale=1]{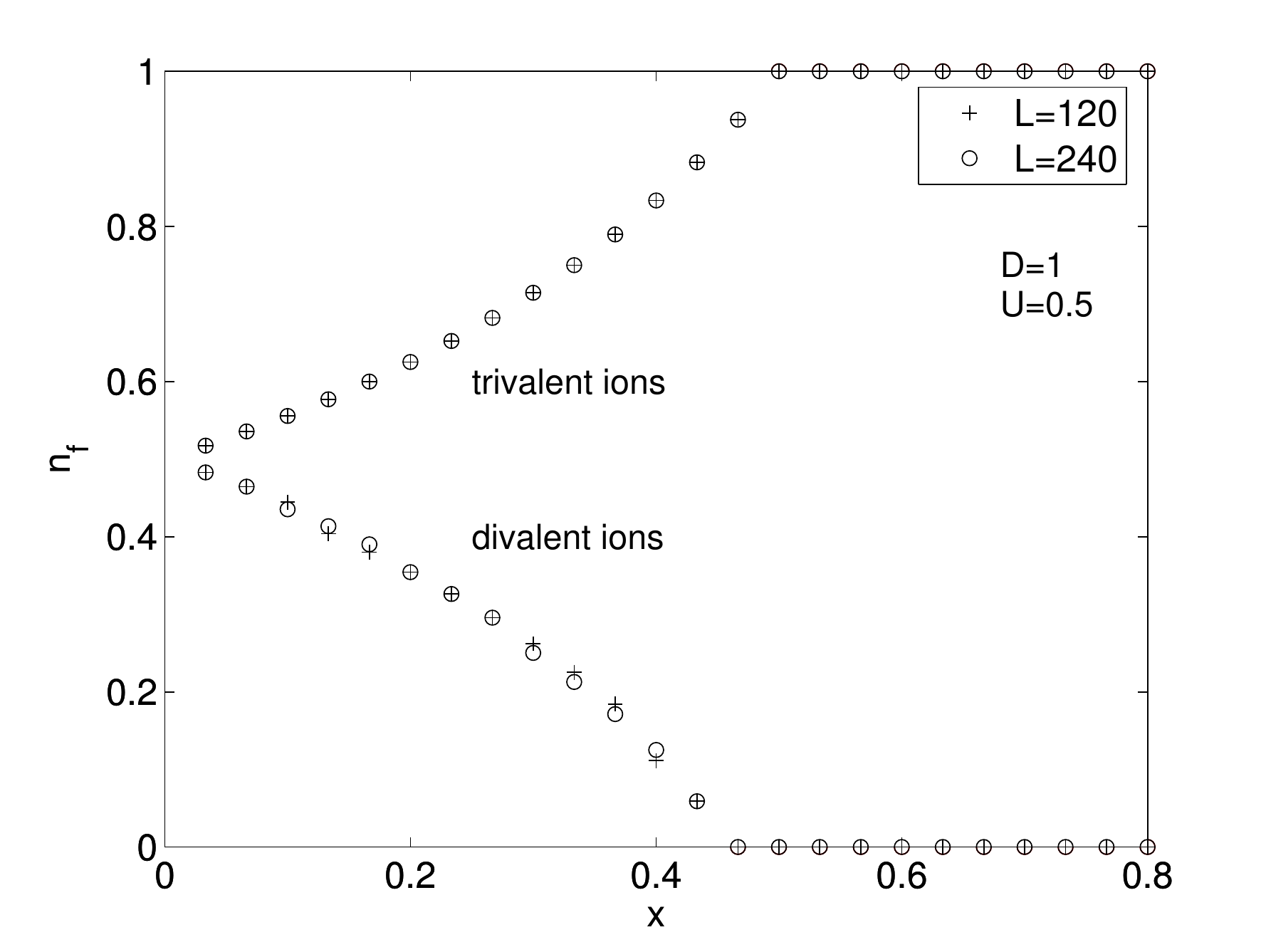}
\end{center}
\vspace{-0.5cm}
\caption{ The average occupancy of $f$ orbitals as a function of $x$ for divalent and trivalent dopants. The one-dimensional case~\cite{Farky57}.}
\label{fig2_x}
\end{figure}

In our previous papers~\cite{Farky57, Farky58} we have examined theoretically both cases: (i) the substitution of rare-earth Sm ions  by non-magnetic trivalent ions (e.g. Y$^{3+}$) which introduce conduction electrons into the $d$-conduction band (one electron per dopant) and (ii) the substitution of rare-earth ions by non-magnetic divalent ions (e.g. Sr$^{2+}$), which play a dilution role and reduce the number of conduction electrons in the $d$-conduction band
(no additional electrons are introduced to the system). We have started our study with the first case which is slightly simpler. In this case the total number of electrons $N=N_f+N_d$ is equal to the total number of lattice sites $L$ (the case of one electron per rare-earth atom is considered) and does not depend on the number of dopants $X$. Consequently, the ground-state energy  $E_{\rm g}(N_f)=\sum^{L-N_f}_{k=1}\lambda_k$  is also independent of $X$ and can be calculated directly using our AM for arbitrary $N_f$ from the interval $[0, L-X]$. The results of numerical calculations for the one dimensional lattices of $L=120$ and $L=240$ sites are summarized in figure~\ref{fig1_x}~(a) for representative model parameters ($E_f=0, U=0.5$).
It is seen that finite-size effects are negligible and thus these results can be  satisfactorily used to represent the behavior of macroscopic systems.  From the $N_f$ dependence of the ground-state energy one can easily deduce (we remember that $N_f=0,1,\ldots,L-X$) that $E_{\rm g}$ has a minimum  at $N^0_f = L/2$ for $N_f \geqslant L/2$ and at  $N^0_f = L-X$ for $N_f < L/2$. Thus, the average occupancy of the $f$-electron orbitals $n_f=\frac{N^0_f}{L-X}$ can be finally written as
\begin{eqnarray}
 n_f =\left\{\begin{array}{ll}
             \frac{L}{2(L-X)}  & \mbox{for $X < L/2$}\\
             1                 & \mbox{for $X \geqslant L/2$.}
            \end{array}
      \right.
\end{eqnarray}
\begin{figure}[!b]
\begin{center}
\hspace*{-0.5cm}
\mbox{\includegraphics[width=12cm,angle=0]{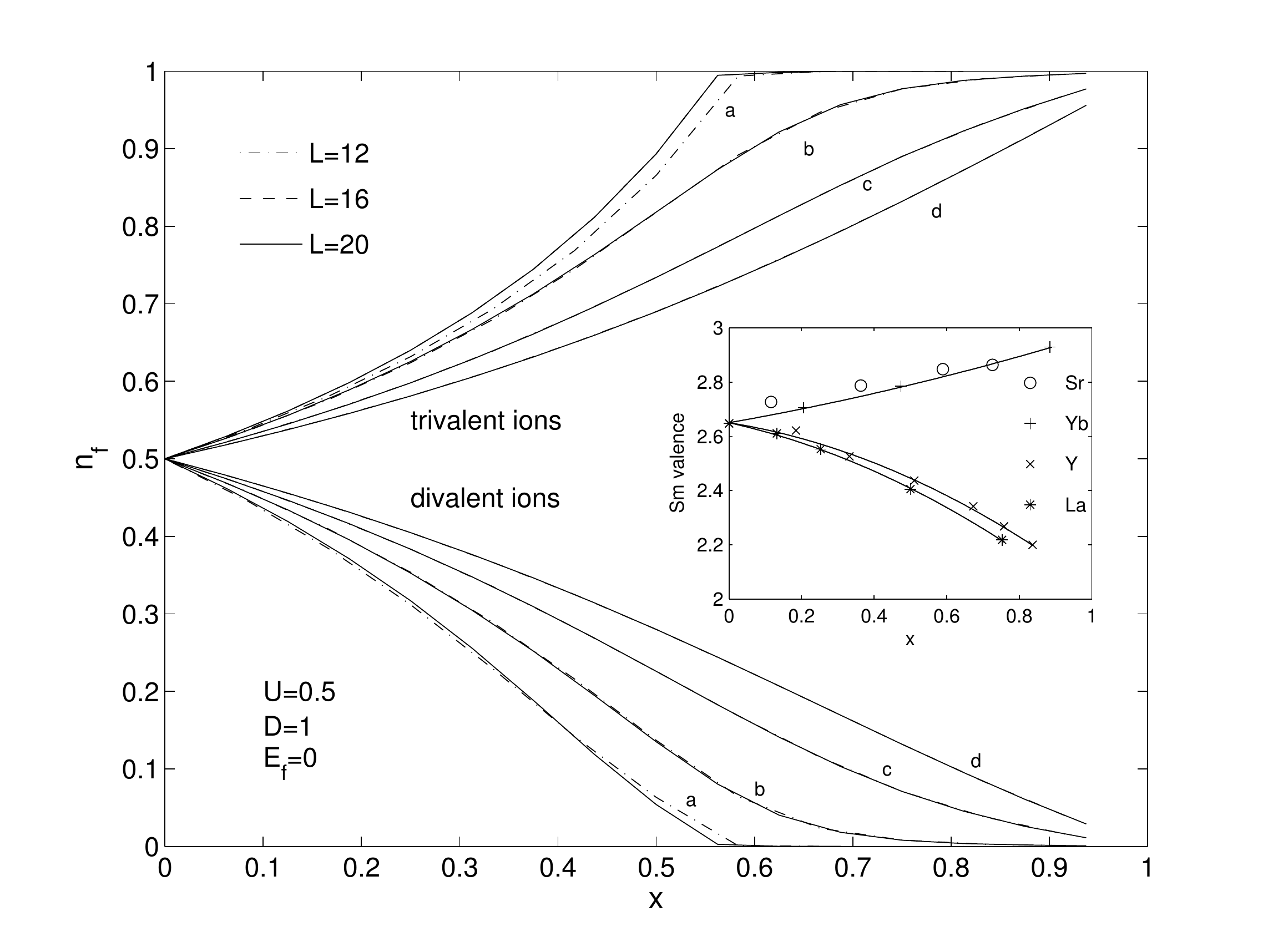}}
\end{center}
\vspace*{-0.7cm}
\caption{ The average occupancy of $f$ orbitals as a function of $x$
for doping by divalent and trivalent ions calculated for
$L=12,16,20$ and several different temperatures $\tau$.
Curve $a$, $\tau=0.1$; curve $b$, $\tau=0.3$;
curve $c$, $\tau=0.6$; curve $d$, $\tau=1$~\cite{Farky58}.
Inset: The average samarium valence in the Sm$_{1-x}$M$_x$B$_6$ systems
(M=Y$^{3+}$, La$^{3+}$, Sr$^{2+}$, Yb$^{2+}$) measured at 300~K (reference~\cite{Tarascon}).}
\label{fig1_xt}
\end{figure}
The situation is more complicated for the substitution of rare-earth ions  by non-magnetic divalent ions that yield no additional electrons in the $d$-electron conduction band. In this case the total number of  electrons $N=L-X$ as well as the number of $d$-electrons $N_d=L-X-N_f$ depend on the number of dopants $X$ and thus the
ground-state energy $E_{\rm g}(N_f,X)=\sum^{N_d}_{k=1}\lambda_k$ has to be calculated many times for the selected values of $X$. The numerical results obtained in the one-dimensional case for several selected values of  $n=N/L$ are presented in figure~\ref{fig1_x}~(b). The finite-size effects are small again and so the results can be  satisfactorily extrapolated to the thermodynamic limit ($L \rightarrow \infty$). Comparing these results with the previous case (doping by trivalent ions) one can see a fully different behavior. For small values of $N=L-X$, the ground-state energy
$E_{\rm g}(N_f,X)$ has s minimum at $N^0_f=0$ and then monotonously increases. This type of behavior holds up to the critical value $N_{\rm c}=L-X_{\rm c} \sim L/2$. Above this value $E_{\rm g}$ reaches a minimum at $N^0_f>0$ and with increasing $N$ (decreasing $X$) the position of this minimum shifts to $L/2$. The resultant behavior of the average occupancy of the $f$-electron orbitals  $n_f=\frac{N^0_f}{L-X}$ on the concentration of dopants $x=X/L$ is plotted  in figure~\ref{fig2_x}. For comparison in figure~\ref{fig2_x} we have displayed the dependence  of $n_f$ on $x$ for doping by trivalent ions.
One can see that the substitution of rare-earth ions by  trivalent and divalent ions produces altogether different effects. While in the first case the average occupancy of $f$-orbitals increases, in the second case $n_f$ decreases. These results are qualitatively in agreement with experimental measurements~\cite{Tarascon} of the average samarium valence  $v=3-n_f$ performed on Sm$_{1-x}$M$_x$B$_6$ (M=Y$^{3+}$, La$^{3+}$, Sr$^{2+}$, Yb$^{2+}$) despite the fact that our calculations have been done at $T=0$, while the experiments have been done at room temperatures. However, it is not expected that the increasing temperature could change dramatically this picture. From the
theoretical investigation of the spinless Falicov-Kimball model at non-zero temperatures it is known~\cite{Farky7} that the effect of temperature is such that a finite temperature only smears the behavior found for $T=0$. Thus, one can expect that for non-zero temperatures the accordance of theoretical and experimental results could be even better.

\begin{figure}[!h]
\begin{center}
\vspace*{-0.5cm}
\mbox{\includegraphics[width=12cm,angle=0]{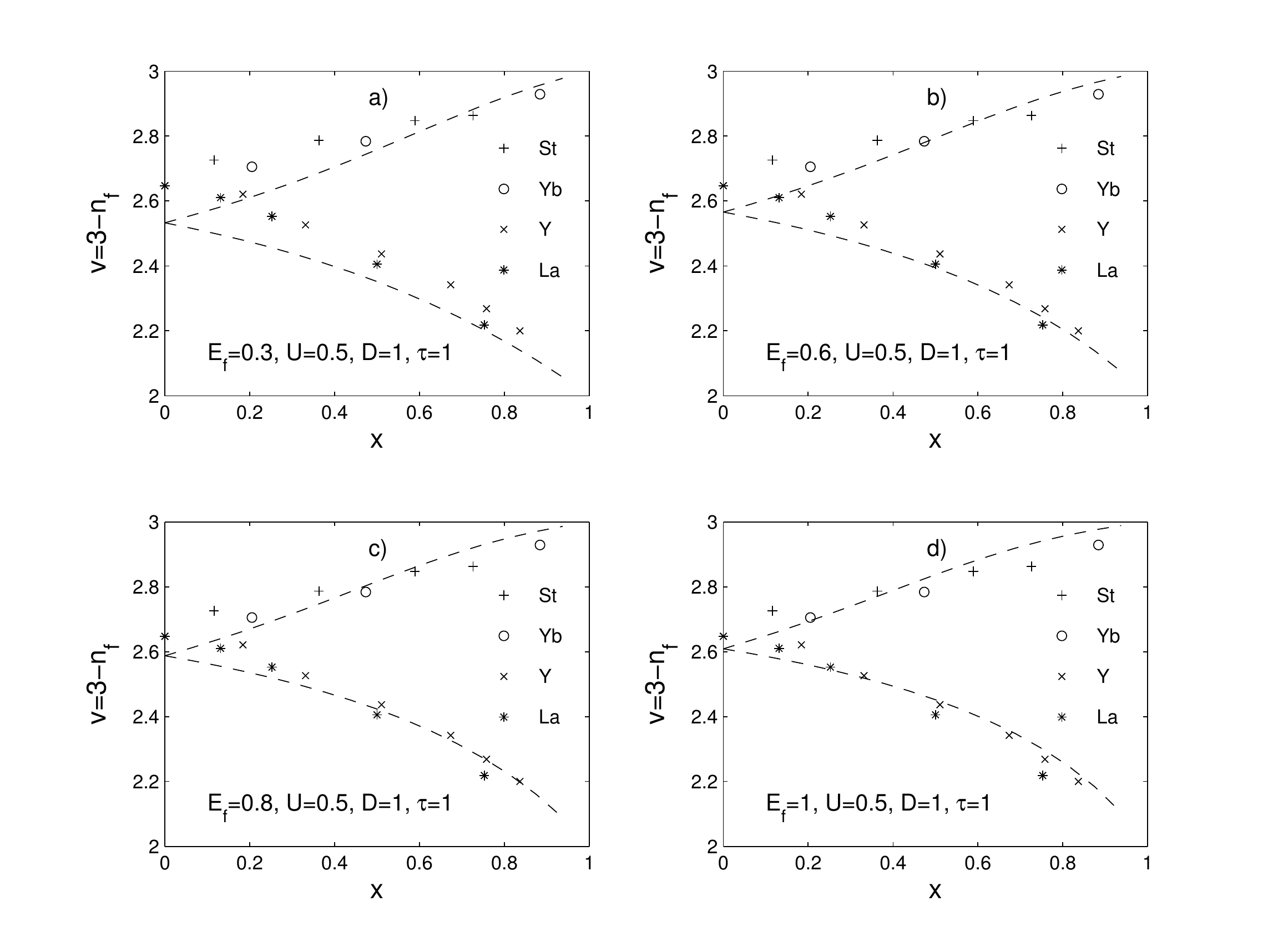}}
\end{center}
\vspace*{-0.8cm}
\caption{ Comparison of the one-dimensional theoretical (dashed lines) and
experimental results for the average valence $v=3-n_f$ as a function of $x$.
The theoretical results have been calculated for $U=0.5, \tau=1$
and several different values of the $f$-level position $E_f$.
The finite-size effects are negligible, and on the linear scale it is not
possible on the drawing to distinguish the behaviors obtained for $L=12, 16$
and 20 over the whole range of $x$ plotted~\cite{Farky58}.
The experimental results have been obtained for T=300 K (reference~\cite{Tarascon}).}
\label{fig2_xt}
\end{figure}
To verify this conjecture, we have  explicitly calculated the dependence of the $f$-state occupancy $n_f$ on the dopant concentration $x$ for nonzero-temperatures using small-cluster exact-diago\-naliza\-tion technique at finite temperatures (this method is described in detail in reference~\cite{Farky7}). The results of our numerical calculations
are summarized in figure~\ref{fig1_xt} for the selected values of temperature $\tau$.
It is seen, at a glance, that the spinless Falicov-Kimball model despite its simplicity can provide a qualitatively correct description of the effects of doping on thermodynamic properties of rare-earth systems for both divalent and trivalent ions. In addition, optimizing the $n_f(x)$ behaviour  with respect to $\tau$ and $E_f$ one can obtain a nice quantitative correspondence between the theoretical and experimental results (see figure~\ref{fig2_xt}).
Since no significant effects of the increasing dimension $D$ ($D=2$, 3) on the behaviour of $n_f(x)$ has been observed for both zero and non-zero temperatures~\cite{Farky57, Farky58}, our results can be satisfactorily used  to describe the behaviour of real three dimensional materials.

\section{Stability of charge and spin ordering at finite temperatures}
\label{Stability of charge and spin ordering at finite temperatures}

Examples of charge and spin ordering discussed in previous sections concerned exclusively the case of  $T=0$. Real experiments are, however, always performed at finite temperatures, and therefore getting an answer to the question about the stability of zero temperature solutions at finite temperatures seems to be the task of fundamental importance. A positive answer to this question is available at least for the chessboard charge ordering that is the ground state of the conventional Falicov-Kimball model at half-filling for all Coulomb interactions $U>0$.  For this case, there exists
an exact proof~\cite{Kennedy_Lieb} of the existence of a phase transition from the
low-temperature ordered phase (the chessboard phase) to the high-temperature
disordered phase at finite critical temperature $\tau_{\rm c}$ (for dimensions
$D\geqslant 2$) which strongly depends on the local Coulomb interaction $U$. In
addition, the numerical simulations within the grand-canonical Monte-Carlo
showed that the phase transitions are of the first order for small and
intermediate values of the Coulomb interaction $U$ and of the second order
for strong interactions~\cite{Maska,Farky46}.
In our paper~\cite{Farky52} we have extended the numerical study of the temperature
induced phase transitions to the case of phase segregated and striped
phases. Moreover, we have considered a more general situation
\begin{equation}
H=H_{\rm FKM}+H_{t'}\, ,
\label{eq:corrhop}
\end{equation}
where $H_{\rm FKM}$ is the conventional  spinless Falicov-Kimball model~(\ref{eq3.1.1.1}) and $H_{t'}$ is the term of the correlated hopping~(\ref{eq4.1.2.2}). As was discussed in section~\ref{Influence of correlated hopping}, all three above mentioned phases (the chessboard phase, the segregated phase and the axial striped phase) are the ground states of the model Hamiltonian~(\ref{eq:corrhop}) at the symmetric band point, where both $N_f$ and $N_d$ are fixed to $L/2$ and, therefore, all
numerical calculations at nonzero temperatures are done exclusively in the canonical
ensemble. In this formalism, the partition function and the internal  energy
corresponding to the model Hamiltonian (\ref{eq:corrhop}) can be written as:
\begin{eqnarray}
Z&=&\sum_{w^f,w^d} \re^{-E/\tau}, \hspace*{1cm} E=\sum_i\varepsilon_i(w^f)w_i^d\, ,
\\
\langle E\rangle&=&\sum_{w^f,w^d}E\re^{-E/\tau},
\vspace{-3mm}
\end{eqnarray}
where $\tau=k_{\rm B}T$ and the summation goes over all possible $\displaystyle\left(\begin{smallmatrix} L\\N_f\end{smallmatrix}\right)$
distributions $w^f$ of $f$ electrons on $L$ lattice sites and $\displaystyle\left(\begin{smallmatrix} L\\N_d\end{smallmatrix}\right)$
distributions $w^d$ of $d$ electrons on $L$ single-particle energy levels
$\varepsilon_i$ corresponding to the matrix $h(w^f)$ with elements
$h_{ij}(w^f)=t_{ij}+t'_{ij}(w^f_i+w^f_j)+Uw^f_i\delta_{ij}$. In the next step, the summation
over all $f$ and $d$ distributions is replaced by the Monte-Carlo summation
with the statistical weight $\re^{-E/\tau}/Z$.

\begin{figure*}[t!]
\begin{center}
\mbox{\includegraphics[width=11.8cm,angle=0]{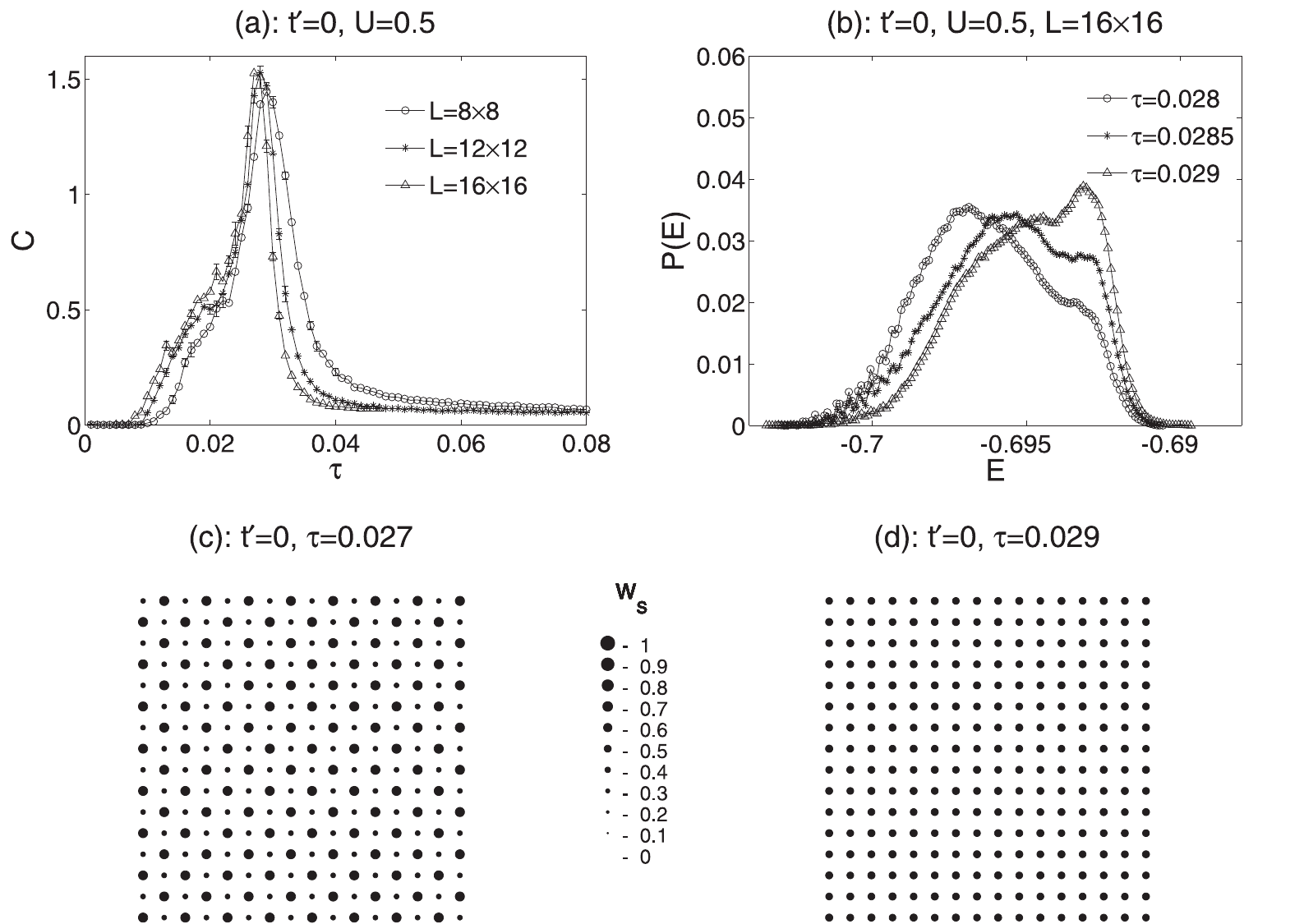}}
\end{center}
\caption{The specific heat (a), the energy distribution (b) and the thermal
average of the $f$-electron occupation (c)--(d) for the conventional
Falicov-Kimball model ($t'=0$) in two dimensions~\cite{Farky52}.}
\label{o_ch1}
\vspace{-3mm}
\end{figure*}
\begin{figure*}[!h]
\begin{center}
\mbox{\includegraphics[width=11.8cm,angle=0]{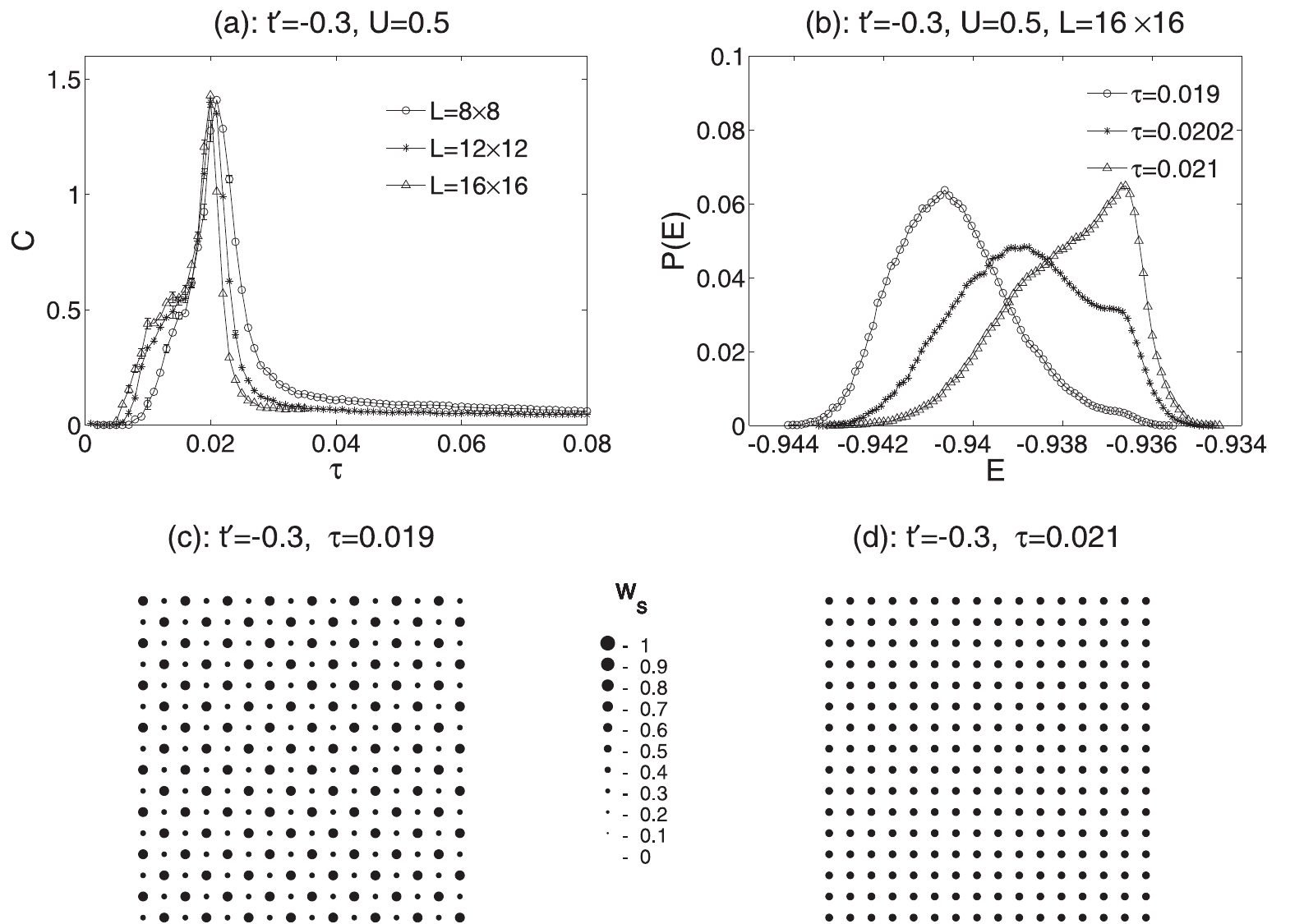}}
\end{center}
\caption{ The specific heat (a), the energy distribution (b) and the thermal
average of the $f$-electron occupation (c)--(d) for the two-dimensional
Falicov-Kimball model with correlated hopping $t'=-0.3$~\cite{Farky52}.}
\label{o_ch2}
\vspace{-3mm}
\end{figure*}
\begin{figure*}[!t]
\begin{center}
\mbox{\includegraphics[width=11.8cm,angle=0]{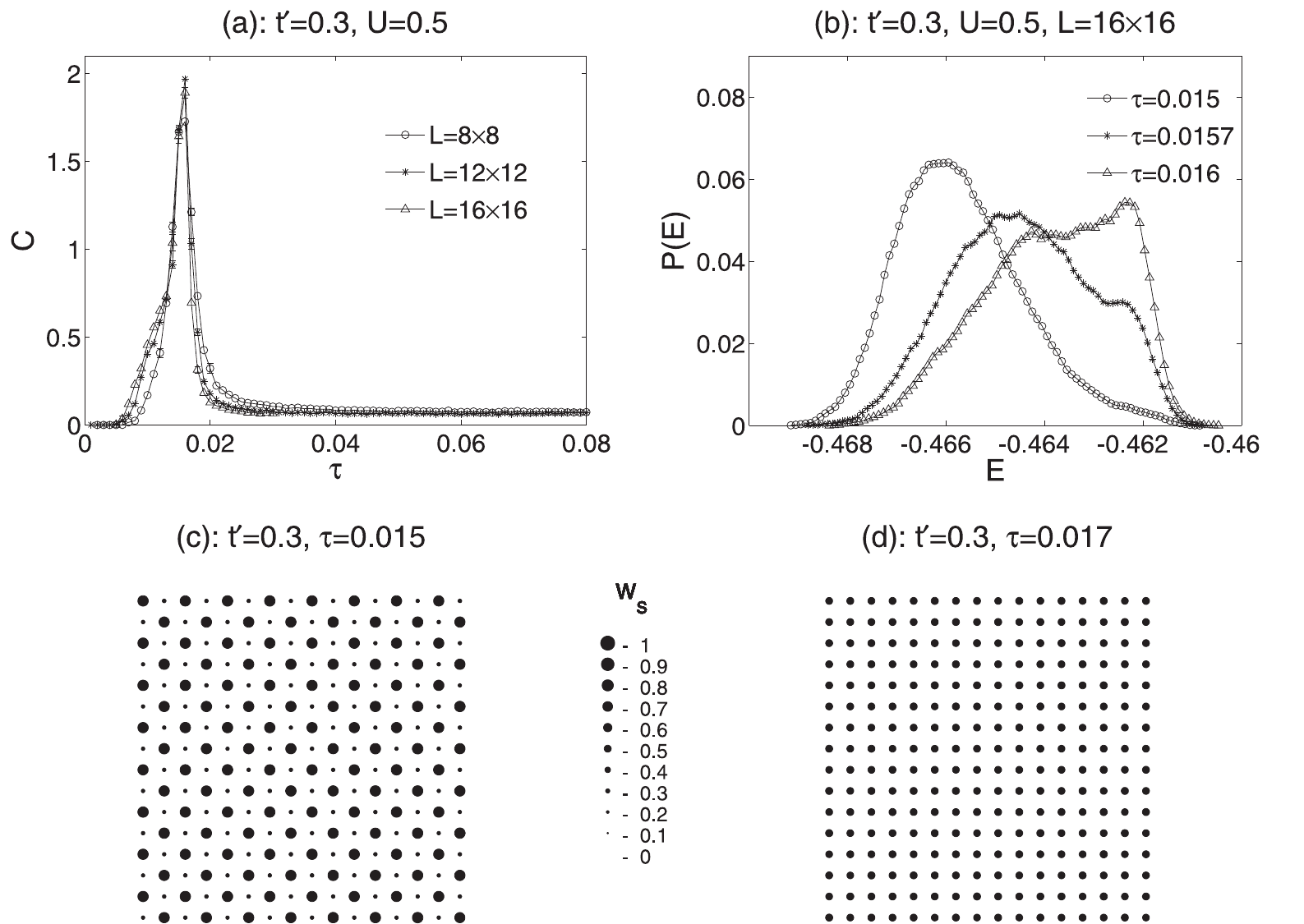}}
\end{center}
\vspace*{2mm}
\caption{ The specific heat (a), the energy distribution (b) and the thermal
average of the $f$-electron occupation (c)--(d) for the two-dimensional
Falicov-Kimball model with correlated hopping $t'=0.3$~\cite{Farky52}.}
\label{o_ch3}
\vspace{-3mm}
\end{figure*}

\looseness=-1To identify the transition temperatures from low-temperature ordered
phases to high-tempera\-ture disordered phase and the type of the phase
transition, we have  numerically calculated the specific heat $C=(\langle
E^2\rangle-\langle E\rangle ^2)/(L\tau ^2)$, the thermal average of
the $f$-electron occupation $w_{\rm s}=\langle w^f\rangle$ and the energy distribution
$P(E)$. The numerical calculations are performed exclusively at $U=0.5$, since the
ground-state phase diagram exhibits a richer spectrum of solutions in the
weak and intermediate coupling regions in comparison to the strong coupling
limit.
To verify the capability of our method to describe the phase transitions at
finite temperatures we have started with the conventional two-dimensional
Falicov-Kimball
model ($t'=0$) at half-filling. As was mentioned above, the physical picture
of temperature-induced phase transitions within this relatively simple
model is quite understandable at present. For all finite Coulomb interaction $U>0$,
the ground state of the model is the chessboard phase that persists up to
critical temperature $\tau_{\rm c}(U)$, where the system undergoes a phase
transition to the homogeneous phase. The phase transition is of the first
order for $U<1$  and of the second order for $U>1$~\cite{Maska}. Our numerical results
obtained within the canonical Monte-Carlo method for $C$, $w_{\rm s}$ and $P(E)$
fully confirm this picture (see figure~\ref{o_ch1})~\cite{Farky52}.
The specific heat curves exhibit a
sharp low-temperature peak at $\tau_{\rm c} \sim 0.028$ that is
obviously connected with the phase transition from the chessboard phase to the
homogeneous phase, as can be seen from the behaviour of the average
$f$-electron occupation $w_{\rm s}$ for temperatures slightly lower or
slightly higher than $\tau_{\rm c}$. Moreover, the energy
distribution function $P(E)$ exhibits an apparent two-peak structure near
the critical point $\tau_{\rm c}$ (it can be considered as a superposition of two
Gaussians), which in accordance with the theory of Challa, Landau and
Binder~\cite{Challa} points
on the first order phase transition at $\tau_{\rm c}$.
Let us now discuss how this picture is changed when the correlated hopping
term is added. Firstly, we have examined the case of small values of
$|t'|$ for which the ground state of the model is still the chessboard
phase~\cite{Farky21}. The typical examples of $C$, $w_{\rm s}$ and $P(E)$ from the positive and
negative region of $t'$ are displayed in figure~\ref{o_ch2} and in figure~\ref{o_ch3} for $t'=-0.3$ and $t'=0.3$.
One can see that the correlated hopping term (in the limit of
small $|t'|$) does not qualitatively change the picture of temperature induced
phase transitions found for $t'=0$. For both, positive and negative $t'$,
there is the first order phase transition from the low-temperature ordered
phase to the high-temperature disordered phase, similarly to $t'=0$, and
the only difference between these cases is that the correlated hopping term
reduces slightly the critical temperature $\tau_{\rm c}$ of the phase transition.

\begin{figure*}[!h]
%\begin{center}
%\vspace*{-0.5cm}
\centerline{\mbox{\includegraphics[width=11.8cm,angle=0]{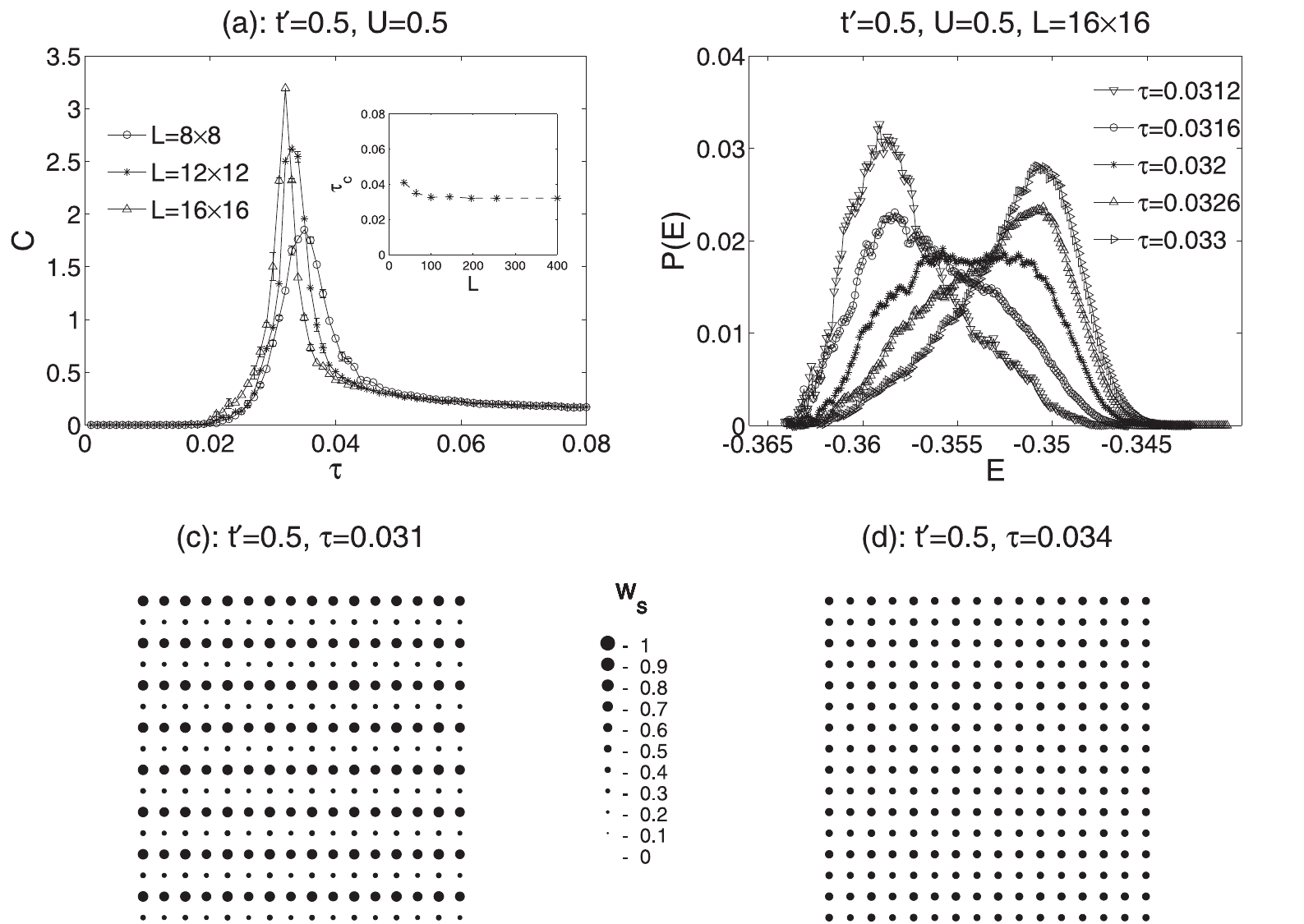}}}
%\end{center}
%\vspace*{-0.5cm}
\caption{ \looseness=-1The specific heat (a), the energy distribution (b) and the thermal
average of the $f$-electron occupation (c)--(d) for the two-dimensional
Falicov-Kimball model with correlated hopping $t'=0.5$. The inset shows the
critical temperature $\tau_{\rm c}$ as a function of the cluster size~$L$~\cite{Farky52}.}
\label{o_ch4}
\end{figure*}
Therefore, in the next step we have turned
our attention to the physically much less explored type of
configurations, namely, the axial striped configurations that are ground
states of the Falicov-Kimball model for the intermediate values of $t'$
($|t'| \sim 0.5$). Note, that for the axial striped phase even the fundamental
question
concerning the temperature stability of this phase has remained unanswered so far.
This is due to the fact that it is very difficult to find this
phase in a pure form. For example, in the conventional Falicov-Kimball
model ($t'=0$), the axial striped phases are stable for a relatively wide
range of model parameters~\cite{Lemanski_Freericks}, but solely in mixtures
with other phases  (e.g., the empty configuration).
In addition, strong finite-size effects have been observed on the stability
of these mixtures and therefore it is practically  impossible to make any
conclusions concerning their stability at finite temperatures from the
numerical calculations on finite clusters. However, in the Falicov-Kimball
model with correlated hopping, the axial striped phase exists in a pure
form for a wide range of model parameters $t'$ and $U$, the finite-size
effects on the stability of this phase at $\tau=0$ are negligible, and so
the corresponding numerical study of the temperature stability of the axial
striped phase can be performed straightforwardly.
\begin{figure*}[t!]
\begin{center}
%\vspace*{-0.5cm}
\mbox{\includegraphics[width=11.8cm,angle=0]{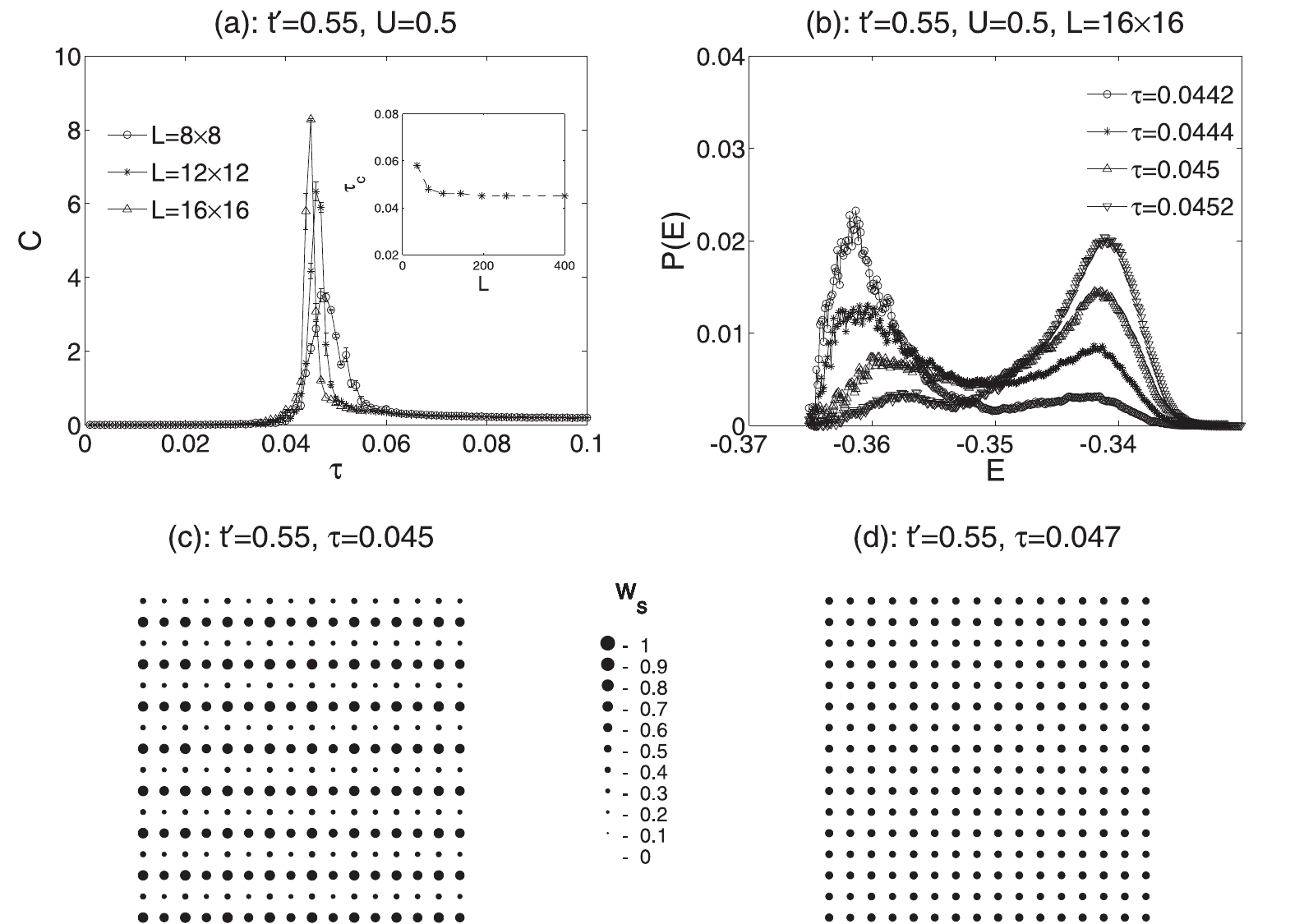}}
\end{center}
\caption{ The specific heat (a), the energy distribution (b) and the thermal
average of the $f$-electron occupation (c)--(d) for the two-dimensional
Falicov-Kimball model with correlated hopping $t'=0.55$. The inset shows the
critical temperature $\tau_{\rm c}$ as a function of the cluster size~$L$~\cite{Farky52}.}
\vspace{-3mm}
\label{o_ch5}
\end{figure*}

In figure~\ref{o_ch4} and figure~\ref{o_ch5} we present our canonical Monte-Carlo results for $C$, $w_{\rm s}$ and
$P(E)$ obtained for two different values of $t'$ ($t'=0.5$ and $t'=0.55$)
from the region where the ground-state of the model is just the axial
striped phase. Again, the specific heat curves exhibit a sharp
low-temperature peak, the existence of which indicates a phase transition
from the axial striped phase to the homogeneous phase. This was independently  verified
by calculating the average $f$-electron occupation $w_{\rm s}$ and
the energy distribution $P(E)$ near the transition point $\tau_{\rm c}$, which
clearly demonstrates the presence of the first order phase transition at
$\tau_{\rm c}$. Since the critical temperature $\tau_{\rm c}$ of the phase transition
for both values of $t'$ shifts to smaller values with increasing $L$, we
have performed a detailed finite-size scaling analysis of the $\tau_{\rm c}(L)$
dependence to exclude the possibility of $\tau_{\rm c}$ vanishing in the
thermodynamic limit $L\rightarrow \infty$. The resultant $\tau_{\rm c}(L)$
dependencies are plotted as insets in figure~\ref{o_ch4} and in figure~\ref{o_ch5}. It is clearly
seen that the critical temperatures $\tau_{\rm c}$ for both $t'=0.5$ and $t'=0.55$ also persist
in the thermodynamic limit, meaning that the axial striped phase remains
stable at finite temperatures as well. Moreover, our numerical results show
that critical temperatures for an axial striped phase are
considerably higher in comparison with critical temperatures for a
chessboard phase. We have observed the same behaviour for negative
values of $t'$ ($t'=-0.7$). However, the critical temperature in
this case was only slightly larger than the one corresponding to $t'=0$.

With increasing $t'$, the half-filled Falicov-Kimball model with correlated
hopping exhibits (at $\tau=0$) a phase transition from the axial striped
phase to the segregated phase~\cite{Farky21} that takes place at $t' \sim 0.6$. Since both the
chessboard phase and the axial striped phase are insulating while
the segregated phase is metallic~\cite{Farky21}, one can expect a fully
different thermodynamic
behaviour of the model for the last case. To verify this conjecture we have
performed exhaustive numerical studies of the temperature dependence of
$C$, $w_{\rm s}$ and $P(E)$ for $t'=1$. This study is also important from the
point of view that the thermodynamics of the metallic phase has been so far examined
only in a few cases~\cite{Czajka,Farky24}, while for the insulating
phase
(usually the chessboard phase) there is a number of analytical and numerical
results~\cite{Kennedy_Lieb,Farky4,Macedo}.

\begin{figure*}[!h]
\begin{center}
\hspace*{-0.5cm}
\mbox{\includegraphics[width=11.8cm,angle=0]{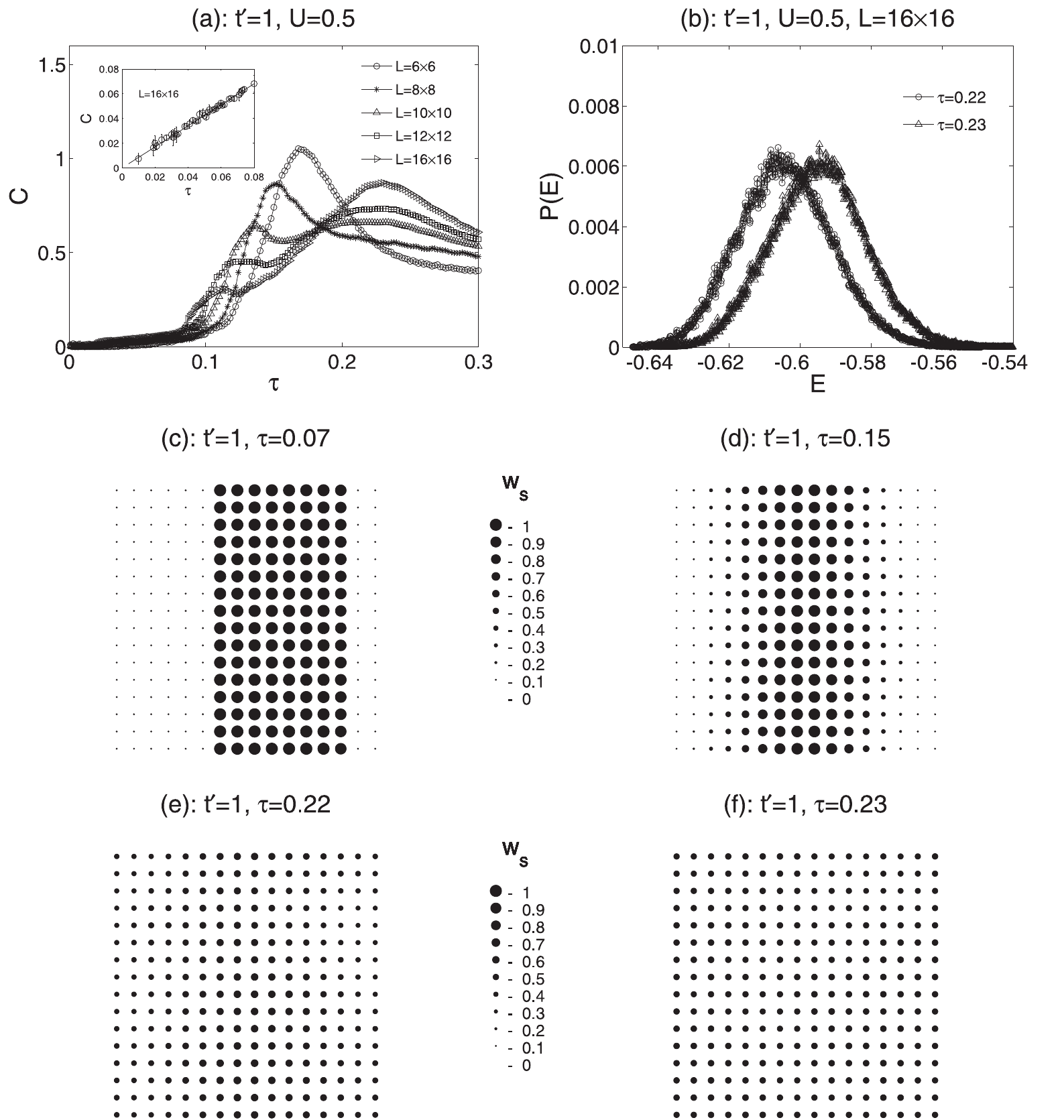}}
\end{center}
\caption{ The specific heat (a), the energy distribution (b) and the thermal
average of the $f$-electron occupation (c)--(f) for the two-dimensional
Falicov-Kimball model with correlated hopping $t'=1$.  The inset shows the
specific heat $C$ in the low-temperature region for $L=16\times16$~\cite{Farky52}.}
\label{o_ch6}
\end{figure*}

The results of our numerical calculations obtained for the specific heat $C$
are shown in figure~\ref{o_ch6}.
To reveal the finite-size effects, the calculations for $C$ have been
done on several different clusters of $L$=$6\times 6$,
$8\times 8$, $10\times 10$,  $12\times 12$ and  $16\times 16$
sites. We have found that the specific heat curves, in the low-temperature
region, strongly depends on the cluster sizes, and, therefore, a very careful
analysis has to be performed to find the correct behaviour
of the model in the thermodynamic limit $L\rightarrow \infty$.
On small finite clusters ($L=6\times 6$ and $L=8\times 8$), the specific heat exhibits only
one-peak structure in the low-temperature region ($\tau \sim 0.15$). With the
increasing cluster size $L$, an additional peak is stabilized at slightly higher temperatures
($\tau \sim 0.23$), while the first peak is gradually suppressed and probably fully disappears in
the thermodynamic limit. The behaviour of the average $f$-electron
occupation shows (see figure~~\ref{o_ch6}) that the second peak in the specific heat corresponds to
the phase transition from the low-temperature ordered (segregated) phase to
the high-temperature disordered phase.

\begin{figure*}[t!]
\begin{center}
\hspace*{-0.3cm}
\mbox{\includegraphics[width=4.2cm,angle=0]{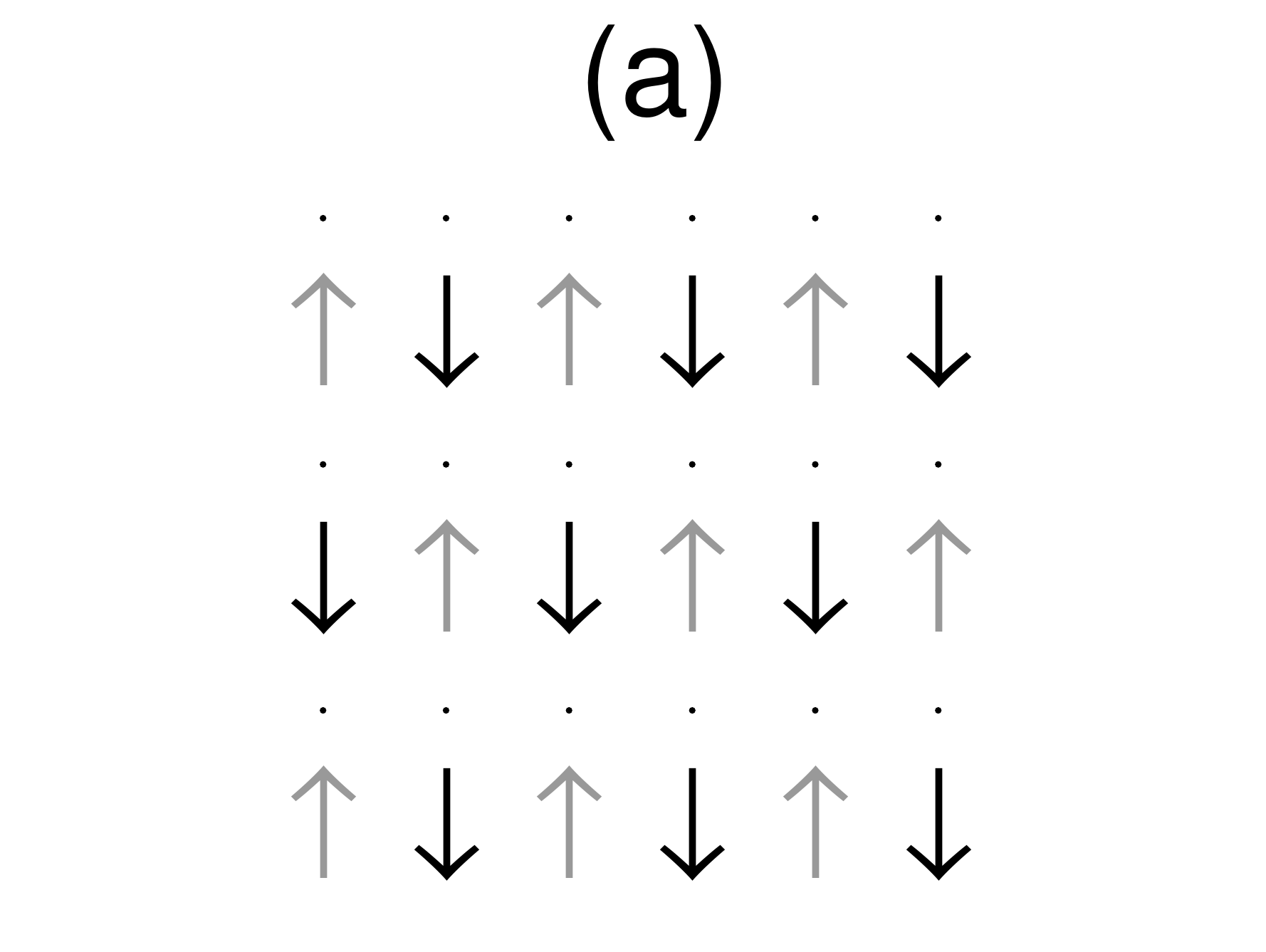}}\hspace*{-0.3cm}
\mbox{\includegraphics[width=4.2cm,angle=0]{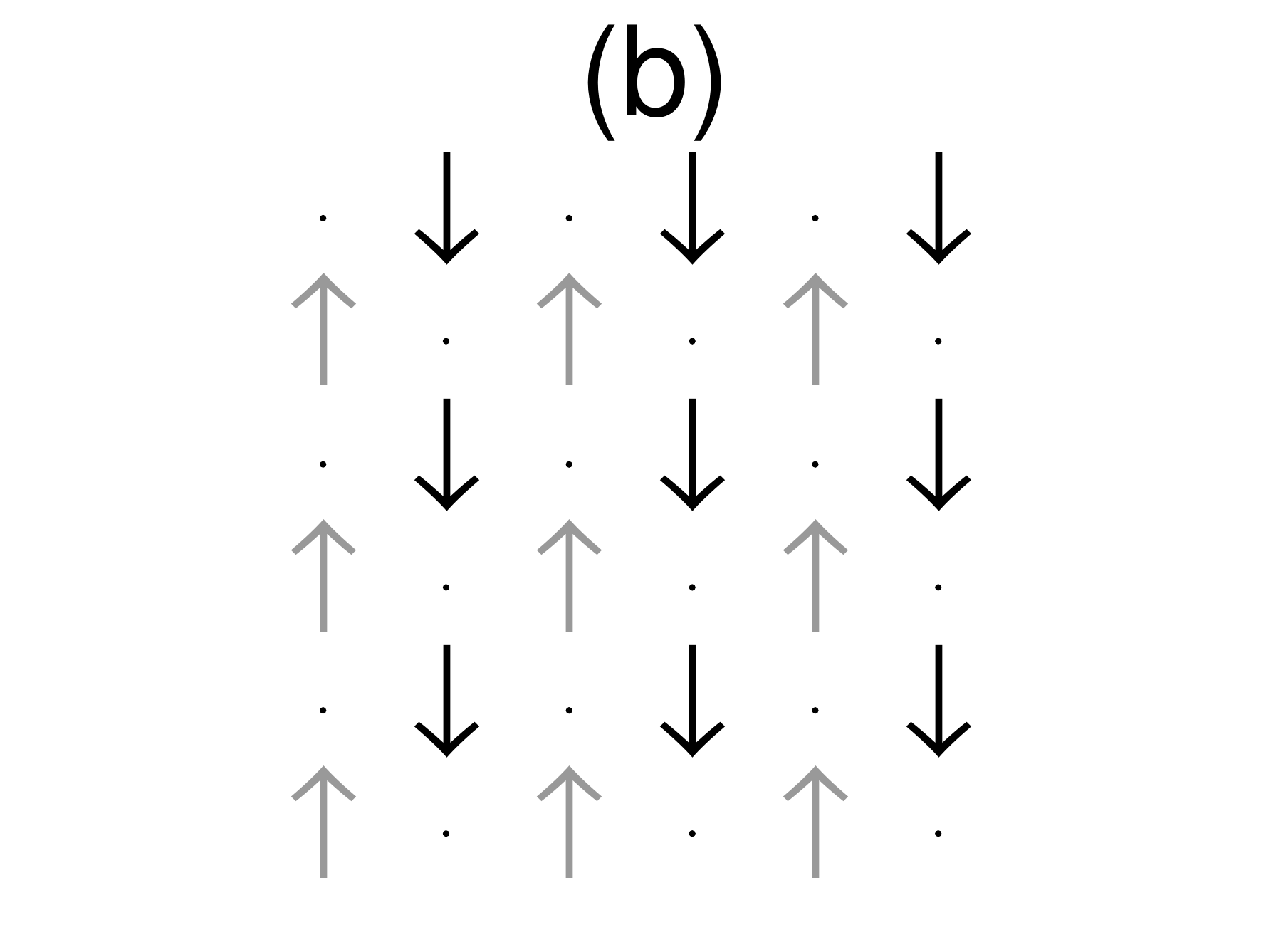}}
\end{center}
\vspace{-2mm}
\caption{ Ground states of the spin-1/2 Falicov-Kimball model extended by spin-dependent interaction $J$ and local Coulomb interaction  $U_{ff}$: (a) for $U=2$, $J=1$, $U_{ff}=4$, $E_f=0$ and $N=2L$, (b) for $U=4$, $J_z=0.8$, $U_{ff}=8$, $E_f=-2$ and $N=3L/2$~\cite{Farky49}.}
\label{o_chs1}
\vspace{-3mm}
\end{figure*}
\begin{figure*}[b!]
\begin{center}
\hspace*{-0.6cm}
\mbox{\includegraphics[width=13cm,angle=0]{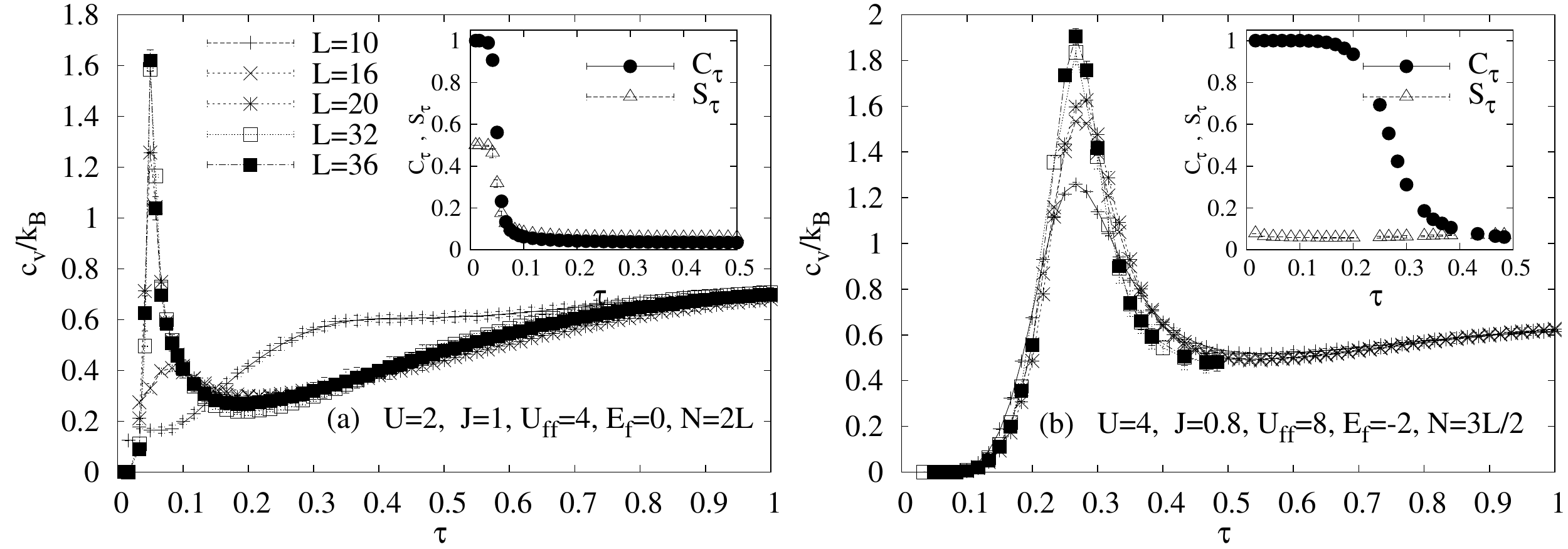}}
\end{center}
\caption{ Specific heat as a function of $\tau=k_{\rm B}T/t$ for the phase (a) and phase (b) from figure~\ref{o_chs1}. Insets present  the temperature dependences of  $C_\tau$ and $S_\tau$ defined in the text~\cite{Farky49}.}
\label{o_chs2}
\end{figure*}

The nature of this phase transition is, however, different in comparison to the
previous cases. While the energy distribution function $P(E)$ is double
peaked for the chessboard and the axial striped phase near the transition
temperature $\tau_{\rm c}$ (the first order phase transition), $P(E)$ exhibits a
single-peak structure for the segregated phase, which points to the second
order phase transition at $\tau_{\rm c}$. Comparing the thermodynamic behaviour of
the model in the chessboard, axial striped and segregated regions one can
find two other important differences, namely, (i) the critical
temperature of the second order phase transition is approximately ten times
higher than critical temperatures of the first order phase transitions,
and (ii) the specific heat (in the low-temperature region) exponentially decreases
for the chessboard and axial striped phase, while in the
segregated phase the specific heat $C(\tau)$ seems to show a linear
behaviour indicating the Fermi-liquid behaviour for $\tau<0.08$ (see the
inset in figure~\ref{o_ch6}~(a)).
The observation of the linear contribution to the specific heat in the
low-temperature region ($\tau<0.08$) is consistent with the behaviour of the
average $f$-electron occupation in this region (see figure~\ref{o_ch6}~(c)). One can see
that despite the increasing temperature (from 0 to 0.08) the $f$-electrons
preferably occupy only one half of the lattice leaving another part
empty. Due to the on-site Coulomb interaction between the $f$ and $d$
electrons, the itinerant $d$ electrons preferably occupy  the empty part
of the lattice, where they can move as free particles yielding the linear
contribution to the specific heat.

We have performed the same numerical study for the spin-1/2 Falicov-Kimball model extended by the spin-dependent Coulomb interaction $J$ between the localized $f$ and itinerant $d$ electrons as well as the on-site Coulomb interaction ($U_{ff}$) between the localized $f$-electrons~\cite{Farky49}. The coexistence of charge and spin degrees of freedom for such type of a system evokes the question concerning a possible sequence of two phase transitions corresponding to the breaking of different types of charge/spin ordering. We have tested the possibility of such a scenario independently for two configuration types displayed in figure~\ref{o_chs1}, which are the ground states of the generalized Falicov-Kimball model for  $U=2$, $J=1$, $U_{ff}=4$, $E_f=0$ and $N=2L$ or $U=4$, $J=0.8$, $U_{ff}=8$, $E_f=-2$ and $N=3L/2$.
To identify the type of phase transition, we have used the the specific heat and the structure factor of the charge and spin ordering defined by~\cite{Farky49}:
\begin{eqnarray}
C_{\tau}=\frac{1}{L^2}\sum_{j,k}^L\exp(\ri\vec{Q}(\vec{R}_j-\vec{R}_k))\langle (w_{j\uparrow}+w_{j\downarrow})(w_{k\uparrow}+w_{k\downarrow})\rangle
\label{eq_chs1}
\end{eqnarray}
and
\begin{eqnarray}
S_{\tau}=\frac{1}{L^2}\sum_{j,k}^L\exp(\ri\vec{Q}(\vec{R}_j-\vec{R}_k))\langle (w_{j\uparrow}-w_{j\downarrow})(w_{k\uparrow}-w_{k\downarrow})\rangle\, .
\label{eq_chs2}
\end{eqnarray}
The temperature behaviours of these quantities are displayed in figure~\ref{o_chs2}.
It is seen that in the first case both types of ordering are  simultaneously  destroyed at the same critical temperature  $\tau_{\rm c} \sim 0.05$, while in the second case the spin ordering disappears already at very low temperatures  ($\tau_{\rm c}< 0.01$), and the charge ordering persists up to $\tau_{\rm c} \sim 0.25$\footnote{It should be noted that similar calculations of finite temperature properties of the model have been recently performed by Wrzodak and Lemanski~\cite{Wrzodak_lemanski2} on the square $4\times4$ cluster at half-filling.}.

%\clearpage

\section{Conclusion}
\label{Conclusion}

In this review we have presented the results of our theoretical study of charge and spin ordering in strongly correlated electron systems obtained within various generalizations of the Falicov-Kimball model.  The primary goal of this study was to identify crucial interactions that lead to the stabilization of various types of charge ordering in these systems such as the axial striped ordering, diagonal striped ordering, phase separated ordering, phase segregated ordering, etc. Among the major interactions that come into account, we have examined the effect of local Coulomb interaction between localized and itinerant electrons,   long-range and correlated hopping of itinerant electrons,  long-range Coulomb interaction between localized and itinerant electrons, local Coulomb interaction between itinerant electrons,  local Coulomb interaction between localized electrons,  spin-dependent interaction between localized and itinerant electrons, both for zero and nonzero temperatures, as well as for doped and undoped systems.

\looseness=-1We have started our study of charge ordering in the strongly correlated systems  with the spinless Falicov-Kimball model. First, we have focused on a description  of the ground-state properties of the model at $T=0$. Since  the previous theoretical studies have shown  that the ground-state properties of the  Falicov-Kimball model are very sensitive to the type of approximation used, to study the charge ordering, valence and metal-insulator  transitions in the Falicov-Kimball model we have used the  EDM on finite clusters with subsequent extrapolation of the results to the thermodynamic limit (infinitely large system). Our results showed that the spinless Falicov-Kimball model is capable of describing homogeneous as well as phase-separated distributions of $f$ electrons and thereby discontinuous valence and metal-insulator transitions induced by pressure. The major driving  interaction of these transitions is the Coulomb interaction between itinerant and localized electrons.

In order to elaborate  the most realistic description of charge ordering and  valence  and metal-insulator transitions in real materials (rare-earth and transition-metal compounds), we have generalized the original type of electron hopping (only to the  nearest neighbours) to a much more realistic type of hopping (the long-range hopping  with power decreasing hopping amplitudes) and in addition we have taken into account the term  of correlated hopping, which modifies the hopping amplitudes of  itinerant electrons from one site to another  according to the occupancy of these sites by localized electrons. We have found that the term of correlated hopping has a significant effect  on the dynamics of valence and metal-insulator transitions in the spinless Falicov-Kimball model and should, therefore, be taken into account for the correct description of these phenomena. Moreover, we have shown that  the correlated hopping term leads to the stabilization of the axial charge stripes, suggesting an alternate and a much simpler mechanism (than the one which was previously considered within  the Hubbard and $t-J$  model) of forming an inhomogeneous charge ordering in  strongly correlated systems. Regarding the effect of long-range hopping term on metal-insulator transitions, we have  found that it is not very significant, at least for $q\leqslant 0.3$, which physically turns out to be the most interesting case.

The next  step towards a comprehensive description of cooperative phenomena in the Falicov-Kimball model was to study the model for dimensions $D=2$ and $D=3$. To fulfill this  task, it was necessary to develop a new numerical method, since the EDM is capable of  analysing only clusters up to  $L\sim 40$ sites, which  does not offer any  possibility to extrapolate the results to the thermodynamic limit ($L=\infty$). This goal has been fully fulfilled. Our   numerical method (based on a modification of an exact diagonalization algorithm) is very accurate, even for cluster sizes of several hundred sites (this has been tested  in the limiting cases where the exact solutions are known). Moreover, it is very flexible,  and so it permits to study  various generalizations of the model.  The study of the Falicov-Kimball model by this method showed that the basic picture of charge ordering and valence and metal-insulator transitions remains unchanged for higher dimensions. The only important difference is  that the stability area of metallic phases shifts to higher values of the Coulomb interaction  $U$. This fact is very important from the practical point of view, since real materials are just in this limit.

The fact that we were able to analyse the model for cluster sizes of several hundred lattice sites  allowed us to study the formation of charge ordering in two- and three-dimensional systems for arbitrary values of $U$, which was very valuable, since  the previous results were limited either to the area of $U\ll 1$ or $U\gg 1$,
or were limited to the type of the investigated configurations (e.g., the periodic configurations with the small periods and their mixtures). We have found that the spinless  Falicov-Kimball model is capable (in $D=2$  and $D=3$) of describing a very wide range of charge  orderings involving periodic charge ordering, phase-segregated and phase-separated ordering as well as axial and diagonal  striped ordering. In the the three-dimensional case, our results represent the first and so far the only  attempt to  systematically
describe the formation of inhomogeneous charge ordering in the spinless Falicov-Kimball model, which was to a large extent possible due to a new elaborated  numerical method.

Moreover, the inclusion of spins and the spin interactions of the Hubbard type between itinerant electrons and of the Ising type between itinerant and localized electrons clearly demonstrated the enormous potential of the model in describing different types of charge and spin superstructures, including the segregated phase and phase-separated orderings, as well as axial and diagonal striped (band) orderings with ferro-, ferri-, or antiferromagnetic ground state. Furthermore, the spin model in the presence of interband Coulomb interaction between localized and itinerant electrons maintains advantages of a spinless model, namely, the capability of describing valence and metal-insulator transitions induced by pressure, temperature and alloying.

\section*{Acknowledgements}
This work was supported by Slovak Grant Agency VEGA under Grant No.~2/0175/10 and by the ERDF EU grant, under the contract No.~ITMS26220120005.

%\newpage

\ukrainianpart

\title{Формування зарядового і спінового впорядкування в сильно скорельованих
 електронних системах}

\author{Г. Ченчарікова, П. Фаркашовскі}

\address{Інститут експериментальної фізики, Словацька академія наук, Кошиці, Словаччина}

\makeukrtitle

\begin{abstract}
\tolerance=3000

У цьому огляді ми представляємо результати наших теоретичних досліджень
 зарядового і спінового впорядкування в сильно скорельованих електронних
 системах, що отримані в рамках різних узагальнень моделі Фалікова-Кімбала.
 Основною метою цього дослідження було ідентифікувати вирішальні взаємодії, що
 приводять до стабілізації різних типів зарядового впорядкування в цих
 системах, таких як осьове стрічкове впорядкування, діагональне стрічкове
 впорядкування, фазове розшарування, фазова сегрегація і т.п. З поміж основних
 взаємодій, які враховують, нами розглядався вплив локальних
 кулонівських взаємодій між локалізованими і колективними електронами,
 далекосяжного і скорельованого переносу колективних електронів, далекосяжної
 кулонівської взаємодії між локалізованими і колективними електронами,
 локальної кулонівської взаємодії між колективними електронами, локальної
 кулонівської взаємодії між локалізованими електронами, спінозалежної взаємодії
 між локалізованими і колективними електронами, як при нульовій так і при
 ненульовій температурах, а також для легованих і нелегованих систем. На
 завершення, обговорюється застосовність отриманих розв'язків для
 опису сполук рідкісноземельних і перехідних елементів.
\keywords зарядове і спінове впорядкування, переходи метал-діелектрик,
 переходи зі зміною валентності, модель Фалікова-Кімбала,
 сильно скорельовані системи

\end{abstract}

\end{document}